\documentclass[11pt]{article}
\usepackage{a4wide,epsfig,amsmath,amssymb,cite,scalefnt,axodraw,graphicx,feynarts}

\renewcommand{\thefootnote}{\fnsymbol{footnote}}
\renewcommand {\theequation}{\arabic{section}.\arabic{equation}}
\renewcommand {\thefigure}{\arabic{section}.\arabic{figure}}
\renewcommand {\thetable}{\arabic{section}.\arabic{table}}
\def\semi{;\hfil\break}
\newcommand{\api}{\frac{\alpha_s}{\pi}}

\newcommand{\Lmu}{L_\mu}

\newcommand{\apiSU}{\frac{\alpha^{\rm SU(5)}(\mu)}{4 \pi}}

\newcommand{\eps}{$\varepsilon$}

\newcommand{\epscalar}{$\varepsilon$-scalar}
\newcommand{\reference}[1]{Ref.~\cite{#1}}
\newcommand{\abbrev}{\scalefont{.9}}
\newcommand{\msbar}{$\overline{\mbox{\abbrev MS}}$}
\newcommand{\mususy}{\mu_{\rm SUSY}}
\newcommand{\MUH}{\mu_{\rm SUSY}}

\newcommand{\msbarmath}{\overline{\rm\abbrev MS}}
\newcommand{\mgut}{M_{\rm G}}
\newcommand{\betaDRbar}{\beta^{\overline{\rm DR}}}
\newcommand{\betaMSbar}{\beta^{\overline{\rm MS}}}
\newcommand{\gammaDRbar}{\gamma^{\overline{\rm DR}}}
\newcommand{\gammaMSbar}{\gamma^{\overline{\rm MS}}}

\newcommand{\mhc}{M_{\rm H_c}}
\newcommand{\mugut}{\mu_{\rm GUT}}

\newcommand{\apiDR}{\frac{\asDRbar}{ \pi}}
\newcommand{\apiMS}{\frac{\asMSbar}{ \pi}}
\newcommand{\aspi}{\frac{\as}{ \pi}}

\newcommand{\aepi}{\frac{\alpha_e}{\pi}}
\newcommand{\asDRbar}{\alpha_s^{\overline{\rm DR}}}
\newcommand{\asMSbar}{\alpha_s^{\overline{\rm MS}}}

\newcommand{\qcd}{{\abbrev QCD}}
\newcommand{\susy}{{\abbrev SUSY}}

\newcommand{\dreg}{{\abbrev DREG}}
\newcommand{\dred}{{\abbrev DRED}}
\newcommand{\nsvz}{{\abbrev NSVZ}}
\newcommand{\Mt}{m_t}
\newcommand{\afourpi}{\frac{\alpha_s}{4\pi}}
\newcommand{\mDRbar}{m^{\overline{\rm DR}}}
\newcommand{\mMSbar}{m^{\overline{\rm MS}}}

\newcommand{\mqDRbar}{m_{q}^{\overline{\rm DR}}}
\newcommand{\mqMSbar}{m_{q}^{\overline{\rm MS}}}
\newcommand{\cA}{C_A}
\newcommand{\cR}{C_R}

\newcommand{\IR}{I_{2}(R)}
\newcommand{\cf}{C_R}
\newcommand{\ca}{C_A}

\newcommand{\Nq}{N_q}
\newcommand{\Nt}{N_t}
\newcommand{\TF}{I_2(R)}
\newcommand{\NTF}{T_f}
\newcommand{\tad}[1]{\ensuremath{t_{\phi_{#1}}}}
\newcommand{\Sighat}{\ensuremath{\hat{\Sigma}}}
\newcommand{\sw}{\sin{\vartheta_W}}

\newcommand{\lmumt}{L_{{\mu}t}}

\newcommand{\lmmtMS}{L_{tS}}
\newcommand{\lmmtMsq}{L_{t\tilde{q}}}
\newcommand{\Msusy}{\ensuremath{m_{\rm SUSY}}}
\newcommand{\z}[1]{\ensuremath{\zeta({#1})}}

\newcommand{\lnMsq}{L_{\tilde{b}_i}}
\newcommand{\Msb}{m_{\tilde{b}_i}}

\newcommand{\hthreel}{{\tt H3m}}
\newcommand{\code}{\tt}

\newcommand{\drbar}{$\overline{\mbox{\abbrev DR}}$}
\newcommand{\drbarmod}{$\overline{\mbox{\abbrev MDR}}$}
\newcommand{\drbarprime}{$\overline{\mbox{\abbrev DR}}^{\,\prime}$}

\newcommand{\Mes}{M_{\varepsilon}}
\newcommand{\mes}{m_{\varepsilon}}
\newcommand{\Msq}{M_{\tilde{q}}}
\newcommand{\msq}{m_{\tilde{q}}}
\newcommand{\mgluino}{m_{\tilde g}}

\newcommand{\Msti}{m_{\tilde{t}_i}}

\newcommand{\mstop}[1]{m_{\tilde t_{#1}}}

\newcommand{\msquark}{m_{\tilde q}}
\newcommand{\Ms}{M_{\rm S}}

\newcommand{\lMstu}{L_{\mu\tilde{t}_1}}
\newcommand{\lMstd}{L_{\mu\tilde{t}_2}}
\newcommand{\lMt}{L_{\mu t}}
\newcommand{\Ls}{L_{\mu S}}
\newcommand{\Lt}{L_{\mu t}}

\newcommand{\lnMh}{L_{\mu h}}

\newcommand{\Mstu}{m_{\tilde{t}_1}}
\newcommand{\Mstd}{m_{\tilde{t}_2}}
\newcommand{\Msbu}{m_{\tilde{b}_1}}
\newcommand{\Msbd}{m_{\tilde{b}_2}}

\newcommand{\Mgl}{m_{\tilde{g}}}
\newcommand{\Smt}{s_{2t}}
\newcommand{\Cmt}{c_{2t}}
\newcommand{\dms}{m_{\tilde{t}_1}^2-m_{\tilde{t}_2}^2}
\newcommand{\stl}{\tilde{t}_L}
\newcommand{\str}{\tilde{t}_R}
\newcommand{\stu}{\tilde{t}_1}
\newcommand{\std}{\tilde{t}_2}
\newcommand{\Sq}{\sin 2 \theta_b}
\newcommand{\St}{\sin 2 \theta_t}
\newcommand{\Sqq}{\sin^2 2 \theta_b}
\newcommand{\Stq}{\sin^2 2 \theta_t}

\newcommand{\lMsq}{L_{\mu\tilde{q}}}
\newcommand{\lMsqgl}{L_{\tilde{q}\tilde{g}}}
\newcommand{\lMgl}{L_{\mu\tilde{g}}}
\newcommand{\lnMgl}{L_{\tilde g}}
\newcommand{\daa}{D_4(AA)}
\newcommand{\daaa}{D_4(AAA)}
\newcommand{\daf}{D_4(RA)}
\newcommand{\lp}{\left(}
\newcommand{\rp}{\right)}

\newcommand{\eqn}[1]{Eq.~(\ref{#1})}

\def\vep{\varepsilon} 
\def\be{\begin{equation}}
\def\ee{\end{equation}}
\def\besub{\begin{subequations}}
\def\eesub{\end{subequations}}
\def\bea{\begin{eqnarray}}
\def\eea{\end{eqnarray}}
\def\Tr{{\rm Tr }}
\def\frak#1#2{{\textstyle{\frac{#1}{#2}}}}

\def\GeV{\hbox{GeV}}

\def\nn{\nonumber\\}

\def\as{\alpha_s}
\def \qq{\qquad}  
\def\pa{\partial}
  
\def\de{\delta}  
\def\ep{\epsilon}
\def \la{\lambda}

\def\Hbar{{\overline{H}}}

\def\Dslash{D\!\!\!\! /}

\def\sy{supersymmetry}
 
\def\psib{\overline{\psi}}

\def\semi{;\hfil\break}

\newcommand{\note}[1]{\marginpar {\scriptsize #1}}

\begin{document}

\title{\vskip-3cm{\baselineskip14pt
    \begin{flushleft}
      \normalsize SFB/CPP-13-02 \\
      \normalsize TTP13-001
  \end{flushleft}}
  \vskip1.5cm
Precision Calculations in  Supersymmetric Theories
}
\author{
  L. Mihaila$$\\[1em]
  {\small\it  Institute for Theoretical Particle Physics,}
   \\
  {\small\it Karlsruhe Institute of Technology (KIT)}\\
}

\date{}

\maketitle

\thispagestyle{empty}

\begin{abstract}
In this review article we report on the newest developments in
 precision calculations in supersymmetric theories. An important issue
 related to this topic  is the construction of a regularization scheme 
preserving simultaneously  gauge invariance and supersymmetry. In
this context, we discuss in detail  dimensional reduction in component
field formalism as it is currently the preferred framework employed in
the literature. Furthermore,   we set special
 emphasis on  the application of multi-loop calculations 
 to  the analysis of gauge coupling unification, the prediction of the
 lightest Higgs boson mass  and the computation of the
 hadronic Higgs production and decay rates in supersymmetric models.
 Such precise theoretical calculations up to the fourth order in
 perturbation theory  are required in order to cope with the expected
 experimental accuracy on the one hand, and to enable us to
 distinguish between the predictions of the  Standard Model and   
those of supersymmetric theories on the other hand. 
\medskip


\end{abstract}
\vspace{2em}

%\centerline{(Submitted to xxx)}

\thispagestyle{empty}
\newpage

\setcounter{page}{1}

\tableofcontents

\renewcommand{\thefootnote}{\arabic{footnote}}
\setcounter{footnote}{0}

%%%%%%%%%%%%%%%%%%%%%%%%%%%%%%%%%%%%%%%%%%%%%%%%%%%%%%%%%%%%

\section{Introduction}
\setcounter{equation}{0}
\setcounter{figure}{0}
\setcounter{table}{0}
%usual stuff
%application of qcd in supergravity and hot qcd\\
Today we know that the Standard Model(SM) of particle
physics~\cite{Glashow:1961tr,Weinberg:1967tq,Salam:1968rm,GellMann:1981ph,Gross:1973id},  
which is a renormalizable gauge theory for the group $SU(3)_C\times
SU(2)_W\times U(1)$, is extremely successful at short distances of the order
of $10^{-16}$ cm. Up to now, all experiments verify it
without any conclusive hint towards new physics. On the other hand, 
 Einstein's gravitational theory based on the same
concept of gauging the symmetries gives a very good classical theory
for long distances. However, the classical theory of gravity could not
be quantized due to its abundant number of singularities. There seems
to be a deep conflict between the    classical theory of gravity and quantum
field theory. Thus, it arises naturally the question whether
gauging is the only organizing principle or there is a deeper connection
between space-time and 
internal space symmetries. In a long
series of ``no-go theorems'' among which the Coleman-Mandula
theorem~\cite{Coleman:1967ad} is 
the most important one, it was shown that the only possible symmetry group of a
consistent four-dimensional quantum field theory is the direct product of
the internal symmetry group and the Poincar\'{e} group. Precisely, it states
that  internal symmetries cannot interact non-trivially with space-time
symmetry. Surprisingly, there is a unique way of combining non-trivially
 space-time and inner space symmetries, namely Supersymmetry
(SUSY). It was shown by Haag, Lopuszanski, and
Sohnius~\cite{Haag:1974qh} that weakening 
the assumptions of the Coleman-Mandula theorem by allowing both
commuting and anti-commuting symmetry generators, there is a nontrivial
 extension of the Poincar\'{e} algebra, namely, the supersymmetry
 algebra. The supersymmetry generators
 transform bosonic particles into fermionic ones and vice versa,
 but the commutator of two such transformations yields a translation in
 space-time. In case of four-dimensional space-time, the algebra generated by
 the SUSY generators will contain the algebra of Einstein's general relativity.

The first attempts to construct  physical models respecting
SUSY can be traced back in the early seventies to
the works by  Golfand and Likhtman~\cite{Golfand:1971iw} and  Volkov and
Akulov \cite{Volkov:1972jx}. However, the first known example of a
renormalizable supersymmetric four-dimensional quantum field theory 
is the Wess-Zumino model~\cite{Wess:1974tw}. Within SUSY it is very
natural to extend the concept of space-time to the concept of
superspace~\cite{Salam:1974yz}. Along with the four-dimensional
Minkowsky space  there are 
also two new "anti-commuting" coordinates $\theta_\alpha$ and
$\overline{\theta}_{\dot{\alpha}}$, that are labeled in
Grassmann numbers rather than real numbers
\begin{eqnarray}
&&\{\theta_\alpha,\theta_\beta\} = 0\,,\quad
\{\overline{\theta}_{\dot{\alpha}},\overline{\theta}_{\dot{\beta}}\} =
0\,,\nonumber\\
&&\,\theta_\alpha^2= 0\,,\qquad\quad\,\,\,\,
\overline{\theta}_{\dot{\alpha}}^2=0\,, \qquad\quad 
\mbox{with}\quad \alpha,\beta,\dot{\alpha}, \dot{\beta} = 1,2\,.
\end{eqnarray}
 The ordinary space
dimensions correspond to bosonic degrees of freedom, the anti-commuting
dimensions to fermionic degrees of freedom.  The fields are now
functions of the superspace variables $(x_\mu, \theta_\alpha,
\overline{\theta}_{\dot{\alpha}})$ and they are organized into 
supersymmetric multiplets in a  natural
way~\cite{Salam:1974yz}. Expanding the multiplets in Taylor series over the
Grassmannian variables, one obtains the  components of the superfield as
the coefficients of the expansion. They are ordinary functions of the
space-time coordinates and can be identified with the usual fields.
 Furthermore, in the superfield notation the manifestly
supersymmetric Lagrangians are polynomials of the superfields. In the
same way, as the ordinary action is the integral over the space-time of
the Lagrangian density, in the supersymmetric case the action 
may  be expressed  as an integral over the
whole superspace.

As quantum field theories the supersymmetric theories are less divergent as they
would be in the absence of SUSY. These properties can be traced back to
the cancellation of diagrams containing bosonic or fermionic
particles, as for example the  cancellation of quadratic
 divergences present in the radiative corrections to the Higgs boson
 mass. Even more, it was 
shown~\cite{Zumino:1974bg,West:1976wz,Grisaru:1979wc}
 that there are parameters of the
theory that do not get any radiative corrections, that is a very special
feature in  quantum field theories. The most important consequence 
for the particle phenomenology is the fact that in a supersymmetric theory there
should be  an
equal number of bosons and fermions with equal masses. In other words,
for every SM particle there should exist a supersymmetric partner with an
equal mass. But in  Nature we do not observe such a situation. An
elegant solution to  break  SUSY in such a way that its
 renormalization properties remain valid (in particular the
 non-renormalization theorems and the cancellation of quadratic
 divergences) is to introduce the so-called soft
 terms~\cite{Girardello:1981wz}.  In this way, the mass difference
 between  supersymmetric partners can become  of the order of SUSY breaking
 scale. Moreover, there will also be  parameters that  receive only finite
 radiative corrections of the  order of magnitude of SUSY breaking
 parameters. This is the case of the Higgs masses and Higgs
 couplings. Accordingly, the
 SUSY partners of the SM particles
  should not be very heavy in order to account for the smallness of
 the Higgs mass and couplings. For example, requiring for consistency of
 the perturbation theory that the radiative corrections to the Higgs
 boson mass do not exceed the mass itself gives~\cite{Gladyshev:2012xq} 
\begin{eqnarray}
\delta M_h^2\approx g^2 M_{\rm SUSY}^2\approx M_h^2\,,
\end{eqnarray}
where $M_{\rm SUSY}$ denotes the mass scale of SM superpartners.
Thus, for $M_h\approx 100$~GeV and $g\approx 10^{-1}$ one obtains $
M_{\rm SUSY}\approx 1000$~GeV. This
 feature is one of the great achievements of  supersymmetric theories,
 namely, the  
solution to the hierarchy problem in particle physics.

The very old concept of the existence of an organizing principle that allows the
unification of all interactions present in Nature is nowadays embedded
in the so-called  Grand Unified Theories (GUT). The predictions of such
theories can be even precisely tested with the help of the experiments
conducted at modern particle colliders.
The most prominent example concerns, for sure, the prediction of gauge
coupling unification.  Once the gauge couplings for the electroweak
and strong interactions had been precisely measured at the Large
Electron-Proton Collider (LEP)~\cite{lepwg},  we could verify this 
hypothesis with high precision. The amazing result of evolving the
low-energy values of the gauge couplings according to the SM
predictions~\cite{Ellis:1990wk,Amaldi:1991cn,Langacker:1991an} 
is that  unification is excluded by more than eight standard
deviations. This means that, unification can be achieved only if new
physics occurs between the electroweak and the Planck scales. If one
considers that a supersymmetric theory describes the new physics, one
obtains that unification at an energy scale of about $10^{16}$~GeV can
be realized if the typical supersymmetric mass scale is of the order of
$10^3$~GeV. This observation was interpreted as first ``evidence'' for
SUSY, especially because the supersymmetric mass scale was in the same range
as that derived from the solution to the hierarchy problem.

Another virtue of SUSY is that it provides a candidate for the cold dark
matter.  Nowadays, it is well established that the visible matter
amounts to only about 4\% of the matter in the Universe. A considerable
fraction of 
the energy is made up from the so-called dark matter. The direct
evidence for the existence of  dark matter are the flat rotation
curves of spiral galaxies (see, for example, Ref.~\cite{Sofue:2000jx}
and  references cited therein), 
%To explain them,one has to asume
%the existence of a galactic halo made of dark-matter that takes part in
%the gravitational interaction. 
%tOther manifestations of the existence of the
%tdark matter are the so-called
the  gravitational lensing caused by invisible
gravitating matter in the sky\cite{Kaiser:1992ps,Kochanek:1994vw}, and
the formation of large 
structures like clusters of galaxies. 
The dark matter is classified in terms of the mass of the constituent
particle(s)  and its (their) typical velocity:
The hot dark matter, consisting of light
relativistic particles and the cold one, made of massive
weakly interacting particles (WIMPs)\cite{Eidelman:2004wy}.
 The hot dark
matter might 
consist of neutrinos; however, this hypothesis cannot explain galaxy
formation. For the cold dark matter, there is obviously no candidate
within the SM. Nevertheless, SUSY provides an excellent candidate for
WIMP, namely , the neutralino as the  lightest supersymmetric particle. 

These three fundamental predictions of SUSY makes it one of the
preferred candidates for physics beyond the SM. This explains the
enormous efforts devoted to  searches for
SUSY   in particle physics experiments at accelerators, in the deep sky
with the help of telescopes, and with the help of underground
facilities, that last already for four decades. 
 The exclusion
bounds on the supersymmetric mass spectrum is in general model
dependent. In the case of the constrained MSSM (CMSSM),  the current
status is  as follows: If one
combines the excluded regions from the direct searches at the
LHC~\cite{atlas:susy,atlas:susy}, the 
stringent lower bound on the mass of the pseudo-scalar Higgs from
XENON100~\cite{Angle:2011th}, the  constraints from the relic density
from WMAP~\cite{wmap} and  
those from muon anomalous magnetic moment~\cite{Bennett:2006fi}, one can set
a lower limit on the WIMP mass of $230$~GeV and on strongly interacting
supersymmetric particles of about $1300$~GeV. If in
addition, the mass of the lightest Higgs boson of
$125$~GeV in agreement with the   recent measurement at the
LHC~\cite{atlas,cms} is imposed,
one can exclude strongly interacting superpartners below
$2$~TeV. Nevertheless, such exclusion bounds concern the gluinos and
mainly the 
first two generation of squarks. On the other hand, for the third
generation of squarks, masses of the order of few hundred GeV are still
allowed. 
%Such light third
%generation squarks are also consistent with the recently measured 
%Higgs boson mass of about $125$~GeV. (At present,  analyses searching for the third
%generation of squarks are  ongoing at the LHC.) \\

In this context, the question whether low-energy SUSY is still 
a valid candidate for physics beyond the SM arises 
naturally. Despite the slight tension that appears in particular models,
as for example the constrained MSSM,
\footnote{The constrained MSSM model
  is based on the universality hypothesis and is described by a set of
  five free parameters defining the mass scale for the Higgs potential
  and the scalar and fermion masses.}
 the supersymmetric parameter
space is large enough to accommodate all the experimental data known at
present. However, the main prediction of low energy SUSY, i.e.  the
existence of supersymmetric particles at the TeV scale, is falsifiable
at the LHC at the full energy run of 14~TeV. If no supersymmetric particle
will be found at the TeV scale, we have to give up the main arguments in
favour of SUSY, namely, the gauge coupling unification and the solution
to the hierarchy problem. 
%In this case, the need of SUSY becomes
%questionable and the possibility to test it beyond the reach of the
%current experiments.

 To draw such powerful conclusions, one definitely
needs an accurate   comparison of the experimental data with the theory
predictions based on SUSY models. There are various possibilities to
perform such comparisons, one of them being high precision analyses,
that requires  precision data both at the experimental and theoretical
level.  On the theory side, the observables for which 
precise theoretical predictions up to the next-to-next-to leading order
in perturbation theory are required, are the electroweak precision
observables (EWPO)~\cite{Heinemeyer:2004gx}, the muon
anomalous magnetic moment~\cite{Stockinger:2006zn}, the lightest Higgs boson
mass~\cite{Heinemeyer:2004gx}, the decay rate for the rare decay of a bottom
quark into a strange quark and a photon $\Gamma(\overline{B}\to
X_s\gamma)$~\cite{delAguila:2008iz}  
and, of course, the production and decay rates of the Higgs boson at
hadron colliders~\cite{Dittmaier:2011ti}. Details about the various topics
can be found in the excellent review articles cited above.  In this paper
we report on the newest developments in
 precision calculations within   SUSY models and set
special emphasis on the recent calculations at the three-loop order
involving several different mass scales. The latter constitute 
in many cases essential ingredients for the state of the art analyses of
the experimental data taken currently at the LHC. 

This article is organized as follows:  In the next section we briefly
review the main results concerning the renormalizability of
supersymmetric theories that can be derived from their holomorphic
properties. In sections~\ref{sec:dred} and \ref{sec:susyqcd} we describe
the regularization 
method based on dimensional reduction  applied to non-supersymmetric and
supersymmetric theories up to the fourth order in perturbation
theory. In the second part of the paper we present the phenomenological
applications of such precision calculations. Namely, in
sections~\ref{sec:3lsm} we concentrate on computation of the three-loop
gauge beta functions within the SM that allows us to predict  
 the  gauge couplings at high energies with very high accuracy. Furthermore,
in section~\ref{sec:running} we report on the gauge coupling unification
within SUSY models taking into account the most precise theoretical
predictions and experimental measurements. Section~\ref{sec:mh} is
devoted to the computation of the lightest Higgs boson mass within SUSY
models with three-loop accuracy. In section~\ref{sec:hdecay}, the
hadronic Higgs production and decay in SUSY models are  reviewed and the
required computations up to the third order in perturbation theory are
presented. Finally, we draw our conclusions and present our perspective on
precision calculations in SUSY models in
section~\ref{sec:conclusions}. In the Appendix~A we give details about
the computation of the  group invariants required in multi-loop
calculations. Appendix~B contains the
main renormalization constants needed for three-loop calculations in
supersymmetric quantum chromodynamics (SUSY-QCD) within the modified minimal
subtraction, that has been employed for the computations reviewed in
sections~\ref{sec:mh} and \ref{sec:hdecay}.

%%%%%%%%%%%%%%%%%%%%%%%%%%%%%%%%%%%%%%%%%%%%%%%%%%%%%%%%%%%%
%%%%%%%%%%%%%%%%%%%%%%%%%%%%%%%%%%%%%%%%%%%%%%%%%%%%%%%%%%%%
\section{\label{sec:holomorphy} Holomorphy and exact 
beta functions in   supersymmetric theories }
\setcounter{equation}{0}
\setcounter{figure}{0}
\setcounter{table}{0}

%Importance of beta functions and RGEs.

In the last decades, enormous progress has been made in understanding the
dynamics of supersymmetric gauge theories. For many models even exact
Renormalization Group Equations (RGEs) for the gauge couplings  have
been derived. However, the connections between the exact results and
those obtained in perturbation theory are still not completely elucidated.   
Shifman and Vainshtein~\cite{Shifman:1986zi} were the first to
propose a solution to this 
puzzle. They based their argumentation  on the difference between the
quantities involved in the exact beta functions derived within the Wilsonian
renormalization approach and those adopted in the common perturbative
framework. A different derivation of the  exact beta functions was presented
in Ref.~\cite{ArkaniHamed:1997mj}, where only the Wilsonian
renormalization approach was used but the authors distinguished between the
holomorphic and canonical normalization of the gauge kinetic term in the
bare Lagrangian.

Within the Wilsonian framework~\cite{Wilson:1973jj} any field theory is
defined by the fundamental Lagrangian, the bare couplings and the
cutoff parameter. Varying the cutoff 
parameter and the bare couplings in a concerted way so that the low-energy
physics remains fixed, one finds the dependence of the bare couplings on the
cutoff parameter 
which is encoded in the Wilsonian Renormalization Group Equations
(WRGEs). The transition from 
a fundamental Lagrangian to an effective Lagrangian involves  
integrating out the high momentum modes of the quantum fields
(i.e.  degrees of freedom with momenta between  some large cutoff
scale $\Lambda$ and some renormalization scale $\mu$). 
 The coefficients of
the resulting operators play the role of renormalized couplings 
and we will call  them Wilsonian effective couplings.
The virtue of this approach is the lack of any infrared effects, since none of
the calculations involves infrared divergences. 
%By separating the infrared
%physics from the discussion, it was possible to derive exact predictions on
%the ultraviolet structure of supersymmetric gauge theories.

Let us consider as an example  supersymmetric electrodynamics
(SQED). The vector superfield in the Wess-Zumino gauge has the following
Grassmannian expansion
\begin{eqnarray}
V(x,\theta,\overline{\theta}) &=& -\theta
\sigma^\mu\overline{\theta}v_\mu(x) + i
\theta\theta\overline{\theta}\overline{\lambda}(x) -i
\overline{\theta}\overline{\theta}\theta \lambda(x) +
\frac{1}{2}\theta\theta\overline{\theta}\overline{\theta} D(x)\,,
\end{eqnarray}
where the physical degrees of freedom correspond to the vector gauge
field $v_\mu$ and the Majorana spinor field $\lambda$, known also as
gaugino field. The field $D$ is an auxiliary field without any physical
meaning and can be eliminated with the help of equations of motion for
the physical fields.
 
The Lagrangian of the model at an energy scale $\mu$ can be written as follows
\begin{eqnarray}
L &=& 
\frac{1}{4 g^2(\mu)}\int {\rm d}^4 x\, {\rm d}^2 \theta\,  {\cal
  W}^{\alpha }{\cal W}_{\alpha 
  } 
+\frac{1}{4} Z(\mu) \int {\rm d}^4 x \,{\rm d}^2 \theta (\bar{T}e^V
T+\bar{U} e^{-V}U)\,,
\label{eq::susyl}
\end{eqnarray}
 where the superfield strength tensor  is defined through the following 
relation
\begin{eqnarray}
{\cal W}_{\alpha } &=& \frac{1}{8} \overline{D}^2D_\alpha V=i
\lambda_\alpha(x)-\theta_\alpha D(x) -i\theta^\beta F_{\alpha \beta}(x)
  +\theta^2\partial_{\alpha \dot{\alpha}}\overline{\lambda}^{\dot{\alpha}}(x)\,,
\end{eqnarray}
with
\begin{eqnarray}
F_{\mu\nu}= \partial_\mu v_\nu - \partial_\nu v_\mu\,.
\end{eqnarray}
Here $D$ and $ \overline{D}$ are the supercovariant derivatives. The superfields
 $T(x,\theta,\overline{\theta})$ and $U(x,\theta,\overline{\theta})$ are
 chiral matter superfields with 
charges $1$ and $-1$, respectively. $g(\mu)$ stands for the gauge
coupling and $Z(\mu)$ denotes ...\note{finish} \\
The maximal value of $\mu$ is equal
to $\Lambda$, the ultraviolet cutoff parameter. At this point the
Lagrangian (\ref{eq::susyl}) is just the original SQED Lagrangian and the
coefficients $1/g^2(\Lambda)$ and $Z(\Lambda)$ are bare
parameters.

 Because the  momentum
integrals are performed in $d=4$ dimension and the regularization is
introduced through the cutoff parameter, 
the Wilsonian renormalization
procedure preserves SUSY.  
Thus, if one calculates the Wilsonian effective Lagrangian, it is manifestly
supersymmetric. As a consequence, the resulting  
 effective superpotential (the part of the Lagrangian density that does
 not contain any derivative)    must be a holomorphic function of
the couplings~\cite{Zumino:1974bg,West:1976wz,Grisaru:1979wc}. This
constraint restricts the running of the Wilsonian 
couplings to just the one-loop order.  \\ 
For example, let us assume that we
integrate out the matter superfields  passing to the
low-energy limit of the theory. The low-energy  
 effective coupling at the low-energy $\mu$ is given through the
 following relation 
\begin{eqnarray}
\frac{\pi}{\alpha_{W}(\mu)}=\frac{\pi}{\alpha_{W,0}(\Lambda)}-2
b_0\ln\frac{\Lambda}{\mu}\,, 
\label{eq::effg}
\end{eqnarray}
where  $\alpha_W(\mu)$ denotes the renormalized or the Wilsonian
low-energy effective coupling constant and $\alpha_{W,0}(\Lambda)$ is the 
cutoff-dependent bare coupling constant. $b_0$ is the coefficient of the
one-loop beta function of the 
underlying theory, where the beta function is defined through
% The beta function for the gauge coupling $\alpha_W$ is defined through
\begin{eqnarray}
\beta(\alpha)=\mu^2\frac{{\rm d}}{{\rm
    d}\mu^2}\frac{\alpha}{\pi}=-\left(\frac{\alpha}{\pi}\right)^2\sum_{n\ge
  0}\left(\frac{\alpha}{\pi}\right)^n b_n \,,\quad \mbox{with}\quad
\alpha=\frac{g^2}{4\pi}\,. 
\label{eq::beta}
\end{eqnarray}
Let us emphasize that Eq.~(\ref{eq::effg}) is exact at all orders. The
two- and higher-loop RGEs involve at least $\ln(\ln( \alpha_{W,0}))$
which is a 
non holomorphic function of the bare coupling and thus,
 cannot contribute to
Eq.~(\ref{eq::effg}). In Ref.~\cite{Shifman:1986zi}, it was  proved through a direct
calculation using the supergraphs  method  that the two-loop contributions to
the running of the effective coupling vanishes. The generalization of
this assertion to  higher loops is based on the extension of the
non-renormalization theorem for $F$-terms in
supersymmetric theories~\cite{Zumino:1974bg,West:1976wz,Grisaru:1979wc}.
% For renormalization scales $\mu\gg m_0$ the question of holomorphic mass
%dependence translates into the question whether the $\mu$-evolution of
%$g_{eff}$ can be written in a holomorphic form. The answer is positive
%because the renormalization is exhausted at one-loop, {\it i.e.} one has
%to replace $m_0$ with $\mu$ in Eq.~(\ref{eq::effg}). This fact is of
%course in contrast with what we know 
%about the running of the gauge couplings in perturbation theory.
%According to general theorems valid in supersymmetric gauge theories,
%the Wilsonian 
%effective action $S_W$ 
%must be expressible as holomorphic function of any chiral superfield
%present in the theory. This means that the
% superpotential and the gauge kinetic term  must be
%holomorphic functions of any chiral superfiled. The  coefficient of the
%gauge kinetic term $W_\alpha W^\alpha$ with $W_\alpha$  the chiral superfield
%generalizing the gluon field strength tensor, is called gauge function
%and is   inverse
%proportional with the gauge coupling in the Wilsonian action.
%Holomorphy
%constraints similar to those giving rise to the non-renormalization
%theorem for F-terms restrict the Wilsonian gauge coupling  to be
%renormalized only 
%at the one-loop. 

As mentioned above, 
%the authors of  Ref.~\cite{Shifman:1986zi} noticed that
%at the perturbative level,
 one has to distinguish between the
holomorphic Wilsonian gauge couplings and the physically-measurable
momentum-dependent effective gauge couplings present in the one-particle
irreducible generating functional.
 Unlike the Wilsonian couplings, the physical couplings do not depend on the
ultraviolet cutoff scale but on momenta of the particles involved. The
dependence  of the physical couplings on the overall momentum scale is
governed by the Gell-Mann-Low equations~\cite{GellMann:1954fq}, which
have different physical 
meaning as the WRGEs and have
different $\beta$-functions~\cite{Callan:1970yg, Symanzik:1970rt} beyond
one loop. 
Going from the effective Lagrangian in the Wilsonian approach to the
classical effective action $\Gamma$ means to
integrate out all of the degrees of freedom down to zero
momentum, that will generate non-holomorphic corrections. 
$\Gamma$ is often
interpreted as a sort of effective Lagrangian, but in general it does not have
the form of a supersymmetric Lagrangian with holomorphic coefficients.

The connection between the Wilsonian gauge coupling $\alpha_W$ and a 
physical gauge coupling $\alpha_{ph}$ was derived in the so called
Novikov-Shifman-Vainshtein-Zakharov renormalization scheme
 ({\abbrev NSVZ})~\cite{Shifman:1986zi}. This scheme requires  a
manifestly supersymmetric 
regularization procedure. In addition, the definition of
the physical couplings is close to that in the momentum subtraction
scheme (MOM).
 The conversion relation reads 
\begin{eqnarray}
\frac{\pi}{\alpha_{W}(\mu)}=\frac{\pi}{\alpha_{ph}(\mu)} +  T(R)
\ln Z(\mu) \,, 
\label{eq::holo}
\end{eqnarray}
where $Z(\mu)$ is the renormalization constant of the matter superfield
and  
the coefficient $T(R)$ is the Dynkin index of the 
representation $R$ of the matter superfield.  
 The  factor $Z(\mu)$
is related to the mass renormalization constants of the matter
superfield through the non-renormalization theorems,  provided 
SUSY is preserved.  However, in general   
the $Z$ factors are not restricted  by any holomorphic constraints and
thus are not 
known analytically. They have to be computed order-by-order in
perturbation theory.
Combining Eqs.~(\ref{eq::holo}) and (\ref{eq::effg}) we get 
\begin{eqnarray}
\frac{\pi}{\alpha_{ph}(\mu)}=\frac{\pi}{\alpha_{ph}(\Lambda)} -  T(R)
\ln\left( \frac{Z(\mu)}{Z(\Lambda)}\right) -2 b_0\ln\frac{\Lambda}{\mu} \,. 
\label{eq::holoph}
\end{eqnarray}
Using Eq.(\ref{eq::beta}) we obtain for the beta function of the
physical coupling in the 
{\abbrev NSVZ} scheme the following relation:
\begin{eqnarray}
\beta^{\rm  NSVZ}_{\rm SQED}(\alpha_{ph})=\left(\frac{\alpha_{ph}}{\pi}\right)^2
\frac{1}{2}T(R)(1-\gamma)\,, 
\label{eq::sqed}
\end{eqnarray}
where  we have specified the value of the coefficient  $b_0$ for the
SQED case and the superfield anomalous dimension is defined through
\begin{eqnarray}
\gamma=-\mu\frac{{\rm d} \ln Z(\mu)}{{\rm
    d}\mu} \,.
\label{eq::gamma}
\end{eqnarray}
Because Eq.(\ref{eq::effg}) is exact at all orders, also the relation between the
beta-function 
of   $\alpha_{ph}$ and the anomalous dimension of the matter
superfields $\gamma$ is valid at all orders.  
 Let us remark however, that this relation holds only in the 
{\abbrev NSVZ} scheme. Unfortunately, it is  
   highly non trivial to fulfill  the requirements of the {\abbrev NSVZ}
   scheme  in practice.

In  supersymmetric
non-abelian models with several matter supermultiplets,
Eq.~(\ref{eq::holo}) becomes 
\begin{eqnarray}
\frac{\pi}{\alpha_{W}(\mu)}=\frac{\pi}{\alpha_{ph}(\mu)}
+\frac{1}{2}C(G)\ln \alpha_{ph}(\mu)+\sum_i
T(R_i)\ln Z_i(\mu)\,,
\label{eq::holononab}
\end{eqnarray}
where $C(G)$ is the quadratic Casimir operator of the adjoint
representation and $T(R_i)$ is the Dynkin index of the 
representation $R_i$ of the matter field $i$. The second term 
stands for the gaugino contribution, while the third one for
contributions generated by the matter superfields.
A simple calculation 
provides us with the exact relation between the gauge beta function and
the  anomalous dimension of the matter  superfields
\begin{eqnarray}
\beta^{NSVZ}=-\left(\frac{\alpha_{ph}}{2\pi}\right)^2\frac{3
  C(G)-2\sum_i T(R_i)(1-\gamma_{i})}{1-C(G)\,\alpha_{ph}/(2\pi)}\,.
\label{eq::betansvz}
\end{eqnarray}
From Eq.~(\ref{eq::betansvz}) it is easy to see that for the
derivation of the $L$-loop beta functions in the {\abbrev NSVZ} scheme one needs
the matter anomalous dimensions $\gamma_i$ at the $(L-1)$-loop order. As will be shown
below this feature was intensively exploited in the literature.

In the case of
SUSY-Yang-Mills theories  the matter superfields are absent, so
$T(R_i)=0$, and an    exact formula for the gauge
coupling beta function  can be derived
\begin{eqnarray}
\beta^{NSVZ}=-\left(\frac{\alpha_{ph}}{2\pi}\right)^2\frac{3
  C(G)}{1-C(G)\,\alpha_{ph}/(2\pi)}\,.
\label{eq::betansvz_sy}
\end{eqnarray}

Similar relations can also be  derived for models with softly-broken SUSY.
 The line of reasoning is as follows:
The powerful supergraph method is also applicable for models with softly-broken
SUSY  by using the ``spurion'' external
field method~\cite{Girardello:1981wz,Grisaru:1985tc}.
Perhaps, one of the most prominent example is
 the relation that can be established  between
the gaugino mass $m_{\tilde {g}}$ and the gauge beta function. In the
presence of the SUSY-breaking gaugino 
mass term, the coefficient of the gauge kinetic term in the Wilsonian action
 becomes
\begin{eqnarray}
\left(\frac{1}{g^2}\right)_W\to\left(\frac{1-2
    m_{\tilde{g}}^2\theta^2}{g^2}\right)_W\,,
\quad\mbox{where $\theta$ is
  the Grassman variable}\,.
\end{eqnarray}
Using the same arguments based on  holomorphy,  it was
shown~\cite{Hisano:1997ua,ArkaniHamed:1998kj} that
 a renormalization group invariant  (RGI)
 relation for the gaugino mass can be derived within {\abbrev NSVZ} scheme
\begin{eqnarray}
\frac{ m_{\tilde{g}} \alpha}{\beta(\alpha)} = \mbox{RGI}\,.
\label{eq::rgi}
\end{eqnarray}

Moreover, it was shown  with the help of the spurion formalism
  that the renormalization constants of softly broken
SUSY gauge theory can be related 
to the renormalization constants of the underlying exact supersymmetric
model~\cite{Yamada:1994id,Avdeev:1997vx,Kazakov:2000ih}. Even more, the
connecting 
formulas are valid at all orders in perturbation theory.  The only 
necessary assumption for their derivation is the existence of a gauge and
SUSY invariant regularization scheme. Thus, such relations are valid
only in {\abbrev NSVZ}-like regularization schemes.

%Using same arguments based on the holomorphy properties of the gauge
%function in the wilsonian action, important relations for the softly
%broken SUSY theories can be made. 

At this point, a few remarks are in order  to comment on the results
discussed above.  The authors of Ref.~\cite{Novikov:1985rd} state that in
$d=4$ dimensions  the
only known   regularization to conserve SUSY is the Pauli-Villars scheme for
matter superfields  and the higher derivative  scheme for the gauge
superfields. Technically 
this construction is rather complicated and hardly applicable to multi-loop
computations. In Ref.~\cite{Grisaru:1985tc}, an attempt  was made to
apply the ``supersymmetric dimensional regularization'' or
``regularization by dimensional reduction''(DRED)~\cite{siegel} within
the supergraph 
formalism. However, as  pointed out by Siegel himself\cite{siegelb},  this
scheme is mathematically inconsistent in its original formulation and a
consistent formulation will break supersymmetry in higher orders of
perturbation theory. A similar situation occurs also for the application
of DRED in component field
formalism~\cite{Avdeev:1981vf,Avdeev:1982xy}\footnote{A detailed analysis of
this issue will be done in the next section.}. Thus, the exact formulas of
the {\abbrev NSVZ} scheme are not valid, in general, for  calculations
based on DRED since they do not involve a
regularization scheme  supersymmetric at all orders. 
% On the other hand, the practical implementation of the {\abbrev NSVZ}
% scheme proposed in   Ref.~\cite{Novikov:1985rd} is rather inaccessible. 
For particle
phenomenology, it means that the powerful predictions of  
Eqs.~(\ref{eq::betansvz}) 
cannot be  tested through experiments, 
since the beta functions are scheme dependent beyond two loops.

The breakthrough regarding  this situation was obtained in
Refs.~\cite{Jack:1997pa,
  Jack:1998iy,Jack:1996qq,Jack:1996vg,Jack:1996cn}, where it is stated 
that, if the
 {\abbrev NSVZ} scheme exists it can be perturbatively related to
schemes based on DRED. Such arguments follow from the equivalence
of  different
renormalization schemes in perturbation theory~\cite{Jack:1994bn}.
% The
%connection between the two schemes was derived for a non-abelian
%supersymmetric gauge theory up to three loops.  
%The calculations of Refs~\cite{Jack:1996qq,Jack:1996vg,Jack:1996cn}  were
%performed with the help of supergraphs formalism together with DRED as 
%regularization method.  
Precisely, the computation of the three-loop mass
anomalous dimension for the chiral matter superfield in a general
non-abelian supersymmetric theory and of the three-loop gauge beta function
in the abelian case allowed the derivation of the three- and four-loop
gauge beta function for a general supersymmetric theory. Remarkably
enough, the derivation (up to a numerical coefficient) of the four-loop
gauge beta function  was based  on a three-loop calculation
and theoretical considerations about special relations valid in $N=2$
supersymmetric theories 
and one-loop finite supersymmetric theories.\\
 Let us mention at this
point also  the calculation of the three-loop gauge beta function
for  supersymmetric Yang-Mills (SYM) theories of
Ref.~\cite{Avdeev:1981ew}. For this calculation, DRED was employed in
component field formalism rather than superfield formalism, and hence a
manifestly not 
supersymmetric gauge was used. The computations of
Refs~\cite{Jack:1996vg,Jack:1996cn} 
and ~\cite{Avdeev:1981ew} coincide as a consequence of gauge invariance
of the gauge beta function.\\  
Moreover, the authors of  Refs.~\cite{Jack:1997pa, Jack:1998iy} 
  noticed that the differential operators
relating the beta functions for soft SUSY breaking parameters to the beta
functions of the gauge and Yukawa couplings are form invariant under
change of scheme ({\it i.e.} from \nsvz{} to DRED scheme).  Thus,
similar relations 
for the soft SUSY breaking parameter
valid to all orders of 
perturbation theory hold also in a DRED like scheme\footnote{ Actually, the
scheme proposed by the authors of  Refs.~\cite{Jack:1997pa, Jack:1998iy}
is the so called DRED$^\prime$, for which  beta
functions of  SUSY-breaking parameters do not depend on 
the unphysical $\epsilon$-scalar mass parameter. For more details about
the DRED$^\prime$ scheme see section~\ref{sec:dred}.}.

In the next section we will discuss in detail the
application of DRED in component field formalism and give some example of
important calculations that can be done within this
approach. Nevertheless,already now  we want to mention the coincidence of all
results obtained with  DRED in component field formalism and those 
derived via  DRED in supergraphs formalism.

%The first phenomenological application of the results obtained in
%Refs.~\cite{Jack:1996qq,Jack:1996vg,Jack:1996cn}
%was the derivation  of the three-loop beta-functions for the minimal
%supersymmetric standard model~\cite{Ferreira:1996ug,Avdeev:1997vx}. It
%was followed 
%by a series of papers concerning the effects of the three-loop beta
%functions on the soft-SUSY breaking
%parameters~\cite{Kolda:1996ea,Jack:2003sx} and on the gauge coupling
%unification~\cite{...} . 

%The classical gene rating functional obtained when integrating out all
%(high and low-momentum) degrees of freedom is not SUSY invariant with
%holomorphic coefficients.\\

%At the perturbative level, we have to distinguish between the
%holomorphic wilsonian gauge couplings and the physically-measurable
%momentum-dependent effective gauge couplings.

%Meanwhile, the RG in particle physics had been reformulated in more
% practical terms by C. G. Callan and K. Symanzik in 1970. 
%%%%%%%%%%%%%%%%%%%%%%%%%%%%%%%%%%%%%%%%%%%%%%%%%%%%%%%%%%%%
%%%%%%%%%%%%%%%%%%%%%%%%%%%%%%%%%%%%%%%%%%%%%%%%%%%%%%%%%%%%

\section{\label{sec:dred}   Dimensional reduction in
 the component field   formalism} 
\setcounter{equation}{0}
\setcounter{figure}{0}
\setcounter{table}{0}

The precision of many present or forthcoming experiments in particle physics
requires inevitably higher order perturbative calculations in the SM
 or its extensions like the  Minimal Supersymmetric Standard Model
 (MSSM). Regularization of the 
divergent loop diagrams arising in the higher order calculations 
 is commonly performed  employing  Dimensional
Regularization (DREG) or  its variants, due to its nice feature to
respect gauge invariance. 
  Higher order
calculations within the SM  predominantly  use \dreg{} in
its original form~\cite{dreg1,'tHooft:1972fi}, while for calculations within
supersymmetric theories \dred{} as defined in Ref.~\cite{siegel} is
commonly employed.
%another technique compatible with the supersymmetry, the Dimensional Reduction
%(DRED), was proposed by Siegel~\cite{siegel}.
It is not {\it a priori} known whether
SUSY as a symmetry of a given Lagrangian is still a symmetry of the full
quantum theory in any particular case. Nevertheless, a detailed formal
renormalization 
program has been pursued in Ref.~\cite{Piguet:1980fa} including a proof that
SUSY is not anomalous. 
%\\The question of anomaly is related to the question of regularization.
 If the regularized theory does not respect SUSY, the
finite amplitude will not satisfy the Ward identities required by SUSY,
giving rise to an apparent anomaly. If SUSY is not anomalous, it is
possible to restore the invariance  by introducing finite
counter-terms. 

 In practice, 
 the choice of regularization scheme is of considerable significance
 for the extraction of physical predictions. This
is the case for the {\abbrev NSVZ} scheme we alluded in the previous
section, that 
rarely   found direct practical applicability. It rather  provides
important checks for results predicted within DRED.
In this section we discuss in detail the application of DRED in the
component field formalism and its application to practical calculations.

\subsection{\label{sec:framework}Framework}
 DRED consists of continuing the number of space
dimensions from $4$ to $d$, where $d$ is less than $4$, but keeping the
dimension of all the fields fixed. In component field language, this
means that the vector bosons and fermions preserve their
four-dimensional character. Furthermore, it is assumed that all  
fields depend on $d$ rather than $4$ space-time coordinates, so that the
derivatives $\partial_\mu$ and momenta $p_\mu$ become
$d$-dimensional. It is the four-dimensional nature of the fields that is
supposed to restore the supersymmetric Ward-Takahashi~\cite{wt} or
Slavnov-Taylor~\cite{st} identities, while the $d$-dimensional
space-time coordinates cures, as 
in \dreg{}, the singularities of the loop integrals.\\
 However, potential
inconsistencies of \dred{}, arising from the use of purely four-dimensional
relations between the Levi-Civita tensor and the metric tensor,  have been
pointed out by Siegel himself~\cite{siegelb}. Even more, inconsistencies
of \dred{} arising without the direct use of Levi-Civita tensors  have been
revealed in Ref.~\cite{Avdeev:1982xy}. The authors have correlated them
with the impossibility of decomposing the finite four-dimensional space
into a direct sum of infinite-dimensional spaces.
The solution proposed by the same authors is to
%renounce to the index counting and
introduce a formal space, called
quasi-four-dimensional space ($Q_4$), with 
``non-integer valued'' vector and spinor indices (thus, the two types of
indices range over an infinite set of values),
 obeying certain algebraic identities
inspired from the   
%retaining  the essential 
properties of the four-dimensional Minkowski space.  
The existence of such
a space was demonstrated by  construction~\cite{ds} starting from similar
arguments as those used to prove the existence of the formal
$d$-dimensional space of \dreg{}~\cite{coll}. In this way the
consistency of the  calculation rules is  guaranteed. 
By construction,  $Q_4$ is represented as the direct sum of two
infinite-dimensional 
spaces: $Q_d$ which is formally $d$-dimensional and is identical with the
one of \dreg{} and $Q_{2 \ep}$  which is formally
$2\ep=4-d$-dimensional\footnote{One needs to perform twice the
  construction of  $n$-dimensional 
integrals and metric tensors for $n=d$ and $n=2\ep$.  The
$d$-dimensional integral 
is the momentum integral in \dred{},  
while $2\ep$ integral is involved
only in the definition of the  $2\ep$-dimensional metric tensor.} 
\begin{equation}
\begin{split}
Q_4=Q_{d} \oplus Q_{2 \ep} \,.
\end{split}
\end{equation}
According to the properties of the three formal spaces at hand
$Q_4$, $Q_d$, $Q_{2 \ep}$
one can derive the following relations for the corresponding  metric
tensors $g^{\mu \nu},\, \hat{g}^{\mu\nu},\, \bar{g}^{\mu\nu}$~\cite{Avdeev:1981vf,ds}:
\begin{equation}
\begin{split}
 g^{\mu \nu} &= \hat{g}^{\mu\nu}+ \bar{g}^{\mu\nu}\,,\quad \,
 g^{\mu \mu} =4\,,\qquad \quad\quad \hat{g}^{\mu \mu} =d\,,\qquad
 \bar{g}^{\mu \mu} =2\ep\,,
\\
 g^{\mu \nu} \hat{g}_{\nu}^{\,\,\,\rho}&=\hat{g}^{\mu\rho}\,,\qquad\quad
 g^{\mu \nu} \bar{g}_{\nu}^{\,\,\,\rho}=\bar{g}^{\mu\rho}\,,\qquad
\hat{g}^{\mu \nu} \bar{g}_{\nu}^{\,\,\,\rho}=0\,.
\end{split}
\end{equation}
Furthermore, any quasi-four dimensional vector can be decomposed with
the help of the projectors $ \hat{g}^{\mu\nu},\, \bar{g}^{\mu\nu}$
 \begin{equation}
\begin{split}
t^\mu&=\hat{t}^\mu+\bar{t}^\mu\,,\qquad \hat{t}^\mu=\hat{g}^{\mu\nu}
t_\nu\,,\qquad  \bar{t}^\mu=\bar{g}^{\mu\nu} t_\nu\,.
\label{eq::vect}
\end{split}
\end{equation}
 Imposing  the Dirac algebra for the $\gamma$-matrices defined in $Q_4$ 
 \begin{equation}
\{\gamma^\mu,\gamma^\nu\}=2 g^{\mu\nu}\mathbf{1}\,, 
\end{equation}
 we can derive similar commutation relations for the components in $Q_d$
 and $Q_{2 \ep}$ 
  \begin{equation}
\begin{split}
\{\hat{\gamma}^\mu,\hat{\gamma}^\nu\}=2\hat{g}^{\mu\nu}\mathbf{1}
\,,\qquad
\{\bar{\gamma}^\mu,\bar{\gamma}^\nu\}=2\bar{g}^{\mu\nu}\mathbf{1}\,,\qquad
\{\bar{\gamma}^\mu,\hat{\gamma}^\nu\}=0\,.  
\end{split}
\label{eq::comm}
\end{equation}
These relations together with the trace condition 
\begin{equation}
{\rm Tr}\mathbf{1} =4
\label{eq::tr}
\end{equation} 
are sufficient for computing Feynman diagrams. Eq.~(\ref{eq::tr}) is
particularly useful in supersymmetric theories, because it ensures that
the numbers of degrees of freedom for fermions and bosons is equal. 

 For  practical computations, it is useful to note  that the fermion
 traces that contain both types of $\gamma$-matrices can be factored out as follows 
\begin{equation}
\rm{Tr}(\hat{\gamma}^{\mu_1}\cdots\hat{\gamma}^{\mu_n}\bar{\gamma}^{\nu_1}
\cdots\bar{\gamma}^{\nu_l})=\frac{1}{4}\rm{Tr}(\hat{\gamma}^{\mu_1}\cdots\hat{\gamma}^{\mu_n})  
\rm{Tr}(\bar{\gamma}^{\nu_1}\cdots\bar{\gamma}^{\nu_l})\,. 
\end{equation}
This relation can be derived from Eqs.~(\ref{eq::tr}) and ~(\ref{eq::comm}) and the
algebra of
Dirac matrices in $d$ dimensions.
 Thus, the Dirac algebra can be performed separately in $d$ and in
$4-d=2\ep$ dimensions.
  
Once we introduced ``non-integer valued'' spinor indices, we need
infinite-dimensional $\gamma$-matrices to represent the Dirac algebra. Thus,
the Fierz
identities valid in the 
genuine four-dimensional space do not hold
 anymore in $Q_4$. Their use  was identified with
one of the sources of \dred{} inconsistencies. Moreover,  within $Q_4$ the
 invariance of the original Lagrangian under SUSY transformations might
 be broken. This feature can be directly correlated with the lack of Fierz
identities that would ensure the cancellation of Lagrangian variation
under SUSY transformations in the genuine four-dimensional space. 
However, it has been
shown~\cite{Avdeev:1982xy,ds} 
that such inconsistencies   become active only in the higher
orders of perturbation theory, when, for example, traces over at least
ten $\gamma$-matrices and anti-symmetrization over
five indices are involved. Thus, \dred{} also breaks SUSY, but starting
from  higher orders    of perturbation theory.
%It is worth mentioning that, although \dred{} was originally constructed
%for applications in supersymmetric models, 
This explains, why one- and even
two-loop calculations of
QCD corrections  within \dred{} ~\cite{Korner:1993pv, Misiak:1993es,
  Kunszt:1993sd, Smith:2004ck, 
  Altarelli:1980fi} based on genuine  four-dimensional  
 Dirac algebra and even Fierz rearrangement provided  correct results.
Even the
supersymmetric character of \dred{} at low orders has been exploited in
the context of QCD with massless quarks in Ref~\cite{Kunszt:1993sd}. 
However, beyond the one-loop level   
the distinction between $g^{\mu \nu}$ resulting from contractions of the
quasi-four-dimensional vector fields and $\hat{g}^{\mu\nu}$ resulting
from momentum integrals is difficult to follow. 
It turned out~\cite{Capper:1979ns} that for higher order computations it
is useful to decompose the 
quasi-four-dimensional vector fields 
according to Eq.~(\ref{eq::vect}). As we shall see in the next section,
in the case of  gauge 
theories the $d$-dimensional 
components behave as  vectors under the gauge transformations whereas
the $2\ep$ components as  scalars, usually called \epscalar{}s. 
%The former ones, usually called
%\eps-scalars, have also the role to convert the  results obtained within
%\dreg{} to \dred{}.

Representing the underlying space of \dred{} $Q_4$ as a formal
infinite-dimensional space renders the extension of $\gamma_5$  as
subtle as  in \dreg{}. The consistent procedure proposed by 't Hooft-Veltman 
(HV)~\cite{'tHooft:1972fi} for defining 
$\gamma_5$ as in  four dimensions  $\bar{\gamma}_5=i \gamma^0 \gamma^1
\gamma^2 \gamma^3$ has in the context of SUSY theories two drawbacks. 
On the
 one hand, it is the fact that the  mathematically consistent treatment
 of $\gamma_5$ in \dreg{}  requires $d > 4$, whereas for \dred{} $d < 4$ is
 needed. However, it has been shown up to
 two-loops~\cite{Nicolai:1980km, Jones:1982zf} that the Adler-Bardeen
 theorem~\cite{Adler:1969er} could  still be satisfied in \dred{}
 with HV scheme, if relations like   
\begin{equation}
\hat{\gamma}^i\bar{\gamma}_5 \hat{\gamma}_i = (d-8)\bar{\gamma}_5\,, 
\end{equation}
which follow in $d>4$ are assumed to hold also for $d<4$.
On the other hand, the use of a not anti-commuting $\gamma_5$
leads to the breakdown of symmetries, {\it e.g.} chiral symmetry of the
SM or supersymmetry in case of the MSSM  already at
the one-loop level. These ``spurious anomalies'' would spoil the
renormalizability and  they have to be cured by 
%imposing  "by hand'' 
introducing  appropriate counter-terms
   to restore Ward-Takahashi and Slavnov-Taylor
identities order by order in perturbation theory (see
Ref~\cite{Trueman:1995ca}). This approach was 
successfully applied for SM predictions within \dreg{} up to
three-loops~\cite{Larin:1993tq,Chetyrkin:1998mw}. However,
for the MSSM 
it becomes much more 
involved due to the complexity introduced by supersymmetric conditions
and it rarely has been employed in practice~\cite{Jones:1982zf}.\\
The implementation  of $\gamma_5$ in \dred{} commonly used in practice
 is inspired by the naive scheme (NS) of \dreg{}.
 Namely, it is treated rather like a
formal object which is not well-defined mathematically but anti-commutes
with all $\gamma$-matrices
\begin{equation}
\begin{split}
\{\hat{\gamma}^\mu,\tilde{\gamma}_5\}&=
\{\hat{\gamma}^\mu,\tilde{\gamma_5}\}\,=\,0\,,\quad 
\mbox{and}\\ (\tilde{\gamma_5})^2&=\mathbf{1}\,.
\label{eq::gam5} 
\end{split}
\end{equation}
Nevertheless, one has to correct the false result that arises from
Eqs.~(\ref{eq::gam5}), that  the
trace of $\gamma_5$ and four or more  $\gamma$-matrices vanishes.
 Paying attention  that now two types
of $\gamma$-matrices occur, the additional constraints read 
\begin{equation}
\begin{split}
    {\rm Tr}(\Gamma^\alpha\Gamma^\beta\Gamma^\gamma\Gamma^
\delta\tilde{\gamma}_5)
  = 4i\,\tilde{\varepsilon}^{\alpha\beta\gamma\delta} + {\cal O}(\epsilon) \,,\quad 
\mbox{with}\quad \Gamma^\mu =
  \hat{\gamma}^\mu \,\,\mbox{or} \,\,\bar{\gamma}^\mu\,.
  \label{eq::trgamma5_2}
\end{split}
\end{equation}
 The tensor $\tilde{\varepsilon}^{\alpha\beta\gamma\delta}$
 has some similarities
 with the four-dimensional Levi-Civita tensor:
i) it is completely antisymmetric in all indices;  ii) when contracted
with a second one of its kind gives the following
result
\begin{equation}
\begin{split}
  \tilde{\varepsilon}^{\alpha\beta\gamma\delta}
  \tilde{\varepsilon}_{\alpha'\beta'\gamma'\delta'} =
  G^{[\alpha\phantom{]}}_{[\alpha'\phantom{]}}
  G^{\phantom{[}\beta\phantom{]}}_{\phantom{[}\beta'\phantom{]}}
  G^{\phantom{[}\gamma\phantom{]}}_{\phantom{[}\gamma'\phantom{]}}
  G^{\phantom{[}\delta\,]}_{\phantom{[}\delta']}\,,\quad
  G^{\mu\nu}=\hat{g}^{\mu\nu}\,\,\mbox{or}\,\,\bar{g}^{\mu\nu}\,,
\label{eq::leci}
\end{split}\end{equation}
depending on the nature of Dirac matrices $\Gamma^\mu$ in
Eq.~(\ref{eq::trgamma5_2}).
Here the square brackets denote complete anti-symmetrization. 
When taking the limit $d\to 4$,
 $\tilde{\varepsilon}^{\alpha\beta\gamma\delta}$ converts into the
four-dimensional Levi-Civita tensor and
Eqs.~(\ref{eq::trgamma5_2}) and (\ref{eq::leci}) ensure that the
correct four-dimensional results are reproduced. 
 This last
constraint is needed in order to correctly compute fermion triangle
diagrams containing an axial vector   current, {\it i.e.} to 
cope with the Adler-Bardeen-Jackiw
anomaly~\cite{Adler:1969gk}.

 At this point a comment on
Eqs.~(\ref{eq::trgamma5_2}) is in order. When we combine it with the
cyclic property of traces, it necessary follows that other traces are
not well defined in $d\ne 4$ dimensions. It turns out that there is an
unavoidable ambiguity of order ${\cal O}(d-4)$ when fixing
the trace condition in Eq.~(\ref{eq::trgamma5_2}). 
%, which can be interpreted~\cite{Chanowitz:1979zu} 
%as a consequence of the ambiguity of the Adler-Bardeen-Jackiw anomaly. 
Even if one does not use the cyclic property of the trace, an
ambiguity in the distribution of the anomaly between the vector and the
axial vector currents shows up~\cite{Nicolai:1980km}. The occurrence of
the ambiguity is  a 
characteristic  of the extension of  $\gamma_5$
away from $d=4$ dimensions.  't Hooft and 
Veltman  have  pointed out in their original paper~\cite{'tHooft:1972fi}
that an 
ambiguity related to the location of $\gamma_5$ shows up in HV scheme,
too.

The use of an anti-commuting $\gamma_5$ in $d\ne 4$ dimensions
was  applied for the first time to the evaluation of fermion traces with
an even number of $\gamma_5$'s 
in Ref.~\cite{Bardeen:1972vi}, and a few years later extended also to odd
$\gamma_5$ fermion traces in Ref.~\cite{Chanowitz:1979zu}.  The
method\footnote{For more details see~\cite{Jegerlehner:2000dz} and
  references cited therein.}
proved to be effective for SM
calculations involving chiral fermions up to two-loop
order\cite{Fleischer:1993ub,Freitas:2000gg}.  The consistency of this 
 $\gamma_5$  prescription has been verified even in  three-loop
 QCD-electroweak calculations~\cite{Avdeev:1994db, Chetyrkin:1995ix}.
Within \dred{}, it  has been successfully employed
in   MSSM calculations at the two- and three-loop
order~\cite{Heinemeyer:2004yq,Pickering:2001aq,Harlander:2009mn}. However, let us
mention at this point that, for these
calculations at most the finite parts of two-loop and the
divergent parts of  three-loop diagrams are required. For the
calculation of finite parts of three-loop diagrams containing two fermion
triangle sub-diagrams, the HV scheme has to be applied as the naive
scheme does not provide correct results.
%Currently, it is
%not clear how should be modified the 
% NS for computations involving
% finite parts of three-loop diagrams that are affected
% by the triangle anomaly.

 Through the consistent formulation of \dred{} we gain a
 regularization scheme which proves
 to be supersymmetric only in the lower orders of perturbation theory. 
 Due to the violation of 
 Fierz identities, SUSY invariance will be broken
 at higher orders. The first consequence of SUSY breaking is that the
 all order relations between different anomalous dimension valid in the
 NSVZ scheme do not hold in DRED. However,
 although \dred{} consistently formulated is not a supersymmetric
 scheme at all orders,  it provides so far the best option for
 computations within SUSY theories.

%%%%%%%%%%%%%%%%%%%%%%%%%%%%%%%%%%%%%%%%%%%%%%%%%%%%%%%%%%%%
\subsection{Minimal Subtraction  Schemes  \msbar{} and \drbar{}}
%\subsection{Minimal Subtraction Scheme \msbar{} }
The common renormalization schemes used for multi-loop calculations are 
the minimal subtraction (MS),
momentum subtraction  and on-shell schemes.   
Minimal subtraction scheme has the 
advantage of involving  the simplest computations, but it is
 non-physical in the sense that it does not take into account mass threshold
 effects for  heavy particles. Nevertheless, it is the main scheme used  in
 renormalization group (RG)
analyses relating the predictions of a given theory at different energy
scales. 
The other two options are computationally much more involved but
 indispensable for the determination of the parameters of a
theory from the quantities measured experimentally. We focus in this
section on the minimal  subtraction methods.

 Minimal subtraction scheme with \dreg{} as regulator~\cite{'tHooft:1973mm} or
 the  
modified \msbar{} scheme~\cite{Bardeen:1978yd} and 
its variant for \dred{} --- the \drbar{} scheme --- are in particular  well
suited for  higher order calculations in perturbation theory. The
advantage of these schemes is that all ultra-violet (UV) counter-terms
are polynomial both in external momenta and
masses~\cite{Collins:1974bg}. This allows to set to zero certain masses or
external momenta, provided no spurious infrared
divergences are introduced. This simplifies substantially the calculations
of the Feynman integrals. It has been shown~\cite{Chetyrkin:1984xa}  by
means of the 
infrared rearrangement (IRR)
procedure~\cite{Vladimirov:1979zm,Chetyrkin:1980pr,Chetyrkin:1984xa}
that the  
renormalization constants within the \msbar{} scheme can be reduced to
the calculation of only massless propagator
diagrams. This method was used for the
first three-loop calculation of the QCD
$\beta$-function~\cite{Tarasov:1980au}, applying it to each individual
diagram. But the most effective approach
is its use in combination with
multiplicative renormalization. This amounts in general to
solve recursively the equation
\begin{eqnarray}
Z_a &=& 1-K_\ep[\Gamma_a(p^2) Z_a]\,,
\label{eq::rec}
\end{eqnarray}
where $K_\ep [f(\ep)]$ stands for the singular part of the Laurent
expansion of $f(\ep)$ in $\ep$ around $\ep=0$.  $\Gamma_a(p^2)$
denotes the renormalized Green function with only one external momentum $p^2$
kept non-zero.  $Z_a$ denotes the renormalization constant associated
with the Green function  $\Gamma_a$. In this case, the renormalization
of $\Gamma_a$  through $(l+1)$-loop
order requires  the
renormalization of the Lagrangian parameters like couplings, masses,
gauge parameters, mixing angles, etc. up to $l$-loop order.  The method
was successfully applied  to the
three-loop calculations of anomalous dimensions within \msbar{} or
\drbar{} schemes~\cite{Larin:1993tp,Larin:1993tq,Pickering:2001aq,Harlander:2006rj,Harlander:2009mn,Chetyrkin:2012rz,Mihaila:2012fm}         
 using the  package {\tt
  MINCER}~\cite{Larin:1991fz} written in  {\tt
  FORM}~\cite{Vermaseren:2000nd}, which computes   
analytically  massless propagator diagrams up to three loops.

Apart from that, a second method was proposed in
Ref.~\cite{Chetyrkin:1997fm}, which has been used for the
calculation of the three- and even four-loop anomalous dimensions of
QCD~\cite{Chetyrkin:1997dh,vanRitbergen:1997va,Vermaseren:1997fq,Czakon:2004bu}  
and the 
beta-function of the quartic coupling of the Higgs boson in the
SM~\cite{Curtright:1979mg,Jones:1980fx, Chetyrkin:2012rz}. It deals
with the IRR
by introducing an artificial mass for all propagators. Expanding in all
 particles masses and external momenta,  one can reduce the
evaluation of the Feynman integrals to massive tadpoles. The analytic
evaluation of the  massive tadpoles up to three-loop order can be
obtained with the help of the package {\tt
  MATAD}~\cite{Steinhauser:2000ry}. 
%For the
%calculations~\cite{vanRitbergen:1997va,Vermaseren:1997fq,Czakon:2004bu}
% even the divergent part of the four-loop massive tadpoles were
% required. 

A third method was introduced for the evaluation of the
 renormalization constants for the quark mass~\cite{Chetyrkin:1997dh}
 and the 
 vector~\cite{Chetyrkin:1996ez}
  and quark scalar current correlators~\cite{Chetyrkin:1996sr} through
  four-loops. It is based on  global IRR properties and
  amounts essentially to set to zero the external momentum and let an
  arbitrary subset of the internal lines to be massive. After non
  trivial manipulations, the four-loop integrals can  be reduced
  to three-loop massless two-point integrals and one-loop massive vacuum
  integrals.

 The three-loop accuracy  for the anomalous dimensions of theories
involving not only vector but also Yukawa and quartic scalar 
interactions (for example  the SM~\cite{Chetyrkin:2012rz,Mihaila:2012fm}) 
  was achieved only very recently.
% restricted, for the time being, to two-loop
%order~\cite{Machacek:1983tz,Jack:1984vj,Ford:1992pn} even in the 
%minimal subtraction schemes. 
Remarkably, for supersymmetric and  softly broken
supersymmetric theories like the MSSM the three-loop anomalous
dimensions were computed  
long before~\cite{Jack:1996vg,Jack:1996qq,Ferreira:1996ug}. Their
derivations used intensively  
 the exact relations 
established between the various anomalous dimensions in the
{\abbrev NSVZ} scheme\footnote{For more details see
  section~\ref{sec:holomorphy}} 
as well as the observation that  the {\abbrev NSVZ} scheme and  \dred{}
can be perturbatively connected. 
%With the help of these observations,
%one can reduce the computation to genuine three-loop order propagator
%diagramms. 

%For the calculation of the finite parts of the renormalization constants
%for Green's functions involving three or more external particles
%additional computational methods have to be employed. Neverthe\\
%{\it ass. expansion\\
%os 2 and 3-loop os propagator\\
%2- and 3-loop vertices h->hadrons, ee->hadrons, form factors\\ }
% However for the coputation of the four-loop anomalous dimensions in QCD
% two new methods were proposed.

%The approch to avoid them, usually called
%infrared rearrangements, can become very much involved in practical
%calculations~\cite{Vladimirov:1979zm,Chetyrkin:1980pr}. In order to get
%rid of this difficulty, the $R^\star$ operation to deal with both UV
%and IR 
%divergences was elaborated~\cite{Chetyrkin:1984xa}.

%\subsection{On Shell Scheme}

%%%%%%%%%%%%%%%%%%%%%%%%%%%%%%%%%%%%%%%%%%%%%%%%%%%%%%%%%%%%

\subsection{\label{sec:qcd} DRED applied to non-supersymmetric
 theories}

Although \dred{} was originally proposed as a candidate for an invariant
regularization in supersymmetric theories, it proved  to be
useful also in non-supersymmetric theories. Its use in
SM calculations up to three-loop orders was motivated  either by the
possibility to apply four-dimensional algebra and even Fierz
rearrangements~\cite{Altarelli:1980fi,Misiak:1993es}\footnote{The
  mathematical inconsistencies alluded to above do not occur at the
  two-loop level in this calculations.}, or
by the possibility to  easily convert a 
non-supersymmetric gauge theory into a SUSY-Yang-Mills theory and use
non-trivial Ward identities as checks of complicated
calculations~\cite{Kunszt:1993sd,Avdeev:1994db,Bern:2002zk}. 
 Apart from
the 
computational advantages, \dred{} applied to non-supersymmetric theories,
in particular to QCD, provides us with a powerful tool to verify its
consistency up to three-loop order via the connection that can be
established with  \dreg{}\footnote{\dred{} and \dreg{} are also
  perturbatively connected.}. Finally, it is motivated by the MSSM, as a
softly broken supersymmetric theory, or by various
models derived from the MSSM which feature lower symmetries (for 
example, the intermediate energy theory obtained by integrating out the
squarks and sleptons). DRED 
  applied to effective field theories, such that  QCD extended to
  include the Higgs-top Yukawa coupling,  was  useful for the
  calculation of the production rate for the Higgs boson in gluon fusion
  channel within MSSM~\cite{Pak:2010cu,Pak:2012xr}.

%\subsubsection{\label{sec::rev}Gauge theory with fermions}

In the following, we consider a non-abelian gauge theory with $n_f$ Dirac 
fermions $\psi_f$ transforming according to  a 
representation $R$ of the gauge group ${\cal G}$.  For the moment we do
not take into account any genuine scalar field.  

The Lagrangian density (in terms of bare fields) reads 
\be
{\cal L}_B = -\frac{1}{4}F^2_{\mu\nu} - \frac{1}{2
  (1-\xi)}(\pa^{\mu}W_{\mu}^a)^2 +  
\pa^{\mu}\bar{c}^{a}(\pa_{\mu} c^a -g f^{abc} c^b W_{\mu}^c) + 
i\sum_{f=1}^{n_f}\psib_f \Dslash\, \psi_f 
\ee
where the field strength tensor is defined through
\be
F^a_{\mu\nu} = \pa_{\mu}W_{\nu}^a - \pa_{\nu}W_{\mu}^a 
+ gf^{abc}W_{\mu}^b W_{\nu}^c
\label{eq::fmunu}
\ee
and 
\be
D_{\mu} = \pa_{\mu} 
- ig (R^a )W_{\mu}^a
\label{eq::dmu}
\ee
is the covariant derivative. $W_{\mu}$ is the gauge field, $c^a$ is the 
Fadeev-Popov-ghost field,
$f^{abc}$ are the structure constants of 
the gauge group ${\cal G}$, $\xi$ is the gauge parameter and $g$ is the
gauge coupling.

For the case when the theory admits a gauge invariant fermion mass 
term we will have $L_B \to L_B + L_B^m$, where 
\be
L_B^m = -m_f \psib_f\psi_f \,.
\ee

DRED amounts to  imposing that all field variables
depend only  on a subset of the total number of space-time dimensions;
in this case $d$ out of $4$ where  $d = 4 - 2\ep$. We can then make the
decomposition 
\be
W_{\mu}^a(x^j ) =  \hat{W}_\mu^a (x^j )+ \bar{W}_\mu^a(x^j )
\label{eq::wdec}
\ee
where
\be
 \hat{W}_\mu^a= \hat{g}_{\mu\nu} W^{\nu,a}\qq \bar{W}_\mu^a=
 \bar{g}_{\mu\nu} W^{\nu,a}\qq \hbox{and}\qq  \hat{g}_{\mu\mu} = d \,. 
\ee
It is then easy to show that~\cite{Kant:2009zza}
\be
L_B = L _B^d + L_B^{\ep} 
\ee
where
\be
L _B^d = -\frac{1}{4} \hat{F}^2_{\mu\nu}
-\frac{1}{2(1-\xi)}(\pa^{\mu}\hat{W}_{\mu})^2 + 
\pa^{\mu}\bar{c}^{a}(\pa_{\mu} c^a -g f^{abc} c^b \hat{W}_{\mu}^c) +
\sum_{f=1}^{n_f}i\psib_f 
\hat{\gamma}^\mu \hat{D}_\mu \psi_f
\label{eq::AD}
\ee
and 
\be
 L_B^{\ep} = \frac{1}{2}(\hat{D}_\mu\bar{W}_{\nu})^2 
- \sum_{f=1}^{nf} g \psib_f \bar{\gamma}^\mu R^a \psi_f \bar{W}_{\mu}^a
-\frac{1}{4}g^2 f^{abc}f^{ade}\bar{W}^b_{\mu}
\bar{W}^c_{\nu}\bar{W}^{d,\mu}\bar{W}^{e,\nu}.
\label{eq::AE}
\ee
$\hat{F}_{\mu\nu}$ and $\hat{D}_\mu$ denote the projection of the field
strength and covariant derivative given in Eqs.~(\ref{eq::fmunu}) and
~(\ref{eq::dmu}) onto $Q_d$, obtained  with the help of the operator
$\hat{g}^{\mu\nu}$. 
Conventional dimensional regularization (\dreg) amounts to using 
Eq.~(\ref{eq::AD}) and discarding Eq.~(\ref{eq::AE}). 

Note that under the gauge transformations: 
\besub
\bea 
\de \hat{W}^a_\mu &=& \pa_\mu\Lambda^a +
gf^{abc}\hat{W}^b_\mu\Lambda^c\label{eq:AFa}\,,\\ 
 \de \bar{W}^a_{\mu}&=& gf^{abc}\bar{W}^b_{\mu}\Lambda^c\label{eq:AFb}\,,\\ 
\de\psi^{\alpha} &=& ig(R^a)^{\alpha\beta}\psi^{\beta}\Lambda^a 
\label{eq:AFc} 
\eea
\eesub 
 each term in
Eq.~(\ref{eq::AE}) is separately invariant.
The $\bar{W}_{\mu}$ fields behave exactly like scalar  fields, and are
hence 
known as \eps{} scalars.  There is
therefore no  reason to expect the $\psib\psi \bar{W}$ vertex to
renormalize in the same  way as the $\psib\psi \hat{W}$ vertex (except in
the case of supersymmetric theories). 
%In the  case of the quartic
%\eps-scalar interaction it is evident that  
%more than one such coupling is permitted by Eq.~(\ref{eq:AFb}).
%In  other words, we cannot in general expect the $f-f$~tensor structure 
%present in Eq.~(\ref{eq::AE}) to be preserved under renormalization.
 The couplings associated with the  $\psib\psi
\bar{W}$  vertex or with the  quartic
\eps-scalar interaction are called {\it evanescent couplings}. They were
first described in Ref.~\cite{Tim}, and later independently discovered
by  van Damme
and 't~Hooft~\cite{hvand}. The vertices $\hat{W} \bar{W} \bar{W}$ and 
$\hat{W}\hat{W} \bar{W} \bar{W}$, on the other hand, are renormalized in
the same way as $\hat{W}\hat{W}\hat{W}\,,\bar{C}C\hat{W}$, etc. because of
the gauge invariance~\cite{Jack:1993ws}.  Thus we can conclude that
$\hat{W}$ is the  gauge particle, while $\bar{W}$ acts as
matter field transforming according to the adjoint representation.
In order to avoid confusion, we denote in the following the gauge
particles with $G^a_\mu$ and the \eps{} scalars with $\vep^a_\mu$
\be
\hat{W}^a_\mu\,\to\, G^a_\mu\qquad \bar{W}^a_\mu\,\to\, \vep^a_\mu\,.
\ee

 Since \eps{} scalars are present only on
internal lines  
we could, in fact, choose the
wave function renormalization of $\vep_{\mu}$ and $G_\mu$ to be the same.
However, such a renormalization prescription will break
unitarity~\cite{hvand}. The  
crucial point is the correct renormalization  of sub-divergences, which
requires that vertices involving \eps{} scalars renormalize in a different way
as their 
gauge counterparts. Thus, to renormalize the \eps{} scalars
one has to treat them as new fields present in the theory.\\
For the renormalization of the theory we distinguish two new types of
couplings: a  
Yukawa like coupling $g_e$ associated with the vertex $\psib\psi\vep$ and
a set of $p$  
quartic couplings $\la_r$ associated with vertices containing four
\eps{} scalars. 
The number $p$ is given  by the number of independent rank four tensors
$H^{abcd}$ which  are non-vanishing when symmetrized with respect to
$(ab)$ and $(cd)$ interchange. We address the issue of the quartic
vertex renormalization in more detail in the next section.\\
The renormalization constants for the couplings, masses as well as
fields and vertices are defined as
 \begin{align}
  g^0 &= \mu^{\epsilon}Z_g g\,,\qquad &
  g_e^0   &= \mu^{\epsilon}Z_e g_e\,,\qquad &
\sqrt{\lambda_r^0} &= \mu^\epsilon Z_{\lambda_r}
\sqrt{\lambda_r}\,,\qquad & 
  \nonumber\\
  1-\xi^0 &= \left(1-\xi\right) Z_3\,,\qquad &
   m_f^0   &= m_f Z_{m_f}\,,\qquad &
   m_\vep^0   &= m_\vep Z^\vep_m\,,\qquad &
\nonumber\\
  \psi^{0} &= \sqrt{Z_2}\,\psi\,,\qquad &
  G_\mu^{0,a}    &= \sqrt{Z_3}\,G_\mu^a\,,\qquad &
  \varepsilon^{0,a}_\mu &=
  \sqrt{Z_3^\varepsilon}\,\varepsilon^a_\mu\,, \qquad &
\nonumber\\
  c^{0,a}        &= \sqrt{\tilde{Z}_3}\,c^a\,,\qquad &
  \Gamma_{\psib\psi G}^{0}        &= Z_1 \Gamma_{\psib\psi G}\,, \qquad &
  \Gamma_{\psib\psi\vep}^{0} &= Z_1^\vep
  \Gamma_{\psib\psi\vep}\,,
  \qquad &
\nonumber\\
 \Gamma_{\bar{c}cG}^{0}              &= \tilde{Z}_1
  \Gamma_{\bar{c}cG}\,,\qquad &
 \Gamma_{\vep\vep G}^{0} &= Z_1^{\vep\vep G}
  \Gamma_{\vep\vep G}\,,
  \qquad &
\Gamma_{\vep\vep G G}^{0} &= Z_1^{\vep\vep G G}
  \Gamma_{\vep\vep G G}\,,
  \qquad &
\nonumber\\
\Gamma_{G G G G}^{0} &= Z_1^{4 G}
  \Gamma_{GG G G}
\,,
  \qquad &
\Gamma_{\vep\vep\vep\vep }^{0} &= Z_1^{4\vep}
  \Gamma_{\vep\vep\vep\vep }
\,,
  \qquad & & 
\qquad & 
  \label{eq::renconst}
\end{align}
where $\mu$ is the renormalization scale and the bare quantities are
marked by the superscript ``0''. $\Gamma{xyz(w)}$ stands for one-particle
irreducible   Green functions involving the external particles $x, y, z,
(w)$. Eq.~(\ref{eq::AD}) takes under renormalization the usual
expression  in terms of renormalized parameters as in  \dreg{} scheme. 
The renormalized Lagrangians $L^{\ep}$ is the new term that
distinguishes \dred{} from \dreg{} and it is 
given by
\begin{eqnarray}
L^{\ep} & =&\frac{1}{2}Z^{\varepsilon\varepsilon}(\pa_\mu \vep_{\nu}^a)^2
+ Z^{\varepsilon\varepsilon G} gf^{abc}\pa_\mu \vep_{\nu}^a G^{b,\mu}
\vep^{c,\nu} 
+ Z^{\varepsilon\varepsilon GG}g^2f^{abc}f^{ade}G_\mu^b \vep_{\nu}^c 
G^{d,\mu} \vep^{e,\nu}\nonumber\\
&-&Z^{\psi\psi\varepsilon}g_e\psib R^a \bar{\gamma}^\mu\psi \vep^a_{\mu}
           - \frac{1}{4} \sum_{r=1}^p Z_{\lambda_r}\la_r
            H^{abcd}_r \vep^a_{\mu}\vep^c_{\nu}\vep^{b,\mu}\vep^{d,\nu}\,.
\label{eq::lren}
\end{eqnarray}
Strictly speaking, Eq.~(\ref{eq::lren}) should also have a mass term
for the \eps{} scalars; but since this mass term does not affect
renormalization of the couplings and fermion masses we omit it here. We
discuss this issue in more detail in section~\ref{sec::ep4mass}.

 The charge renormalization constants are obtained from the
Slavnov-Taylor identities.  For example, if one
computes the $N$-point Green function with external fields
$\phi_1,\cdots,\phi_n$ and denote its coupling constant by $g$, one
obtains
\begin{eqnarray}
Z_g = \frac{Z_{\phi_1\cdots\phi_N}}{\sqrt{Z_{\phi_1}\cdots Z_{\phi_N}}}\,,
\label{eq::ZZZ}
\end{eqnarray}
where the $Z_{\phi_i}$ are the wave function renormalization constants
for the $\phi_i$, $Z_{\phi_1\cdots\phi_N}$ is the corresponding vertex
renormalization constant, and $Z_g$ the charge renormalization. Within
the minimal subtraction scheme, one is free to choose any masses and  
external momenta, as long as infra-red divergences are avoided.  One can set
all masses to zero, as well as one of the  two independent
external momenta in  three-point functions. In this case, one  arrives at
three-loop integrals with one non-vanishing external momentum $q$ which
can be calculated with the help of {\tt MINCER}. One can also calculate
the  three-point functions  setting a common mass $m$ to all
  particles and expanding the Feynman integrals in the limit
$m^2/q^2\ll 1$ with the help of asymptotic
expansions~\cite{Chetyrkin:1997fm}. This approach is much more tedious, but 
possible infra-red singularities would  manifest in $\ln m^2/q^2$
terms. If such terms are absent
 in the final expression, the limit
  $m\to 0$ can be taken and the result should coincide with the one obtain
  with the massless set-up\footnote{ For a comprehensive overview about
    the multi-loop techniques within \dreg{} see the
review article~\cite{Steinhauser:2002rq}.}.

Precisely, the charge renormalization of
the gauge coupling can be derived from the ghost-gauge boson,
fermion-gauge boson, \eps-scalar-gauge boson vertices or from the  gauge
boson self interaction
  \begin{eqnarray}
  Z_g &=& \frac{\tilde{Z}_1}{\tilde Z_3\sqrt{Z_3}}
  \,\, = \,\, \frac{Z_1}{Z_2\sqrt{Z_3}}\,\, = \,\, 
\frac{Z_1^{\vep\vep G}}{Z_3^\vep\sqrt{Z_3}} \,\, = \,\,\mbox{etc.}
  \label{eq::Zg}
\end{eqnarray}
as a consequence of  gauge invariance.\\
Similarly, for the charge renormalization constants of the evanescent
couplings it holds the following relations
 \begin{eqnarray}
  Z_e &=& \frac{Z_1^\varepsilon}{Z_2\sqrt{Z_3^\varepsilon}}\qquad
 Z_{\lambda_r} = \frac{Z_1^{4\vep}}{(Z_3^\varepsilon)^2}
  \,.
  \label{eq::Zh}
\end{eqnarray}
In general,  $Z_g\not= Z_e$ even at one-loop order. However, in
supersymmetric theories $Z_g= Z_e$ should hold at all orders because of
SUSY. This can be understood following the same line of reasoning as for
the derivation of the equality of the
charge renormalization constants for the interactions involving gluons
and those involving gluinos. 
%Furthermore, both $Z_g$ and $Z_e$ depend on $g$,
%$g_e$, and $\lambda_r$~\cite{Jack:1993ws};
%note, however, that $Z_g$ depends on $g_e$ and $\lambda_r$
%only starting from three- and four-loop order, respectively,
%while $Z_e$ depends on $g_e$ and  $\lambda_r$ already at
%one- and two-loop order, respectively. 
%%We detail on this point in Sections~\ref{sec::qcd3}.

%%%%%%%%%%%%%%%%%%%%%%%%%%%%%%%%%%%%%%%%%%%%%%%%%%%%%%%%%%%%

\subsubsection{\label{sec::ep4}The \epscalar{} self couplings}

Let us discuss the structure  of the quartic \epscalar{} couplings
for an  arbitrary gauge group. These interactions are  invariant under the 
symmetry ${\cal G}\otimes O(2\epsilon)$, where only   ${\cal G}$ is 
gauged. The number of independent quartic \epscalar{} couplings  
  is given by the number of independent rank $n=4$ tensors
$H^{abcd}$ invariant with respect to $(a,b)$ and $(c,d)$ exchange,
because of the $O(2\epsilon)$ invariance. It has been shown that for
${\cal G}=SU(N) , SO(N), SP(N)$ with $N\ge 4$ there are four such
tensors~\cite{Jack:2007ni}.  For the case $N=3$ only three independent tensors
can be built~\cite{predrag}, while for $N=2$ their number reduces to
two~\cite{Jack:1993ws}. The answer to the general
question concerning rank $n$ tensors is not yet known.
For the explicit construction of the set of tensors $H^{abcd}$ we
consider first   the   $SU(N)$  group and then generalize the results for
the other two groups. 

A natural choice for a basis for  rank $n=4$ tensors when $N \geq 4$ 
is  
given by~\cite{Dittner:1972hm}\footnote{An alternative way to define a
  basis which has the virtue of  
being immediately generalizable to any group~\cite{predrag} is in terms
of traces of products  
of the generators in the defining representation, thus 
$\Tr\left(T^a T^b T^c T^d\right)$, 
$\Tr\left(T^a T^b\right)\Tr\left(T^c T^d\right)$
 etc.} 
\bea
K_1 &=& \delta^{ab}\delta^{cd}
\quad K_4 = d^{abe}d^{cde}\quad K_7 = d^{abe}f^{cde}\nn  
K_2 &=& \delta^{ac}\delta^{bd}
\quad K_5 = d^{ace}d^{bde}\quad K_8 = d^{ace}f^{bde}\nn  
K_3 &=& \delta^{ad}\delta^{bc}
\quad K_6 = d^{ade}d^{bde}\quad K_9 = d^{ade}f^{bce}.
\eea
Here $d^{abc}$ stands for the completely symmetric rank $n=3$  tensors.
The dimension of the basis reduces to $8$ in the case of $SU(3)$. This is 
achieved via the relation~\cite{msw,Dittner:1972hm} 
\be
K_4 + K_5 + K_6 = \frac{1}{3}(K_1 + K_2 + K_3)
\label{eq:su3compl}
\ee
which is not valid for $N \geq 4$.\\ 
To describe the \epscalar{} quartic interactions one needs 
to construct rank $n=4$ tensors invariant w.r.t. exchange of pairs of
indices. Thus, one has to take linear combinations of the basis tensors
and symmetrized them with respect to the pair
of indices $(ab)$ and $(cd)$.
A possible choice for $N \geq 4$ is given by
\bea
H_1 &=& \frak{1}{2}K_1\,,\nn
H_2 &=& \frak{1}{2}(K_2 + K_3)\,,\nn
H_3 &=& \frak{1}{2}K_4\,,\nn
H_4 &=& \frak{1}{2}(K_5 + K_6).
\label{eq::basis1}
\eea
Note that the absence of a $d-f$ type term from Eqs.~(\ref{eq::basis1}) follows from the
identity~\cite{Dittner:1972hm}
\be
K_8 + K_9 = -f^{abe} d^{cde}.
\ee
However, for practical purposes a basis constructed with the help of the
structure constants $f^{abc}$ and avoiding the use of the $d$-tensors is
more suited. For example, it would allow to 
explore more easily the supersymmetric case and to generalize to other
groups. It is natural to consider the alternative
choice~\cite{Tim,Harlander:2006rj} 
\bea
\Hbar_1 &=& \frak{1}{2}(f^{ace}f^{bde} + f^{ade} f^{bce})\,,\nn
\Hbar_2 &=&\delta^{ac}\delta^{bd} +  \delta^{ad}\delta^{bc} 
+  \delta^{ab}\delta^{cd}\,,\nn
\Hbar_3 &=& \frak{1}{2}( \delta^{ac}\delta^{bd} +  \delta^{ad}\delta^{bc})
-  \delta^{ab}\delta^{cd}\,,\nn
\Hbar_4 &=& \frak{1}{2}(f^{aef}f^{bfg}f^{cgh}f^{dhe}
+f^{aef}f^{bfg}f^{dgh}f^{che}).
\label{eq:hbarbasis}
\eea

Let us introduce the coupling constants
\begin{eqnarray}
  \alpha_s = \frac{g^2}{4\pi}\,,
  \quad
  \alpha_e = \frac{g_e^2 }{4\pi}
  \quad\mbox{and}\quad 
  u_r = \frac{\lambda_r}{4\pi}\,.
\end{eqnarray}
Then we can write the last term in Eq.~(\ref{eq::lren})
\be
\sum_{r=1}^4 Z_{\lambda_r}\la_r H^{abcd}_r = 4\pi  \sum_{r=1}^4 Z_{u_r}  u_r H^{abcd}_r=
4\pi \sum_{r=1}^4 Z_{\eta_r} \eta_r 
\Hbar^{abcd}_r\,,
\label{ref::lbasis}
\ee
%Feynman rules necessary for the computations with \eps{} scalars in the
%component field approach can be  derived from Eq.~(\ref{eq::lbasis})
%using the conventional methods. They are reproduced for completeness in
%Appendix~\ref{sec::eprules}.\\
where $\eta_r$ denote the quartic \epscalar{} couplings in the basis $\Hbar^{abcd}$.
The renormalization constants $Z_\eta,\,Z_u,\,\mbox{etc.}$ have been computed
through one loop in the \drbar{} scheme for a general gauge group  in
Ref.~\cite{Tim,Jack:2007ni} and in Ref.~\cite{Harlander:2006rj} for
$SU(3)$. The calculation performed in Ref.~\cite{Harlander:2006rj} has
employed the method of Ref.~\cite{Chetyrkin:1997fm} to
introduce an artificial mass for all propagators in order to avoid
spurious infrared divergences. For the calculation of
the results in terms of group invariants the package {\tt
  color}~\cite{RSV} has been used. For completeness, we reproduce here
the one-loop results for the couplings $\eta_r$. 
{\allowdisplaybreaks
\begin{align}
%\begin{eqnarray}
Z_{\eta_1}&=1 +\frac{1}{\ep}\bigg[- \apiDR\cA\frac{3}{2}
    + \frac{\eta_1}{\pi}\cA\frac{1}{2}+\frac{\eta_2}{\pi}\cA 2
-\frac{\eta_3}{\pi}\frac{7}{2}
-\frac{\eta_2}{\pi}\frac{\eta_4}{\eta_1}\cA
-\frac{\eta_3}{\pi}\frac{\eta_4}{\eta_1}\cA\frac{1}{2}
\nonumber\\
&+\frac{\eta_4}{\pi}\frac{\cA^4(-61+7N_A)+48D_4(AA)(N_A-1)/N_A}{36\cA^2(N_A-3)}
+ \aepi T_f 
\nonumber\\
&- \aepi\frac{\alpha_e}{\eta_1}  
  \frac{4\cA(2 + N_A) D_4(RA)/I_2(R) + 5\cA^3(7\cA - 2\cR)N_A  -16(2 +
    N_A) D_4(AA)  }{2(25\cA^4 N_A - 12 D_4(AA)(2 + N_A))}T_f \nonumber\\
&-    \frac{\eta_4}{\pi} \frac{\eta_4}{\eta_1}
     \frac{1 }{54\cA N_A(N_A-3)(25\cA^4N_A  - 12 D_4(AA)(2 + N_A))}
\nonumber\\
&
\bigg(-144 D_4(AA)^2(2 + N_A)(1 +2 N_A) + 216\cA^2 D_4(AAA)N_A (2 +
N_A)( N_A-3)\nonumber\\
&
 -12        \cA^4 D_4(AA)N_A(-191-56 N_A+N_A^2)-25\cA^8N_A^2(4N_A+23)\bigg)
\bigg]\,,
\displaybreak[2]\nonumber\\
%\end{eqnarray}
Z_{\eta_2}&=1 +\frac{1}{\ep}\bigg[ -\apiDR\cA\frac{3}{2} 
- \frac{\eta_1}{\pi}\cA\frac{1}{6} 
+ \frac{\eta_3}{\pi}\frac{  N_A-1}{6}
+\frac{\eta_4}{\pi}\cA^2\frac{13}{12}  
+ \frac{\eta_2}{\pi}\frac{2(8 + N_A)}{3} \nonumber\\
&+ \frac{\eta_3}{\pi}\left(\frac{\eta_1}{\eta_2}  \cA\frac{1}{6}
+ \frac{\eta_4}{\eta_2}\cA^2\frac{1}{6} 
- \frac{\eta_3}{\eta_2}\frac{( N_A-1)}{12}\right) \nonumber\\
&-\frac{\eta_4}{\pi}\frac{\eta_4}{\eta_2}\frac{2}{9}\frac{72
  D_4(AA)^2/N_A - 90\cA^2 D_4(AAA)  
+25\cA^4 D_4(AA) }{25\cA^4N_A  - 12 D_4(AA)(2 + N_A)} \nonumber\\
&+      \aepi\left( T_f -
   \frac{\alpha_e}{\eta_2}2\frac{5\cA^2 D_4(RA)/ I_2(R) + (\cA - 6\cR)
     D_4(AA) }{25\cA^4 N_A - 12 
D_4(AA)(2 + N_A)}T_f \right)
\bigg]\,,
\displaybreak[2]\nonumber\\
Z_{\eta_3}&=1 +\frac{1}{\ep}\bigg
[ (-\apiDR\cA \frac{3}{2} 
+ \frac{\eta_4}{\pi}\cA^2\frac{5}{12}
+\frac{\eta_2}{\pi}\frac{2(2 + N_A)}{3}
 +\frac{\eta_3}{\pi}\frac{-26 + 5 N_A}{12} \nonumber\\
&+ \frac{\eta_4}{\pi}\frac{\eta_4}{\eta_3}\frac{7}{108}\frac{12\daa -
  5\cA^4 N_A}{( N_A-3)N_A}
- \frac{\eta_2}{\pi}\left(\frac{\eta_4}{\eta_3}\cA^2\frac{2}{3}
  +\frac{\eta_2}{\eta_3}\frac{(2 + N_A)}{3}\right)
 \nonumber\\
&+ \frac{\eta_1}{\pi}\left(-\cA\frac{5}{6} -
  \frac{\eta_2}{\eta_3}\frac{2\cA}{3}  
+\frac{\eta_4}{\eta_3}\frac{12\daa - 5\cA^4N_A}{ 9 \cA N_A( N_A - 3)}\right) +
   \aepi T_f
\bigg]\,,
\displaybreak[2]\nonumber\\
Z_{\eta_4}&=1 + \frac{1}{\ep}\bigg[-\apiDR\cA\frac{3}{2} -
\frac{\eta_1}{\pi}\cA\frac{1}{4} 
+ \frac{\eta_2}{\pi} 8   - \frac{\eta_3}{\pi}\frac{1}{2} +
\apiDR\frac{\asDRbar}{\eta_4}\frac{3}{4} - 
   \frac{\eta_1}{\pi}\frac{\eta_1}{\eta_4}\frac{1}{4} \nonumber\\
&+ \frac{\eta_4}{\pi}\frac{-1152\daaa(2 + N_A) +
      5\cA^2(125\cA^4N_A + 4\daa(98 + N_A))}{
    48(25\cA^4 N_A - 12\daa(2 + N_A))}\nonumber\\
& + \aepi \left( T_f +
   \frac{\alpha_e}{\eta_4}\frac{5\cA^2(\cA - 6\cR)N_A + 12 (2 +
     N_A)\daf/I_2(R) } 
    {25\cA^4N_A - 12\daa(2 + N_A)}T_f\right)
\bigg]\,,
\label{eq::renquart}
\end{align}
}
$\!\!$with the  group invariants defined in Appendix~A and the abbreviation
$T_f=I_2(R) n_f$, where $n_f$ denotes the number of active
fermions. Let us notice at this point the presence 
of negative power of couplings in the expressions of the renormalization
constants. This results into beta functions that are not proportional
to the coupling itself.
 This feature is  specific to scalar couplings and it implies
  that, even if we set such a coupling to zero at a given scale, it
 will receive non-vanishing radiative corrections due to the other
 couplings present in the theory.
%Although the results have been derived for an $SU(N)$ group, they
%hold also for the case of $SO(N)$ and $SP(N)$ groups with $N\ge 4$ . One
%has only to specify the group invariants according to Appendix~A.

The above results have been computed using an $SU(N)$ gauge
group. However, they are parametrized in terms of group invariants. Thus
they are also valid for other physically interesting groups like $SO(N)$
and   $SP(N)$. The explicit values of the group invariants for the three
groups can be found in Appendix~A. \\
In the case of $SU(3)$ group, the invariant
$\overline{H}_4$ becomes a 
linear combination of $\overline{H}_{i}\,, i=1,2,3$ because of relation~(\ref{eq:su3compl}).The same is also
true for the coupling $\eta_4$ that can be expressed in terms of the
other three couplings. Thus in this case one can ignore $\eta_4$. 

 Actually, the one- and two-loop
renormalization constants for totally symmetric quartic scalar couplings
with scalars in an 
arbitrary representation  have been known for long time~\cite{Cheng:1973nv}.   
However, these results cannot be directly applied to \eps-scalar self
interactions, due to their particular symmetry with respect to exchange
between pairs of indices.

\subsubsection{\label{sec::3loop}Three-loop renormalization constants
  for a non-supersymmetric theory}
In this section we report on the explicit computation of the charge
$Z_g,\, Z_e$ 
and mass $Z_m^q,\, Z_m^\ep$ renormalization constants to 
three-loop order within \drbar{} scheme. This  requires the calculation of
divergent parts of logarithmically divergent integrals. One can exploit
the fact that such contributions are independent of the masses and
external momenta. 
Precisely, one  sets all internal masses to zero and keeps only one external
momentum different from zero and then solve recursively
Eq.~(\ref{eq::rec}). In practice, use of the automated programs {\tt
  QGRAF}~\cite{Nogueira:1991ex}, {\tt q2e} and {\tt
  exp}~\cite{Seidensticker:1999bb,Harlander:1997zb} and {\tt MINCER} are
essential due to the  large number of diagrams that occur.

%The quantities  $Z_g^{\overline{\rm DR}}$ and $Z_e$ have been computed to
% four-loop order~\cite{Jack:2007ni}. The results
%have been presented in terms of the corresponding $\beta$ functions.
The analytical form of
$Z_g^{\overline{\rm DR}}$ up to two-loop order is identical to the
corresponding result in 
the $\overline{\rm MS}$ scheme. This has been shown by an explicit
calculation for the first time in Ref.~\cite{Capper:1979ns} and is a
consequence of the minimal renormalization.  The three- and four-loop
results for a general 
theory have been derived in Refs.~\cite{Harlander:2006rj,Harlander:2006xq,Jack:2007ni}.
For completeness we present in the following the three-loop results 
{\allowdisplaybreaks
\begin{align}
%\begin{eqnarray}
  Z_g^{\overline{\rm DR}} &= 1 +
  \apiDR\frac{1}{\ep}\left(-\frac{11}{24} C_A + \frac{1}{6}
   T_f\right)
  +\left(\apiDR \right)^2\bigg[
    \frac{1}{\ep^2}\left( \frac{121}{384} C_A^2 - \frac{11}{48} C_A
     T_f
    \right.\nonumber\\&\left.\mbox{}
    + \frac{1}{24} T_f^2 \right)
    + \frac{1}{\ep}\left(-\frac{17}{96} C_A^2 +
    \frac{5}{48} C_A  T_f + \frac{1}{16} C_R  T_f\right)
    \bigg]
\nonumber\\&\mbox{}
+\left(\apiDR \right)^3\bigg[\frac{1}{\ep^3}\left(
\frac{-6655}{27648}\cA^3 + \frac{605}{2304}\cA^2  T_f -
     \frac{55}{576}\cA T_f^2 + \frac{5}{432} T_f^3\right)
\nonumber\\&\mbox{}
+\frac{1}{\ep^2}\left(\frac{2057}{6912}\cA^3 - \frac{979}{3456}\cA^2
T_f + \frac{11}{288}\cR T_f^2 -\frac{121}{1152}\cA\cR T_f
      + \frac{55}{864}\cA T_f^2\right)
\nonumber\\&\mbox{}
+\frac{1}{\ep}\left(
-\frac{3115}{20736}\cA^3 + \frac{1439}{10368}\cA^2 T_f +
\frac{193}{3456}\cA\cR T_f - \frac{79}{5184}\cA T_f^2
\right.\nonumber\\&\left.\mbox{}
-\frac{1}{192}\cR^2 T_f - \frac{11}{864}\cR T_f^2  \right)
  \bigg]
+ \left(\apiDR
\right)^2\aepi\frac{1}{\ep}\left(\frac{1}{32}\cR^2 T_f\right) 
\nonumber\\&\mbox{}
+
\apiDR\left(\aepi \right)^2\frac{1}{\ep}\left(
\frac{1}{96}\cA\cR T_f - \frac{1}{48}\cR^2 T_f -
\frac{1}{96}\cR T_f^2\right) \,.
\label{eq::zas3loop}
\end{align}
}
$\!\!$The one-loop result for $Z_e$
can be found in Ref.~\cite{Jack:1993ws}. For the particular case of QCD,
{\it i.e.} ${\cal G} = SU(3)$ and $\eta_4=0$,
the two-, three- and four-loop  results have been computed in
Refs.~\cite{Harlander:2006rj,Harlander:2006xq}. The two-, three- and four-loop
results for a general 
theory have been derived in Ref.~\cite{Jack:2007ni}. Because of the
complexity of the results,
 we reproduce below only the
two-loop contributions that are however enough for most of the practical applications
{\allowdisplaybreaks
\begin{align}
% \displaybreak[2] \\
  Z_e &= 1 + \apiDR \frac{1}{\ep} \left(-\frac{3}{4} C_R\right)
  + \aepi
  \frac{1}{\ep} \left(-\frac{1}{4} C_A + \frac{1}{2} C_R  +
  \frac{1}{4}  T_f \right)
  \nonumber\\&\mbox{}
  + \left(\apiDR \right)^2\bigg[
    \frac{1}{\ep^2}\left(\frac{11}{32} C_A C_R
    + \frac{9}{32} C_R^2  -
    \frac{1}{8}  C_R  T_f  \right)
    +\frac{1}{\ep}\left( \frac{7}{256} C_A^2 - \frac{55}{192} C_A C_R
    \right.\nonumber\\&\left.\mbox{}
    - \frac{3}{64} C_R^2-\frac{1}{32} C_A  T_f  +
    \frac{5}{48} C_R  T_f  \right)
    \bigg]
  + \apiDR \aepi \bigg[
    \frac{1}{\ep^2} \left(\frac{3}{8} C_A C_R  - \frac{3}{4} C_R^2
    \right.\nonumber\\&\left.\mbox{}
    - \frac{3}{8}  C_R  T_f  \right)
    +\frac{1}{\ep} \left( \frac{3}{32} C_A^2 - \frac{5}{8} C_A C_R
    + \frac{11}{16} C_R^2  + \frac{5}{32} C_R  T_f  \right)
    \bigg]
  \nonumber\\&\mbox{}
  + \left(\aepi\right)^2\bigg[
    \frac{1}{\ep^2} \left(\frac{3}{32} C_A^2  - \frac{3}{8} C_A C_R
    + \frac{3}{8} C_R^2  - \frac{3}{16} C_A T_f  +
    \frac{3}{8} C_R  T_f +\frac{3}{32}  T_f^2
    \right)
    \nonumber\\&\mbox{}
    +\frac{1}{\ep} \left( -\frac{3}{32} C_A^2 + \frac{5}{16} C_A C_R
    -\frac{1}{4} C_R^2  + \frac{3}{32} C_A  T_f -\frac{3}{16} C_R
     T_f \right)
    \bigg]
  \nonumber\\&\mbox{}
  +  \aepi \frac{1}{\ep} \bigg[\frac{\eta_1}{\pi} \left(\frac{1}{32}\cA^2\right)
    +\frac{\eta_2}{\pi}\left(\frac{1}{16}\cA - \frac{3}{8}\cR\right)
   +\frac{\eta_3}{\pi} \left(-\frac{1}{16}\cA\right) 
  \nonumber\\&\mbox{}
+\frac{\eta_4}{\pi}
   \left(\frac{1}{192}\cA^3 - \frac{1}{8}D_4(RA)\IR\right)\bigg]
  +\left(\frac{\eta_1}{\pi} \right)^2 \frac{1}{\ep}
  \left(-\frac{3}{256}\cA^2\right) 
  + \left(\frac{\eta_2}{\pi} \right)^2\frac{1}{\ep}
  \left(\frac{3}{32}(N_A+2)\right)  
 \nonumber\\&\mbox{}
  +\frac{\eta_1}{\pi} \frac{\eta_3}{\pi}
  \frac{1}{\ep} \left( \frac{3}{64}\cA\right)
  + \left(\frac{\eta_3}{\pi} \right)^2 \frac{1}{\ep}
  \left(-\frac{3}{128}(N_A-1)\right)
   +\frac{\eta_1}{\pi} \frac{\eta_4}{\pi}
  \frac{1}{\ep} \left( -\frac{1}{256}\cA^3\right)
\nonumber\\&\mbox{}
 +\frac{\eta_2}{\pi} \frac{\eta_4}{\pi}
  \frac{1}{\ep} \left( \frac{5}{32}\cA^2\right)
+\frac{\eta_3}{\pi} \frac{\eta_4}{\pi}
  \frac{1}{\ep} \left( \frac{1}{128}\cA^2\right)
  + \left(\frac{\eta_4}{\pi} \right)^2 \frac{1}{\ep}
  \left(-\frac{1}{3072}\cA^4 + \frac{1}{32}D_4(AA)\right) 
  \,.
\end{align}
}
$\!\!$The group invariants $C_A, C_R, I_2(R), D_4(XY)$ occurring in the above
equations are defined in Appendix~A and we used the abbreviation $T_f=
I_2(R) n_f$.

 There is also an
indirect way to  derive the three-loop gauge beta function in the
\drbar{} scheme
starting from the knowledge of the three-loop gauge beta function in the
\msbar{} scheme and the fact that the gauge couplings defined in the  two
schemes can be perturbatively related to each other. This method will be
discussed in more detail in the next section. Let us mention however
that, using the expression for the   three-loop gauge beta function in
the 
\msbar{} scheme $\beta_s^{\overline{\rm MS}}$ and the two-loop conversion
relation of $\alpha_s$ given 
in Eq.~(\ref{eq::asMS2DR_2}) one obtains exactly the same results for 
$\beta_s^{\overline{\rm DR}}$ as given in  Eq.~(\ref{eq::zas3loop}). This
is a powerful consistency check for the calculation reviewed in this
section.   It is interesting to mention that the equality of the two
results can be obtain only if one keeps $\asDRbar\ne\alpha_e$ during the
calculation and renormalize them differently. The
identification of $\asDRbar$ and $\alpha_e$ leads to 
inconsistent results. In case of $\betaDRbar_s$  the
error is a finite, gauge parameter independent
term~\cite{Bern:2002zk}. For quark mass renormalization, this
identification 
(precisely the identification of the renormalization constants for the
two couplings)
generates much more severe problems. Namely, the renormalization
constant for the quark mass $Z_m^{\overline{\rm DR}}$ will contain non-local
terms at three-loop order and the mass anomalous dimension will
erroneously become divergent at this loop order.

 The renormalization constant for the fermion masses 
  $Z_m^{\overline{\rm DR}}$ has been computed in
Ref.~\cite{Harlander:2006rj} to three- and in
Ref.~\cite{Harlander:2006xq,Jack:2007ni} even to four-loop
order. Whereas in~\cite{Harlander:2006rj,Harlander:2006xq,Jack:2007ni} only the
anomalous dimensions were
given we want to present the explicit three-loop result for the renormalization
constant, that reads
{\allowdisplaybreaks
\begin{align}
%\begin{eqnarray}
  Z_m^{\overline{\rm DR}} &= 1 + \apiDR  \frac{1}{\ep} \left(-\frac{3}{4}
  C_R \right)
  + \left(\apiDR\right)^2\bigg[
    \frac{1}{\ep^2} \left(\frac{11}{32} C_A C_R + \frac{9}{32} C_R^2
    -\frac{1}{8} C_R  T_f  \right)
    \nonumber\\&
    +\frac{1}{\ep}\left(-\frac{91}{192} C_A C_R - \frac{3}{64} C_R^2
    + \frac{5}{48} C_R  T_f \right)
    \bigg]
  + \apiDR \aepi
  \left(\frac{3}{16}\frac{1}{\ep} C_R^2\right)
  \nonumber\\ &
  + \left(\aepi\right)^2 \frac{1}{\ep}\left(
  \frac{1}{16} C_A C_R - \frac{1}{8} C_R^2  - \frac{1}{16} C_R  T_f
  \right)
  +\left(\apiDR\right)^3\bigg[
    \frac{1}{\ep^3} \bigg(
    -\frac{121}{576} C_A^2 C_R
    \nonumber\\&
    - \frac{33}{128} C_A C_R^2
    - \frac{9}{128} C_R^3
    + \frac{11}{72} C_A C_R  T_f
    + \frac{3}{32} C_R^2 T_f - \frac{1}{36} C_R  T_f^2
    \bigg)
    \nonumber\\&
    +\frac{1}{\ep^2} \bigg(
    \frac{1613}{3456}C_A^2 C_R + \frac{295}{768} C_A C_R^2
    + \frac{9}{256} C_R^3
    -\frac{59}{216} C_A C_R  T_f - \frac{29}{192} C_R^2 T_f
    \nonumber\\&
    + \frac{5}{216} C_R  T_f^2
    \bigg)
+\frac{1}{\ep} \bigg(
    -\frac{10255}{20736} C_A^2 C_R + \frac{133}{768} C_A C_R^2
    - \frac{43}{128} C_R^3
    +\left(\frac{281}{2592}
    \right.\nonumber\\&\left.
    + \frac{1}{4} \zeta_3\right) C_A C_R  T_f
    + \left(\frac{23}{96}- \frac{1}{4} \zeta_3 \right)C_R^2  T_f
    + \frac{35}{1296} C_R   T_f^2
    \bigg)
    \bigg]
  \nonumber\\&
  +\left(\apiDR\right)^2\aepi\bigg[
    \frac{1}{\ep^2} \left(
    -\frac{11}{192} C_A C_R^2 - \frac{15}{64} C_R^3 + \frac{1}{48} C_R^2
     T_f \right)
    +\frac{1}{\ep}\left(
    \frac{5}{256} C_A^2 C_R
    \right.\nonumber\\&\left.
    + \frac{7}{32} C_A C_R^2 + \frac{9}{64} C_R^3
    - \frac{3}{32} C_R^2  T_f \right)
    \bigg]
  +\apiDR \left(\aepi\right)^2 \bigg[
    \frac{1}{\ep^2} \left(
    -\frac{9}{64} C_A C_R^2
    + \frac{9}{32} C_R^3
    \right.\nonumber\\&\left.
    + \frac{9}{64} C_R^2
     T_f \right)
    +\frac{1}{\ep} \left(
    -\frac{1}{64} C_A^2 C_R  + \frac{7}{32} C_A C_R^2  - \frac{3}{8} C_R^3
    - \frac{1}{64} C_A C_R T_f
    \right.\nonumber\\&\left.
    - \frac{1}{8} C_R^2  T_f
    \right)
    \bigg]
+\left(\aepi\right)^3 \bigg[
    \frac{1}{\ep^2} \bigg(
    -\frac{1}{48} C_A^2 C_R
    + \frac{1}{12} C_A C_R^2
    -\frac{1}{12} C_R^3
    \nonumber\\&
    + \frac{1}{24} C_A C_R T_f
    -\frac{1}{12} C_R^2  T_f  - \frac{1}{48} C_R T_f^2
    \bigg)
    +\frac{1}{\ep} \bigg(
    \frac{1}{32} C_A^2 C_R - \frac{1}{8} C_A C_R^2
    + \frac{1}{8} C_R^3
    \nonumber\\&
    - \frac{1}{24} C_A C_R  T_f
    +\frac{5}{48} C_R^2  T_f + \frac{1}{96} C_R T_f^2
    \bigg)
    \bigg]
   +\left(\aepi\right)^2\frac{1}{\ep} \bigg[ \frac{\eta_1}{\pi}\left(-
   \frac{1}{96}\cA^2\cR\right) 
\nonumber\\&
  +\frac{\eta_2}{\pi}\left(-\frac{1}{48}\cA\cR +\frac{1}{8}\cR^2\right)  
  +\frac{\eta_3}{\pi}    \left(\frac{1}{48}\cA\cR\right)   
 +\frac{\eta_4}{\pi}\left(-\frac{1}{576}\cA^3\cR + \frac{1}{12}\cR
   D_4(RA) \right)    \bigg]
\nonumber\\&
  + \aepi \frac{1}{\ep} \bigg[\left(\frac{\eta_1}{\pi}\right)^2
    \left(\frac{1}{256}\cA^2\cR\right) 
  +  \left(\frac{\eta_2}{\pi}\right)^2\left( - \frac{1}{32}\cR (N_A+2)\right) 
  +  \left(\frac{\eta_3}{\pi}\right)^2\left(\frac{1}{128}\cR(N_A-1)  \right)
  \nonumber\\&
  +  \left(\frac{\eta_4}{\pi}\right)^2
\left(\frac{1}{9216}\cA^4\cR - \frac{1}{96}\cR D_4(AA) \right)
  + \frac{\eta_1}{\pi}\frac{\eta_3}{\pi}
\left(-\frac{1}{64}\cA\cR \right)
 \nonumber\\&
  + \frac{\eta_1}{\pi}\frac{\eta_4}{\pi} 
\left(\frac{1}{768}\cA^3\cR \right)
  + \frac{\eta_2}{\pi}\frac{\eta_4}{\pi}
\left(-\frac{5}{96}\cA^2\cR \right)
 + \frac{\eta_3}{\pi}\frac{\eta_4}{\pi}
\left(-\frac{1}{384}\cA^2\cR \right)\bigg]
  \,.
  \label{eq::ZmDR}
%\end{eqnarray}
\end{align}
}
$\!\!$where $\z3$ is Riemann's zeta function with $\zeta(3)= 1.20206\ldots$.

Again, the consistency of the above results can be proved using the
indirect method alluded above. To derive the three-loop quark mass
anomalous dimension in the \drbar{} scheme $\gamma_m^{\overline{\rm
    DR}}$, one needs the three-loop result for $\gamma_m^{\overline{\rm
    MS}}$ and the two-loop conversion relation fro the quark mass as
given in Eq.~(\ref{eq::mMS2DR_2}). Full agreement 
has been found between the two methods~\cite{Harlander:2006rj}, that
provides a further consistency check of the calculation.

%ze=zg discussion\\Results for $Z_g$ and $Z_e$ in a general theory...
%%%%%%%%%%%%%%%%%%%%%%%%%%%%%%%%%%%%%%%%%%%%%%%%%%%%%%%%%%%%
\subsubsection{\label{sec::ms2dr}The general four-loop order results in the \drbar{} scheme}
The direct way to compute the renormalization constants in minimal
subtraction schemes as \msbar{} or \drbar{} requires the calculation of
divergent parts of logarithmically divergent integrals. 
Up to three loops there are well established methods and
automated programs exist to perform such calculations (see, e.g.,
Refs.~\cite{Larin:1991fz,Steinhauser:2000ry}).
Also at four-loop order a similar approach  is applicable. Nevertheless, it is
technically much more involved~\cite{Chetyrkin:1997dh,vanRitbergen:1997va,
  Czakon:2004bu,Baikov:2008jh,Baikov:2010iw}. There is however an
indirect method discussed  
in Refs.~\cite{Bern:2002zk,Harlander:2006rj} to derive the
renormalization constants in the \drbar{} scheme starting from their
\msbar{} expressions. It relies on the perturbative relation that can be
established between the couplings and masses defined in the two
schemes and takes into account that the four-loop results in the \msbar{}
scheme are known~\cite{Chetyrkin:1997dh,vanRitbergen:1997va,  Czakon:2004bu}.
For example,  to derive the beta-function for the gauge coupling to
four-loop order in \drbar{} scheme one needs the relation between the
gauge couplings defined in the \msbar{} and \drbar{} schemes up to
three-loop order. The latter can be determined using the following
arguments.

 To compute the relations between  running parameters 
defined in two different renormalization schemes, one has to relate them
 to physical observables which cannot depend on the choice of 
 scheme. For example, the relationship between the strong
 coupling constant defined in the \msbar{} and \drbar{} schemes can be obtained
 from the S-matrix amplitude of a physical process involving the gauge
 coupling computed in the two schemes. However, beyond one-loop  the
 computation of the physical amplitudes 
becomes very much  involved and requires the computation of multi-loop
and multi-scale  on-shell Feynman integrals that is a highly
non-trivial task. Nevertheless, one can avoid the use of on-shell kinematics
introducing a physical 
renormalization scheme defined through convenient kinematics, for which
the renormalization constants 
can be computed applying the  ``large-momentum'' or the ``hard-mass''
procedures. Up to three loops, there are 
 well established methods  ( for a details see previous sections) to
 compute the divergent as well as finite pieces  of the  Feynman integrals  and
 automated programs exist to perform
 such calculations. Once the renormalization constants in the physical
 renormalization scheme are determined, one uses the constraint that 
the {\it effective} gauge coupling constant defined in such a scheme is
unique and thus,
independent on the  regularization procedure. Furthermore, one relates
the running 
gauge couplings defined in the two regularization schemes through  the
following relations
\begin{equation}
\begin{split}
  \alpha_s^{\rm ph} = \left(z_s^{\rm ph,X}\right)^2 \alpha_s^{\rm X}\,,\qquad
  z_s^{\rm ph,X} &= Z_s^{\rm X}/Z_s^{\rm ph,X} \,,\qquad
  {\rm X} \in \{\overline {\rm MS},\overline {\rm DR}\}\\
  \Rightarrow
  \asDRbar &= \left(\frac 
           {Z_s^{\rm ph,\overline{\rm DR}}\,Z_s^{\overline{\rm MS}}}
           {Z_s^{\rm ph,\overline{\rm MS}}\,Z_s^{\overline{\rm DR}}}
           \right)^2
  \,\asMSbar\,,
  \label{eq::asDRMSderiv}
\end{split}
\end{equation}
where $Z_s^{{\overline {\rm MS}}/{\overline {\rm DR}}}$ are the charge
renormalization constants using minimal
subtraction in \dreg{}/\dred{}, as defined above.  
%For $Z_s^{{\rm ph},\overline{\rm MS}/\overline{\rm DR}}$,
%on the other hand, one can use \dreg{}/\dred{} combined with a {\it physical}
%renormalization condition. One observes that the ratio in
%Eq.\,(\ref{eq::asDRMSderiv}) is momentum independent, such that the
%calculation amounts to keeping the constant finite pieces in the charge
%renormalization constants
%$Z_s^{{\rm ph},\overline{\rm MS}/\overline{\rm DR}}$.
Note that the various $Z_s$ in Eq.~(\ref{eq::asDRMSderiv}) 
depend on differently renormalized
$\alpha_s$, so that the equations have to be used iteratively at higher
orders of perturbation theory. Working out these considerations for the
gauge coupling and for the fermion mass up to the three-loop order, one obtains
\begin{eqnarray}
\asDRbar &=& \asMSbar\left[1+\frac{\asMSbar}{\pi} \frac{1}{12}C_A
  +\Bigl(\frac{\asMSbar}{\pi}\right)^2
  \frac{11}{72}C_A^2
  - \frac{\asMSbar}{\pi} \aepi
  \frac{1}{8}C_R T_f
  + \delta_\alpha^{(3)}  \Bigr]
  \,,
\label{eq::asMS2DR_2}\\
%  \nonumber\\
  \mDRbar &=& \mMSbar\Bigg[1 -\aepi\frac{1}{4}C_R +
  \left(\apiMS\right)^2 \frac{11}{192}C_AC_R 
-\apiMS\aepi
  \frac{1}{32}C_R(3C_A+8C_R) 
  \nonumber\\
  &&\mbox{} + \left(\aepi\right)^2 \frac{1}{32}[3C_R+T_f] 
      + \delta_m^{(3)} 
  \Bigg]
%  \asDRbar &=& \asMSbar\left[1+\frac{\asMSbar}{\pi} \frac{1}{4}
%  +\left(\frac{\asMSbar}{\pi}\right)^2
%  \frac{11}{8} 
%  - \frac{\asMSbar}{\pi} \aepi
%  \frac{1}{12} n_f
%  + \delta_\alpha^{(3)}  \right]
%  \,,
%  \nonumber\\
%  \mDRbar &=& \mMSbar\Bigg[1 -\aepi\frac{1}{3} +
%  \left(\apiMS\right)^2 \frac{11}{48} -\apiMS\aepi
%  \frac{59}{72}
%  \nonumber\\
%  &&\mbox{} + \left(\aepi\right)^2 \left( \frac{1}{6}
%      +  \frac{1}{48} n_f\right)   
%      + \delta_m^{(3)} 
%  \Bigg]
  \,,
  \label{eq::mMS2DR_2}
\end{eqnarray}
where we have suppressed the explicit dependence on the renormalization
scale $\mu$.  $\delta_\alpha^{(3)}$ and $\delta_m^{(3)}$ denote the
three-loop terms  
  and they are obtained from
the finite parts of three-loop diagrams (see
\reference{Harlander:2006rj} for details). They
read~\cite{Tim,Jack:2007ni}
%\bea
{\allowdisplaybreaks
\begin{align}
\pi^3\delta^{(3)}_{\alpha}
 &= \frac{1}{96}\asMSbar \alpha_e^2 T_f[2C_A^2-3C_AC_R+2C_R^2
-C_A T_f+7C_R T_f]\nn
&
-\frac{1}{192}(\asMSbar)^2\alpha_e T_f(5C_A^2+60C_AC_R+6C_R^2) \nn
&+\frac{1}{9216}\asMSbar (36\cA^3\eta_1^2 -576\cA\eta_2^2-144\cA^2
\eta_1\eta_3-72\cA\eta_3^2\nn
&+12\cA^4\eta_1\eta_4-480\cA^3\eta_2\eta_4-24\cA^3\eta_3\eta_4
+\cA^5\eta4^2-288\cA 
N_A\eta_2^2\nn
&+72\cA N_A\eta_3^2
)
-\frac{1}{96N_A}\asMSbar \eta_4^2C_A D_4 (AA)
+\frac{1}{48}(\asMSbar)^2\eta_4D_4 (AA)\nn
&+\frac{1}{4608}(\asMSbar)^2(-6C_A^3\eta_1+240 
C_A^2\eta_2+12C_A^2\eta_3
-C_A^4\eta_4)\nn
&+\frac{1}{10368}(\asMSbar)^3[3049C_A^3-416C_A^2 T_f
-138C_AC_R T_f] \,,
\label{eq:deltaa}
 \displaybreak[2] \\
%\eea
%\bea
\pi^3\delta^{(3)}_m &= -\frac{1}{384}\alpha_e^3 C_R
[-10C_A^2+14C_AC_R+27C_R^2-7C_A T_f\nn
&+39C_R T_f
-10I_2(R)^2 T_f^2+12C_A^2\zeta_3-36C_AC_R\zeta_3+24C_R^2\zeta_3]\nn
&-\alpha_e^2 C_R\Bigl(\frac{1}{322}[(6C_R-C_A)\eta_2]
+\frac{1}{16I_2(R)N_A}D_4 (RA)\eta_4
+\frac{1}{384}\asMSbar[47C_A^2
+10C_R^2\nn
&-3 C_A T_f
-19 C_R T_f-165C_AC_R+144C_R^2\zeta_3\nn
&-48
C_A T_f\zeta_3
+48 C_R T_f\zeta_3
+72C_A^2\zeta_3
-216C_AC_R\zeta_3]\Bigr)\nn
&+\alpha_e C_R
\Bigl(\frac{1}{12288}[
-36\cA^2\cR\eta_1^2+1728\cR\eta_2^2+144\cA\cR\eta_1\eta_3
+72\cR\eta3^2-12\cA^3\cR\eta_1\eta_4
\nn
&
+1440\cA^2\cR\eta_2\eta_4+24\cA^2\cR\eta_3\eta_4-\cA^4\cR\eta4^2+864\cR
N_A\eta_2^2-72\cR N_A\eta3^2] 
\nn
&+\frac{1}{3072}(\asMSbar)^2[2880C_R^2\zeta_3-168C_A T_f
-1544C_AC_R-52C_R^2\nn
&-128 C_R T_f
+1440C_A^2\zeta_3-4320C_AC_R\zeta_3
-79C_A^2]\Bigr)\nn
&+\frac{1}{20736}(\asMSbar)^3 C_RC_A[4354C_A+135C_R+304
T_f]+\frac{3}{128N_A}D_4 (AA)\eta_4^2.
\label{eq:deltam}
%\eea
\end{align}
}
$\!\!$Inserting Eqs.~(\ref{eq::asMS2DR_2}) and (\ref{eq::mMS2DR_2}) into the definition of the beta
function for the gauge coupling Eq.~(\ref{eq::beta}) and the mass
anomalous dimension Eq.~(\ref{eq::gamma}), one
can show that 
\begin{eqnarray}
  \betaDRbar_s
  &=& \mu^2 \frac{{\rm d}}{{\rm d}\mu^2} \apiDR\nonumber\\
&=&\betaMSbar_s
  \frac{\partial \asDRbar}{\partial \asMSbar} + 
  \beta_e \frac{\partial \asDRbar}{\partial \alpha_e} +
  \sum_r \beta_{\eta_r} \frac{\partial \asDRbar}{\partial \eta_r}
  \,,
  \nonumber\\
  \gammaDRbar_m &=&\frac{\mu^2}{\mDRbar} \frac{{\rm d}}{{\rm d}\mu^2} \mDRbar
\nonumber\\
 &=&  \gammaMSbar_m \frac{\partial \ln \mDRbar}{\partial \ln \mMSbar}
  + \frac{\pi \betaMSbar_s}{\mDRbar} 
  \frac{\partial \mDRbar}{\partial \asMSbar}
  + \frac{\pi \beta_e}{\mDRbar}   
  \frac{\partial \mDRbar}{\partial \alpha_e}
  + \sum_r \frac{\pi \beta_{\eta_r}}{\mDRbar}   
  \frac{\partial \mDRbar}{\partial \eta_r}
  \,,
  \label{eq::DRED-DREG}
\end{eqnarray}
where the first equality is due to the definition of $\betaDRbar_s$ and
$\gammaDRbar_m$, and 
the second one is a consequence of the chain rule.
Let us briefly discuss the order in perturbation theory
up to which the individual building blocks are needed. Of course, the
\msbar{} quantities are needed to four-loop order; they can be found in
Refs.~\cite{vanRitbergen:1997va,Chetyrkin:1997dh,
Vermaseren:1997fq,Czakon:2004bu}.  The dependence of $\asDRbar$ and
$\mDRbar$ on $\alpha_e$ starts at two- and one-loop
order~\cite{Harlander:2006rj}, respectively. Thus, $\beta_e$ is needed
up to the three-loop level (cf. Eq.~(\ref{eq::DRED-DREG})).  On the other
hand, both $\asDRbar$ and $\mDRbar$ depend on $\eta_r$ starting from
three loops and consequently only the one-loop term of $\beta_{\eta_r}$
enters in Eq.~(\ref{eq::DRED-DREG}).
The \drbar{} four-loop results were derived for QCD in
Ref.~\cite{Harlander:2006rj} and for a general theory in
Ref.~\cite{Tim,Jack:2007ni}. The explicit four-loop results are too
lengthy to be presented 
in this review and we refer to the original papers for the explicit
results. We discuss however their supersymmetric limit in the next section.
%%%%%%%%%%%%%%%%%%%%%%%%%%%%%%%%%%%%%%%%%%%%%%%%%%%%%%%%%%%%
\subsubsection{\label{sec:4lsusy}The four-loop supersymmetric case}

An important check of the complicated formulas derived in the previous
sections can be obtained by converting them to a supersymmetric
Yang-Mills theory. For this case, one has to replace the fermions by the
supersymmetric partner of the gauge bosons, the so-called gauginos. Technically,
this amounts to set the fermions in the adjoint representation of the
gauge group. In addition, closed fermion loops have to be multiplied by
an extra factor $1/2$ in order to take into account the Majorana
character of the gauginos. Explicitly, for the derivation of the three-
and four-loop results one needs the replacements
 \bea
C_R &\to& C_A\nn
I_2 (R) &\to& C_A\nn   
n_f &\to& \frak{1}{2}\nn
D_4 (RR) &\to& D_4 (AA)\nn
D_4(RA) &\to& D_4 (AA)\nn 
D_4 (RAA) &\to& D_4 (AAA)\,.
\label{susycase}
\eea
Furthermore, SUSY requires that the gauge coupling $\alpha_s$ equals the
evanescent coupling $\alpha_e$ to all orders of perturbation theory,
and therefore, the $\beta$ functions are also equal $\beta_e^{\rm
  SYM}=\beta_s^{\rm SYM}$. Moreover, SUSY also requires
that the  \epscalar{} quartic interaction containing
the structure constants   is equal to the gauge
coupling to all orders of perturbation theory. In this case, the other
three quartic couplings
have to vanish, so that the decomposition Eq.~(\ref{eq::wdec})  holds 
to all orders of perturbation theory.
  Indeed, using Eqs.~(\ref{eq::renquart})
one can  easily derive the corresponding one-loop beta functions for
supersymmetric theories and
obtains  
 \bea
&&\beta_{\eta_1}^{\rm SYM}=\beta_{e}^{\rm SYM}=\beta_{s}^{\rm SYM}\quad
\mbox{and}\quad
\beta_{\eta_2}^{\rm SYM}=\beta_{\eta_3}^{\rm SYM}=\beta_{\eta_4}^{\rm
  SYM}=0\,,
\label{eq::symbeta}
\eea
when the SUSY restrictions 
\bea
&& \eta_1=\alpha_3=\alpha_s\quad
\mbox{and}\quad \eta_2=\eta_3=\eta_4=0\,.
\label{eq::symcoupl}
\eea
are imposed. It is also interesting to notice that the terms in the
renormalization 
constants Eqs.~(\ref{eq::renquart}) that contain negative power of
couplings cancel out in the SUSY limit, so that the limit
$\eta_2=\eta_3=\eta_4\to 0$ can be computed trivially. 
Thus, if relations (\ref{eq::symcoupl}) are imposed at the tree-level,
they will not be spoiled by the renormalization at the one-loop
order. Checks of this statement at two- and three-loop orders  are
available so far only for 
the evanescent coupling $\alpha_e$~\cite{Harlander:2006rj,Harlander:2006xq}. 

Applying the substitutions given in Eq.~(\ref{susycase}) and
(\ref{eq::symcoupl}) 
one can obtain the four-loop results for the gauge beta-function
$\beta_{s}^{\rm SYM}$~\cite{Harlander:2006rj,Jack:2007ni} and compare it
with the expression  derived in 
Ref.~\cite{Jack:1998uj}
\begin{equation}
\begin{split}
\beta_s^{\rm SYM} = -\left(\api\right)^2\,\left[
  \frac{3}{4}C_A
  + \frac{3}{8}C_A^2\api
  + \frac{21}{64}C_A^3\left(\api\right)^2
  + \frac{51}{128}C_A^4\left(\api\right)^3
\right] + {\cal O}(\alpha_s^6)\,.
\end{split}
 \label{eq::betag4}
\end{equation}
The method employed in \reference{Jack:1998uj} to obtain the four-loop result 
was very indirect, in particular relying on the existence of the \nsvz{} 
formula for $\beta_s^{\rm SYM}$~\cite{Jones:1983ip,
  Novikov:1983ee}\footnote{For more details see section~\ref{sec:holomorphy}.}.
It is therefore a remarkable check on both calculations that indeed   
precise agreement was obtained.

Turning now to the case of softly-broken \sy, 
there exists an exact result relating $\beta_s$ and
$\gamma_m$~\cite{Hisano:1997ua,Jack:1997pa} within the \nsvz{} scheme:
%\footnote{For more details
%  see section\ref{sec:holomorphy}.}: 
\begin{equation}
\gamma_m^{\rm SYM} = \pi \alpha_s\frac{\rm d}{\rm d\alpha_s}
\left[\frac{\beta_s^{\rm SYM}}{\alpha_s}\right],
\label{eq::exactgamma}
\end{equation}
that nevertheless holds in \drbar{} scheme too.
Hence, it follows that
\begin{equation}
\begin{split}
\gamma_m^{\rm SYM} = -\left(\api\right)\,\left[
  \frac{3}{4}C_A
  + \frac{3}{4}C_A^2\api
  + \frac{63}{64}C_A^3\left(\api\right)^2
  + \frac{51}{32}C_A^4\left(\api\right)^3
\right] + {\cal O}(\alpha_s^5)\,.
\end{split}
 \label{eq::gamma4}
\end{equation}
Inserting \eqn{susycase}\ in \eqn{eq::DRED-DREG},
one can easily  reproduce \eqn{eq::gamma4}.   

The invariant $D_4 (AAA)$ does not occur in either  calculation, and the
dependence on $D_4 (AA)$, $N_A$, $\zeta_3$,  $\zeta_4$ and $\zeta_5$ all 
 cancel although they appear in individual terms. It is tempting  to
speculate that this absence of higher order invariants and
transcendental  numbers (other than $\pi$) is related to the existence
of the \nsvz{} scheme, in which  the gauge $\beta$-function
for any simple gauge group  is given (in the supersymmetric case without
matter fields)  by the expression  in Eq.~(\ref{eq::betansvz_sy}),
%\be \beta_s^{\rm
%NSVZ} = -\frac{3}{4}C_A\left(\api\right)^2
%\left(1-\frac{C_A\alpha_s}{2\pi}\right)^{-1} 
%\ee 
which is manifestly
free of transcendental numbers to all orders. It is natural  to conjecture that the same
property holds in the \dred{} scheme, too.   
%For discussion of the
%relationship between $\beta_s^{\rm NSVZ}$ and  $\betaDRbar_s$
%see~\reference{cjn}.

%%%%%%%%%%%%%%%%%%%%%%%%%%%%%%%%%%%%%%%%%%%%%%%%%%%%%%%%%%%%

\subsubsection{\label{sec::ep4mass} \eps-scalar mass }
\begin{figure}[t]
  \begin{center}
    \begin{tabular}{c}
      \leavevmode
      \epsfxsize=.95\textwidth
      \epsffile{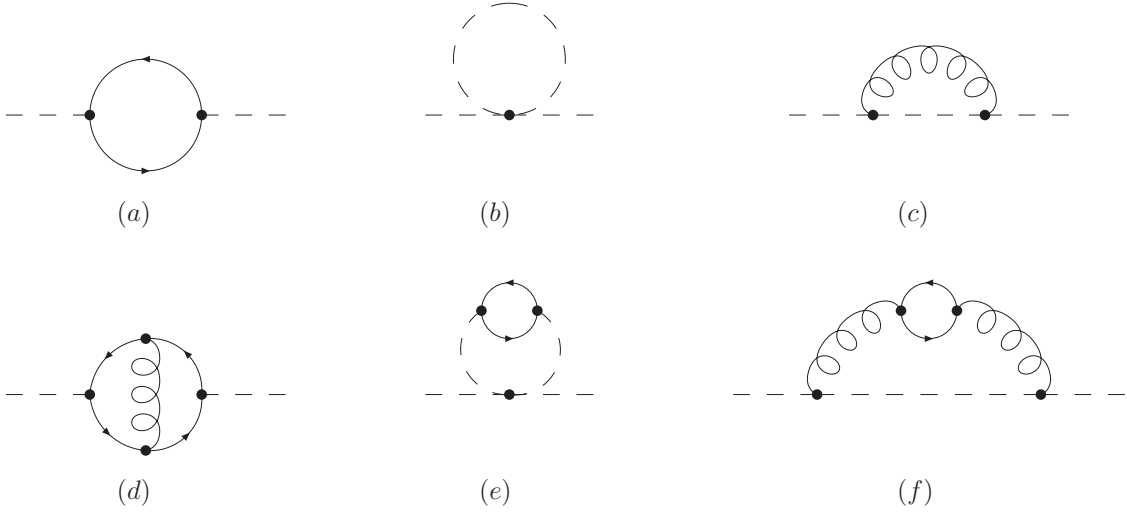}
    \end{tabular}
  \end{center}
  \caption{\label{fig::eps_prop}
    One- and two-loop Feynman diagrams contributing to the
    $\varepsilon$-scalar propagator. Dashed lines denote
    $\varepsilon$ scalars, curly lines denote the gauge bosons and solid lines
    denote massive fermions with mass $M_f$.
          }
\end{figure}

%In order to achieve the
% finite result for the relation between the
%pole and the \drbar{} quark mass it is necessary to fix a
%renormalization scheme also for the mass of the $\varepsilon$ scalar,
%$m_\varepsilon$. 
Although
there is in general no tree-level term in the Lagrangian for the mass of
the $\varepsilon$ scalar 
%$m_\varepsilon$,
 there are loop induced  contributions to it
  that require the introduction of the corresponding
counter term. Let us introduce first the renormalization constant for
the \eps-scalar mass 
\begin{eqnarray}
\left(m_\varepsilon^{0}\right)^2 = Z_{m_\varepsilon} m_\varepsilon^2\,.
\end{eqnarray}
 The relevant Feynman diagrams contributing to the
$\varepsilon$-scalar propagator show  quadratic divergences and
 therefore, one  needs to consider
only contributions from massive particles. Thus, in this case, only
diagrams involving  massive fermions have to be taken into account,
since they are the only particles allowed by the gauge invariance to
have non zero masses. Sample diagrams are
shown in Fig.~\ref{fig::eps_prop}.

It is advantageous in  practice to renormalize $m_\varepsilon$ on-shell and
require that the renormalized mass is zero to each order in
perturbation theory. In this scheme  the
$\varepsilon$-scalar mass completely decouples from the physical
observables. 
%In supersymmetric theories the contributions to
%$m_\varepsilon$ from fermions and bosons sitting in the same
%super-multiplet cancel each other, so that no explicit renormalization
%for it has to carried out.
% Nevertheless, there are also other
%options for the renormalisation of the \eps-scalar{} mass, that we
%allued to in the next section. 

At the one-loop order there is only one relevant diagram
(cf. Fig.~\ref{fig::eps_prop}$(a)$) which has to be evaluated
for vanishing external momentum.
A closer look at the two-loop diagrams shows that they
develop infrared divergences in the limit $m_\varepsilon\to 0$
(cf., e.g., Fig.~\ref{fig::eps_prop}$(e)$). They can be regulated by
introducing a small but non-vanishing mass for the
$\varepsilon$-scalars. After the subsequent application
of an asymptotic expansion~\cite{Smirnov:2002pj}
in the limit $q^2=m_\varepsilon^2\ll M_f^2$ the infra-red divergences
manifest themselves as $\ln(m_\varepsilon)$ terms.
Furthermore, one-loop diagrams like the ones in
Fig.~\ref{fig::eps_prop}$(b)$ and $(c)$ do not
vanish anymore and have to be taken into account as well.
Although they are proportional to $m_\varepsilon^2$, after
renormalization they induce two-loop contributions which
are proportional to $M_f^2$, partly multiplied by
$\ln(m_\varepsilon)$ terms. It is interesting to note that
in the sum of the genuine two-loop diagrams and the counter-term
contributions the limit $m_\varepsilon\to 0$ can be taken
which demonstrates the infra-red finiteness of the on-shell mass of the
$\varepsilon$ scalar.
 The two-loop renormalization constant within QCD has been computed in
Ref~ \cite{Marquard:2007uj}. It 
 is given by
{\allowdisplaybreaks
\begin{align}
%\begin{eqnarray}
  \frac{m_\varepsilon^2}{M_f^2} (Z_{m_\varepsilon}^{\rm OS}-1) &=
   - \aepi n_h I_2(R)
\bigg[
  \frac{2}{\ep} + 2 + 2 \Lmu + \ep\left(2 + \frac{1}{6} \pi^2  + 2 \Lmu
  + \Lmu^2
\right)
%% \nonumber\\
%% &&
%%  +
%%  \ep^2\bigg[1+ \frac{1}{2} \zeta_2  - \frac{1}{3} \zeta_3 + \left(1
%%   + \frac{1}{2} \zeta_2 \right) \Lmu + \frac{1}{2} \Lmu^2  \bigg]
\bigg]
\nonumber\\
& - \left(\apiDR\right)^2 n_h I_2(R)
\bigg(\frac{3}{4} \frac{1}{\ep}  + \frac{1}{4} +
  \frac{3}{2} \Lmu
%%  + \ep \left(\frac{13}{4}+ \frac{1}{2}\Lmu +
%%   \frac{3}{2}  \Lmu^2 + \frac{1}{8} \pi^2\right)
\bigg) C_A
 + \apiDR\aepi n_h I_2(R) \bigg\{
\frac{1}{\ep^2} \left( \frac{3}{8} C_A + \frac{3}{2} C_R\right)
\nonumber\\
& +\frac{1}{\ep}\bigg[ \frac{7}{8} C_A + \frac{3}{2} C_R +\left(
 \frac{3}{4} C_A  + \frac{3}{2} C_R\right)  \Lmu \bigg]
+ \left(\frac{15}{8} + \frac{1}{16}\pi^2 \right) C_A
\nonumber\\
& + \left(\frac{3}{2} + \frac{1}{8}\pi^2\right) C_R + \left(\frac{7}{4}  C_A
 +\frac{3}{2}  C_R\right)\Lmu +\left(\frac{3}{4} C_A + \frac{3}{4}
  C_R\right) \Lmu^2
\bigg\}
\nonumber\\
&
+ \left(\aepi\right)^2 n_h I_2(R) \bigg\{
\frac{1}{\ep^2} \left(
\frac{1}{4} C_A  - \frac{1}{2} C_R  - \frac{1}{2} T_f \right)
+\frac{1}{\ep}  \bigg[
\frac{1}{2} C_R
\nonumber\\
& - \frac{1}{2} \left(1+\Lmu \right)  T_f
\bigg]
- \frac{1}{2} C_A  + \frac{5}{2} C_R - \left(\frac{1}{2} +\frac{1}{24}
  \pi^2 \right) T_f
\nonumber\\
&
- \left(\frac{1}{2} C_A - 2 C_R  + \frac{1}{2}  T_f \right) \Lmu
- \left(\frac{1}{4} C_A  -\frac{1}{2} C_R  + \frac{1}{4}  T_f
  \right) \Lmu^2
\bigg\}
\nonumber\\
&
+ \aepi\frac{\eta_1}{\pi} n_h \bigg[
  \frac{3}{16} \frac{1}{\ep^2} +\frac{1}{\ep}\left(
\frac{3}{16} + \frac{3}{8} \Lmu \right) +\frac{3}{16}+\frac{1}{32} \pi^2
+ \frac{3}{8} \Lmu + \frac{3}{8} \Lmu^2
\bigg]
\nonumber\\
&
- \aepi\frac{\eta_2}{\pi} n_h \bigg[
 \frac{5}{4} \frac{1}{\ep^2} + \left(5 + \frac{5}{2} \Lmu \right )\frac{1}{\ep}
+\frac{25}{2}  + \frac{5}{24} \pi^2 + 10 \Lmu + \frac{5}{2} \Lmu^2
\bigg]
\nonumber\\
&
- \aepi\frac{\eta_3}{\pi} n_h \bigg[
 \frac{7}{16}\frac{1}{\ep^2}  + \left( \frac{7}{16} + \frac{7}{8} \Lmu
  \right )\frac{1}{\ep}
+\frac{7}{16} + \frac{7}{96} \pi^2  + \frac{7}{8} \Lmu + \frac{7}{8}
  \Lmu^2
\bigg]
  \,,
%\end{eqnarray}
\end{align}
}
$\!\!$where $\Lmu=\ln(\mu^2/M_f^2)$,  $T_f =n_f I_2(R)$, where $n_f$ and $n_h$
denotes the number of fermions and  heavy
fermions, respectively. The overall factor $n_h$ in front of the one- and two-loop
corrections 
shows that the renormalization of $m_\varepsilon$ is only influenced by
those diagrams which contain a closed heavy fermion loop.

It is also possible to renormalize  $m_\varepsilon$ so that
$m_\varepsilon^{\rm OS}\ne 0$ or  adopt the \drbar{} scheme  for it. In
the latter case, the physical observables will 
depend on $m_\varepsilon$. In order to get rid of this unphysical
dependence, one has to introduce additional finite shifts in the
renormalization constants of the physical parameters. This new
renormalization scheme is called  \drbarprime{} and it will be discussed
in more detail in the next section. Nevertheless, in context of
QCD, the  \drbarprime{} has rarely been used~\cite{Martin:2001vx}.
%%%%%%%%%%%%%%%%%%%%%%%%%%%%%%%%%%%%%%%%%%%%%%%%%%%%%%%%%%%%
\section{\label{sec:susyqcd} Dimensional Reduction applied to 
SUSY-QCD at   three loops}
\setcounter{equation}{0}
\setcounter{figure}{0}
\setcounter{table}{0}

%%SUSY is a strong candidate for an extension of the SM. Apart from the
%%solution to the dark Matter, it provides
%%an appealing solution to the fine tuning problem by canceling the
%%quadratic divergences in the Higgs self-energies and explains
%%electroweak symmetry breaking by a simple evolution of the parameters in
%%the Higgs potential. Through its extended particle spectrum it also allows
%%the unification of the gauge couplings at scales compatible with proton
%%decay.

All the appealing features of supersymmetric theories have to be
confirmed by an  accurate comparison with the experimental data like those
measured in collider experiments~\cite{tevatron,atlas,cms}. 
Such an ambitious task
  requires precision data as well as precision
calculations. But, precise predictions for observables 
 implies computations of
higher order radiative corrections. Thus, it
necessarily  rises the question of constructing  regularization and
renormalization schemes 
that are gauge and SUSY invariant. As discussed in the previous
sections,  DRED scheme was proposed
as a solution, although it
could violate SUSY at higher orders of perturbation theory. Currently, it
is believed that  DRED preserves SUSY at three-loop order as was
explicitly checked in Refs.~\cite{Pickering:2001aq,Harlander:2009mn} and
that it breaks  SUSY at 
four-loop order, taking into consideration formal
arguments~\cite{Avdeev:1982xy,ds}. 
Nevertheless, renormalization by combining DRED with minimal subtraction
(the \drbar{} scheme) or the  on-shell scheme has become the
preferred schemes in higher order 
supersymmetric calculations~\cite{Heinemeyer:2004yq,Heinemeyer:2004ms,Degrassi:2009yq,Kant:2010tf,Pak:2010cu}.      

\subsection{Renormalization of the gauge
 coupling and fermion masses at   three loops}

%Renormalization group functions, governing the energy dependence of
%masses and couplings, are among the simplest quantities to compute in
%perturbative quantum field theory. They are also among the main
%ingredients for the analyses revealing the posibility of gauge and
%yukawa coupling 
%unification around the Grand Unification Scale (GUT) scale.

As was  already reviewed in section~\ref{sec:holomorphy}, for supersymmetric gauge
theories one can devise a particular
renormalization scheme, the so-called {\abbrev NSVZ}
scheme~\cite{Novikov:1983uc}, 
where an all-order relation between the gauge $\beta$ function and the
anomalous dimension of the chiral supermultiplet is valid. So, in the
absence of the matter supermultiplet, {\it i.e.} for \susy{}-Yang-Mills
theory, the $\beta$ function is known to all orders in the coupling
constant. Applying the same method based on the connection between the
holomorphic 
and   the {\abbrev NSVZ} scheme to softly broken \susy{} gauge theory,
the authors of 
Ref.\,\cite{Hisano:1997ua} derived the renormalization group equation
governing the running of the gaugino and sfermion masses as functions of
the gauge and Yukawa coupling $\beta$ functions, valid
to all orders in perturbation theory.  Actually, all these
calculations received  important 
phenomenological applications only after the authors of
Ref.\,\cite{Jack:1996vg} found the three-loop conversion formula between
the   {\abbrev NSVZ} and \drbar{}. This  allowed the derivation of
three-loop order beta-functions for the parameters of the MSSM in the
\drbar{} scheme~\cite{Jack:2003sx}. 

The goal of this section is to report on 
another confirmation of the results for the anomalous
dimensions of  SUSY-QCD parameters, that is based on a direct
calculation of 
relevant three-loop  Feynman diagrams implementing the \dred{}
approach in the component field formalism. 
The agreement of the two
independent and conceptually completely different calculations is a very
important check of the two methods on the one side, and on the other side
it establishes the \dred{} as a consistent framework for computations of
radiative 
corrections in supersymmetric theories.

%%As already explained in the previous section,
% the  implementation of \dred{} is
%done by decomposing the quasi-four-dimensional 
%vector fields into $d$- and $2\ep$-dimensional components:
%\begin{eqnarray}
%A_{\mu} = A_{\hat{\mu}} + A_{\bar{\mu}}\,.
%\label{eq::epscalar}
%\end{eqnarray}
%For convenience, $A_{\hat{\mu}}$ will be simply called the ``gluon field'' in
%what follows, while  
% $A_{\mu}$ will be explicitely referred to as ``four-dimensional gluon
% field''.$A_{\bar{\mu}}$ is the so-called
% $A_{\mu}$ will be explicitely referred to as ``four-dimensional gluon
% field''.\epscalar.
%the bare coupling constants for the gluon and the \epscalar{} are
%identical by construction. However, {\it a priori} it is not clear
%whether this holds also for the renormalized couplings. In principle,
%all \epscalar{} couplings could be different without violating gauge
%invariance\footnote{See section~\ref{sec:dred} for details.}.
The renormalized Lagrangian of a supersymmetric theory will obey
\susy{} constraints, only if the 
decomposition of \eqn{eq::wdec}  hold  at all  orders of perturbation theory.
Therefore, the renormalized gluon and \epscalar{} coupling constants
must be equal, i.e., their $\beta$ functions must be the same.  
 An all order proof of this statement is currently not
available. However, it was
explicitly shown~\cite{Harlander:2009mn}  that the coupling
constant arising from the vertices: $gc\bar{c}$, $ggg$, $gq\bar{q}$,
$\tilde{g}q\bar{q}$ and that from 
 the vertices $q\bar q\ep$, 
$\tilde g\tilde 
g\ep$, and  $g\ep\ep$  are equal through three loops.\\
Even more:
in order to renormalize the quartic \epscalar{} vertex, one has to take
into account all possible colour structures for it, and attribute  to
each one a separate coupling constant\footnote{For details see
sections~\ref{sec::ep4} and \ref{sec:4lsusy}.}. For \susy{}-\qcd{}, it
has been 
explicitly checked~\cite{Jack:2007ni} that at the one-loop order only
the $\beta$ function 
associated with the usual colour structure of the four-gluon interaction,
{\it i.e.} $f_{abe} f_{cde}$\footnote{$f_{abe}$ denotes the structure
  constants of the gauge group.}, does not vanish and it equals the
one-loop gauge $\beta$ function.  Thus, through one-loop, one can identify
the coupling constant of the corresponding \epscalar{} quartic
interaction with the strong coupling constant and set to zero the other
three quartic couplings. This order of accuracy is sufficient for the results
discussed here, as the \epscalar{} quartic interactions
contribute to the anomalous dimensions starting from the two-loop order.
 A similar observation was made also in the previous
section when the SUSY-Yang-Mills theory was discussed at four-loop accuracy.
 All these tests confirm the 
consistency of \dred{} with \susy{} at 
  next-to-next-to-next-to-leading order (NNNLO) of perturbation theory.

\begin{figure}
\begin{center}
\newcommand{\qvertex}[2]{%
\ifnum#1=-1 \ifnum#2=1 0 \else\ifnum#2=2 0\fi\fi\else
\ifnum#1=6
    \ifnum#2=1 75
\else\ifnum#2=2 10
\fi\fi\else
\ifnum#1=4
    \ifnum#2=1 75
\else\ifnum#2=2 140
\fi\fi\else
\ifnum#1=1
    \ifnum#2=1 130
\else\ifnum#2=2 130
\fi\fi\else
\ifnum#1=3
    \ifnum#2=1 130
\else\ifnum#2=2 20
\fi\fi\else
\ifnum#1=7
    \ifnum#2=1 75
\else\ifnum#2=2 75
\fi\fi\else
\ifnum#1=2
    \ifnum#2=1 20
\else\ifnum#2=2 75
\fi\fi\else
\ifnum#1=5
    \ifnum#2=1 110
\else\ifnum#2=2 75
\fi\fi\else
\fi\fi\fi\fi\fi\fi\fi\fi }
\SetScale{.7}
\begin{picture}(120,120)
\SetWidth{0.5}
\ArrowLine(\qvertex{4}{1},\qvertex{4}{2})(\qvertex{2}{1},\qvertex{2}{2})
\SetWidth{0.5}
\ArrowLine(\qvertex{2}{1},\qvertex{2}{2})(\qvertex{6}{1},\qvertex{6}{2})
\SetWidth{0.5}
\ArrowLine(\qvertex{5}{1},\qvertex{5}{2})(\qvertex{3}{1},\qvertex{3}{2})
\SetWidth{0.5}
\DashArrowLine(\qvertex{1}{1},\qvertex{1}{2})(\qvertex{4}{1},\qvertex{4}{2}){3}
\SetWidth{0.5}
\DashArrowLine(\qvertex{6}{1},\qvertex{6}{2})(\qvertex{5}{1},\qvertex{5}{2}){3}
\SetWidth{0.5}
\Line(\qvertex{1}{1},\qvertex{1}{2})(\qvertex{5}{1},\qvertex{5}{2})
\SetWidth{0.5}
\Gluon(\qvertex{1}{1},\qvertex{1}{2})(\qvertex{5}{1},\qvertex{5}{2}){6}{5}
\SetWidth{0.5}
\Line(\qvertex{7}{1},\qvertex{7}{2})(\qvertex{4}{1},\qvertex{4}{2})
\SetWidth{0.5}
\Gluon(\qvertex{7}{1},\qvertex{7}{2})(\qvertex{4}{1},\qvertex{4}{2}){6}{5}
\SetWidth{0.5}
\Line(\qvertex{6}{1},\qvertex{6}{2})(\qvertex{7}{1},\qvertex{7}{2})
\SetWidth{0.5}
\Gluon(\qvertex{6}{1},\qvertex{6}{2})(\qvertex{7}{1},\qvertex{7}{2}){6}{5}
\SetWidth{0.5}
\Gluon(\qvertex{7}{1},\qvertex{7}{2})(\qvertex{3}{1},
\qvertex{3}{2}){6}{7.73702731645053} 
\SetWidth{0.5}
\ArrowLine(160,150)(\qvertex{1}{1},\qvertex{1}{2})
\SetWidth{0.5}
\DashLine(-10,75)(\qvertex{2}{1},\qvertex{2}{2}){2}
\SetWidth{0.5}
\ArrowLine(\qvertex{3}{1},\qvertex{3}{2})(160,0)
\Vertex(\qvertex{6}{1},\qvertex{6}{2}){3}
\Vertex(\qvertex{4}{1},\qvertex{4}{2}){3}
\Vertex(\qvertex{1}{1},\qvertex{1}{2}){3}
\Vertex(\qvertex{3}{1},\qvertex{3}{2}){3}
\Vertex(\qvertex{7}{1},\qvertex{7}{2}){3}
\Vertex(\qvertex{2}{1},\qvertex{2}{2}){3}
\Vertex(\qvertex{5}{1},\qvertex{5}{2}){3}
\end{picture} 
\caption[]{\label{fig::qqep}Sample diagram for the three-loop $q\bar
  q\ep$ vertex where a non-vanishing trace with a single $\gamma_5$
  matrix occurs. Solid lines are quarks, dashed lines
  are squarks, slashed springy lines are gluinos, and the external
  dashed line depicts an \eps{} scalar. The arrows on the lines denote the charge flow.}
\end{center}
\end{figure}
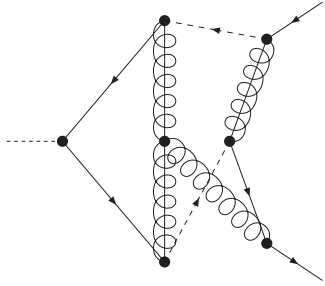
For the calculation of  renormalization constants within
supersymmetric theories 
 one can apply the same methods as the ones
discussed in section~\ref{sec::3loop} in the context of non-SUSY
theories. Let us however mention at this point, a technical subtlety related to the
implementation of  $\gamma_5$ matrix. Traces with a single $\gamma_5$ and at
least four $\gamma$-matrices do not
contribute to any of the 
two-point functions\footnote{For a detailed
  discussion about this aspect see Ref.~\cite{Mihaila:2012pz}.}. They do
contribute for some of 
the three-point functions though, in particular the $q\tilde q\tilde g$, the
$\tilde g\tilde g\ep$, and the $q\bar q\ep$ vertex. An example
diagram for the latter vertex is shown in Figure~\ref{fig::qqep}. Such diagrams
contribute  (among others) a colour factor $d_R^{abcd}d_A^{abcd}$ (for
the notation, see Appendix~A), but they cancel against
the same factors from other sources in the final result for the
renormalization constants and the $\beta$ functions. Precisely, the
naive scheme for the implementation of the $\gamma_5$ give rise to
incorrect results. One has to supplement it with the relations given in
Eqs.~(\ref{eq::trgamma5_2}) and (\ref{eq::leci}). The first equation
takes into
account  the contributions arising in  triangle diagrams containing
Dirac fermions, whereas the
second one generalizes the contraction properties of the {\it pseudo}
Levi-Civita tensors defined away from $d=4$ dimensions.

The results for the three-loop renormalization constants of the gauge
coupling constant $\alpha_s$  are very compact and are given by
{\allowdisplaybreaks
\begin{align}
%\begin{eqnarray}
Z_{s}&=1 + \frac{\alpha_s}{4\pi}\frac{1}{ \ep}[-3\ca + 2\NTF] \nonumber\\
%        & 
&
+ 
 \left(\frac{\alpha_s}{4\pi}\right)^2\bigg\{\frac{1}{   \ep^2} [9  \ca^2
 - 12  \ca  \NTF + 4  \NTF^2] 
+ \frac{1}{  \ep} [-3  \ca^2 +
 2  \ca  \NTF + 4  \cf \NTF]\bigg\}
\nonumber\\
%&
&
+\left(\frac{\alpha_s}{4\pi}\right)^3\bigg\{
\frac{1}{  \ep^3}
    [-27  \ca^3 + 54  \ca^2  \NTF - 36  \ca  \NTF^2 + 8
   \NTF^3]
\nonumber\\
%&
&
 + \frac{7}{3  \ep^2} 
   [9  \ca^3 - 12  \ca^2  \NTF - 12  \ca  \cf  \NTF + 4  \ca  \NTF^2 +
   8  \cf  \NTF^2]
\nonumber\\
%&
&
+ \frac{1}{3 \ep}  [-21
\ca^3 + 20  \ca^2  \NTF + 52  \ca  \cf  \NTF - 16  \cf^2  \NTF - 
     4  \ca  \NTF^2 - 24  \cf  \NTF^2]
\bigg\}\,,
\label{eq::as3l}
%\end{eqnarray}
\end{align}
}
$\!\!$where we have introduced the notation $T_f= I_2(R) n_f$, with $n_f$ the number of
active fermions of the theory and the invariants $\cf, \ca, \TF$ are explicitly
given in the  Appendix A. 

The case $T_f=0$ corresponds to \susy{}--Yang-Mills theory that
 has been treated in detail 
in sections~\ref{sec:holomorphy} and \ref{sec:4lsusy}. Full agreement
has been found between the two methods up to three-loop order.

The  three-loop renormalization constants for the gluino 
mass read 
{\allowdisplaybreaks
\begin{align}
%\begin{eqnarray}
Z_{\Mgl}&=1 + \frac{\alpha_s}{4\pi}\frac{1}{\ep}[-3\ca +
2\NTF]
\nonumber\\
%&
&
 + \left(\frac{\alpha_s}{4\pi}\right)^2
\bigg\{\frac{1}{\ep^2}[9\ca^2 - 12\ca\NTF + 4\NTF^2]
+\frac{2}{\ep}[-3\ca^2 + 2\ca\NTF + 4\cf\NTF]
\bigg\}
\nonumber\\
%&
& + \left(\frac{\alpha_s}{4\pi}\right)^3\bigg\{
\frac{1}{\ep^3}[-27\ca^3 + 54\ca^2\NTF - 36\ca\NTF^2 + 8\NTF^3]
\nonumber\\
%&
&
+ \frac{4}{\ep^2}(9\ca^3 - 12\ca^2\NTF - 12\ca\cf\NTF + 4\ca\NTF^2 + 
     8\cf\NTF^2)
\nonumber\\
%&
&
+\frac{1}{\ep}(-21\ca^3 + 20\ca^2\NTF + 52\ca\cf\NTF - 16\cf^2\NTF -
4\ca\NTF^2 -  
     24\cf\NTF^2)
\bigg\}\,.
%\\
%\nonumber\\
\label{eq::mgl3l}
%\end{eqnarray}
\end{align}
}
$\!\!$The \drbar{} quark mass renormalization constant is also independent of
any mass  parameter and is given by the following formula
{\allowdisplaybreaks
\begin{align}
%\begin{eqnarray}
Z_{m_q}&=1 - \frac{\alpha_s}{4\pi}\frac{1}{\ep} 2\cf
+ \left(\frac{\alpha_s}{4\pi}\right)^2\bigg\{
\frac{1}{\ep^2}[3\ca\cf +2 \cf^2 - 2\cf\NTF]
+ \frac{1}{\ep}[-3\ca\cf + 2\cf^2 + 2\cf\NTF]
\bigg\}
\nonumber\\
%&
&
 + \left(\frac{\alpha_s}{4\pi}\right)^3
\bigg\{\frac{1}{\ep^3}\bigg[-6\ca^2\cf - 6\ca\cf^2 -\frac{4}{3} \cf^3 + 
     (8\ca\cf + 4\cf^2)\NTF - \frac{8}{3}\cf\NTF^2
\bigg]
\nonumber\\
%&
&\frac{1}{\ep^2}\bigg[
10\ca^2\cf + 2\ca\cf^2 - 4\cf^3 + (-\frac{32}{3}\ca\cf -
\frac{20}{3}\cf^2)\NTF +  
     \frac{8}{3}\cf\NTF^2\bigg]
\nonumber\\
%&
&\frac{1}{\ep}\bigg[
-4\ca^2\cf + 4\ca\cf^2 - \frac{16}{3}\cf^3 + 
     \frac{8}{3}\cf\NTF^2 
\nonumber\\
%&
&
+ \NTF(\cf^2(\frac{32}{3} - 16\z3) +
     \ca\cf(-\frac{4}{3} +16 \z3)\bigg]
\bigg\}\,,
\label{eq::mq3l}
%\end{eqnarray}
\end{align}
}
$\!\!$where $\z3$ is Riemann's zeta function with $\zeta(3)= 1.20206\ldots$.
The results of Eqs.~(\ref{eq::as3l},\ref{eq::mq3l},\ref{eq::mgl3l}) are in
agreement with 
  Refs.~\cite{Jack:1996vg,Jack:1996qq}. Using Eqs. (\ref{eq::as3l}) and
  (\ref{eq::mgl3l}), it is an easy exercise to confirm the relation derived
in Ref.~\cite{Jack:1997pa} between 
   the anomalous dimension of  gluino mass and the
  gauge $\beta$-function that holds also in \dred{}. This result  is  similar to the  
NSVZ relation given in Eq.~(\ref{eq::rgi}) and holds to all orders in perturbation
theory. It reads
\begin{equation}
  \gamma^{{\tilde g}}_n  =(n+1) \beta_{n}\,
  \,.
\end{equation}
where $n$ denotes the number of loops. 
%It is a consequence of 
%\dred{} properties to respect SUSY in the lower orders of perturbation theory.

\subsection{\label{sec:drbarprime}Renormalization of the squark sector at three loops}

In this section, we report
 on the renormalization of the squark sector of SUSY-QCD up to three loop order
 within the \drbar{} scheme in the component field approach
 ~\cite{Hermann:2011ha}. 
These results are on the one side  important for the phenomenological
analyses aiming to predict the squark masses at the TeV scale with an
accuracy of the order of ${\cal O}(50~\mbox{GeV})$,  that is
required by the precision achieved in the current experimental searches
at the LHC. On the other side, they have also  genuine theoretical
significance, since they provide an independent confirmation of the
three-loop results obtained with the help of  the NSVZ
scheme~\cite{Avdeev:1997vx,Kazakov:2000ih,Jack:2003sx}.     

The calculations presented  in this section are performed in the framework of
SUSY-QCD with $n_q=5$ massless quarks and a massive top quark ($m_t$).
The scalar super partners of the latter has two mass eigenstates 
($m_{\tilde{t}_1}$ and $m_{\tilde{t}_2}$) which may
have different masses and thus a non-vanishing mixing angle occurs.
The super partners of the $n_q$ light quarks are assumed to have degenerate
masses ($m_{\tilde{q}}$) and vanishing mixing angles. A generalization to
a non-degenerate 
spectrum is possible in a straightforward way from the formalism for the top
squark sector which is discussed in detail in the following.

Unless stated otherwise all parameters in the following derivation are
\drbar{} quantities which depend on the renormalization scale $\mu$. For the
sake of compactness the latter is omitted. Bare quantities are marked by
a superscript ``(0)''. 
To define the framework, we 
start from the bare Lagrangian containing the  kinetic energy and the 
mass terms for the top squarks
\begin{eqnarray}
{\cal L}_{\tilde{t}}^{(0)} = \frac{1}{2}\partial_{\mu}(\stl^\ast,
\str^\ast)^{(0)}\partial^{\mu}
\left(\begin{array}{c}
\stl\\
\str
\end{array}\right)^{\!\!(0)}-\frac{1}{2}(\stl^\ast,
\str^\ast)^{(0)} ( {\cal M}_{{\tilde t}}^{2})^{(0)}
\left(\begin{array}{c}
\stl\\
\str
\end{array}\right)^{\!\!(0)}\,,
\label{eq::lbare}
\end{eqnarray}
where  $\stl$ and $\str$
denote the interaction eigenstates. The  top squark mass matrix
is given by 
\begin{align}
  {\cal M}^2_{\tilde{t}} &= 
  \left(
    \begin{array}{cc}
      m_t^2 + M_Z^2\big( \frac{1}{2} - \frac{2}{3} \sin^2\vartheta_W \big)
      \cos 2\beta  + M^2_{\tilde{Q}} & m_t \big( A_t - \mu_{\rm SUSY}
      \cot \beta \big) \\ 
      m_t \big( A_t - \mu_{\rm SUSY} \cot \beta \big) &  m_t^2 +
      \frac{2}{3} M^2_Z 
      \sin^2\vartheta_W \cos 2\beta  + M^2_{\tilde{U}} \\ 
    \end{array}
  \right) \nonumber \\  \nonumber \\
  &\equiv \left(
    \begin{array}{cc}
      m^2_{\tilde{t}_L} & m_t X_t \\
      m_t X_t  & m^2_{\tilde{t}_R} \\
    \end{array}
  \right)
  \label{eq::Mtil}
\end{align}
with $X_t = A_t - \mu_{\rm SUSY} \, \cot{\beta}$.
$A_t$ is the soft SUSY breaking trilinear coupling, and $M_{\tilde{U}}$ and
$M_{\tilde{Q}}$ are the soft SUSY breaking masses.

The  top squark mass eigenstates are  related to the interaction
eigenstates through the unitary transformation
\begin{eqnarray}
\left(\begin{array}{c}
\stu\\
\std
\end{array}\right)^{\!\!(0)} = {\cal R}_{{\tilde t}}^{(0)\, \dag}
\left(\begin{array}{c} 
\stl\\
\str
\end{array}\right)^{\!\!(0)}\,.
\label{UnitTransSquark} 
\end{eqnarray} 
 The unitary matrix ${\cal R}_{{\tilde t}}$ is defined through the 
diagonalization relation for the mass matrix ${\cal M}^2_{\tilde{t}}$
\begin{align}
  \left(
    \begin{array}{cc}
      m^2_{\tilde{t}_1} & 0\\
      0 & m^2_{\tilde{t}_2}\\
    \end{array}
  \right) = R^{\dagger}_{\tilde{t}}\, {\cal M}^2_{\tilde{t}}\,
  R_{\tilde{t}} \,. 
  \label{eq::Mtildiag}
\end{align}
The eigenvalues are the masses of the eigenstates $\tilde{t}_1$ and
$\tilde{t}_2$. They read
\begin{align}
  m^2_{\tilde{t}_{1,2}} = \frac{1}{2}\Bigg[ m^2_{\tilde{t}_L} +
  m^2_{\tilde{t}_R} \mp \sqrt{\Big( m^2_{\tilde{t}_L} 
    - m^2_{\tilde{t}_R} \Big)^2 + 4 m_t^2 X_t^2} \Bigg] \,.
\end{align}
The unitary transformation can be parameterized by the mixing angle 
\begin{align}
  R_{\tilde{t}} =
  \left(
    \begin{array}{cc}
      \cos\theta_t & -\sin\theta_t\\
      \sin\theta_t & \cos\theta_t\\
    \end{array}
  \right)\,,
\end{align}
with
\begin{align}
  \sin\big(2\theta_t \big) = \frac{2 m_t \, \big(A_t - \mu_{\rm SUSY}
    \, \cot \beta 
    \big)}{m^2_{\tilde{t}_1} - m^2_{\tilde{t}_2}} \,. 
\end{align}
The renormalization constants connected to the top squark are extracted
from the  
top squark propagator. At tree-level it is a diagonal
$2\times 2$ matrix which receives non-diagonal entries at loop-level.
In order to be able to write
down the renormalized top squark propagator we define the
renormalization constants 
as follows: The wave function renormalization constant is introduced through
the relation
\begin{eqnarray}
\left(\begin{array}{c}
\stu\\
\std
\end{array}\right)^{\!\!(0)} ={\cal Z}_{\tilde t}^{1/2} \left(\begin{array}{c}
\stu\\
\std
\end{array}\right)\,,\quad\mbox{with} \quad {\cal Z}_{\tilde t}^{1/2}=
\left(\begin{array}{cc} 
Z_{11}^{1/2}& Z_{12}^{1/2}\\
Z_{21}^{1/2}& Z_{22}^{1/2}
\end{array}\right)\,,
\end{eqnarray}
where it holds
${\cal Z}_{\tilde t}^{1/2} = \mathbf{I} +{\cal O}(\alpha_s)$. 
%Thus, 
%$ Z_{11}^{1/2} = 1 + {\cal  O}(\alpha_s)$,
%$Z_{22}^{1/2} = 1 + {\cal  O}(\alpha_s)$, 
%$Z_{12}^{1/2} = {\cal O}(\alpha_s)$
%and $Z_{21}^{1/2} =  {\cal  O}(\alpha_s)$. \\

In case of SUSY-QCD, the matrix ${\cal Z}_{\tilde t}^{1/2}$ has a
particularly symmetric form. This can be derived from the observation that the
left- and right-handed components of the top squark fields have the same
 renormalization constant for their  wave functions within SUSY-QCD
\begin{eqnarray}
\left(\begin{array}{c}
\tilde{t}_L
\\
\tilde{t}_R
\end{array}\right)^{(0)} = \tilde{Z}^{1/2}_2
\left(\begin{array}{c}
\tilde{t}_L\\
\tilde{t}_R
\end{array}\right)
\,.
\label{eq:wfqcd}
\end{eqnarray}
 Furthermore, if we introduce the
renormalization constant for the mixing angle via 
\begin{align}
  \theta^{(0)}_t = \theta_t + \delta\theta_t \label{dThetat}
  \,.
\end{align}
and make use of Eq.~(\ref{UnitTransSquark}) we obtain
\begin{align}
 {\cal Z}^{1/2}_{\tilde{t}} = \tilde{Z}^{1/2}_2
 \left( \label{squarkWFMaxtrix}
 \begin{array}{cc}
  \cos\delta\theta_t & \sin\delta\theta_t \\
  -\sin\delta\theta_t & \cos\delta\theta_t\\
 \end{array}
 \right) \,.
\end{align}

When supersymmetric electroweak (SUSY-EW) corrections are taken into account, Eq.~(\ref{eq:wfqcd})   
becomes
\begin{eqnarray}
\left(\begin{array}{c}
\tilde{t}_L
\\
\tilde{t}_R
\end{array}\right)^{(0)} = 
\left(\begin{array}{cc}
\tilde{Z}_L^{1/2} &0\\
0&\tilde{Z}_R^{1/2}
\end{array}\right)
\left(\begin{array}{c}
\tilde{t}_L\\
\tilde{t}_R
\end{array}\right)
\,.
\label{eq:wfew}
\end{eqnarray}
This assignment takes into account supersymmetric 
constraints~\cite{Hollik:2002mv} and is 
sufficient to absorb all divergences. As a consequence also the matrix
${\cal Z}_{\tilde t}^{1/2}$ 
has a more complicated structure and additional renormalization
conditions are required.

Furthermore, the mass matrix Eq.~(\ref{eq::Mtil}) has to be
renormalized. It can be parameterized as follows
\begin{align}
  \left( %\label{squarkMassMatrix}
    \begin{array}{cc}
      (m^{(0)}_{\tilde{t}_1})^2 & 0 \\
      0 & (m^{(0)}_{\tilde{t}_2})^2 \\
    \end{array}
  \right) \rightarrow 
  \left(
    \begin{array}{cc}
      m^2_{11}Z_{m_{11}} & m^2_{12}Z_{m_{12}}\\
      m^2_{21}Z_{m_{21}} & m^2_{22}Z_{m_{22}}\\
    \end{array}
  \right) \equiv {\cal M} \,,
  \label{eq::Mt}
\end{align}
where we require that the off-diagonal elements in the renormalized mass matrix
vanish. This ensure that the renormalized fields are the true mass
eigenstates. As a consequence, the counter-term $\delta\theta_t$ takes
care of the 
divergences in the self-energy contribution where a $\tilde{t}_1$ transforms
into a $\tilde{t}_2$ or vice versa. This can be seen in the explicit formulae
given below.
The diagonal elements of Eq.~(\ref{eq::Mt}) can be identified with the
renormalization constants of the masses 
\begin{equation}
  (m^{(0)}_{\tilde{t}_i})^2 
  = m_{ii}^2 Z_{m_{ii}}
  = m^2_{\tilde{t}_i} Z_{m_{\tilde{t}_i}} 
  \,.
\end{equation}

In order to formulate the renormalization conditions it
is convenient to consider the renormalized inverse top squark
propagator given by  
\begin{align}
  i{\cal S}^{-1}(p^2) = p^2\left({\cal Z}^{1/2}_{\tilde{t}}\right)^{\dagger}
  {\cal Z}^{1/2}_{\tilde{t}} 
  - \left({\cal Z}^{1/2}_{\tilde{t}}\right)^{\dagger}\left[ {\cal
      M} - \Sigma(p^2) \right] {\cal
    Z}^{1/2}_{\tilde{t}} 
  \label{squarkProp} 
\end{align}
where 
\begin{align}
  \Sigma(p^2) =
  \left(
    \begin{array}{cc}
      \Sigma_{11}(p^2) & \Sigma_{12}(p^2)\\
      \Sigma_{21}(p^2) & \Sigma_{22}(p^2)\\
    \end{array}
  \right) \,,
\end{align}
stands for the matrix of the squark self energies in the mass
eigenstate basis.

In the \drbar{} scheme the renormalization conditions read
\begin{equation}
  {\cal S}^{-1}_{ij}(p^2)\bigg|_{\rm pp} = 0 
  \,,
  \label{squarkRenoCondition}
\end{equation}
where ``pp'' stands for the ``pole part''.

In order to obtain explicit formulae for the evaluation of the
renormalization constants it is convenient to define perturbative expansions 
of the quantities entering Eq.~(\ref{squarkRenoCondition}). Up to three-loop
order we have
\begin{align}
  Z_k &= 1 + \left (\frac{\alpha_s}{\pi} \right)\delta Z_k^{(1)} + \left
  (\frac{\alpha_s}{\pi} \right)^2\delta Z_k^{(2)} 
  + \left (\frac{\alpha_s}{\pi} \right)^3\delta Z_k^{(3)} + {\cal
    O}(\alpha_s^4) \nonumber \,,\\ 
  \delta\theta_t &= \left (\frac{\alpha_s}{\pi} \right)\delta \theta_t^{(1)}
  + \left (\frac{\alpha_s}{\pi} \right)^2\delta\theta_t^{(2)} 
  + \left (\frac{\alpha_s}{\pi} \right)^3\delta \theta_t^{(3)} + {\cal
    O}(\alpha_s^4) \,,\nonumber\\ 
  \Sigma_{ij} &= \left (\frac{\alpha_s}{\pi} \right) \Sigma_{ij}^{(1)} +
  \left (\frac{\alpha_s}{\pi} \right)^2\Sigma_{ij}^{(2)} 
  + \left (\frac{\alpha_s}{\pi} \right)^3\Sigma_{ij}^{(3)} + {\cal
    O}(\alpha_s^4)\,,
\end{align}
where $i,j\in\{1,2\}$ and $k\in\{2,m_{\tilde{t}_1},m_{\tilde{t}_2}\}$.
Inserting these equations into~(\ref{squarkProp}) one can solve 
Eq.~(\ref{squarkRenoCondition}) iteratively order-by-order in $\alpha_s$.
At one-loop order one gets
\begin{eqnarray}
  \bigg\{\Sigma_{ii}^{(1)} - m^2_{\tilde{t}_{i}}\left(\delta \tilde{Z}^{(1)}_2
  + \delta Z^{(1)}_{m_{\tilde{t}_i}} \right)  + p^2\delta
  \tilde{Z}^{(1)}_2\bigg\} \Bigg|_{\rm pp} &=& 0  
  \,, \quad i=1,2  \,,\nonumber\\
  \bigg\{ \Sigma_{12}^{(1)} - \delta\theta_t^{(1)} \left( m^2_{\tilde{t}_{1}} -
  m^2_{\tilde{t}_{2}} \right) \bigg\} \Bigg|_{\rm pp} &=& 0 \,. 
  \label{squarkRenBeding1L}
\end{eqnarray}

The terms proportional to $p^2$ in the first equation
of~(\ref{squarkRenBeding1L}) are used to compute the wave function
renormalization constant which is independent of all occurring masses. Thus
they can be set to zero and one obtains
\begin{equation}
  \delta \tilde{Z}^{(1)}_2 
  = -\frac{1}{p^2} \Sigma_{11}^{(1)}(p^2) \Bigg|_{\rm pp}
  = -\frac{1}{p^2} \Sigma_{22}^{(1)}(p^2) \Bigg|_{\rm pp}
  \,.
\end{equation}
Once $\delta \tilde{Z}^{(1)}_2$ is known Eq.~(\ref{squarkRenBeding1L}) is used
to obtain $\delta Z^{(1)}_{m_{\tilde{t}_i}}$ keeping the mass dependence in
$\Sigma_{ii}^{(1)}$ (see below for more details).
The second equation of~(\ref{squarkRenBeding1L}) is used to obtain the 
renormalization constant of the mixing angle via
\begin{equation}
 \delta \theta_t^{(1)} = \frac{\Sigma_{12}^{(1)}}{ m^2_{\tilde{t}_{1}} -
   m^2_{\tilde{t}_{2}}} \Bigg|_{\rm pp} \,. 
\end{equation}

Proceeding to two loops we obtain the equations
\begin{align}
  &\Bigg[ \Sigma_{ii}^{(2)} + \delta \tilde{Z}_2^{(1)} \Sigma_{ii}^{(1)} -
  m^2_{\tilde{t}_{i}} \Big( \delta \tilde{Z}_2^{(2)}  
  +  \delta \tilde{Z}_2^{(1)}\delta Z^{(1)}_{m_{\tilde{t}_i}} +  \delta
  Z^{(2)}_{m_{\tilde{t}_i}}\Big) 
  + \delta \tilde{Z}^{(2)}_2 p^2 
  \nonumber  \\
  &+ (-1)^{(i+1)} \delta \theta_t^{(1)} \Big( -2 \Sigma_{12}^{(1)} + \delta
  \theta_t^{(1)} \big( m^2_{\tilde{t}_{1}} - m^2_{\tilde{t}_{2}} \big) \Big) 
  \Bigg] \Bigg|_{\rm pp} = 0 \, , \quad i=1,2 \,, 
  \label{squarkRenBeding2L}
\end{align}
\begin{align}
  &\Bigg[
  -\delta \theta_t^{(2)} \Big( m^2_{\tilde{t}_{1}} - m^2_{\tilde{t}_{2}} \Big)
  - \delta \theta_t^{(1)} \delta \tilde{Z}_2^{(1)} \Big( m^2_{\tilde{t}_{1}} 
  - m^2_{\tilde{t}_{2}} \Big) - \delta \theta_t^{(1)} \delta
  Z^{(1)}_{m_{\tilde{t}_1}}  m^2_{\tilde{t}_{1}} 
  + \delta \theta_t^{(1)} \delta Z^{(1)}_{m_{\tilde{t}_2}}
  m^2_{\tilde{t}_{2}} 
  \nonumber \\
  &+ \delta \theta_t^{(1)} \Sigma_{11}^{(1)} - \delta \theta_t^{(1)}
  \Sigma_{22}^{(1)} + \delta \tilde{Z}_2^{(1)} \Sigma_{12}^{(1)} +
  \Sigma_{12}^{(2)} 
  \Bigg] \Bigg|_{\rm pp} = 0 \,,
\end{align}
which are solved for $\tilde{Z}_2^{(2)}$, $\delta Z^{(2)}_{m_{\tilde{t}_i}}$
and $\delta \theta_t^{(2)}$ using the same strategy as at one-loop level. 

Similarly, at three-loop order we have
\begin{align}
  &\Bigg[ (-1)^{i+1} \, \bigg\{ \,  \Big(\delta \theta_t^{(1)}\Big)^2 \,
  \bigg( \delta \tilde{Z}_2^{(1)} \big( m^2_{\tilde{t}_{1}} -
  m^2_{\tilde{t}_{2}} \big) + 
  \delta Z^{(1)}_{m_{\tilde{t}_1}} m^2_{\tilde{t}_{1}} - \delta
  Z^{(1)}_{m_{\tilde{t}_2}} m^2_{\tilde{t}_{2}} - \Sigma_{11}^{(1)}
  +\Sigma_{22}^{(1)} \bigg) \nonumber \\ 
  & \quad \quad \quad + \delta \theta_t^{(1)} \bigg( 2\, \delta \theta_t^{(2)}
  \big( m^2_{\tilde{t}_{1}} - m^2_{\tilde{t}_{2}} \big)  
  - 2 \delta \tilde{Z}_2^{(1)} \Sigma_{12}^{(1)} - 2 \Sigma_{12}^{(2)} \bigg) 
  -2 \delta \theta_t^{(2)} \Sigma_{12}^{(1)} \bigg\}\nonumber \\
  &+\delta \tilde{Z}_2^{(1)} \bigg( \Sigma_{ii}^{(2)} - \delta
  Z^{(2)}_{m_{\tilde{t}_i}} m^2_{\tilde{t}_{i}} \bigg)  
  - \delta \tilde{Z}_2^{(2)} \delta Z^{(1)}_{m_{\tilde{t}_i}}
  m^2_{\tilde{t}_{i}} + \delta \tilde{Z}_2^{(2)} \Sigma_{ii}^{(1)} -  
  \delta \tilde{Z}_2^{(3)} m^2_{\tilde{t}_{i}} + \delta \tilde{Z}_2^{(3)} p^2
  \nonumber \\ 
  & - \delta Z^{(3)}_{m_{\tilde{t}_i}} m^2_{\tilde{t}_{i}} + \Sigma_{ii}^{(3)} 
  \Bigg] \Bigg|_{\rm pp} = 0 \, , \quad i=1,2 \, ,
  \label{eq::ren3l_1}
\end{align}
\begin{align}
  &\Bigg[
  \,  \delta \theta_t^{(1)} \, \bigg(- \delta \tilde{Z}_2^{(1)} \delta
  Z^{(1)}_{m_{\tilde{t}_1}} m^2_{\tilde{t}_{1}}  
  + \delta \tilde{Z}_2^{(1)} \delta Z^{(1)}_{m_{\tilde{t}_2}}
  m^2_{\tilde{t}_{2}}  
  + \delta \tilde{Z}_2^{(1)} \Sigma_{11}^{(1)} - \delta \tilde{Z}_2^{(1)}
  \Sigma_{22}^{(1)} \nonumber \\ 
  & \quad \quad \quad - \delta \tilde{Z}_2^{(2)} \big( m^2_{\tilde{t}_{1}} -
  m^2_{\tilde{t}_{2}} \big)  
  - \delta Z^{(2)}_{m_{\tilde{t}_1}} m^2_{\tilde{t}_{1}} + \delta
  Z^{(2)}_{m_{\tilde{t}_2}} m^2_{\tilde{t}_{2}} 
  + \Sigma_{11}^{(2)} - \Sigma_{22}^{(2)} \bigg) \nonumber \\
  & + \delta \theta_t^{(2)} \bigg( - \delta \tilde{Z}_2^{(1)} \big(
  m^2_{\tilde{t}_{1}} - m^2_{\tilde{t}_{2}} \big)  
  - \delta Z^{(1)}_{m_{\tilde{t}_1}} m^2_{\tilde{t}_{1}} + \delta
  Z^{(1)}_{m_{\tilde{t}_2}} m^2_{\tilde{t}_{2}} + \Sigma_{11}^{(1)} -
  \Sigma_{22}^{(1)} \bigg) \nonumber \\ 
  &- \delta \theta_t^{(3)} \big( m^2_{\tilde{t}_{1}} - m^2_{\tilde{t}_{2}}
  \big)  
  + \delta \tilde{Z}_2^{(1)} \Sigma_{12}^{(2)}+ \delta \tilde{Z}_2^{(2)}
  \Sigma_{12}^{(1)} + \Sigma_{12}^{(3)} 
  +\frac{2}{3} \, \Big(\delta \theta_t^{(1)}\Big)^3 \, \big(
  m^2_{\tilde{t}_{1}} - m^2_{\tilde{t}_{2}} \big) \nonumber \\ 
  &-2 \Big(\delta \theta_t^{(1)}\Big)^2 \Sigma_{12}^{(1)}
  \Bigg] \Bigg|_{\rm pp} = 0 \,.
  \label{eq::ren3l_2}
\end{align}

\begin{figure}[t]
  \centering
  \includegraphics[width=\linewidth]{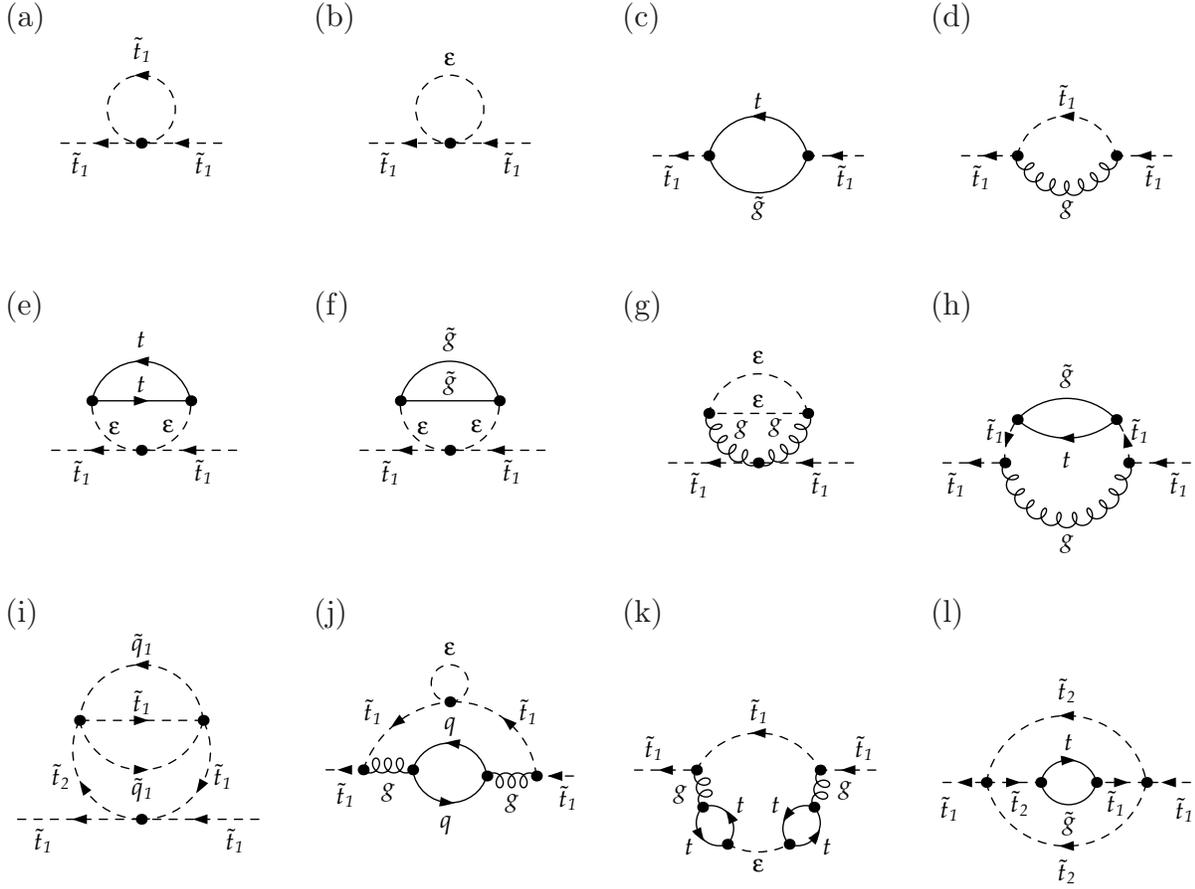}
  \caption[]{\label{fig::dias}Sample diagrams contributing to 
    $\Sigma_{11}$ at one, two and three loops.
    The symbols $t$, $\tilde{t}_i$, $g$, $\tilde{g}$ and
    $\epsilon$ denote top quarks, top squarks, gluons, gluinos,
    and $\epsilon$ scalars, respectively.
    }
\end{figure}

Sample diagrams contributing to $\Sigma_{11}$ up to three loops can be found
in Fig.~\ref{fig::dias}; the contributions to $\Sigma_{12}$ and $\Sigma_{22}$
look very similar.  Once the quantities $\Sigma_{11}$, $\Sigma_{12}$ and
$\Sigma_{22}$ are known to three-loop order it is possible to extract the
renormalization constants for the squark wave function and mass and the mixing
angle from Eqs.~(\ref{eq::ren3l_1}) and~(\ref{eq::ren3l_2}).

As compared to the corresponding self-energy contributions for fermions or
gauge bosons, which after proper projection only lead to logarithmically
divergent integrals, the quantities in the above equations have mass dimension
two. As a consequence the renormalization constants of the squark masses and
the mixing angles depend on the occurring masses, even in a minimal
subtraction scheme like \drbar{}.  At three-loop order an exact evaluation of
the corresponding integrals is not possible. It is nevertheless possible to
reconstruct the complete dependence on the occurring masses using repeated
asymptotic expansions and in addition some knowledge about the structure of
the final result. Thus, one has to keep during the calculation non
vanishing squark, gluino and the top quark masses and chose convenient
hierarchies between them.
  For the asymptotic expansion (see, e.g.,
Ref.~\cite{Smirnov:2002pj}) one can use  {\tt
  exp}~\cite{Harlander:1997zb,Seidensticker:1999bb}. As a result only
one-scale integrals up to three 
loops appear which can be evaluated with the packages {\tt
  MINCER}~\cite{Larin:1991fz} and {\tt
  MATAD}~\cite{Steinhauser:2000ry}.

After the calculation of the bare self energies one has to renormalize all
occurring parameters in the \drbar{} scheme. For the three-loop
calculation one needs the counter-terms for 
$\alpha_s$, $m_t$, $m_{\tilde{g}}$, $m_{\tilde{t}_i}$, $\theta_t$ and
$m_{\epsilon}$ to two-loop order and the one  for $m_{\tilde{q}}$ to
one-loop approximation. Furthermore, also the QCD gauge parameter
has to be renormalized to two loops since it appears in the results for the
wave function anomalous dimensions.

At this point some comments on the treatment of the $\epsilon$ scalar mass,
$m_{\epsilon}$, are in order. In practice there are two
renormalization schemes for $m_{\epsilon}$ which are frequently used, the
\drbar{} and on-shell scheme. In the latter one requires that the renormalized
mass vanishes in each order in perturbation theory whereas in the \drbar{}
prescription only the pole parts are subtracted by the renormalization
constant. In the \drbar{} scheme it is important to keep $m_{\epsilon}$
different from 
zero since the renormalization group equations for the squark masses and
$m_{\epsilon}$ are coupled. A non-vanishing $\epsilon$-scalar
mass in intermediate steps is also required for the computation of the 
anomalous dimensions in the \drbarprime{} scheme~\cite{Jack:1994rk} (see
below) which was constructed in order to disentangle the running of
$m_{\epsilon}$ from the one of the squark parameters.

 In the following, we present only the results derived in the scheme where the  $\epsilon$
scalar mass is renormalized in \drbar{} scheme. 
%Afterwards,  the difference to the
%on-shell& and \drbarprime{} schemes will be discussed.
The two-loop results for the renormalization constants of the top  squark mass
$\Mstu$ read
%The results for the  renormalization constants of the top  squark mass
%$\Mstu$ up to two loops   read
{\allowdisplaybreaks
\begin{align}
  \Mstu^2\delta Z_{{m}_{\tilde{t}_1}}^{(1)} &= \cf
  \left(-\Mgl^2 - \Mt^2 + \Mgl\Mt\Smt +
    \frac{\Mstd^2-\Mstu^2}{4}\Smt^2\right)\frac{1}{\ep}\,,\nonumber\\
  \Mstu^2\delta Z_{{m}_{\tilde{t}_1}}^{(2)} &= \bigg\{
  \cf^2\bigg[\frac{\Cmt^2\Mgl^2\Mt^2}{\dms} + \frac{(1 + \Cmt^2)\Smt^2(\dms) +
    8\Mt^2}{16} -  \frac{(1 + \Cmt^2)\Mgl\Mt\Smt}{2} \bigg] 
\nonumber\\
&+
\ca\cf \bigg[\frac{9\Mgl^2}{8} + 3\frac{\Smt^2(\dms) + 4\Mt^2}{32}
  - \frac{3\Mgl\Mt\Smt}{4}\bigg]
\nonumber\\
&+
\cf T_f\bigg[\frac{-3\Mgl^2}{4} - \frac{\Smt^2(\dms) + 4\Mt^2}{16} 
+ \frac{\Mgl\Mt\Smt}{2} \bigg]
\bigg\}\frac{1}{\ep^2}
\nonumber\\
&+ \bigg\{
\cf^2\bigg[\frac{3\Mgl^2}{4} + \frac{\Smt^2(\dms) + 4\Mt^2}{16} 
- \frac{\Mgl\Mt\Smt}{2} \bigg] 
\nonumber\\
&+
\ca\cf \bigg[\frac{-11\Mgl^2}{8} - 3\frac{\Smt^2(\dms) +
    4\Mt^2}{32} + \frac{3\Mgl\Mt\Smt}{4}\bigg] 
\nonumber\\
&+\cf T_q\bigg[\frac{3\Mgl^2}{4} + \frac{\Smt^2(\dms) + 8\msquark^2
    + 4\Mt^2}{16} - \frac{\Mgl\Mt\Smt}{2}\bigg] 
\nonumber\\
&+\cf T_t
\bigg[
\frac{3\Mgl^2}{4} + \frac{\Smt^2(\dms) + 4\Mstu^2 + 4\Mstd^2 - 4\Mt^2}{16} - 
 \frac{\Mgl\Mt\Smt}{2}
\bigg]
\bigg\}\frac{1}{\ep}
\nonumber\\
&+\mes^2 \left(-\ca\cf\frac{3}{8} + \cf T_f \frac{1 }{4} 
\right)\frac{1}{\ep}\,, 
\label{eq::mst1}
\end{align}
}
$\!\!$where we have introduced the abbreviations $T_l=n_l I_2(R)$, with
$l=f,q,t$ and  $c_{nt} = \cos(n\theta_t)$ and $s_{nt} = \sin(n\theta_t)$.
$n_q$ denotes the number of light quark flavours and takes in this case
the value 
$n_q=5$. $n_t=1$ has been introduced for convenience and it holds $n_f=n_q+n_t$. Furthermore
   $\mes$ denotes the \drbar{} renormalized \epscalar{} mass. 
The corresponding results for $\Mstd$ can be derived
from Eq.~(\ref{eq::mst1})
by interchanging $\Mstu$ and $\Mstd$ and changing the sign of
$\theta_t$.

Finally, for the mixing angle we have
{\allowdisplaybreaks
\begin{align}
(&\dms)\delta\theta_t^{(1)}=\cf \Cmt
\left(\Mgl\Mt - \frac{\Smt(\Mstu^2 - \Mstd^2)}{4}\right)\frac{1}{\ep}\,,
\nonumber\\
(&\dms)\delta\theta_t^{(2)}=\bigg\{
\cf^2\Cmt\bigg[(\Smt^2 - \Cmt^2)\left(\frac{\Mgl\Mt}{2} -
  \frac{\Smt(\dms)}{16}\right) - \frac{2\Smt\Mgl^2\Mt^2}{\dms} 
\bigg]\nonumber\\
&+
\cf\ca\Cmt\bigg[\frac{-3\Mgl\Mt}{4} + \frac{3\Smt(\dms)}{32}
\bigg]
+
\cf T_f \Cmt\bigg[\frac{\Mgl\Mt}{2} - \frac{\Smt(\dms)}{16}
\bigg]
\bigg\}\frac{1}{\ep^2}\nonumber\\
&+\bigg\{
\cf^2\Cmt\bigg[-\frac{\Mgl\Mt}{2} + \frac{\Smt(\dms)}{16}\bigg]+
\cf\ca\Cmt\bigg[\frac{3\Mgl\Mt}{4} - \frac{3\Smt(\dms)}{32}
\bigg]\nonumber\\
&+
\cf T_f \Cmt\bigg[-\frac{\Mgl\Mt}{2} + \frac{\Smt(\dms)}{16}
\bigg]
\bigg\}\frac{1}{\ep}\,.
\label{eq::mixa2l}
\end{align}
}
$\!\!$The three-loop results are also available in electronic
form~\cite{Hermann:2011ha}, 
but they are  too lengthy to be explicitly given in this review.

In the case of degenerate squark masses, one can take naively the limit   
$m_{{\tilde t}_2}\to m_{{\tilde t}_1}$ in Eqs.~(\ref{eq::mst1}).
 Furthermore one has to nullify the mixing angle.
The quantities $\delta\theta_t^{(1,2)}$ are not defined in the  mass-degenerate
case which is reflected by the fact that the limit $m_{{\tilde t}_2}\to
m_{{\tilde t}_1}$ does not exist in 
Eqs.~(\ref{eq::mixa2l}).
         
For completeness let us also provide the three-loop result for mass-degenerate
squarks which is given by
{\allowdisplaybreaks
\begin{align}
%\begin{eqnarray}
\msq^2 Z_{\msq}&= \msq^2 - \frac{\alpha_s}{\pi}\frac{1}{\ep} \cf \Mgl^2 + 
\left(\frac{\alpha_s}{\pi}\right)^2\frac{1}{16}\bigg\{\frac{2}{\ep^2} 
 \cf ( 9  \ca   - 6    \NTF )\Mgl^2
\nonumber\\
%&
&
+ \frac{1}{\ep}\bigg[
   4 \cf  [2 T_q \msq^2 + T_t   (\Mstu^2 + \Mstd^2 - 2  M_t^2) ]
\nonumber\\
%&
&
+ 
       [2\cf (-11 \ca + 6 \cf) + 12 \cf \NTF]  \Mgl^2
+ 
       (-6 \ca \cf + 4\cf \NTF) \mes^2 
\bigg]
\bigg\}
\nonumber\\
%&
&
\left(\frac{\alpha_s}{\pi}\right)^3\frac{1}{64}\bigg\{\frac{8}{\ep^3}\cf \bigg[
  -9 \ca^2  + 12 \ca  \NTF - 
      4 \NTF^2\bigg]\Mgl^2
%\nonumber\\
%&&
+\frac{1}{\ep^2}\bigg[
8\cf T_q(-3 \ca + 2\NTF)\Msq^2
\nonumber\\
%&
&
+ 4\cf\NTF (-3\ca + 2\NTF) (\Mstu^2 + \Mstd^2 - 2M_t^2)
%\nonumber\\
%&&
+2\cf (3\ca - 2\NTF)^2  \mes^2\bigg] + 
\nonumber\\
%&
&
+\frac{1}{\ep}\bigg[\cf(5\ca - 2\cf + 2\NTF)[T_q \Msq^2
%\nonumber\\
%&&
+4 T_t(\Mstu^2 + \Mstd^2 - 2M_t^2)]
\nonumber\\
%&
&
+8\cf\Mgl^2[-10\ca^2 + 7\ca\cf - 8\cf^2
 + 4\NTF^2 + 
   6\NTF\cf(3 - 4\z3) + 24\NTF\ca\z3]
\nonumber\\
%&
&
+2\cf(3\ca - 2\NTF)(-5\ca + 2\cf + 2\NTF)\mes^2
%\nonumber\\
%&&
\bigg]
\bigg\}
\,,
\label{eq::msq3l}
%\end{eqnarray}
\end{align}
}
$\!\!$where we have used the above mentioned abbreviations  $T_l=n_l I_2(R)$, with
$l=f,q,t$. The terms that do not involve $T_t$ can be obtained from
$Z_{{m}_{\tilde{t}_1}}$
by setting $m_{\tilde{t}_2}=m_{\tilde{t}_1}$, $m_t=0$ and $\theta_t=0$. 

As mentioned before, the $\epsilon$
scalar mass needs to be renormalized at two loops within the 
\drbar{} scheme, in order to obtain the three-loop renormalization
constants for squark masses and  mixing angles. The corresponding
renormalization constant is given by 
{\allowdisplaybreaks
\begin{align}
%\begin{eqnarray} 
Z_{\mes}&=
1 + \aspi \frac{1}{\ep}\bigg\{\frac{-3}{4} \cA +\frac{1}{2} T_f + 
    \bigg[-\frac{\cA}{2} \Mgl^2 + 2 T_q \msq^2  + T_t (\Mstu^2 
+ \Mstd^2 - 2 m_t^2) 
    \bigg ]\frac{1}{2\mes^2}\bigg\}
\nonumber\\ 
%&
&+ 
 \left(\aspi\right)^2 \bigg\{
\frac{1}{\ep^2}\bigg[
\frac{9}{16} \cA^2 - \frac{3}{4} \cA  T_f + 
     \frac{1}{4} T_f^2 
+ \bigg(\frac{3 \cA^2 -2\cA T_f -2 \cR T_f}{4}\Mgl^2
\nonumber\\ 
%&
&
 -  \frac{3 \cA T_q -2 T_f T_q}{4} \msq^2 -
\frac{3\cA T_t-2T_f T_t}{8} (\Mstu^2 + \Mstd^2 - 2 m_t^2)\bigg)\frac{1}{\mes^2}
\bigg]
\nonumber\\ 
%&
&
\frac{1}{\ep}\bigg[-\frac{3}{8}\cA^2+\frac{1}{4}\cA T_f
 +\bigg(
-\frac{5\cA^2-2 \cA T_f -4 \cR T_f}{8}\Mgl^2 
\nonumber\\ 
%&
&
+\frac{\cA T_q}{2}\msq^2+
\frac{\cA T_t}{4}(\Mstu^2 + \Mstd^2 - 2 m_t^2)
\bigg)\frac{1}{\mes^2}
\bigg]
\bigg\}\,.
\label{eq::zmes}
%\end{eqnarray}
\end{align}
}
$\!\!$Let us detail at this point on  the choice of scheme. When
 computing the
anomalous dimensions for the physical parameters, one has 
to consider the combined set of differential equations of all \drbar{}
parameters appearing in the corresponding renormalization constants.
This concerns
in particular the unphysical $\epsilon$-scalar mass which means
that although $m_\epsilon$ is set to zero at one scale it is
different from zero once this scale is changed. A way out from this
situation is to renormalize the $\epsilon$ scalar 
mass on-shell and set the renormalized mass $\Mes$ to zero. However, 
this scheme might become quite involved in  practice, because of the on shell
two-loop diagrams that have to be computed. Alternatively one could
shift the squark masses by a finite term which is chosen such that the
$\epsilon$ scalar decouples from the system of differential equations.
The resulting renormalization scheme is called \drbarprime{} scheme
and has been suggested in Ref.~\cite{Jack:1994rk}. For this calculation
the finite shift is needed up to two loops and  is given
by~\cite{Jack:1994rk,Martin:2001vx}
\begin{align}
  m^2_{\tilde{f}} \rightarrow  m^2_{\tilde{f}} - \frac{\alpha_s}{\pi} \,
  \frac{1}{2} \, C_F \, m^2_{\epsilon} 
  + \left( \frac{\alpha_s}{\pi} \right)^2 \, C_F \, m^2_{\epsilon} \, \left(
    \frac{1}{4} \, T_f \, \left( n_q + n_t \right) + \frac{1}{4} \, C_F
 - \frac{3}{8} \, C_A
  \right)
  \,,
\label{eq::fshift}
\end{align}
where $f=t$ or $f=q$.

At the end of this section we want to discuss briefly the numerical
impact of the higher order corrections on the squark masses. If one
chooses the SUSY mass parameters  of the order of ${\cal
  O}(1~\mbox{TeV})$, one observes a moderate shift of a few GeV when
going from one to two loops. After switching on the three-loop terms, however,
the squark masses are decreased by about 40~GeV which is approximately an
order of magnitude larger than the two-loop corrections. Nevertheless it
corresponds to a shift in the masses of about 3\% which is a reasonable amount
for a three-loop SUSY-QCD term.  Our observation coincides with the findings
of Ref.~\cite{Jack:2003sx} where also relatively large three-loop
corrections for
the squarks have been identified.

%%%%%%%%%%%%%%%%%%%%%%%%%%%%%%%%%%%%%%%%%%%%%%%%%%%%%%%%%%%%
%%%%%%%%%%%%%%%%%%%%%%%%%%%%%%%%%%%%%%%%%%%%%%%%%%%%%%%%%%%%

 \section{\label{sec:3lsm} The SM gauge beta functions 
to three loops} 

In this section we report on the recent calculation of the three-loop
gauge beta functions of the SM. In contrast to the supersymmetric
theories, the SM  beta functions to three loops have been computed
only last year. At this point, it becomes probably clear the importance  of 
all order relations for the anomalous dimensions of supersymmetric
theories valid in special regularization schemes. In the absence of
SUSY and its holomorphic properties, one has to derive the anomalous
dimension from a pure diagrammatic computation, which at the three-loop
level becomes computationally quite involved.

The  SM  beta functions are important tools that allow us to  relate theory
predictions for various parameters   at different energy scales. An
important example in this respect is 
the inspection of the gauge coupling unification at high energies, for which 
precise experimental data of the couplings at the electroweak scale combined
with accurate calculations of the RGEs yields
precise predictions. 

The computation of the beta functions of gauge theories has a long
history. The one-loop beta functions in gauge theories along with the discovery
  of asymptotic freedom have been presented in
  Refs.~\cite{Gross:1973id,Politzer:1973fx}. 
The computation of the corresponding two-loop corrections followed a few
years later in a series of papers. Namely, for gauge theories without
fermions the results were computed in
Refs.~\cite{Jones:1974mm,Tarasov:1976ef}, those for gauge theories with
fermions neglecting Yukawa 
    couplings in Refs.~\cite{Caswell:1974gg,Egorian:1978zx,Jones:1981we}
    and considering also Yukawa    couplings in
    Ref.~\cite{Fischler:1981is}. The two-loop gauge coupling beta
    functions in an arbitrary quantum field 
  theory have been considered in
  Ref.~\cite{Machacek:1983tz,Jack:1984vj}. At the three-loop order, the
  first computed contributions to the gauge beta functions were those
  induced through the
  scalar self-interactions in
  Refs.~\cite{Curtright:1979mg,Jones:1980fx}. An important contribution
  to the field was the computation of the three-loop beta function in
  QCD~\cite{Tarasov:1980au,Larin:1993tp}. Yukawa contributions to it
  have been obtained in Ref.~\cite{Steinhauser:1998cm}. The
  generalization of these results to a general quantum field theory based on a
  single gauge group has been achieved in
  Ref.~\cite{Pickering:2001aq}. For QCD, even the four-loop corrections
  are known from Refs.~\cite{vanRitbergen:1997va,Czakon:2004bu}. 
In the following we concentrate on the calculation of the beta functions
for the three gauge couplings of the SM up to three loops in the
\msbar{} scheme. They have been computed for the first time in
Ref~\cite{Mihaila:2012fm} and confirmed by an independent calculation in
Ref.~\cite{Bednyakov:2012rb}.

 Let us in a first step fix the notation.
 We denote the three
gauge couplings by $\alpha_1$, $\alpha_2$ and $\alpha_3$ and adopt a
$SU(5)$-like normalization. They are  related to the quantities
usually used in the SM by the all-order relations
\begin{eqnarray}
  \alpha_1 &=& \frac{5}{3}\frac{\alpha_{\rm QED}}{\cos^2\theta_W}\,,\nonumber\\
  \alpha_2 &=& \frac{\alpha_{\rm QED}}{\sin^2\theta_W}\,,\nonumber\\
  \alpha_3 &=& \alpha_s\,,
\end{eqnarray}
where $\alpha_{\rm QED}$ is the fine structure constant, $\theta_W$ the weak
mixing angle and $\alpha_s$ the strong coupling. 

The SM Yukawa interactions are described by (see, e.g., Chapter~11 of
Ref.~\cite{Nakamura:2010zzi})
\begin{equation}
  {\cal L}_{\rm Yukawa} = - \bar{Q}_i^LY^U_{ij} \epsilon H^{\star} u_j^R -
  \bar{Q}_i^LY^D_{ij} H d_j^R
 - \bar{L}_i^LY^L_{ij} H l_j^R+ \text{h.c.}\,,
\end{equation}
where $Y^{U,D,L}$ are complex $3\times 3$ matrices, $i,j$ are generation
labels, $H$ denotes the Higgs field and $\epsilon$ is the $2 \times 2$
antisymmetric tensor. $Q^L,L^L$ are the left-handed quark and lepton
doublets, and $u^R,d^R,l^R$ are the right-handed up- and down-type
quark and lepton singlets, respectively. The physical mass-eigenstates
are obtained by diagonalizing $Y^{U,D,L}$ by six unitary matrices
$V_{L,R}^{U,D,L}$ as follows
\begin{equation}
 \tilde{Y}^{f}_{\rm diag}=V_L^f Y^f V_R^{f\dagger}\,, \quad f=U,D,L\,.
 \label{eq::yuk1}
\end{equation}
As a result the charged-current $W^{\pm}$  couples to the physical quark
states 
with couplings parametrized by the Cabibbo-Kobayashi-Maskawa (CKM) matrix
$V_{CKM}\equiv V_L^UV_L^{D\dagger}$.
We furthermore introduce the notation
\begin{equation}
  \hat{T} = \frac{1}{4\pi} Y^U {Y^U}^{\dagger},\ \hat{B} =
  \frac{1}{4\pi}  Y^D {Y^D}^{\dagger},\ \hat{L} = \frac{1}{4\pi}  Y^L
  {Y^L}^{\dagger}. 
 \label{eq::yuk2}
\end{equation}
Of course, only traces over products of Yukawa matrices can occur because
they arise from closed fermion loops.  Using Eqs.~(\ref{eq::yuk1})
and~(\ref{eq::yuk2}) it is straightforward to see that  only traces of
diagonal matrices have to be taken 
except for $\text{tr}\hat{T}\hat{B}$ which is given by
\begin{equation}
  \text{tr}\hat{T}\hat{B} = \text{tr}\left[\left(\begin{matrix}
        \alpha_u & 0 & 0 \\
        0        & \alpha_c & 0 \\
        0 & 0 & \alpha_t
      \end{matrix}
    \right) V_{\text{CKM}}
    \left(\begin{matrix}
        \alpha_d & 0 & 0 \\
        0        & \alpha_s & 0 \\
        0 & 0 & \alpha_b
      \end{matrix}
    \right) V_{\text{CKM}}^{\dagger}\right].
  \label{eq:ckm}
\end{equation}
%The addition of a fourth generation of fermions to the SM particle content can
%be also easily accounted for by this general notation. In this case, the
%Yukawa matrices become $4\times 4$ dimensional. 

The Yukawa couplings
are related to the SM parameters via   the tree-level relations
\begin{eqnarray}
\alpha_x &=& \frac{\alpha_{\rm QED} m_x^2}{2 \sin^2\theta_W M_W^2}\,,
\quad \mbox{with} \quad x=t,b,\tau, c, s,\ldots \,,
\end{eqnarray}
where $m_x$ and $M_W$ are the fermion and W boson mass, respectively.

We denote the Higgs boson self-coupling by $\hat{\lambda}$, where the
Lagrange density contains the following term
\begin{eqnarray}
{\cal L}= \ldots -(4\pi\hat{\lambda}) (H^\dagger H)^2+ \ldots
\end{eqnarray}
describing the quartic Higgs boson self-interaction.

The beta functions are obtained by calculating the renormalization constants
relating bare and renormalized couplings via the relation
\begin{eqnarray}
  \alpha_i^{\text{bare}} &=&
  \mu^{2\epsilon}Z_{\alpha_i}(\{\alpha_j\},\epsilon)\alpha_i
  \,.
  \label{eq::alpha_bare}
\end{eqnarray}
Taking into account that $\alpha_i^{\text{bare}}$ does not depend on
$\mu$ and taking into account that $Z_{\alpha_i}$ may depend on all  couplings
leads to the following formula
\begin{eqnarray}
  \label{eq::renconst_beta}
  \beta_i &=& 
  -\left[\epsilon\frac{\alpha_i}{\pi}
    +\frac{\alpha_i}{Z_{\alpha_i}}
    \sum_{{j=1},{j \neq i}}^7
    \frac{\partial Z_{\alpha_i}}{\partial \alpha_j}\beta_j\right]
  \left(1+\frac{\alpha_i}{Z_{\alpha_i}}
    \frac{\partial Z_{\alpha_i}}{\partial \alpha_i}\right)^{-1}
  \,,
\end{eqnarray}
where $i=1,2$ or $3$. We furthermore set $\alpha_4=\alpha_t$,
$\alpha_5=\alpha_b$, $\alpha_6=\alpha_{\tau}$ and
$\alpha_7=\hat{\lambda}$ and neglect the rest of Yukawa couplings.

The first term in the first factor of Eq.~(\ref{eq::renconst_beta}) originates
from the term $\mu^{2\epsilon}$ in Eq.~(\ref{eq::alpha_bare}) and vanishes in
four space-time dimensions. 
The second term in the first factor
contains the beta functions of the remaining six couplings of the SM. Note
that (for the gauge couplings) the one-loop term of $Z_{\alpha_i}$ only
contains $\alpha_i$, whereas at two loops all couplings are present except
$\hat{\lambda}$. The latter appears for the first time at
three-loop level. As a
consequence, it is necessary to know $\beta_j$ for $j=4,5,6$ to
 one-loop order and only the $\epsilon$-dependent term for
  $\beta_7$, namely $\beta_7 = - \epsilon \alpha_7/\pi$.
From the second term in the first factor and the
second factor of Eq.~(\ref{eq::renconst_beta}) one can read
off that three-loop corrections to $Z_{\alpha_i}$ are required 
for the computation of $\beta_i$ to the same loop order.  

In the $\overline{\rm MS}$ scheme the beta functions are mass independent.
This allows us to use the SM in the unbroken phase as a
framework for our calculation. 
In principle each vertex containing the coupling
$g_i=\sqrt{4\pi\alpha_i}$ can be used in 
order to determine the corresponding renormalization constant 
via the relation~(\ref{eq::ZZZ}). In order to compute the individual renormalization constants entering
Eq.~(\ref{eq::ZZZ}) one can proceed as outlined in the previous sections.
%\begin{eqnarray}
%  Z_{\alpha_i} &=& \frac{(Z_{\text{vrtx}})^2}{\prod_k Z_{k,{\text{wf}}}}\,, 
%  \label{eq::Zalpha}
%\end{eqnarray}
%where $Z_{\text{vrtx}}$ stands for the renormalization constant of the
%vertex and $Z_{k,{\text{wf}}}$ for the wave function renormalization constant;
%$k$ runs over all external particles. 
%Sample diagrams can be found in
%Fig~\ref{fig:smbetadiag}.

%\begin{figure}[tb]
%  \begin{center}
%\end{center}
%\end{figure}

% The underlying formula can be written in the
%form
%\begin{eqnarray}
%  Z_{\Gamma} &=& 1 - K_\epsilon\left( Z_{\Gamma} \Gamma \right)
%  \,,
%  \label{eq::Z}
%\end{eqnarray}
%where $\Gamma$ represents the two- or three-point function corresponding to
%the renormalization constant $Z_{\Gamma}$ and the operator $K_\epsilon$ extracts the
%pole part of its argument.
%From the structure of Eq.~(\ref{eq::Z}) it is clear that $Z_{\Gamma}$ is computed
%Forder-by-order in perturbation theory in a recursive way. It
%Fis understood that the bare parameters 
%Fentering $\Gamma$  on 
%Fthe right-hand side are replaced by the renormalized ones
%Fbefore applying $K_\epsilon$. The corresponding counter-terms are only needed to
%Flower loop orders than the one which is requested 
%Ffor $\Gamma$.

A second method that can be used to get an independent result for the
renormalization constants of the gauge  couplings is a calculation in
the background field gauge (BFG)~\cite{Abbott:1980hw,Denner:1994xt}.  The
basic idea of the BFG is the splitting of all gauge fields
in a ``quantum'' and a ``classical''
part where in practical calculations the latter only occurs as external
particle.

The BFG has the advantage that Ward identities guarantee
that renormalization constants for gauge couplings can be obtained from the
exclusive knowledge of the corresponding wave function renormalization
constant. Thus we have the following formula
\begin{eqnarray}
  Z_{\alpha_i} &=& \frac{1}{Z_{A_i,\text{wf}}}
  \,,
  \label{eq::Z_BFG}
\end{eqnarray}
where $A$ denotes the gauge boson corresponding to the gauge coupling
$\alpha_i$.\\
In the BFG calculation it is advisable to adopt Landau gauge in order to
avoid the 
renormalization of the gauge parameters $\xi_i$. However, it
is not possible to choose Landau gauge from the very beginning since some
Feynman rules for vertices involving a background gauge boson contain terms
proportional to $1/\xi_i$ where $\xi_i=0$ corresponds to Landau gauge.  To
circumvent this problem one has to evaluate the bare integrals for
arbitrary gauge 
parameters. In the final result all inverse powers of $\xi_i$ cancel and thus
the limit $\xi_i=0$ can be taken at the bare level.

An important issue in the present 
calculation is the treatment of $\gamma_5$ within dimensional regularization. 
Non-trivial contributions may arise if in the course of the
calculation two fermion traces occur where both of them contain an
odd number of $\gamma_5$ matrices and four or more $\gamma$
matrices. It is straightforward to see that the three-point Green
functions that are required for this computation contain at most
one-loop triangle sub-diagrams.~\footnote{  Three-point Green
functions  involving external fermion lines are not considered here.} This could potentially lead to
contributions where a careful treatment of $\gamma_5$ is required.
However, all these contributions vanish identically due
to anomaly cancellations in the SM (see, e.g., Ref.~\cite{Peskin:1995ev}).
This can also be checked by an explicit calculation using the
“semi-naive” regularization prescription for $\gamma_5$ as discussed in
section~\ref{sec:framework}. Due to the ${\cal O} (\epsilon)$
ambiguity of the Eq.~(\ref{eq::trgamma5_2}), this approach can be directly
applied only to 
diagrams that contain at most simple poles in $\epsilon$. Otherwise,
finite counter-terms have to be introduced in order to restore
Ward-Identities~\cite{Larin:1993tq}. However, the diagrams contributing
to this calculation that contain one-loop triangle sub-diagrams,
have at most  simple poles in $\epsilon$.  This explains why one obtains  correct
results for the three-loop beta functions even without implementing the
't Hooft-Veltman scheme for the regularization of $\gamma_5$. 

From the technical point of view, all the methods and programs
discussed in the previous section can also be applied in this
computation. The main difficulty of this calculation is the enormous
number of diagrams (of about a million diagrams) that contribute to the
individual  renormalization factors. In order to handle such an enormous
amount of diagrams in a reasonable wall-clock time, one needs to
parallelize the calculation. 

We are now in the position to present the 
results for the beta functions of the gauge couplings which are
given by
{\allowdisplaybreaks
\begin{align}
\beta_1 &=
      \frac{\alpha_1^2}{\lp4\pi\rp^2} \bigg\{ \frac{2}{5} + \frac{16 n_G}{3} \bigg\} \notag \\
 &  + \frac{\alpha_1^2}{\lp4\pi\rp^3} \bigg\{ \frac{18 \alpha_1}{25} + \frac{18 \alpha_2}{5} - \frac{34 \text{tr}\hat{T}}{5} - 2 \text{tr}\hat{B} - 6 \text{tr}\hat{L} + n_G \bigg[ \frac{76 \alpha_1}{15} + \frac{12 \alpha_2}{5} + \frac{176 \alpha_3}{15} \bigg] \bigg\} \notag \\
 &  + \frac{\alpha_1^2}{\lp4\pi\rp^4} \bigg\{ \frac{489 \alpha_1^2}{2000} + \frac{783 \alpha_1 \alpha_2}{200} + \frac{3401 \alpha_2^2}{80} + \frac{54 \alpha_1 \hat{\lambda}}{25} + \frac{18 \alpha_2 \hat{\lambda}}{5} - \frac{36 \hat{\lambda}^2}{5} - \frac{2827 \alpha_1 \text{tr}\hat{T}}{200} \notag \\
 &  - \frac{471 \alpha_2 \text{tr}\hat{T}}{8} - \frac{116 \alpha_3 \text{tr}\hat{T}}{5} - \frac{1267 \alpha_1 \text{tr}\hat{B}}{200} - \frac{1311 \alpha_2 \text{tr}\hat{B}}{40} - \frac{68 \alpha_3 \text{tr}\hat{B}}{5} - \frac{2529 \alpha_1 \text{tr}\hat{L}}{200} \notag \\
 &  - \frac{1629 \alpha_2 \text{tr}\hat{L}}{40} + \frac{183 \text{tr}\hat{B}^2}{20} + \frac{51 (\text{tr}\hat{B})^2}{10} + \frac{157 \text{tr}\hat{B}\text{tr}\hat{L}}{5} + \frac{261 \text{tr}\hat{L}^2}{20} + \frac{99 (\text{tr}\hat{L})^2}{10} \notag \\
 &  + \frac{3 \text{tr}\hat{T}\hat{B}}{2} + \frac{339 \text{tr}\hat{T}^2}{20} + \frac{177 \text{tr}\hat{T}\text{tr}\hat{B}}{5} + \frac{199 \text{tr}\hat{T}\text{tr}\hat{L}}{5} + \frac{303 (\text{tr}\hat{T})^2}{10} \notag \\
 &  + n_G \bigg[ - \frac{232 \alpha_1^2}{75} - \frac{7 \alpha_1 \alpha_2}{25} + \frac{166 \alpha_2^2}{15} - \frac{548 \alpha_1 \alpha_3}{225} - \frac{4 \alpha_2 \alpha_3}{5} + \frac{1100 \alpha_3^2}{9} \bigg] \notag \\
 &  + n_G^2 \bigg[ - \frac{836 \alpha_1^2}{135} - \frac{44 \alpha_2^2}{15} - \frac{1936 \alpha_3^2}{135} \bigg] \bigg\},
\end{align}
\begin{align}
\beta_2 &=
      \frac{\alpha_2^2}{\lp4\pi\rp^2} \bigg\{ - \frac{86}{3} + \frac{16 n_G}{3} \bigg\} \notag \\
 &  + \frac{\alpha_2^2}{\lp4\pi\rp^3} \bigg\{ \frac{6 \alpha_1}{5} - \frac{518 \alpha_2}{3} - 6 \text{tr}\hat{T} - 6 \text{tr}\hat{B} - 2 \text{tr}\hat{L} + n_G \bigg[ \frac{4 \alpha_1}{5} + \frac{196 \alpha_2}{3} + 16 \alpha_3 \bigg] \bigg\} \notag \\
 &  + \frac{\alpha_2^2}{\lp4\pi\rp^4} \bigg\{ \frac{163 \alpha_1^2}{400} + \frac{561 \alpha_1 \alpha_2}{40} - \frac{667111 \alpha_2^2}{432} + \frac{6 \alpha_1 \hat{\lambda}}{5} + 6 \alpha_2 \hat{\lambda} - 12 \hat{\lambda}^2 - \frac{593 \alpha_1 \text{tr}\hat{T}}{40} \notag \\
 &  - \frac{729 \alpha_2 \text{tr}\hat{T}}{8} - 28 \alpha_3 \text{tr}\hat{T} - \frac{533 \alpha_1 \text{tr}\hat{B}}{40} - \frac{729 \alpha_2 \text{tr}\hat{B}}{8} - 28 \alpha_3 \text{tr}\hat{B} - \frac{51 \alpha_1 \text{tr}\hat{L}}{8} \notag \\
 &  - \frac{243 \alpha_2 \text{tr}\hat{L}}{8} + \frac{57 \text{tr}\hat{B}^2}{4} + \frac{45 (\text{tr}\hat{B})^2}{2} + 15 \text{tr}\hat{B}\text{tr}\hat{L} + \frac{19 \text{tr}\hat{L}^2}{4} + \frac{5 (\text{tr}\hat{L})^2}{2} + \frac{27 \text{tr}\hat{T}\hat{B}}{2} \notag \\
 &  + \frac{57 \text{tr}\hat{T}^2}{4} + 45 \text{tr}\hat{T}\text{tr}\hat{B} + 15 \text{tr}\hat{T}\text{tr}\hat{L} + \frac{45 (\text{tr}\hat{T})^2}{2} \notag \\ 
 &  + n_G \bigg[ - \frac{28 \alpha_1^2}{15} + \frac{13 \alpha_1 \alpha_2}{5} + \frac{25648 \alpha_2^2}{27} - \frac{4 \alpha_1 \alpha_3}{15} + 52 \alpha_2 \alpha_3 + \frac{500 \alpha_3^2}{3} \bigg] \notag \\
 &  + n_G^2 \bigg[ - \frac{44 \alpha_1^2}{45} - \frac{1660 \alpha_2^2}{27} - \frac{176 \alpha_3^2}{9} \bigg]\bigg\},
\end{align}
\begin{align}
\beta_3 &=
      \frac{\alpha_3^2}{\lp4\pi\rp^2} \bigg\{ - 44 + \frac{16 n_G}{3} \bigg\} \notag \\
 &  + \frac{\alpha_3^2}{\lp4\pi\rp^3} \bigg\{ - 408 \alpha_3 - 8 \text{tr}\hat{T} - 8 \text{tr}\hat{B} + n_G \bigg[ \frac{22 \alpha_1}{15} + 6 \alpha_2 + \frac{304 \alpha_3}{3} \bigg] \bigg\} \notag \\
 &  + \frac{\alpha_3^2}{\lp4\pi\rp^4} \bigg\{ - 5714 \alpha_3^2 - \frac{101 \alpha_1 \text{tr}\hat{T}}{10} - \frac{93 \alpha_2 \text{tr}\hat{T}}{2} - 160 \alpha_3 \text{tr}\hat{T} - \frac{89 \alpha_1 \text{tr}\hat{B}}{10} - \frac{93 \alpha_2 \text{tr}\hat{B}}{2} \notag \\
 &  - 160 \alpha_3 \text{tr}\hat{B} + 18 \text{tr}\hat{B}^2 + 42 (\text{tr}\hat{B})^2 + 14 \text{tr}\hat{B}\text{tr}\hat{L} - 12 \text{tr}\hat{T}\hat{B} + 18 \text{tr}\hat{T}^2 + 84 \text{tr}\hat{T}\text{tr}\hat{B} \notag \\ 
 &  + 14 \text{tr}\hat{T}\text{tr}\hat{L} + 42 (\text{tr}\hat{T})^2 \notag \\ 
 &  + n_G \bigg[ - \frac{13 \alpha_1^2}{30} - \frac{\alpha_1 \alpha_2}{10} + \frac{241 \alpha_2^2}{6} + \frac{308 \alpha_1 \alpha_3}{45} + 28 \alpha_2 \alpha_3 + \frac{20132 \alpha_3^2}{9} \bigg] \notag \\
 &  + n_G^2 \bigg[ - \frac{242 \alpha_1^2}{135} - \frac{22 \alpha_2^2}{3} - \frac{2600 \alpha_3^2}{27} \bigg] \bigg\}.
\end{align}
}
$\!\!$In the above formulas $n_G$ denotes the number of fermion generations. It is
obtained by labeling the closed quark and lepton loops present in the
diagrams.

\begin{figure}
  \begin{center}
  \includegraphics[width=0.7\linewidth]{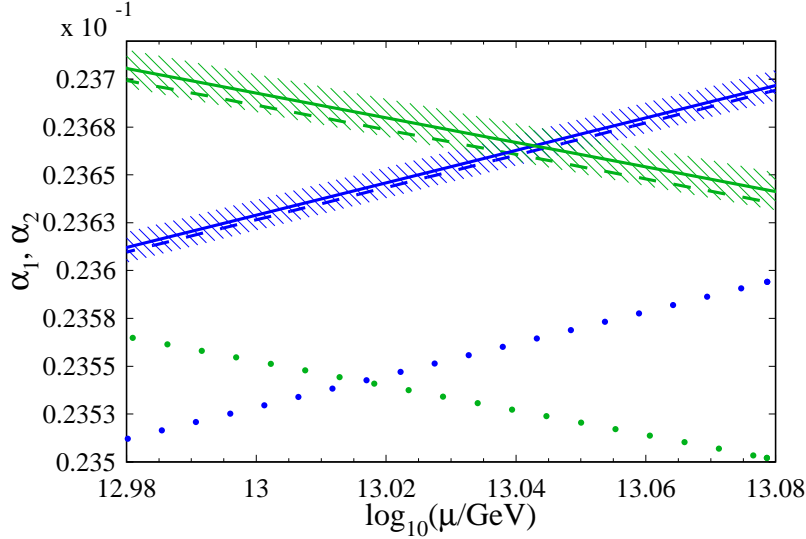}
  \caption{The running of the electroweak gauge couplings in the SM from
    Ref.~\cite{Mihaila:2012fm}. The lines 
    with positive slope correspond to $\alpha_1$, the lines with negative
    slope to $\alpha_2$.  The dotted, dashed and solid lines correspond to
    one-, two- and three-loop precision, respectively.  The bands around the
    three-loop curves visualize the experimental
    uncertainty.}\label{fig::run2}
  \end{center}
\end{figure}

Let us finally briefly discuss the numerical impact of the new three-loop 
corrections. In Fig.~\ref{fig::run2} from Ref.~\cite{Mihaila:2012fm} we reproduce the 
running of $\alpha_1$ and $\alpha_2$ from $\mu=M_Z$ to the energy scales 
where these two couplings become equal. The dotted and dashed lines correspond
to one- and two-loop running, respectively. One observes a significant change
of the curves, which is in particular much bigger than the experimental
uncertainty indicated by the dashed band. Thus in case only one- and two-loop
perturbative corrections are included the theory uncertainty is much bigger
than the experimental one. This changes with the inclusion of the three-loop
terms. The  results are shown as solid lines which are closed to the
corresponding dashed curves. The effect is small, however, still of the order
of the experimental uncertainty, in particular for $\alpha_2$. The
three-loop effects on $\alpha_3$ predictions are, as expected, much
smaller than  the experimental uncertainty. For this reason, the strong
coupling was not displayed in Fig.~\ref{fig::run2}. Let us briefly point
out that the energy scale at which the electroweak couplings meet each
other is of about $10^{13}$~GeV. Coupling unification at such a low energy
scale would imply a too rapid proton decay, in contrast to the
experimental results. Thus, even from this partial analysis, we can
conclude that the statement that 
gauge coupling unification cannot be achieved within the SM remains
valid even after the inclusion of the three-loop radiative
corrections. More details about this topic can be found in the next section.

%%%%%%%%%%%%%%%%%%%%%%%%%%%%%%%%%%%%%%%%%%%%%%%%%%%%%%%%%%%%
%%%%%%%%%%%%%%%%%%%%%%%%%%%%%%%%%%%%%%%%%%%%%%%%%%%%%%%%%%%%

 \section{\label{sec:running} Gauge coupling unification in
   supersymmetric models}
%%\setcounter{equation}{0}
%%\setcounter{figure}{0}
%%\setcounter{table}{0}
%... from the paper with Waldemar

An appealing hint in favour of supersymmetry  is the apparent 
unification of 
gauge couplings at a scale of about
$10^{16}$~GeV~\cite{Ellis:1990wk,Amaldi:1991cn,Langacker:1991an}. 
Gauge coupling unification is highly sensitive to the heavy particle
mass spectrum. This property
allows us to probe unification through precision measurements of
low-energy parameters like the gauge couplings at the electroweak
scale and the supersymmetric mass spectrum.
%A simple algebraic exercise taking into account
%the naive step-function approximation~\cite{Ross:1992tz}
%based on one-loop RGEs provides analytical formulas for the determination
%of the 
%%GUT spectrum as a function of the three gauge couplings measured at the
%$Z$-boson mass scale. 
The current precision of the
experimental data for the relevant input
parameters~\cite{Bethke:2009jm,Nakamura:2010zzi} and the substantial
progress on the theory
side~\cite{Ferreira:1996ug,Jack:2003sx,Harlander:2005wm,Bauer:2008bj,Harlander:2009mn}     
require  renormalization group analyses even at
three-loop accuracy. Within this method, one needs    $n$-loop RGEs and   
$(n-1)$-loop threshold corrections to achieve $n$-loop precision. We
have discussed in detail the derivation of  RGEs in the previous
sections. The first part of this section is devoted to the calculation of
threshold corrections. As an example, the determination of the two-loop
SUSY-QCD threshold corrections for the strong coupling $\alpha_s$ and
the bottom quark mass $m_b$ will be presented. In the second part of this
section, we outline  the phenomenological analysis of gauge coupling
unification  within the minimal SUSY SU(5) model.

\subsection{\label{sec::dec} Effective field theory approach. Decoupling coefficients.}
As already stated above, the underlying motivation for the running analysis is
to relate physical parameters measured at the electroweak scale with the
Lagrange parameters at the GUT scale. The running parameters are most
conveniently defined in mass-independent renormalization schemes such as
\msbar{} for the SM parameters and \drbar{} for  
the MSSM parameters. These schemes have the advantage that the gauge
beta-functions are mass independent and their computation is much easier
than in physical mass dependent schemes. 
%allow a very simple 
% application of the RGEs to the evolution of the
%couplings.
 However, quantum corrections  
to low-energy processes contain logarithmically enhanced
contributions from heavy particles with masses  much greater than the
energy-scale  of the process under consideration. In other words 
in such  ``unphysical''  renormalization schemes the
Appelquist-Carazzone decoupling theorem~\cite{Appelquist:1974tg}  does
not hold in its naive form. An elegant approach to
get rid of this unwanted behaviour in the \msbar{} or \drbar{} scheme  is to
formulate an effective 
theory (ET)~(for more details see Refs.~\cite{Chetyrkin:1997un,
  Steinhauser:2002rq}) integrating out all heavy particles. The
parameters of the ET  must be  modified (``rescaled'') in order to
take into account  the effects of the heavy fields. The ET parameters
are related to the 
parameters  of the full theory by  the so-called matching or decoupling
relations.\\
 They have been computed in QCD including
 corrections up to the four-loop
order  for the strong coupling~\cite{Schroder:2005hy} and 
three-loop order for quark masses~\cite{Chetyrkin:1997un}.
In the MSSM the  two-loop 
 SUSY-QCD~\cite{Harlander:2005wm, Bednyakov:2007vm,Bauer:2008bj}
 and SUSY-EW~\cite{Bednyakov:2009wt,Noth:2008tw} expressions are  known. Very
 recently, even the three-loop SUSY-QCD corrections to decoupling
 coefficient of the strong coupling was computed~\cite{Kurz:2012ff}.

%For the derivation of the coefficient functions one has to compute
%Green functions in the full and effective theory and make use of 
%the decoupling relations  to connect
%them~\cite{Chetyrkin:1997un}.....
% For moderate mass splittings between the  particle masses of a given
%theory
%, {\it i.e.} there are no large   logarithms in the theory that have to
%be resummed by means of RGEs,  
% the decoupling of heavy particles might be performed in one
%step~\cite{Ferreira:1996ug}.  The 
% energy-scale at which the decoupling is performed is not fixed by the
% theory. It is usually chosen to be $\mu \simeq \tilde M $, where
% $\tilde M$ is a typical  heavy particle mass. 
%For example, the MSSM parameters at energies $E\simeq \tilde M$ can be
%determined from the knowledge of the corresponding 
%SM parameters and the associated  decoupling relations. 

In the following, we concentrate on the calculation of the decoupling
coefficients for the strong coupling and the bottom-quark mass within
SUSY-QCD. They are the most interesting quantities from the
phenomenological point of view because they are on the one hand the 
main ingredients for the study of the gauge and Yukawa coupling
unification. On the other hand they are the quantities that receive the
largest radiative corrections, for which next-to-next-to-leading-order
corrections are essential for high precision predictions.
 \begin{figure}[t]
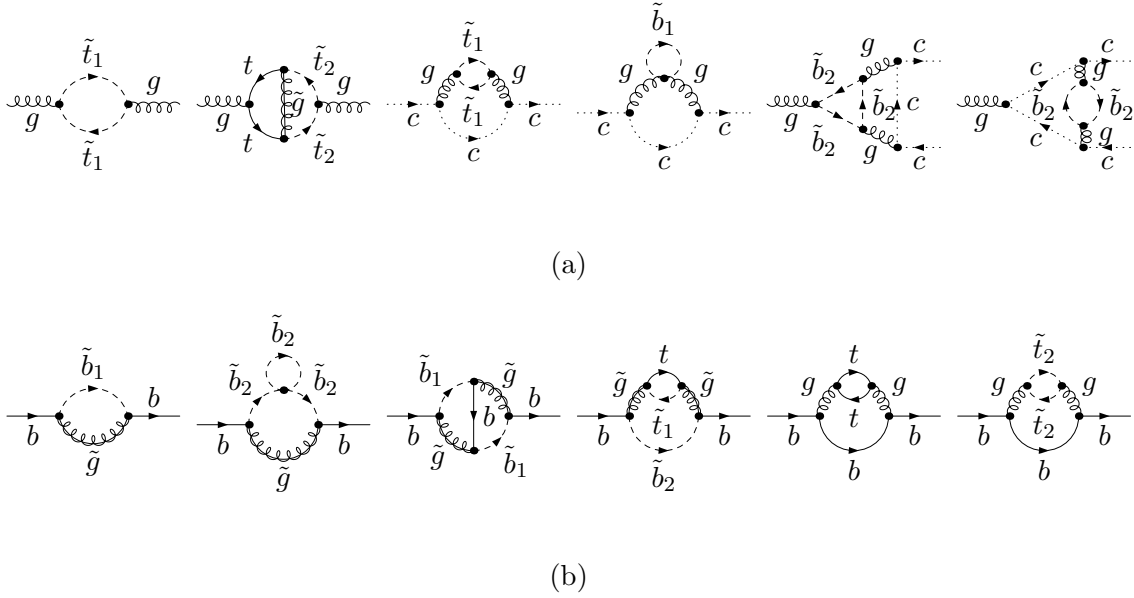

  \begin{center}
    \begin{tabular}{c}
\unitlength=1.bp%
\begin{feynartspicture}(430,100)(6,1)
\input  figs/diagcc
\end{feynartspicture}\\
(a)\\
\unitlength=1.bp%
\begin{feynartspicture}(430,100)(6,1)
\input  figs/diaqq
\end{feynartspicture}\\
(b)
    \end{tabular}
    \parbox{14.cm}{
 \caption[]{\label{fig::diagrams}\sloppy Sample diagrams
        contributing to $\zeta_3$, $\tilde{\zeta}_3$, $\tilde{\zeta}_1$
        and $\zeta_m$ with 
        gluons ($g$), ghosts ($c$),  bottom/top quarks ($b/t$),
        bottom/top squarks
        ($\tilde{b}/\tilde{t}$) and gluinos ($\tilde{g}$).    }
}
  \end{center}
\end{figure}

\subsubsection{Framework}
We consider SUSY-QCD with $n_f$ active quark and $n_s=n_f$ active
squark flavours 
and $n_{\tilde g}=1$ gluinos. Furthermore, we assume that $n_l=5$ quarks
are light (among which the bottom quark) and that the top quark and
all squarks and the gluino are heavy.
Integrating out the heavy fields from the full SUSY-QCD Lagrangian, we
obtain 
the Lagrange density corresponding to the effective QCD with $n_l$
light quarks plus non-renormalizable interactions. The latter are
suppressed by negative powers   of the heavy masses and  will be
neglected here. The effective Lagrangian can be written as follows:

\begin{eqnarray}
{\cal L}_{\rm eff}(g_s^{0}, m_q^{0}, \xi^{0}; q^{0},G_{\mu}^{0,a},
c^{0,a}; \zeta_i^{0}) = {\cal L}^{QCD}(g_s^{0 \prime}, m_q^{0\prime},
\xi^{0\prime}; q^{0\prime},G_{\mu}^{0\prime,a},c^{0\prime,a})\,, 
 \label{eq::lag}
\end{eqnarray} 
where $ q, G_{\mu}^a ,c^a $ denote the light-quark, the gluon and
the ghost 
fields, respectively, $m_q$ stands for the light quark masses, $\xi$ is
the gauge parameter and $g_s=\sqrt{4\pi \alpha_s}$ is the strong
coupling. The index $0$ 
marks bare quantities
${\cal L}^{QCD}$ is the usual QCD Lagrangian from which all heavy
fields have been discarded. As a result the fields, masses and
couplings  associated with light particles have to be rescaled.  They
are  labeled by a prime in Eq.~(\ref{eq::lag}) and are related to the
original parameters through  decoupling relations:

\begin{eqnarray}
  g_s^{0\prime}  =\zeta_g^0 g_s^0   \,,&\quad
  m_q^{0\prime}  =\zeta_m^0m_q^0    \,,&\quad
  \xi^{0\prime}-1=\zeta_3^0(\xi^0-1)\,,
  \nonumber\\
  q^{0\prime} =\sqrt{\zeta_2^0}q^0     \,,&\quad
  G_\mu^{0\prime,a}=\sqrt{\zeta_3^0}G_\mu^{0,a}  \,,&\quad
  c^{0\prime,a}    =\sqrt{\tilde\zeta_3^0}c^{0,a}\,.
  \label{eq::bare_dec}
\end{eqnarray}
Since the decoupling coefficients are universal quantities, they are
independent of the momenta carried by the incoming and outgoing
particles. The authors of 
Refs.~\cite{Chetyrkin:1997un}  showed that the  bare decoupling coefficients
$\zeta_m^0,\, \zeta_2^0,\, \zeta_3^0, 
\tilde\zeta_3^0 $ can be  derived from the quark, the gluon and the ghost
propagators, all evaluated at vanishing external momenta, via the
relations 
\begin{eqnarray}
 \zeta_3^{(0)}=1+\Pi^{0,h}(0)\,,\nonumber\\
 \zeta_2^{(0)}=1+\Sigma_v^{0,h}(0)\,,\nonumber\\
 \zeta_m^{(0)}=\frac{1-\Sigma_s^{0,h}(0)}{1+\Sigma_v^{0,h}(0)}\,.
 \label{eq::letse}
\end{eqnarray} 
%For the derivation of the coefficient $\zeta_g^{(0)}$ one has to
% consider in addition one vertex involving the strong coupling.
%for example $\bar{q}qg$ or $\bar{c}cg$, where $c$ denotes the Faddeev-Popov ghost.
  The
 superscript $h$ indicates that in the framework of DREG or DRED only diagrams
 containing at least one heavy particle inside the loops contribute and that
  only the hard regions in the asymptotic expansion of the diagrams are
  taken into account.\\
 In  Fig.~\ref{fig::diagrams} are shown 
 sample Feynman diagrams contributing to the
decoupling coefficients for the strong coupling (a) and the bottom-quark
mass~(b). 

For the
computation of $\zeta_g$ one has to consider in addition one vertex
involving the strong coupling.  A convenient choice is the relation:
\begin{eqnarray}
  \zeta_g^0 &=& \frac{\tilde{\zeta}_1^0}{\tilde{\zeta}_3^0\sqrt{\zeta_3^0}}
  \,,
  \label{eq::zetag0}
\end{eqnarray}
where $\tilde{\zeta}_1^0$ denotes the decoupling constant for the
ghost-gluon vertex.\\
The finite decoupling coefficients are obtained upon the
renormalization of the bare parameters. They are
given by
\begin{eqnarray}
  \zeta_g = \frac{Z_g}{Z_g^\prime} \zeta_g^0
  \,,\quad  \zeta_m = \frac{Z_m}{Z_m^\prime} \zeta_m^0\,,
  \label{eq::zetagren}
\end{eqnarray}
where $Z_g^\prime$ and $Z_m^\prime$ correspond to the renormalization
constants in the 
effective theory, and  $Z_g$ and $Z_m$ denote the same quantities in the
full theory. Since we are interested in the two-loop results for
$\zeta_i\,,\, i=g,m$, the corresponding renormalization constants for
SUSY-QCD and QCD have to be implemented with the same accuracy. Analytical 
results for the latter up to the three-loop order can be found in the
previous sections and the references 
cited therein, {\it e.g.} 
Refs.~\cite{ Jack:1994kd, Steinhauser:2002rq, Bednyakov:2002sf}.  

 \subsubsection{Renormalization scheme}
Apart from the renormalization constants of the external fields, also
the renormalization of the input parameters is required.  However, for
the renormalization of the gluino and 
squark masses  and the squark mixing angle we choose the on-shell
scheme.   This scheme allows us to use directly
the physical parameters in the running analyses making the implementation
very simple. The explicit formulae at the one-loop order can be found in
Refs.~\cite{Pierce:1996zz,Harlander:2004tp}. The two-loop counterterms
are known analytically only for specific mass hierarchies~\cite{Kant:2010tf}
and numerically for arbitrary masses~\cite{Martin:2005qm}.

For the computation of the decoupling coefficient for the bottom quark
mass at order ${\cal O}(\alpha_s^2)$ one needs to renormalize in
addition the bottom quark mass and the trilinear coupling $A_b$ as well
as the $\epsilon$-scalar mass. 
As  the bottom-quark mass is neglected w.r.t. heavy particle masses, an
explicit dependence of the radiative corrections on $m_b$ can occur only
through bottom Yukawa coupling. In order to avoid the occurrence of
large logarithms of the form 
$\alpha_s^2\log(\mu^2/m_b^2)$ with $\mu\simeq \tilde M$, one has to
renormalize  
the bottom Yukawa coupling in the \drbar{} scheme. In this way, the large
logarithms are absorbed into the running mass and the higher
order corrections are maintained small.

The renormalization prescription  for the trilinear coupling $A_b$ is fixed
by the tree-level relation 
\begin{eqnarray}
\sin 2\theta_b =  \frac{2 m_b (A_b-\mu
  \tan\beta)}{m_{\tilde{b}_1}^2-m_{\tilde{b}_2}^2} \,. 
\label{eq::mixb} 
\end{eqnarray}
The parameters
$\mu$ and $\tan\beta$ do not acquire 
${\cal O}(\alpha_s)$ corrections to the one-loop level.
Generically, the counter-term for $A_b$ can be expressed as
\begin{eqnarray}
\delta A_b = \left(2 \cos 2\theta_b
  \delta\theta_{b}+\sin 2\theta_b \frac{ \delta m^2_{\tilde{b}_1}-\delta
    m^2_{\tilde{b}_2}}{m^2_{\tilde{b}_1}-m^2_{\tilde{b}_2}}-\sin
  2\theta_b\frac{\delta m_b}{m_b}\right)
\frac{m^2_{\tilde{b}_1}-m^2_{\tilde{b}_2}}{2
  m_b} \,,
\label{eq::dab}
\end{eqnarray}
where $\delta m_b$ and $ \delta m^2_{\tilde{b}_{1,2}} $ 
are the counter-terms corresponding to bottom-quark and squark masses,
respectively. 
Due to the use of different renormalization prescriptions for the
bottom quark and squark  masses and mixing angle, the parameter $A_b$ is renormalized
in a {\it mixed} scheme.

Finally, the last parameter to be renormalized is the $\epsilon$-scalar
mass. In softly broken SUSY theories, as it is the case of MSSM or SUSY-QCD, they get
a radiatively induced mass. As already discussed in the previous
sections, there are different approaches in the literature to perform the
renormalization in such a case. 
%In  one scheme, the $\epsilon$-scalar
%mass is renormalized on-shell, requiring that the renormalized mass is equal to 
%zero~\cite{Bauer:2008bj}. This means that the  $\epsilon$-scalar are not
%decoupled, {\it   i.e.} the 
%effective theory is regularized in the \drbar{} scheme. In a   second
%approach~\cite{Bednyakov:2007vm}   the $\epsilon$-scalars are treated as
%massive particles. In this case,
%the $\epsilon$-scalars have to be decoupled together with the  heavy
%particles of the theory. 
To obtain decoupling coefficients
independent of the unphysical parameter $m_\epsilon$, one has to
modify the bottom squark
masses by finite
quantities~\cite{Jack:1994rk,Martin:2001vx} according to the
relation~(\ref{eq::fshift}). Such finite shifts have to be performed
for both renormalization schemes for squark masses \drbar{} and
on-shell.  
%\begin{eqnarray}
%m_{\tilde{b}}^2|_{\mbox{\drbarprime}}=m_{\tilde{b}}^2|_{\mbox{\drbar}}
%-\aspiDR 2 C_F m_{\epsilon}^2 \,.
%\end{eqnarray}
%Thus one has to use the \drbarprime scheme and the indices \drbar{} and
%\drbar{}$^\prime$ in the above  equation just specify the regularization
%scheme.

\subsubsection{Analytical results} 
The exact one-loop results for the decoupling coefficients of the strong
coupling constant $\zeta_s$ 
and  bottom-quark mass $\zeta_m$ can be found in
Refs.~\cite{Pierce:1996zz, Bednyakov:2007vm,Bauer:2008bj}. We list them
below up to order $\cal{O}(\epsilon)$
{\allowdisplaybreaks
\begin{align}
%\begin{eqnarray}
\zeta _s =&
%& 
1+\frac{\alpha_s^{\rm{ (SQCD)}}}{ \pi }
\Bigg[-\frac{1}{6} C_A L_{\tilde{g}}-\frac{1}{6}L_t
-\sum_{q}\sum_{i=1,2}\frac{1}{24} L_{\tilde{q}_i}
  \nonumber \\ 
%&
&-\epsilon  \left(\frac{C_A}{12}
  \left(L_{\tilde{g}}^2+\zeta(2)\right) 
  +\frac{1}{12} \left(
  L_t^2+\zeta(2)\right)-\frac{1}{48}\sum_{q}\sum_{i=1,2}\left( 
    L_{\tilde{q}_i}^2+ \zeta(2)\right)\right)\Bigg]\,,
\label{eq::zetas1l}
\\
\zeta_{m_b} = &
% & 
 1 + \frac{\alpha_s^{\rm{ (SQCD)}}}{ \pi }
C_F\sum_{i=1,2}\bigg\{-\frac{(1 + \lnMsq)}{4}\frac{ 
  \Msb^2}{ (\Msb^2-\Mgl^2)} 
+  \frac{ (3 +
     2 \lnMsq) \Msb^4-(3 + 2 \lnMgl) \Mgl^4 }{16 (\Msb^2-\Mgl^2)^2}
 \nonumber\\
%&
&-
\frac{(-1)^i\, X_b \Mgl }{m_{\tilde{b}_1}^2-m_{\tilde{b}_2}^2}\frac{ 
\Msb^2  \lnMsq  - \Mgl^2  \lnMgl}{2 (\Msb^2-\Mgl^2)}  + 
  \epsilon \bigg[-\frac{\Msb^2 (2 + \lnMsq (2 + \lnMsq) +\zeta(2)
   )}{8 (\Msb^2-\Mgl^2)} 
\nonumber\\
%&
&+
 \frac{ 
     \Msb^4 (7 + 2 \lnMsq (3 + \lnMsq) + 2 \zeta(2))-\Mgl^4 (7 + 2
     \lnMgl (3 + \lnMgl) + 2 \zeta(2)) }{32 (\Msb^2-\Mgl^2)^2}  
 \nonumber\\
%&
&+ 
   \frac{(-1)^i\, X_b \Mgl}{m_{\tilde{b}_1}^2-m_{\tilde{b}_2}^2}
   \frac{ \Mgl^2  \lnMgl (2 
+ \lnMgl)  -  \Msb^2  \lnMsq (2 + \lnMsq) }{4
     (\Msb^2-\Mgl^2)} \bigg] 
       \bigg\}\,,
\label{eq::zetam1l}
\end{align}
%\end{eqnarray}
}
$\!\!$where $\zeta(2)$ is Riemann's zeta function with $\zeta(2)=\pi^2/6$.
In the above equations we have
adopted the abbreviations 
\begin{eqnarray}
L_i &=& \ln\frac{\mu^2}{m_i^2}\,,\quad i \in
\{t,\tilde{g},\tilde{q}_{1,2},\tilde{b}_{1,2}\}\,,\quad \mbox{and} \quad
X_b = A_b - \mu _{\rm SUSY}\tan\beta\,.
%X_q &=& A_q-\mu_{\rm SUSY}\left\{\begin{array}{ll}
%\tan\beta\,, &\mbox{ for  down-type quarks}\\
%\cot\beta\,,  &\mbox{ for  up-type quarks}
%\end{array}\right . \,.
\label{eq::abbv}
\end{eqnarray} 
$\alpha_s^{\rm{ (SQCD)}}$ denotes the strong coupling constant in SUSY-QCD.

The presence of the terms proportional to the parameter $X_b$ is a
manifestation of the supersymmetry breaking. They are generated
by the Yukawa interaction between left- and right-handed bottom squarks
and the CP-neutral Higgs fields. 
 Their contribution to the decoupling
coefficient of the bottom-quark mass can be related through the Low
Energy Theorem~\cite{Shifman:1978zn} to the decay rate of the Higgs
boson to $b\bar{b}$ 
pairs. To one-loop order, the  $X_b$-term of
Eq.~(\ref{eq::zetam1l}) coincides with the SUSY-QCD corrections to the decay
rate $\phi\to b\bar{b}$~\cite{Carena:1999py}. To higher orders, the relation
between the two parameters becomes more involved\footnote{For details
  see section{\ref{sec:hdecay}}}. 
These Yukawa-coupling induced contributions attracted a lot of attention
due to the fact that they  are the dominant 
corrections for large values of $\tan \beta$. They may in general become
comparable with the tree-level bottom-quark mass. Thus, they need to be
resummed even at the two-loop level.

The analytical two-loop results for the decoupling coefficients are
too lengthy to be displayed here. They are available in
Ref~\cite{Bauer:2008bj} together with their expressions  for  some
phenomenologically motivated mass hierarchies. The dominant
two-loop contributions to $\zeta_{m_b}$, i.e. the terms enhanced by $\tan\beta$, 
have been confirmed by the  independent computation of
Ref.~\cite{Noth:2008tw}. Also the dominant SUSY QCD-EW corrections to $\zeta_{m_b}$ at
the two-loop order have been computed in Ref.~\cite{Noth:2008tw}.

\subsubsection{Numerical analysis}

In this section we discuss  briefly the numerical  impact of the
two-loop decoupling coefficients derived above on the
prediction of the strong coupling constant 
at the GUT scale. As already pointed out, the scale $\mu_{\rm dec}$ at which the decoupling of
the heavy  
particles is performed is not fixed by the theory. The dependence of 
physical observables on this unphysical parameter is a measure of the
theoretical uncertainty left over. At fixed order perturbation theory,
it is expected that the 
relations between the running parameters 
evaluated at high-energy scale and their low-energy values become less
sensitive to the choice of $\mu_{\rm dec}$, once higher order radiative
corrections are considered. 

In Ref.~\cite{Harlander:2007wh} a consistent method for the calculation
of the energy evolution of physical parameters was proposed. For example, 
one derives the SM values $\alpha_s^{(5)}(\mu_{\rm dec})$ and
 at the heavy scale $\mu_{\rm dec}$ from the
$n$-loop SM RGEs. Here $\mu_{\rm
  dec}$ denotes the energy 
scale at which the heavy particles are supposed to become active, {\it i.e.}
 the scale where the matching between the SM and the MSSM is
performed. Before the matching procedure, one has to perform also the
change of regularization scheme from \msbar{} to \drbar{}.
For consistency, the  $n$-loop running parameters have to be
folded with 
 $(n-1)$-loop  conversion and decoupling relations. The latter are known
 in SUSY-QCD up to two-loop order~\cite{Mihaila:2009bn} and within MSSM
 to one-loop order~\cite{Stockinger:2011gp}. 
 Above the decoupling
scale, the energy dependence of the running parameters is governed by
the $n$-loop  MSSM  RGEs.

\begin{figure}[t]
\epsfig{figure=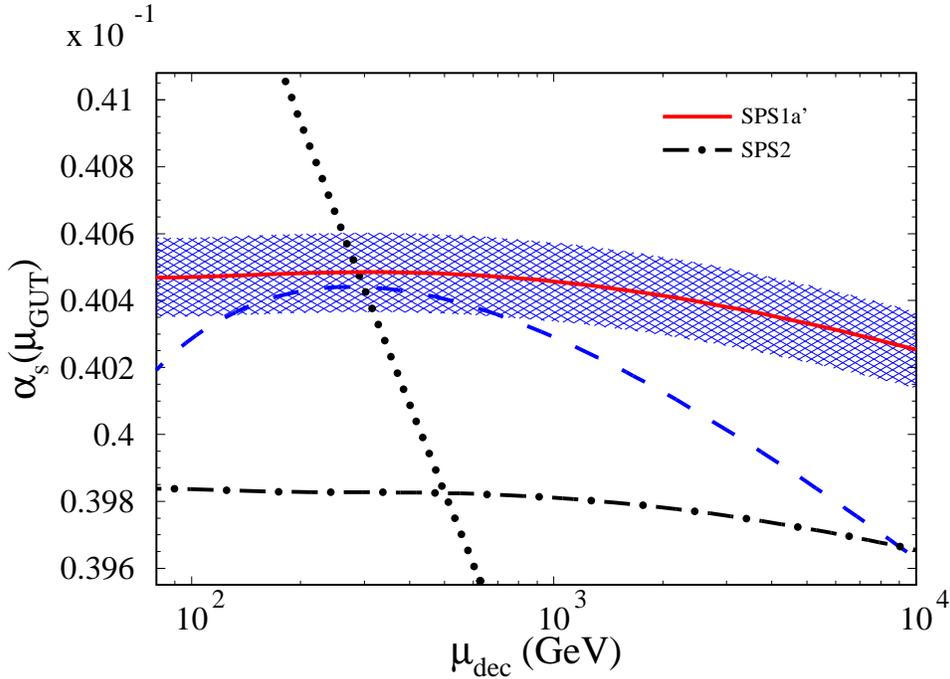,width=36em}
  \caption{\label{fig::asGUT}
    $\alpha_s(\mu_{\rm GUT})$ as a function of $\mu_{\rm dec}$ from Ref~\cite{Bauer:2008bj}. .
    Dotted, dashed and solid lines denote  the  one-, two-, and three-loop
    contributions, respectively, obtained using for the input parameters their
    values for the SPS1a$^\prime$ benchmark point. The dash-dotted line
    shows the 
    three-loop running corresponding to the SPS2 point.
  }
\end{figure}

The dependence of $\alpha_s(\mu_{\rm GUT})$ on the decoupling scale 
 is displayed in Fig.~\ref{fig::asGUT} from Ref~\cite{Bauer:2008bj}. 
 The dotted, dashed and solid lines denote the one-, two-, and
 three-loop running, where the corresponding exact results for the
 decoupling coefficients have been implemented. One can see the improved
 stability of the three-loop results  w.r.t. the decoupling-scale
 variation. The uncertainty  induced by the current experimental
accuracy  on $\alpha_s(M_Z)$, $\delta \alpha_s=
0.001$\cite{Bethke:2006ac},  is indicated by the 
hatched band.\\
In order to get an idea of the effects induced by the SUSY mass parameters
on $\alpha_s(\mugut)$, two different mass spectra are shown. As
reference was chosen the so called Snowmass Point SPS1a$^\prime$
scenario~\cite{AguilarSaavedra:2005pw} for which rather low SUSY mass
parameters are required: $\Mgl=607.1$~GeV,
$m_{\tilde{t}_1}=366.5$~GeV, $m_{\tilde{t}_2}=585.5$~GeV,
$m_{\tilde{b}_1}=506.3$~GeV, $m_{\tilde{b}_2}=545.7$~GeV, $A_t^{\mbox{\drbar}}
(1~{\mbox{TeV}})=-565.1$~GeV, $A_b^{\mbox{\drbar}}
(1~{\mbox{TeV}})=-943.4$~GeV, $\mu = 396.0$~GeV, and $\tan\beta=10.0$
. In addition  the dash-dotted line shows the  three-loop results when
the SUSY parameters  corresponding to the SPS2~\cite{Ghodbane:2002kg}
scenario are adopted. Their explicit values  are $\Mgl=784.4$~GeV,
$m_{\tilde{t}_1}=1003.9$~GeV, $m_{\tilde{t}_2}=1307.4$~GeV,
$m_{\tilde{b}_1}=1296.6$~GeV, $m_{\tilde{b}_2}=1520.1$~GeV, and
$\tan\beta=10.0$.  One clearly notices  the great
impact of the SUSY-mass pattern on the predicted value of the strong
coupling at high energies.  Accordingly, for precision studies
concerning gauge coupling unification 
the explicit mass
pattern of heavy particles must  be taken into account.\\
At this point, a comment on the chosen mass spectrum is in order.
The nature of this plot is rather academic and aims to quantify the mass
dependence of the strong coupling constant at high energies. The two
mass spectra are already excluded by the direct searches at the
LHC. Nevertheless, one can easily estimate that for heavier SUSY
particles $M_{\rm SUSY} > 1.5$~TeV the value of $\alpha_s(\mu_{\rm GUT})$
decreases below the value $0.398$. Its implication on the quality of the
unification will be discussed in the next section.

\subsection{Gauge coupling unification in the minimal 
SUSY SU(5) model} 

The gauge coupling unification might be  predicted,
even under the assumption of a
minimal particle content of the underlying GUT like in  the
so-called minimal SUSY SU(5) model~\cite{Dimopoulos:1981zb,Sakai:1981gr}.
 This is the most predictive model among the currently known candidates for
 SUSY GUTs. However, immediately after its formulation it has been noticed that
 new dimension-five operators may cause a rapid proton decay. Together
 with the requirement of gauge coupling unification this aspect was used
 to even  rule out the SUSY SU(5)
 model~\cite{Goto:1998qg,Murayama:2001ur}. However,  subsequent careful 
analyses have shown that the proton decay rate for the dominant channel
$p\rightarrow 
K^+\overline{\nu}$ can be suppressed either
by sfermion mixing~\cite{Bajc:2002bv} or by taking into account higher
dimensional operators induced at the Planck
scale~\cite{EmmanuelCosta:2003pu,Wiesenfeldt:2004qz,Bajc:2002pg}.

In the following, we review the latest analysis  on the gauge coupling
unification in the renormalizable version of minimal  SUSY SU(5). This 
model is not the best motivated phenomenologically, but it requires the most
severe constraints on the GUT parameters.  More precisely, we outline the
constraints on the mass of the 
colour triplet Higgs 
 $\mhc$ and the grand unification scale\footnote{See 
  below  the exact definition of $\mhc$ and
  $\mgut$.} $\mgut$ 
 taking into
account the latest experimental data for the weak scale parameters and
the most precise theoretical predictions currently available.
The two parameters are predicted in the ``bottom-up'' approach, taking
into account threshold corrections generated by the superpartners of the
SM particles as well as those due to  the super-heavy SUSY-GUT particles.
In addition,   the gauge coupling constants of the SM at the electroweak
scale and the MSSM mass spectrum are required as input parameters. The
predicted values for the two parameters have to be compared with the
constraints derived from the non observation of proton decay.

For completeness, we present below our notation in the framework of
minimal SUSY SU(5). The superpotential of the model~\cite{Dimopoulos:1981zb} is given by 
\begin{eqnarray}
  {\cal W} &=& M_{1}\rm{Tr}(\Sigma^2)+\lambda_1 \rm{Tr}(\Sigma^3)+
  \lambda_2 \bar{H}\Sigma H +M_2 \bar{H} H \nonumber\\
  &&\mbox{}+ \sqrt{2} Y_d^{ij}\Psi_i \phi_j \bar{H} + \frac{1}{4}
  Y_u^{ij} \Psi_i\Psi_j H
  \,,
\end{eqnarray}
where $\Psi_i$ and $\phi_i$ ($i=1,2,3$ is a generation index) are matter
multiplets in the $\mathbf{10}$- and $\mathbf{\overline{5}}$-dimensional 
representation of SU(5). Their decomposition  w.r.t.  the SM gauge
group\\ SU(3)$\times$SU(2)$\times$U(1) reads 
\begin{eqnarray}
\mathbf{5}&=& (3,1,-\frac{1}{3})\oplus (1,2,+\frac{1}{2})\nonumber\\
\mathbf{10}&=&(\bar{3},1,-\frac{2}{3})\oplus (3,2,\frac{1}{6}) \oplus (1,1,1)\,.
\end{eqnarray}
The field $H$ ($\bar{H}$) is
realized in the $\mathbf{5}$ ($\mathbf{\bar{5}}$) representation.
The gauge group SU(5) is broken to the SM gauge group if 
the adjoint Higgs boson
 $\Sigma\equiv\Phi^a T^a$ ($a=1,\ldots,24$) living in the
$\mathbf{24}$ representation
  gets the
vacuum expectation value  $\langle\Sigma\rangle =
V/(2\sqrt{30})\times\rm{diag}(-2,-2,-2,3,3)$,\\ 
with $V=-4\sqrt{30} M_1/(3\lambda_1)$.\footnote{Here, we parametrize as
  usual the
  $\mathbf{24}$ representation like a $5\times 5$ matrix.} 
Its decomposition w.r.t. the SM gauge group reads
\begin{eqnarray}
\mathbf{24}&=& (1,1,0)\oplus (1,3,0) \oplus (8,1,0)\oplus
(3,2,-\frac{5}{6})\oplus (\bar{3},2,\frac{5}{6})\,.
\end{eqnarray}
Choosing $\langle\bar{H}\rangle=\langle H\rangle\ll V$ and in addition
 imposing the (tree-level-) fine-tuning
condition $M_2=-\sqrt{3}\lambda_2 V/\sqrt{40}$,  the iso-doublets in $H$ and
$\bar{H}$ remain massless.
Furthermore, one gets the following super-heavy mass spectrum:
\begin{eqnarray}
  M_X^2=\frac{5}{12}g^2V^2\, ,\quad M_{H_c}^2 =
  \frac{5}{24}\lambda_2^2 V^2\,,\quad M_{\Sigma}^2\equiv 
  M^2_{(8,1)} = M^2_{(1,3)} = 25 M^2_{(1,1)} = \frac{15}{32}\lambda_1^2 V^2 
  \,,
\end{eqnarray} 
where the indices in round brackets refer to the SU(3) and SU(2) 
quantum numbers.
Here $M_{\Sigma}$ denotes the mass of the colour octet part of the
adjoint Higgs boson $\Sigma$ and $M_{H_c}$ stands for the mass of the colour
triplets of $H$ and $\bar{H}$. $M_X$ is the mass of the gauge bosons
and $g$ is the gauge coupling.
The equality $ M^2_{(8,1)} = M^2_{(1,3)}$  holds only
if one neglects operators that are suppressed by $1/M_{\rm Pl}$ as we do here.

In the study of the energy evolution of the gauge couplings up to scales
of the order ${\cal O}(10^{16})$~GeV, one has to apply the effective
field theory (EFT) approach
twice: once at an energy scale comparable with the SUSY particle masses  and
once at the GUT scale. In practice, this translates into the following steps: 
\begin{enumerate}
\item Running within the SM from $\mu=M_Z$ to the
  SUSY scale $\mususy$.\\
In this step,  the three-loop beta function of
QCD~\cite{Tarasov:1980au,Larin:1993tp} 
  and up to three-loop RGEs in the electroweak
  sector~\cite{Mihaila:2012fm, Chetyrkin:2012rz} are necessary
 in order
  to obtain the values of the gauge couplings at $\mususy\approx 1$~TeV.
At this point we want to stress once again that, the value of $\mususy$ is  a free
parameter.  Let us mention that the   top-quark threshold effects are
taken into account in the determination of the input parameters (for
details see next section) and the running analysis  is performed in SM
with six active quark flavours.

\item SUSY threshold corrections.\\
In order to still cure the naturalness problem of the SM, the SUSY mass
spectrum has to be at most in the TeV range. Thus, for energies of this
order of magnitude, it is expected that the SUSY particles
  become active and the proper matching between the SM and the MSSM has to be
  performed. The  one-loop decoupling  relations for $\alpha_1$ and $\alpha_2$~\cite{Yamada:1992kv}
  and the Yukawa couplings~\cite{Pierce:1996zz} are known since long time. The SUSY-QCD
  decoupling effects for 
  $\alpha_3$ and $m_b$ are known to three- and two-loop order,
  respectively, as discussed in the 
  previous section.
  A  fully consistent approach would require
  two-loop threshold corrections not only in the strong but also in
  the electroweak sector. They are  not yet available, nevertheless it is
   expected that their numerical impact is relatively small.

 At this stage also the change of renormalization scheme
  from \msbar{} to \drbar{} has to be taken into account. To establish
  the conversion relations between the running parameters in the two schemes,
  one can use the method discussed in section~\ref{sec::ms2dr}, where such
  relations have been derived in the context of non-supersymmetric
  theories.\footnote{For more details
  see Ref.~\cite{Mihaila:2009bn}.} The conversion relations that are of interest 
for the numerical analysis discussed in this section are those involving
the gauge couplings and the quark masses of the third generation, as
only their Yukawa couplings give sizable effects. They are known up to
the two-loop order in SUSY-QCD~\cite{Mihaila:2009bn}. For the
convenience of the reader we cite them below
\begin{eqnarray}
\asMSbar &=& \asDRbar \bigg[ 1 - \apiDR \frac{\cA}{3} +
  \left(\apiDR\right)^2 \left(-\frac{11}{9} \cA^2  + 2 T_f 
    \cR\right)\bigg]\,,
\label{eq::asms}
\\
\mqMSbar &=& \mqDRbar\bigg[1 + \apiDR \cR +
\left(\apiDR\right)^2 \bigg(
\frac{7}{12}\cA\cR + \frac{7}{4}\cR^2 - \frac{1}{2} T_f \cR
\bigg)\bigg]\,,
\label{eq::mqms}
\end{eqnarray}
 where the group invariants are defined as in Appendix A and $T_f= \TF
 n_f$, with $n_f$ the number of active fermions.

\item Running within the MSSM from $\mususy$ to the high-energy
  scale $\mugut$.\\
 In this step the  three-loop RGEs of the
 MSSM~\cite{Ferreira:1996ug,Harlander:2009mn} are required to evolve the 
  gauge and Yukawa couplings  from $\mususy$ to some very high scale
  of the order of $10^{16}$~GeV, that we denote as 
  $\mugut$. At this energy scale it is expected that SUSY-GUT particles become
  active.

\item SUSY-GUT threshold effects.\\
At the energy scale $\mugut$, threshold corrections induced by the
  non-degenerate SUSY-GUT spectrum have to be taken into account.
   The one-loop formulas of the decoupling coefficients for a general
   gauge group have been known for a long
   time~\cite{Hall:1980kf,Weinberg:1980wa,Einhorn:1981sx}. The
   specification to the minimal SUSY SU(5) reads 
~\cite{Hagiwara:1992ys,Dedes:1996wc}
\begin{eqnarray}
  \zeta_{\alpha_1}(\mu)&=&1+\apiSU\left(-\frac{2}{5}\,L_{\mu H_c}+10\,L_{\mu
      X}  
  \right)
  \,,
  \nonumber\\
  \zeta_{\alpha_2}(\mu)&=& 1+\apiSU\left(-2\,L_{\mu\Sigma}+
  6\,L_{\mu X} \right)
  \,,
  \nonumber\\
  \zeta_{\alpha_3}(\mu)&=& 1+\apiSU\left(-L_{\mu
      H_c}-3\,L_{\mu\Sigma}+4\,L_{\mu X} 
  \right)\,,
  \label{eq::gutdec}
\end{eqnarray}
where $L_{\mu x}=\ln(\mu^2/M_x^2)$ and for simplicity  we  keep
from the list of arguments of the coefficients $\zeta_{\alpha_i}$ only
the decoupling scale. $\apiSU$ is the gauge coupling constant of the
unified theory, {\it i.e.} of the SUSY SU(5) model.

A suitable linear combination of the three equations above leads to the
following two relations
\begin{eqnarray}
    4\pi\left(-\frac{1}{\alpha_1(\mu)}+3\,\frac{1}{\alpha_2(\mu)} - 2\,
      \frac{1}{\alpha_3(\mu)}\right)&=&-\frac{12}{5} L_{\mu H_c}
    \,,\nonumber\\
    4\pi\left(5\,\frac{1}{\alpha_1(\mu)}-3\,\frac{1}{\alpha_2(\mu)} - 2\,
      \frac{1}{\alpha_3(\mu)}\right)&=&-24\left( L_{\mu
      X}+\frac{1}{2}L_{\mu\Sigma}\right)\,,
    \label{eq::gutrel}
  \end{eqnarray}
  where $\alpha^{\rm SU(5)}$ has been eliminated.
  These equations allow  the prediction of the coloured triplet Higgs boson
  mass $\mhc$
  from the knowledge of the MSSM gauge
  couplings at the energy scale $\mu=\mugut$.
 It is furthermore common to define a new mass parameter
  $M_G=\sqrt[3]{M_X^2M_\Sigma}$, the so-called grand unified mass scale.
  It can    
  also be determined from the knowledge of the MSSM gauge
  couplings at $\mugut$ through Eq.~(\ref{eq::gutrel}).
  These observations makes it quite easy to test the
  minimal SUSY SU(5) model once the required
  experimental data are available in combination with a high-order analysis.

\item Running from $\mugut$ to the Planck scale $M_{\rm Pl}$.\\ 
  The last sequence of this approach consists in
  the running within the SUSY-SU(5) model.
  The three-loop RGEs for the gauge~\cite{Jack:1996vg}, and the
  one-loop formulas for the Yukawa and Higgs
  self couplings~\cite{Hisano:1992jj} are available in the literature.
 In addition, the perturbativity constraints ({\it i.e.} all couplings
 of the theory are smaller than unity)  have to be imposed.
\end{enumerate}

\subsubsection{Input parameters}

As can be inferred from the discussion above, to  constrain the GUT
parameters one needs in addition to the precise running analysis also
precise input parameters. Explicitly, one needs the values of weak mixing
angle in the $\overline{\rm MS}$ 
scheme~\cite{Amsler:2008zzb}, the QED coupling constant at zero
momentum transfer and its hadronic
contribution~\cite{Teubner:2010ah} in order to obtain its counterpart at
the $Z$-boson scale, and the strong
coupling constant~\cite{Bethke:2009jm}.\footnote{We adopt the central
  value from Ref.~\cite{Bethke:2009jm}, however, use as our default
  choice for the uncertainty $0.0020$ instead of $0.0007$.} 
Their numerical values and uncertainties are 
\begin{eqnarray}
  \sin^2\Theta^{\overline{\rm MS}} &=& 0.23119 \pm 0.00014
  \,,\nonumber\\
  \alpha &=& 1/137.036
  \,,\nonumber\\
  \Delta\alpha^{(5)}_{\rm had} &=& 0.02761\pm 0.00015
  \,,\nonumber\\
  \alpha_s(M_Z) &=& 0.1184 \pm 0.0020
  \,.
  \label{eq::parin5}
\end{eqnarray}
To determine the value of $\alpha$ in the \msbar{} scheme,
  it is necessary  to take into account 
 the hadronic $\Delta\alpha^{(5)}_{\rm had}$ ,
 leptonic $\Delta\alpha^{(5)}_{\rm lep}$~\cite{Steinhauser:1998rq} 
 and top-quark $\Delta\alpha^{(5)}_{\rm top}$~\cite{Kuhn:1998ze}  
 contributions to the
 on-shell value. In addition, the  conversion formula to the
 \msbar{} scheme has to be taken into account. Thus, one obtains
\begin{eqnarray}
  \alpha^{\overline{\rm MS}}(M_Z) &=&
 \frac{\alpha}{1-\Delta\alpha_{\rm lep}^{(5)} -
    \Delta\alpha^{(5)}_{\rm had} - \Delta\alpha^{(5)}_{\rm top} -
 (\Delta\alpha^{(5),\overline{\rm MS}}
  - \Delta\alpha^{(5),\rm OS}) }\nonumber\\
  &=& \frac{1}{127.960 \pm 0.021}  \,.
\end{eqnarray}
For supersymmetric particle masses of order ${\cal O}(1\,\,\mbox{TeV})$ it is
appropriate to take into account top-quark threshold effects in a
separate step. For convenience, we choose the scale at which we decouple
the top-quark to be $\mu_{\rm
  dec}=M_Z$. The corresponding threshold corrections  are available from the
Refs.~\cite{Amsler:2008zzb,Fanchiotti:1992tu,Chetyrkin:2000yt} and
give the following contributions
\begin{eqnarray}
  \alpha^{(6),\overline{\rm MS}}(M_Z) &=& 1/(128.129\pm 0.021)
  \,,\nonumber\\
  \sin^2\Theta^{(6),\overline{\rm MS}}(M_Z) &=& 0.23138 \pm 0.00014
  \,,\nonumber\\
  \alpha_s^{(6)}(M_Z) &=& 0.1173\pm 0.0020
  \,.
  \label{eq::alphasin}
\end{eqnarray}
Even more, the supersymmetric particles can induce sizeable effects in
the extraction of the weak mixing angle from experimental data. Such
effects are by construction suppressed by the square of the
supersymmetric mass scale~\cite{Dedes:1998hg, Yamada:1992kv}. For
a typical supersymmetric  mass scale $\ge$1~TeV such corrections can
lead to shifts in $M_{H_c}$ of the order of $\le 10$\%.

\subsubsection{Numerical analysis}

For illustration of the numerical effects we adopt the 
{\tt mSUGRA}~\cite{Chamseddine:1982jx} scenario  for the SUSY breaking
mechanism with the following 
initial  parameters 
\begin{eqnarray}
  m_0 &=&m_{1/2}\,\,=\,\,-A_0\,\,=\,\,1000~\mbox{GeV}\,,\nonumber\\
  \tan\beta &=& 3\,,\qquad
  \mu>0\,,
  \label{eq::msugra_par1}
\end{eqnarray}
and generate with the help of the code {\tt SOFTSUSY}~\cite{Allanach:2001kg}
the supersymmetric mass spectrum. This results in squark masses  of
the order of 2~TeV, thus beyond above the exclusion bounds currently
established   by direct searches at the  LHC. 

\begin{figure}[t]
  \centering
  \begin{tabular}{cc}
    \includegraphics[width=.5\linewidth]{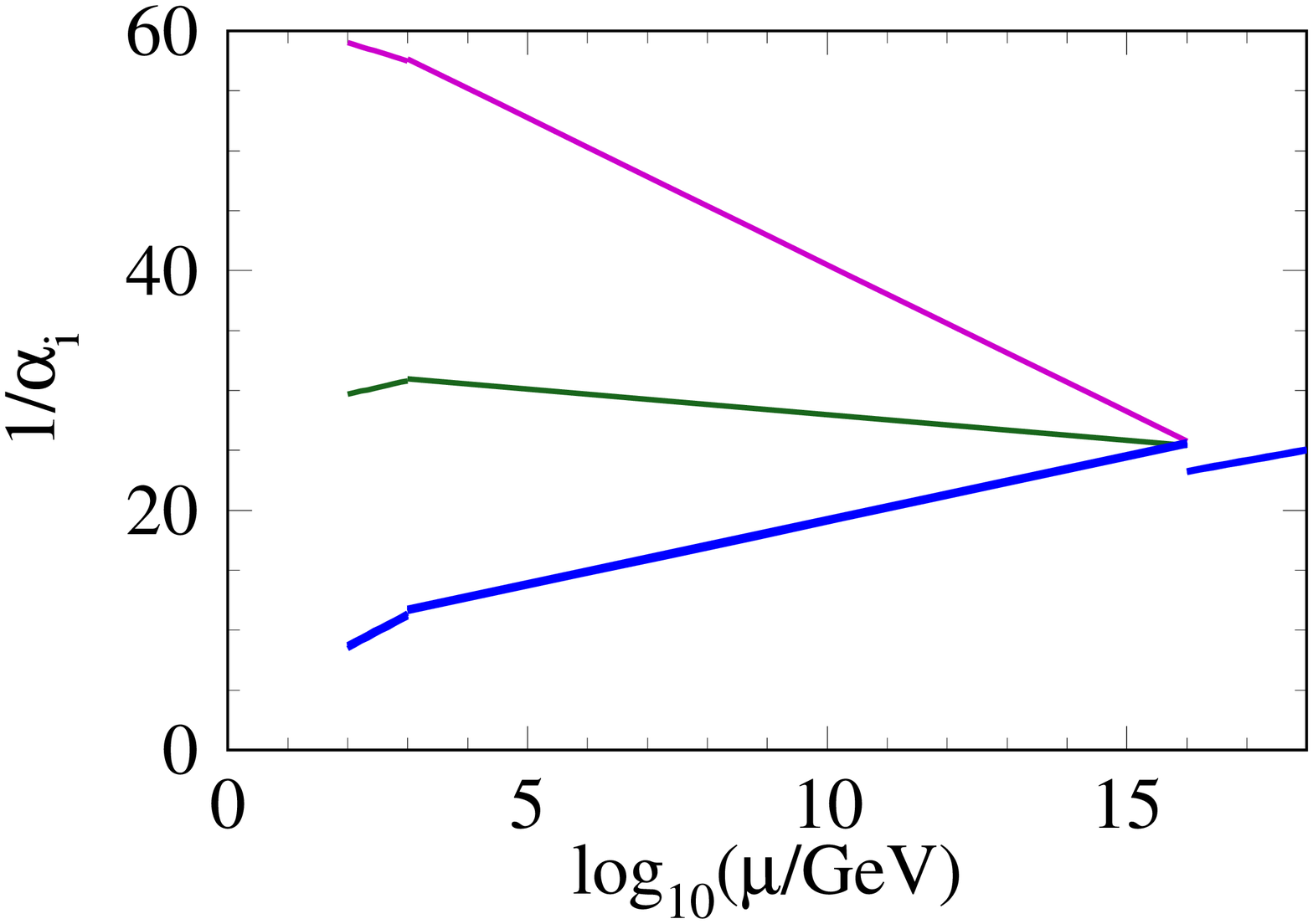}
    &
   \includegraphics[width=.5\linewidth]{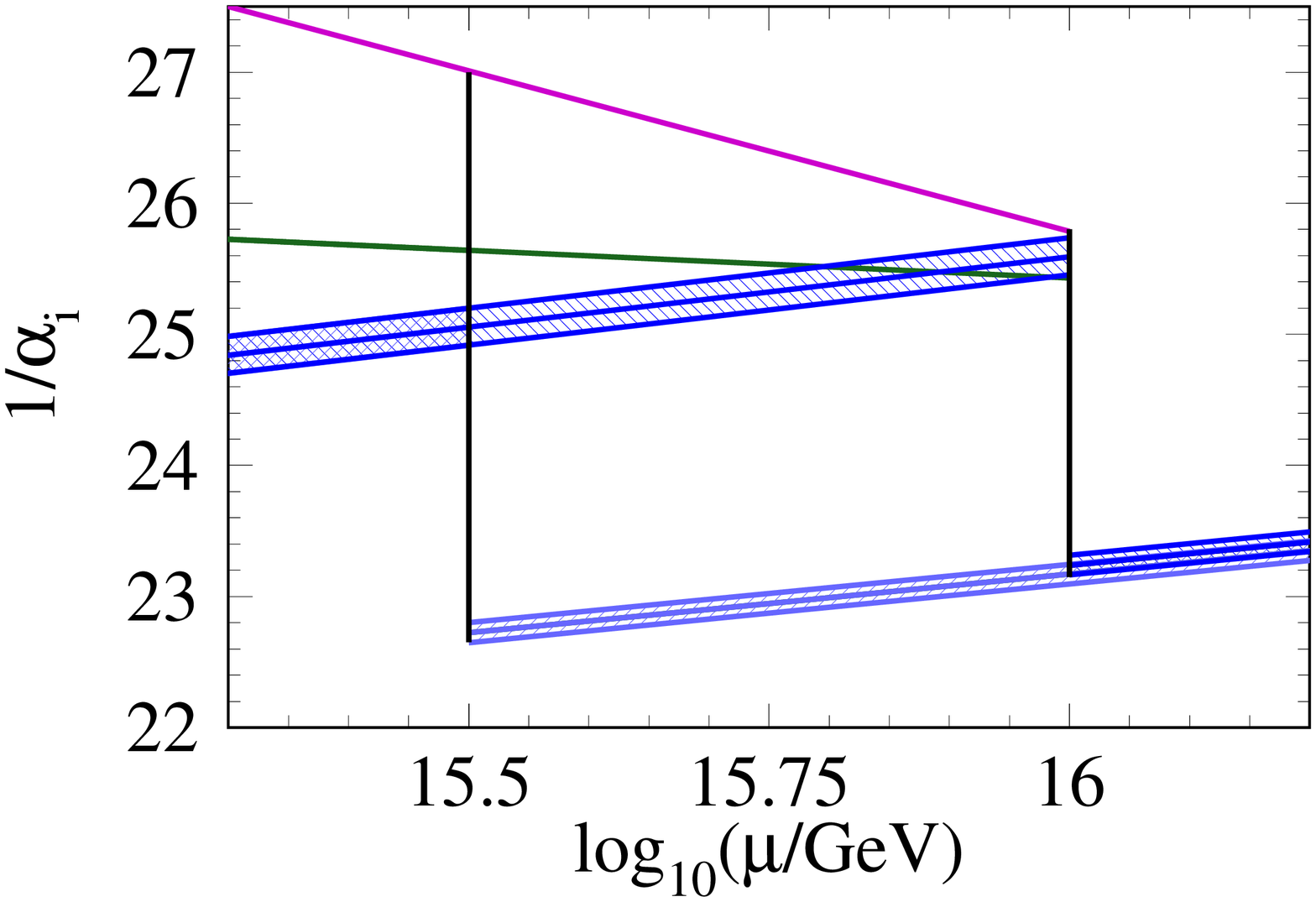}
    \\
    (a) & (b)
  \end{tabular}
  \caption{\label{fig::unification}Running of the gauge couplings from the
    electroweak to the Planck scale from Ref~\cite{Martens:2010nm}. The discontinuity for $\mu=\mususy$ and
    $\mu=\mugut$ are clearly visible. In panel (b) an enlargement of (a) for
    the region around $\mu=\mugut$ is shown where for the decoupling the two
    values $\mugut=10^{15.5}~\mbox{GeV}\approx 3.2\cdot 10^{15}$~GeV and
    $\mugut=10^{16}$~GeV have been chosen.} 
\end{figure}

In Fig.~\ref{fig::unification} from Ref.~\cite{Martens:2010nm}
we visualize the running (and decoupling) of the gauge
couplings where the parameters of 
Eq.~(\ref{eq::msugra_par1}) together with 
$\mususy = 1000$~GeV and $\mugut = 10^{16}$~GeV have been adopted.
In addition we have chosen $M_\Sigma = 1\cdot 10^{15}~\mbox{GeV}
$ which
leads via Eq.~(\ref{eq::gutrel}) to $\mhc = 1.7\cdot 10^{15}~\mbox{GeV}$
and $M_X = 4.6\cdot 10^{16}~\mbox{GeV}$. 
One can clearly see the discontinuities at the matching scales and the change
of the slopes when passing them.
In panel (b) the region around $\mu = 10^{16}$~GeV
is enlarged which allows for a closer look at the unification region.
The bands indicate $1\sigma$ uncertainties of $\alpha_i$ at the
electroweak scale (cf. Eq.~(\ref{eq::alphasin})).
In panel (b) we furthermore perform the decoupling of the super-heavy masses
for two different values of $\mugut$. One observes quite different threshold
corrections leading to a nice agreement of $\alpha^{\rm SU(5)}$ 
above $10^{16}$~GeV.
Fig.~\ref{fig::unification} stresses again that the uncertainty 
of $\alpha_s$
is the most important one for the   constraints that one can set
 on  GUT models from low-energy data. Furthermore, it illustrates the
 size of the GUT threshold corrections 
and emphasizes the importance of precision calculations.
%of the two-loop corrections for the corresponding
%decoupling constants.

\begin{figure}
  \centering
  \begin{tabular}{cc}
    \includegraphics[width=.75\linewidth]{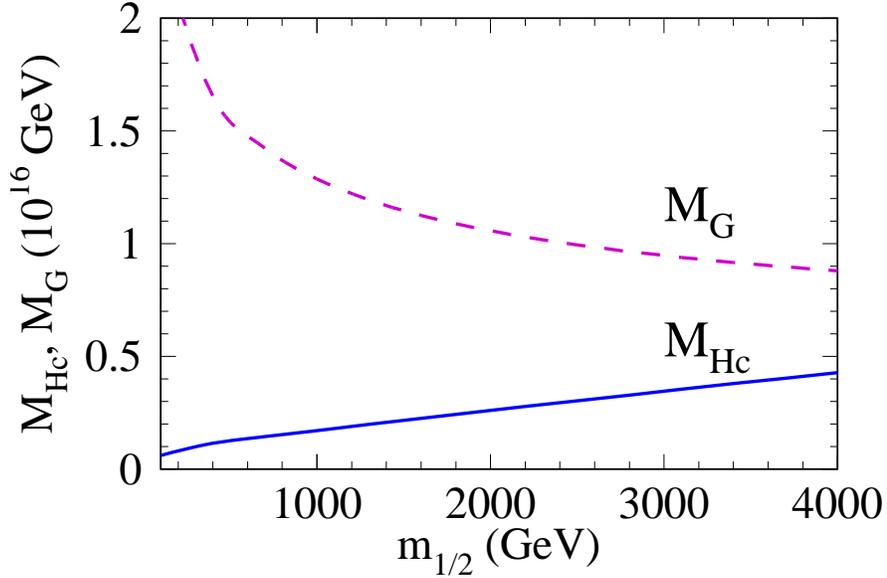}
  \end{tabular}
  \caption{\label{fig::mhc_msugra}
    Dependence of $\mhc$ (solid) and $\mgut$ (dashed) 
    on $m_{1/2}$ from Ref~\cite{Martens:2010nm}.}
\end{figure}

In the following, we discuss the dependence of $\mhc$ and $\mgut$ on various
parameters entering our analysis. We start
with varying the supersymmetric mass spectrum 
and use
Eq.~(\ref{eq::gutrel}) in order to extract  both $\mhc$ and
$\mgut$. The decoupling scales are fixed to $\mususy = 1000$~GeV and $\mugut =
10^{16}$~GeV, respectively, which ensures 
that the three-loop effect is rather small. In
Fig.~\ref{fig::mhc_msugra} the parameter $m_{1/2}$ is varied up to 4~TeV.
The solid and dashed lines correspond to $\mhc$ and $\mgut$, respectively,
which show a substantial variation. On the other hand, $m_0$, $\tan\beta$ and
$A_0$ have only a minor influence on the GUT masses and thus we refrain from
explicitly showing the dependence.

\begin{figure}
  \centering
  \begin{tabular}{c}
    \includegraphics[width=.75\linewidth]{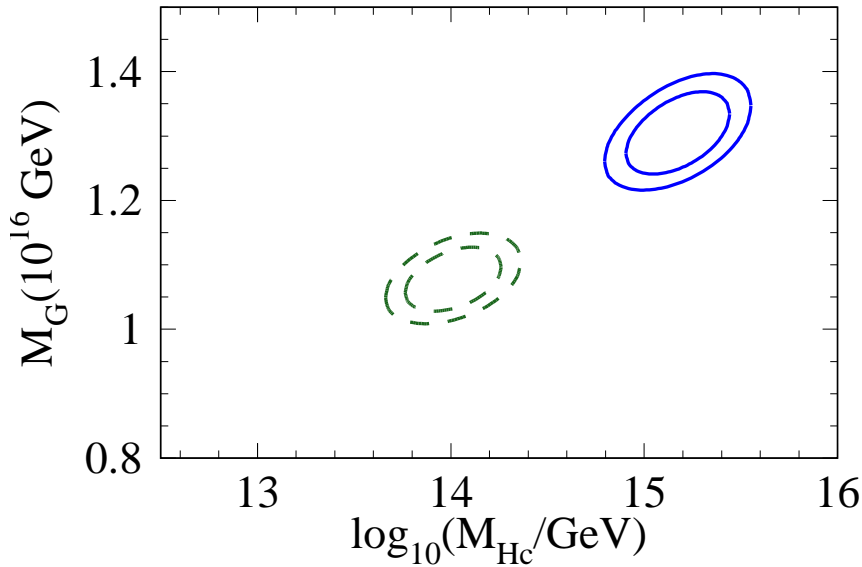}
  \end{tabular}
  \caption{\label{fig::mhc_mgut_mz}
    Ellipses in the $\mhc-\mgut$ plane obtained from the uncertainties
    of the gauge couplings at the electroweak scale.  The input parameters of
    Eq.~(\ref{eq::parin5}) have been used whereas 
    $\delta\alpha_s=0.0010$ has been chosen.
    Dashed and solid lines correspond to the two- and three-loop
    analysis, respectively. 
  }
\end{figure}

An interesting aspect from the phenomenological point of view is the
study of  the effects
of the experimental uncertainties of
$\alpha_i$ (c.f. Eq.~(\ref{eq::parin5})) on the prediction of $\mhc$ and
$\mgut$. For this, we fix
the SUSY spectrum as 
before (see Eq.~(\ref{eq::msugra_par1})) and set $\mu_{\rm SUSY}=M_Z$
which has often been common practice in similar  
analyses (see, e.g., Ref.~\cite{Murayama:2001ur}).   
Taking into account correlated errors and performing a $\chi^2$
analysis leads to ellipses in the $\mhc-\mgut$ plane.
%Let us mention that we can reproduce the results of
%Ref.~\cite{Murayama:2001ur} after adopting their parameters and
%restricting ourselves to the perturbative input used in that publication.
In Fig.~\ref{fig::mhc_mgut_mz}  from Ref~\cite{Martens:2010nm} the results 
for the two- (dashed lines) and
three-loop (continuous lines)  analyses are shown. The two concentric
ellipses correspond to 68\% and 90\% confidence level, respectively, where
only parametric uncertainties from Eq.~(\ref{eq::alphasin}) have
 been taken into account. Let us, however, stress  that an
 optimistic uncertainty of 
$\delta\alpha_s=0.0010$ has been adopted for this plot.
As expected, the uncertainty of $\alpha_s$ induces the
largest contributions to the uncertainties on $\mhc$ and $\mgut$. 
In particular, it essentially determines the semi-major axis of the ellipses.
The three-loop corrections induce a significant shift to higher masses of about 
an order of magnitude  for $\mhc$. In the same time $\mgut$ increases 
by
about $2\cdot 10^{15}$~GeV. This demonstrates the importance of 
the precision calculations in such type of analyses.
As has been  discussed in the original paper~\cite{Martens:2010nm}
they are also essential in order to remove the dependence on 
$\mu_{\rm SUSY}$. In fact, choosing $\mu_{\rm SUSY}$ close
to the supersymmetric mass scale leads to small three-loop effects,
since the two-loop ellipses are essentially shifted on top of the
three-loop ones.

At this point a discussion about the additional constraint on the Higgs
triplet mass 
$M_{H_c}$ that can be derived from the non-observation of the proton
decay is in order. The latest upper bound on the proton 
decay rate for the channel $p\to K^+\bar{\nu}$~\cite{Kobayashi:2005pe}
is $\Gamma_{\rm exp}=4.35\times 10^{-34}/\mbox{y}$. In order to
translate it into a lower bound for the  the Higgs triplet mass, one
needs an additional assumption about the Yukawa couplings that enter
the expression of the decay rate $\Gamma(p\to K^{+}\bar{\nu})$.  As
pointed out in Ref.~\cite{EmmanuelCosta:2003pu} this is because down quark
and lepton Yukawa couplings fail to unify within the minimal renormalizable SUSY
SU(5) model and so a completely consistent treatment is not possible.
Therefore one could either choose\footnote{$Y_{ql}$ is the Yukawa
  coupling of the quark and lepton doublet to the Higgs colour triplet.
  $Y_{ud}$ is the corresponding coupling for the up and down quark
  singlet.} (i) $Y_{ql}=Y_{ud}=Y_d$ or 
(ii) $Y_{ql}=Y_{ud}=Y_e$, which leads to completely different
phenomenological consequences.  Both choices are equally justified
once higher dimensional operators are included. Since these operators
further weaken the bounds presented below, we refrain to include these bounds
into the analysis presented here.
For the case (i) and supersymmetric particle masses around 1~TeV the lower bound
for  the Higgs triplet mass can be read off from Fig.~2 of
Ref.~\cite{EmmanuelCosta:2003pu} 
and amounts to  $M_{H_c}\ge 1.05 \times 10^{17}$~GeV whereas for
the second choice it becomes $M_{H_c}\ge 5.25\times 10^{15}$~GeV.
From our phenomenological analysis 
presented above it turns out that within the minimal
SUSY SU(5) model the upper bound for $M_{H_c}$ is 
of about $10^{16}$~GeV.  Thus, the substantial increase   of about one
order of magnitude for the upper bound on $M_{H_c}$
induced by the three-loop order running analysis attenuates
the tension
between the theoretical predictions made under the assumption (i) and the
experimental data. The choice (ii) for the Yukawa couplings  clearly shows that
the minimal SUSY SU(5) model cannot be ruled out by  the current
experimental data on proton decay rates.   More experimental
information about the SUSY mass spectrum and proton decay rates is
required in order to be able to draw a firm conclusion.

%%%%%%%%%%%%%%%%%%%%%%%%%%%%%%%%%%%%%%%%%%%%%%%%%%%%%%%%%%%%
%%%%%%%%%%%%%%%%%%%%%%%%%%%%%%%%%%%%%%%%%%%%%%%%%%%%%%%%%%%%
%\newpage
\section{\label{sec:mh} The mass of the lightest Higgs
 boson in the   MSSM }
\setcounter{equation}{0}
\setcounter{figure}{0}
\setcounter{table}{0}

\subsection{Higgs boson mass  in the SM}
Spontaneous symmetry breaking was introduced into the particle
physics in the seminal
papers~\cite{Nambu:1960xd,Englert:1964et,Higgs:1964ia} and the 
existence of the Higgs boson was postulated by P.~Higgs in 1964 in
Ref.~\cite{Higgs:1964pj}. 
The next important step was the incorporation of the  spontaneous
symmetry breaking into the unified model of the weak and electromagnetic
interactions~\cite{Weinberg:1967tq,Salam:1968rm}.
% The notion that the
%fermion mass could arise from Yukawa couplings of the Higgs boson to the
%fermions appear for the first time in the paper by S.~Weinberg.
 The breakthrough of these ideas came with the proof of the
renormalizability of spontaneously-broken gauge theories by G.~'t Hooft and
M.~Veltman~\cite{'tHooft:1971rn,'tHooft:1972fi}. 

%Thus, the Higgs boson becomes  essential for the calculability of the
%SM and  its 
%consistency with experimental data.
 
The direct  Higgs boson searches  performed at  LEP 1 in $Z^0\to H+\bar{f}f$  and at LEP 2 in
$e^+e^-\to Z^0+ H$  channels provided us with a lower bound on its mass
of $M_h>114.4$ GeV at the $95$\% CL~\cite{Barate:2003sz}.  In parallel to the direct searches, the
high precision electroweak data obtained at LEP allowed us to estimate the
possible mass range of the Higgs boson within the SM, namely
$M_h=96^{+31}_{\,\,\,\, - 21}$~GeV~\cite{Baak:2012kk}.
 Moreover, the CDF and D0 experiments at the
Tevatron~\cite{:2012nc} excluded a range of Higgs masses between
$156<M_h<177$ GeV, as well as lower masses in the range already excluded by
LEP. Only very recently, 
 the existence of the Higgs boson could have been experimentally
 confirmed by the ATLAS and CMS collaborations at the
 LHC~\cite{atlas,cms}. Its mass  
is around $125-126$~GeV. Currently, dedicated analysis are performed in
order to establish if the observed boson is just the one predicted by
the SM or  hints towards new physics.

\begin{figure}[h]
\begin{center}
  \epsfig{file=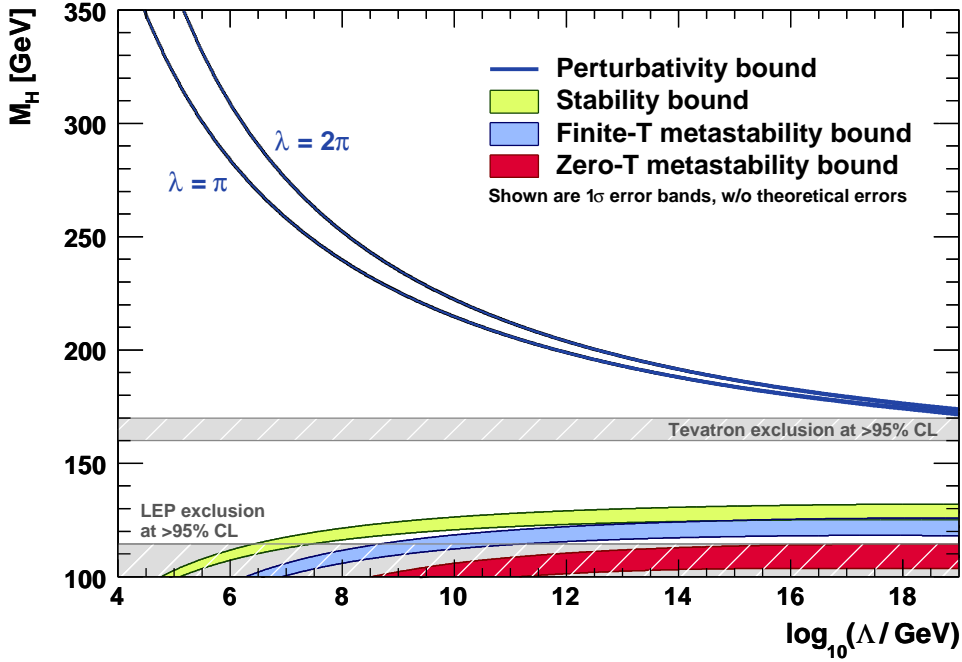, scale=0.65}
\end{center}
  \vspace{-0.5cm}
\caption{ The Higgs boson mass as a function of the scale $\Lambda$
  up to which the SM may remain valid obtained from perturbativity
  (solid dark blue line) and the stability of the electroweak vacuum
  (shaded regions). The figure is taken from Ref.~\cite{Ellis:2009tp}.  
\label{fig:bounds}}
\end{figure}

The Higgs boson mass itself is a fundamental parameter of the SM.
Together with the top quark mass and the strong 
coupling constant, it plays a crucial role in determining the stability
bounds for the SM electroweak vacuum. The usual way to present this
interplay is to display the allowed domains for  $M_h$ as a function of
$\Lambda$, the scale  
up to which the SM may remain valid.   
If $M_h$ is too large , the RGEs
of the SM drive the Higgs self coupling into the non-perturbative regime
at some scale $\Lambda<M_{Planck}$. This is shown as the upper pair bold
lines in Fig.~\ref{fig:bounds} that is taken from
Ref.~\cite{Ellis:2009tp}. In this 
case new physics at a scale  $\Lambda$ 
 will be required in order to prevent the Higgs self-coupling to
 blow-up. On the other hand if $M_h$ is too small, the RGEs drive the
 Higgs self-coupling to a negative value. In this case the Higgs
 potential can develop an instability at high field values $>\Lambda$, 
 unless there is new physics at some scale $<\Lambda$
 that prevents the occurrence of an additional minimum in the potential.
 This is shown as light
 shaded bands in Fig.~\ref{fig:bounds}. Between the blow-up and the
 stability regions, there is a range of intermediate values of $M_h$ for
 which the SM can survive up to the Planck scale.  Taking into account
 the current
 theoretical and experimental 
 errors on $M_h$, $M_t$ and $\alpha_s$, stability up to the Planck scale
 cannot be yet 
 excluded~\cite{EliasMiro:2011aa}.

\begin{figure}[hb]
\begin{center}
\includegraphics[width=0.7\textwidth]{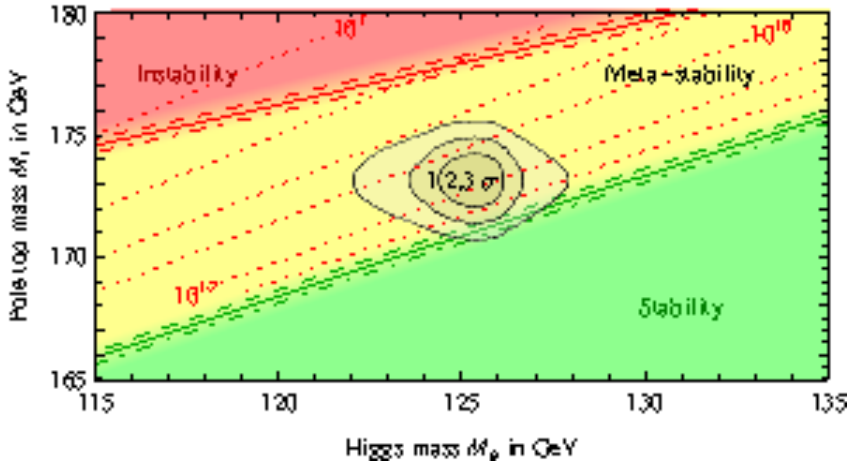}
\end{center}
\vspace{-0.5cm}
\caption{ Measured value of the top mass and preferred range of $m_h$
  as revealed by the present searches at the LHC , compared to the
  regions corresponding to 
absolute stability, meta-stability and instability of the SM
vacuum~\cite{EliasMiro:2011aa}. 
The three  boundaries lines corresponds to $\alpha_s(M_Z)=0.1184\pm
0.0007$, and the grading of the  
colours indicates the size of the theoretical errors.  
The dotted contour-lines show the instability scale $\Lambda$ in $\GeV$
assuming $\alpha_s(M_Z)=0.1184$. 
%  (1.5~GeV in $m_t$). 
\label{fig:regions}}
\end{figure}

Nevertheless, as shown in Fig.~\ref{fig:regions} from
Ref.~\cite{EliasMiro:2011aa} and confirmed by
Ref~\cite{Bezrukov:2012sa}, the range of $M_h$ as 
revealed by the present searches at the LHC lies right at the edge
between electroweak stability and instability regions.
The possibility that the SM potential becomes unstable
at large field values, below the Planck scale, does not contradict any
experimental observation, provided its lifetime is longer than the age
of the universe. Indeed, the authors of Ref.~\cite{Degrassi:2012ry}
found that for $M_h=125$~GeV, the instability scale lies around
$10^{11\pm 1}$~GeV. In this case, tunneling through quantum fluctuations 
is slow enough to ensure at least metastability of the  electroweak vacuum.
 
It is also interesting to note that the SM extrapolation of the Higgs
parameters (the  mass parameter $m^2$ and quartic coupling $\lambda$) corresponds to near
vanishing $\lambda$ and its beta function at the Planck scale. The coupling
$\lambda=0$ is the critical value for the  electroweak stability. Moreover, the
coefficient $m^2$ of the Higgs bilinear in the scalar potential  is also
approximately zero (at the Planck scale). This is again a
critical value that separates the symmetric phase ($m^2>0$) from the
broken phase ($m^2<0$). At present, we do not know if this is just a
numerical coincidence or the consequence of an underlying
symmetry.\\
 There are different interpretations in the literature for the
near-criticality of the SM parameters. For instance, SUSY implies that
$m^2=0$. If SUSY is softly broken, $m^2$ would remain near zero, solving
the hierarchy problem. Nevertheless, the analysis performed in
Ref.~\cite{Degrassi:2012ry} shows that the usual low-scale SUSY
scenario can accommodate a Higgs mass around $125$~GeV only for extreme
values of the parameters, {\it e. g.} large $\tan\beta$, heavy stops, or
maximal stop mixing.  Other explanations of the near-criticality can be
given via interpreting the Higgs as a Goldstone boson (composite Higgs modes) or as a     
consequence of transplanckian dynamics (like in multiverse models). In
the following we concentrate on the SUSY explanation.

\subsection{Higgs boson mass in the  MSSM}

\begin{figure}
  \centering    
    \includegraphics[width=.85\textwidth]{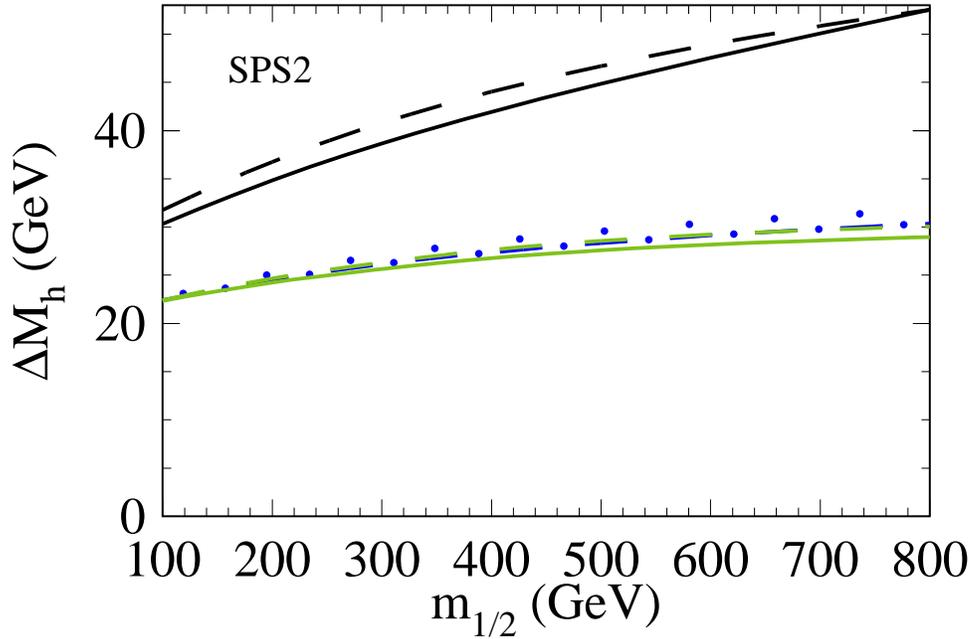}
  \caption{\label{fig::mt4}Comparison of complete and approximate one-
    and two-loop corrections to the Higgs boson mass for 
    SPS2 scenario.  The solid (full result) and dashed lines ($m_t^4$
    approximation) represent the results in the on-shell scheme where
    the black and gray curves correspond to the one- and two-loop
    results, respectively.  For comparison, the two-loop \drbar{} results are shown as
    dash-dotted (full result) and dotted ($m_t^4$ approximation) curves.
  }
\end{figure}

A natural possibility to counterbalance the effects of the top
quark on the evolution of the Higgs self-coupling was found within  SUSY
models, via the opposite effects induced by  the top quark superpartners. The mass of
the Higgs boson within SUSY models is linked to the magnitude of its self-coupling, which
in turn is fixed by SUSY in terms of the electroweak gauge
couplings. Compared to the SM, the MSSM Higgs sector is described by
two additional parameters, usually chosen to be the pseudo-scalar mass
$M_A$ and the ratio of the vacuum expectation values of the two Higgs
doublets, $\tan\beta=v_2/v_1$. The masses of the other Higgs bosons are
then fixed by \susy{} constraints. In particular, the mass of the light
{\abbrev CP}-even Higgs boson, $M_h$, is bounded from above. At
tree-level, this constraint reads $M_h<M_Z$.
 Radiative corrections to the Higgs pole
mass raise this bound substantially to values that were inaccessible
at {\abbrev LEP}~\cite{Ellis:1990nz,Okada:1990vk,Haber:1990aw}.
The
dominant radiative corrections are given by the contribution $\sim \alpha_t m_t^2\sim
m_t^4$ coming from top- and top squark loops ($m_t$ is the top quark
mass and $\sqrt{\alpha_t}$ is proportional to the top Yukawa
coupling). For illustration,
complete and approximate ({\it i.e.} only contributions $\sim m_t^4$ )
one- and two-loop corrections to the lightest 
Higgs boson mass are shown in Fig.~\ref{fig::mt4} from 
Ref.~\cite{Kant:2010tf}. In this figure, the mass difference  $\Delta M_h = M_h^{i--\mbox{loop}}-M_h^{\mbox{tree}}$
is shown as a function of the parameter $m_{1/2}$
in the scenario SPS2~\cite{Allanach:2002nj}.
The small differences between the solid (full result)  and dashed
($m_t^4$ approximation) lines demonstrate that the 
leading  term $\sim m_t^4$ approximates the full result to a high
accuracy. This motivates the computation of higher order corrections
taking into account only the contributions that scale like  $\sim m_t^4$.\\
From the one-loop corrections to the Higgs pole masses, that  are known without any
approximations~\cite{Chankowski:1991md,Brignole:1992uf,
Dabelstein:1994hb,Pierce:1996zz}, one can    show that a second
approximation is appropriate:
The bulk of the
numerical effects can be obtained in the so-called effective-potential
approach, for which the  external momentum of the Higgs propagator is
set to zero.
Most of the   relevant two-loop corrections have  been
evaluated in this approach (for reviews, see
e.g. Refs.\,\cite{Heinemeyer:2004ms,Allanach:2004rh}).
In addition,
 two-loop corrections  including even
CP-violating couplings and improvements from renormalization group
considerations have been computed in
Refs.~\cite{Heinemeyer:2004ms,Allanach:2004rh,Frank:2006yh}. In particular
CP violating phases can lead to a shift of a few GeV in $M_h$, see,
e.g., Refs.~\cite{Heinemeyer:2007aq,Carena:2000yi}.
In Ref.~\cite{Martin:2002wn} a large class of sub-dominant two-loop
corrections to the lightest Higgs boson mass have been considered.
 Furthermore, leading logarithmic corrections at
three-loop order have been computed in  Ref.~\cite{Martin:2007pg}. The first complete three-loop
calculation of the leading quartic top quark mass terms within
  supersymmetric QCD  has been performed in
Refs.~\cite{Harlander:2008ju,Kant:2010tf}.\\ 
There are by now three computer programs publicly available which
include most of the higher order corrections.  {\code FeynHiggs} has
been available already since
1998~\cite{Heinemeyer:1998yj,Degrassi:2002fi,Frank:2006yh} and has been
continuously improved since
then~\cite{Hahn:2009zz}. In particular, it contains
all numerically important two-loop corrections and accepts both real and
complex MSSM input parameters.  The second program, {\code
  CPSuperH}~\cite{Lee:2003nta,Lee:2007gn}, is based on a renormalization
group improved  calculation and allows for explicit CP
violation.  Both programs compute the masses  as well as the decay
widths of the neutral and charged Higgs bosons. The third program,
\hthreel{}~\cite{h3m}, 
contains all currently available  three-loop results. Furthermore,
\hthreel{} constitutes an interface to {\code
  FeynHiggs}~\cite{FeynHiggs} and various 
SUSY spectrum generators which allows for precise predictions of $M_h$ on the
basis of realistic SUSY scenarios.

\subsubsection{\label{subsec:mh3l} Calculation of ${\cal
    O}(\alpha_t\alpha_s^2)$ corrections  in the MSSM }

In this section we focus on details of the calculation of the lightest
Higgs boson 
mass to three-loop accuracy in SUSY-QCD. It was   the first 
calculation of an observable at this order of accuracy in the framework
of  SUSY-QCD and it raised  technical
difficulties specific to higher order calculations. 

At tree-level, the mass matrix of the neutral, {\abbrev CP}-even Higgs
bosons $h$, $H$ has the following form:
\begin{eqnarray}
%  \lefteqn{
  {\cal M}_{H,\rm tree}^2 =
  \frac{\sin 2\beta}{2}\times
%}
%\\&&
  \left(
  \begin{array}{cc}
    M_Z^2 \cot\beta + M_A^2 \tan\beta &
    -M_Z^2-M_A^2 \\
    -M_Z^2-M_A^2 &
    M_Z^2 \tan\beta + M_A^2 \cot\beta
  \end{array}
  \right)
  \,.
%\nonumber
\end{eqnarray}
The diagonalization of ${\cal M}_{H,\rm tree}^2$ gives the tree-level result
for $M_h$ and $M_H$, and leads to the well-known bound $M_h < M_Z$ which
is approached in the limit $\tan\beta\to \infty$.

%Quantum corrections to the Higgs boson masses are incorporated by
%evaluating  the Higgs boson propagators at higher orders. As
%already mentioned, the numerically dominant contributions
%can be obtained in the approximation of zero external momentum (see,
%e.g., Refs.~\cite{Heinemeyer:1998jw:1998kz:1998np:1999be}).
%Furthermore, only corrections of order $\alpha_t\alpha_s^2$
% will be  considered.  

\begin{figure}[ht]
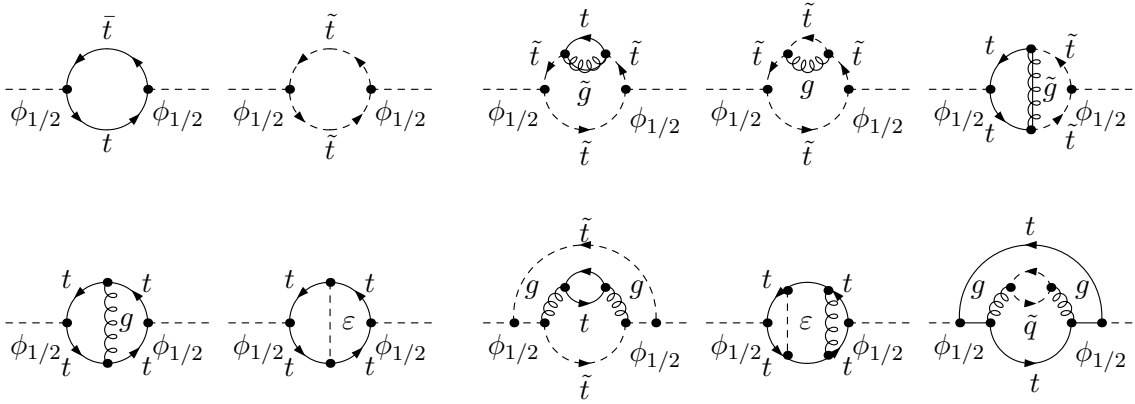

  \centering
  \begin{tabular}{ll}
\unitlength=1.4bp%
\begin{feynartspicture}(120,60)(2,1)
\input  figs/diagramsmh1l
\end{feynartspicture}
&
\unitlength=1.4bp%
\begin{feynartspicture}(180,60)(3,1)
%\begin{feynartspicture}(285,85)(3,1)
\input  figs/diagramsmh2l
\end{feynartspicture}\\
\unitlength=1.4bp%
\begin{feynartspicture}(120,60)(2,1)
\input  figs/diagramsmh2l2
\end{feynartspicture}
&
\unitlength=1.4bp%
\begin{feynartspicture}(180,60)(3,1)
\input  figs/diagramsmh3l
\end{feynartspicture}
  \end{tabular}
  \caption{Sample diagrams contributing to $\Sigma_{\phi_1}$,
   $\Sigma_{\phi_2}$, $\Sigma_{\phi_1\phi_2}$, etc. to one-, two- and three-loops.
%    $\Sigma_A$, $t_{\phi_1}$ and $t_{\phi_2}$.
    Internal solid, dashed, dotted and curly lines correspond to
    top quarks, top squarks, \epscalar{}
    and gluons, respectively. Gluinos are
    depicted with as curly lines with an additional solid line in the middle.
    The external dashed line corresponds to the Higgs bosons.
  }
  \label{fig::diags}
\end{figure}

The mass matrix ${\cal M}_H^2$ is obtained from the quadratic terms in
the Higgs boson potential constructed from the fields $\phi_1$ and
$\phi_2$. They are related to the physical Higgs mass eigenstates via
the mixing angle $\alpha$
\begin{eqnarray}
\left(
\begin{array}{c}
H\\h
\end{array}
\right) =
\left(
\begin{array}{cc}
\cos\alpha&\sin\alpha\\
-\sin\alpha & \cos\alpha 
\end{array}
\right)
\left(
\begin{array}{c}
\phi_1\\
\phi_2
\end{array}
\right)\,.
\label{eq::hmix}
\end{eqnarray}
As usual, $h$ stands for the lightest Higgs boson. 
The mixing angle $\alpha$ is determined at the leading order through
\begin{eqnarray}
\tan 2\alpha= \tan 2 \beta \frac{M_A^2+M_Z^2}{M_A^2-M_Z^2}\,; 
\quad -\frac{\pi}{2}<\alpha<0\, ,
\label{eq::h_alpha}
\end{eqnarray}
where $M_Z$ is the mass of the Z boson and $\tan  \beta=v_2/v_1$. 
Since $\phi_1$ does not couple directly to top
quarks, it is convenient to perform the calculations of the Feynman
diagrams in the $(\phi_1,\phi_2)$ basis.

Including higher order corrections, one obtains the Higgs boson mass
matrix
\begin{eqnarray}
  {\cal M}_{H}^2 &=&
  {\cal M}_{H,\rm tree}^2 -
  \left(
  \begin{array}{cc}
    \hat\Sigma_{\phi_1}       & \hat\Sigma_{\phi_1\phi_2} \\
    \hat\Sigma_{\phi_1\phi_2} & \hat\Sigma_{\phi_2}
  \end{array}
  \right)
  \,,
  \label{eq::MH}
\end{eqnarray}
which again gives the physical Higgs boson masses upon diagonalization.
The renormalized quantities $\hat\Sigma_{\phi_1}$, $\hat\Sigma_{\phi_2}$
and $\hat\Sigma_{\phi_1\phi_2}$ are obtained from the self energies of
the fields $\phi_1$, $\phi_2$, $A$, evaluated at zero external
momentum, as well as from tadpole contributions of $\phi_1$ and $\phi_2$
(see, e.g., Ref.\,\cite{Heinemeyer:2004ms}). 
\begin{align}
  \Sighat_{\phi_1} &= \Sigma_{\phi_1} \nonumber
  \begin{aligned}[t]
    &- \Sigma_A \sin^2\beta\\
    &+ \frac{e}{2M_W\sw} \tad1 \cos\beta\left( 1 + \sin^2\beta \right)\\
    &- \frac{e}{2M_W\sw} \tad2 \cos^2\beta \sin\beta \,,
  \end{aligned}\\
  \Sighat_{\phi_2} &= \Sigma_{\phi_2} \nonumber
  \begin{aligned}[t]
    &- \Sigma_A \cos^2\beta\\
    &- \frac{e}{2M_W\sw} \tad1 \sin^2\beta \cos\beta\\
    &+ \frac{e}{2M_W\sw} \tad2 \sin\beta \left( 1 + \cos^2\beta \right) \,,
  \end{aligned}\\
  \Sighat_{\phi_1\phi_2} &= \Sigma_{\phi_1\phi_2}
  \begin{aligned}[t]
    &+ \Sigma_A \sin\beta \cos\beta\\
    &+ \frac{e}{2M_W\sw} \tad1 \sin^3\beta\\
    &+ \frac{e}{2M_W\sw} \tad2 \cos^3\beta \,.
  \end{aligned}
\end{align}
In this equation, $\vartheta_W$ is the weak mixing angle, $\Sigma_A$
denotes the self energy of the pseudo-scalar Higgs boson and
$t_{\phi_i}$ the tadpole contributions of the field $\phi_i$.
Typical diagrams to the
individual contributions can be found in Fig.~\ref{fig::diags}.

 Considering the many different mass parameters entering the formula for
the Higgs boson mass an exact calculation of the three-loop corrections
is currently not feasible. However, it is possible to apply
expansion techniques~\cite{Smirnov:2002pj} for various limits which
allow to cover a large 
part of the supersymmetric  parameter space. After the application of
the asymptotic expansion the resulting integrals have to be reduced to an
independent set of  master integrals. For the
case of the Higgs mass corrections there will be only three-loop tadpole
integrals that can be handled with the program
MATAD~\cite{Steinhauser:2000ry}.    

\begin{figure}[t]
  \centering
%  \begin{tabular}{cc}
    \includegraphics[width=.45\linewidth]{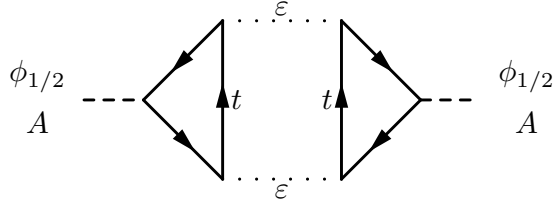}
\caption{Sample diagram contributing  a finite term to $\Sigma_{\phi_1}$,
    $\Sigma_{\phi_2}$, etc. when the  infra-red
divergence are regulated through a small external momentum or a finite
\eps-scalar mass.}
  \label{fig::diageps}
\end{figure}

A technical subtlety arises when calculating  diagrams like those  shown in
Fig.~\ref{fig::diageps}.  If both the external momentum and the
\epscalar{} mass are set to zero from the beginning, an infra-red
divergence occurs and cancels the ultra-violet divergence of the
integral. In effect, the diagram will be of order $(d-4)$ due to the
\epscalar{} algebra. In order to avoid this, one can keep the external
momentum $q$ non-zero, though much smaller than all other scales. The
ultra-violet pole multiplied by the algebraic factor of $(d-4)$ then
produces a finite contribution, while the infra-red divergence leads to
a contribution 
$(d-4)\ln(q^2)$ that vanishes as $d\to 4$.
Instead of the requirement $q\not=0$ one could also introduce a nonzero
mass for the \eps{} scalars in order to regulate the infra-red
divergences. In the final result one  observes that the regulator is
multiplied by an additional factor $(d-4)$ leading to a finite 
result for $\Mes\to 0$. Alternatively, one can  allow a non zero
\eps-scalar mass and shift the squark mass
counter-terms so that all  $\Mes$ dependent 
contributions in the final result  cancel out\footnote{This
  renormalization scheme is equivalent with the  \drbarprime{} 
scheme discussed in section~\ref{sec:drbarprime}, however it is not
identical.}. 
All these renormalization prescriptions
lead to identical results for the corrections to the Higgs boson mass
$M_h$, that is a non trivial check of the calculation.

\begin{figure}[t]
  \centering
  \begin{tabular}{cc}
    \includegraphics[width=.45\linewidth]{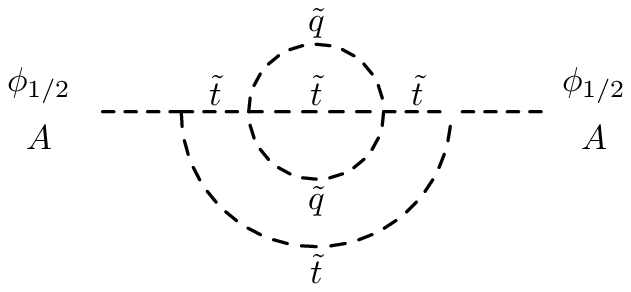}
    &
    \includegraphics[width=.45\linewidth]{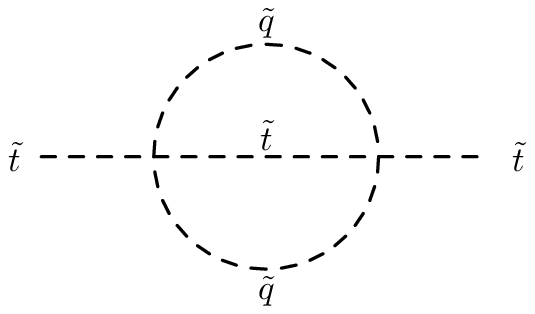}
    \\ (a) & (b)
  \end{tabular}
  \caption{(a) Feynman diagram involving a heavy virtual squark 
    contributing to the Higgs boson self energy. 
    (b) Counter-term diagram related to the diagram in (a).
    The same notation as in Fig.~\ref{fig::diags}
    has been adopted.
  }
  \label{fig::stst_squark}
\end{figure}

Concerning the renormalization, it is well known that the perturbative
series can exhibit a bad 
convergence behaviour in case it is parametrized in terms of the
on-shell quark masses\footnote{For a typical example we refer to the
  electroweak $\rho$ parameter. Using the on-shell top quark mass the
  four-loop
  corrections~\cite{Schroder:2005db,Chetyrkin:2006bj,Boughezal:2006xk}
  are larger by a factor 50 as compared to the $\overline{\rm MS}$
  scheme.}  which is due to intrinsically large contributions related to
the infra-red behaviour of the theory.  Thus, it is tempting to
re-parametrize the results for the Higgs boson mass in terms of
 the top quark mass renormalized in the \drbar{} scheme. Moreover, the two-loop
 renormalization constants for the masses of the
SUSY particles and the top squark mixing angle, that are required for
this calculation, 
are much more complicated
in the on-shell scheme as compared to the \drbar{} ones. Thus, it is
preferable to adopt the \drbar{}  scheme also for these
parameters. The renormalization constants for the gluino and \epscalar{}
masses are needed only at the one-loop order. For them, both schemes are
accessible. Nevertheless, the \epscalar{} mass renormalized in the
on-shell scheme might 
be better suited for this type of calculations. In this case, it can be
set equal to zero in the 
 three-loop diagrams, that makes the calculation less involved. An
 extensive discussion about the calculation of the two-loop 
  renormalization constants required in this
computation as well as explicit formulae can be found in
section~\ref{sec:susyqcd}.  
 In the remainder of this section we will
refer to this renormalization scheme as \drbar{} scheme although it
contains a mixture of on-shell and \drbar{} parameters in order to
distinguish between it and the genuine on-shell scheme.

At this point a comment concerning the minimal  \drbar{}
renormalization constants for the masses of the top squarks is in order.  Due to
diagrams involving heavy squarks $\tilde q$, for example
Fig.~\ref{fig::stst_squark}(a), the squared Higgs boson mass receives
contributions which are proportional to $\msq^2$ and thus can lead to
unnatural large corrections. For this reason the on-shell
scheme for these contributions is better suited, because it avoids the
potentially large terms $\sim m_{\tilde q}^2$ from the three-loop
diagrams.  The
renormalization of the mixing angle is free of such enhanced
contributions and  can be done in the pure \drbar{} scheme.
A similar behaviour is  observed when the gluino  is much heavier than the  
top squarks~\cite{Degrassi:2001yf,Kant:2009zza}.
  In this case,  the two- and three-loop 
corrections to the Higgs masses contain 
terms proportional to  $m_{\tilde{g}}$ and $m_{\tilde{g}}^2$. These 
contributions are canceled when the masses are renormalized in the
on-shell scheme by the finite parts of the  
relevant counter-terms. 
Thus, in order to avoid unnatural large radiative corrections 
to the Higgs mass for scenarios with heavy  gluinos
a modified  non-minimal renormalization scheme for the top squark masses
is required.  The additional finite shifts of top 
squark masses are chosen such that they  cancel the power-like behaviour
of the gluino contributions. Again, the renormalization of the mixing angle
will not be modified as compared to the genuine \drbar{} scheme.
The relevant finite shifts for commonly adopted  scenarios 
are explicitly given in Appendix~B.

\begin{figure}[t]
  \centering
%  \begin{tabular}{cc}
    \includegraphics[width=.85\textwidth]{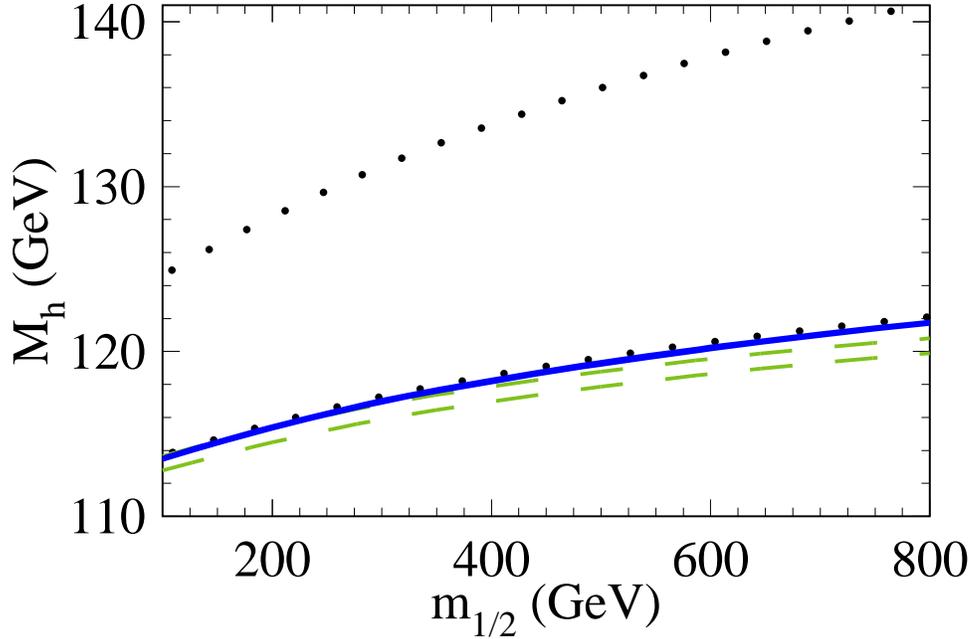}
%    &
%    \includegraphics[width=.49\textwidth]{figs/mh_dros_sps2_magn.eps}
%  \end{tabular}
  \caption{\label{fig::DRvsOS} Renormalization scheme dependence 
    of $M_h$ as a function of $m_{1/2}$ adopting SPS2.
    Dotted, dashed and solid curves correspond to one-, two- and
    three-loop results. The \drbar{} (on-shell) results correspond to the 
    lower (upper)  curves. The three-loop curves obtained in the two
    renormalization schemes lay on top of each other.
  }
\end{figure}

As an illustration of the renormalization scheme issue,  
 we show in Fig.~\ref{fig::DRvsOS} from Ref.~\cite{Kant:2010tf}  the
 renormalization scheme dependence
of $M_h$ as a function of the parameter $m_{1/2}$ for the SPS2 scenario.
In the left panel of Fig.~\ref{fig::DRvsOS}  the upper dotted, dashed and
solid curve correspond to the one-, two- and three-loop prediction of $M_h$ in
the on-shell scheme whereas the corresponding lower three curves 
are obtained in the \drbar{} scheme.
In the on-shell scheme one observes large positive one-loop
corrections which get reduced by 10 to 20~GeV after including the two-loop
terms. The three-loop corrections amount to several hundred MeV. They are
positive or negative  depending on the value of $m_{1/2}$.
The situation is completely different for \drbar{} mass
parameters: the one-loop corrections are significantly smaller and lead
to values of $M_h$ which are already of the order of the two- and
three-loop on-shell prediction. The two-loop term leads to a small shift
of the order of $-1$~GeV and the three-loop term to a positive shift of
about the same order of magnitude. The final prediction for $M_h$ is
very close to the one obtained after incorporating three-loop on-shell
results.\footnote{There are
  regions in the parameter space where the two-loop corrections are
  accidentally small in the \drbar{} scheme leading to relatively large
   three-loop terms. Nevertheless the overall size of the two-
  and three-loop corrections is small.}

The three-loop results have in general very long expressions. However,
for simplifying assumptions about the supersymmetric mass spectrum, like
for example the {\it natural} SUSY, for which the superpartners of the
first and second generation of quarks are much heavier than the gluino
and third generation of squarks, {\it i. e.}
 $m_{\tilde q} \gg m_{\tilde t_1} \approx m_{\tilde t_2} \approx m_{\tilde g}$,
the analytical expressions for the dominant contributions have a quite compact
form. Let us mention that in general, for the case of quasi degenerate
masses a naive Taylor expansion in the mass differences is
sufficient, while for large mass ratios an asymptotic expansion is necessary. 
For illustration, we give below the three-loop results for the
two-point functions contributing to the Higgs boson mass,
%$\hat\Sigma_{\phi_1}$, $\hat\Sigma_{\phi_2}$ and
%$\hat\Sigma_{\phi_1\phi_2}$ 
 where for the renormalization of the stop quark
masses the modified \drbar{} scheme as given in Eq.~(\ref{eq:h4shift})
was adopted.
  % \mbox{Notation wie im PRL aber DRbar!; incl. At and $1/Msq^2$}
  % \,,\nonumber\\
{\allowdisplaybreaks
\begin{align} 
  \hat\Sigma_{\phi_1}
  &= \frac{G_F \Mt^4 \sqrt{2}}{\pi^2 \cos^2\beta}
%  \begin{aligned}[t]
    \left(\afourpi\right)^{\!2}
%    &
    \frac{A_t^2}{\Msusy^2}
    \bigg[
    -\frac{349}{9} 
    + \frac{32}{9} \lmumt 
    + \frac{32}{9} \lmumt^2 
%    \notag\\&
    + \bigg(
    \frac{56}{9} 
    + \frac{64}{9} \lmumt
    \bigg) \lmmtMS 
    + \frac{32}{9} \lmmtMS^2 
  \notag\\&
    + \frac{94}{3} \z{3}
    + \mathcal{O}\left(\frac{\Msusy^4}{\msq^4}\right)
    \bigg]\,,
%  \end{aligned}
  \nonumber\\
%  \displaybreak[2]\\
  \hat\Sigma_{\phi_{12}}
  &= \frac{G_F \Mt^4 \sqrt{2}}{\pi^2 \cos\beta \sin\beta}
  \bigg[
  \afourpi \frac{A_t}{\Msusy}
  \Big(
  -2 
  - 4 \lmumt 
  - 2 \lmmtMS
  \Big)
  \notag\\&
  + \left(\afourpi\right)^{\!2} \negthickspace
%  \begin{aligned}[t]
    \bigg\{{}
%&
    \frac{A_t^2}{\Msusy^2}
 %   \begin{aligned}[t]
      \bigg({}
%&
      \frac{349}{9} 
      - \frac{32}{9} \lmumt 
      - \frac{32}{9} \lmumt^2 
      + \Big(
      -\frac{56}{9} 
      - \frac{64}{9} \lmumt
      \Big) \lmmtMS 
      - \frac{32}{9} \lmmtMS^2 
      - \frac{94}{3} \z{3}
      \bigg)
%    \end{aligned}
    \notag\\&
    + \frac{A_t \Msusy}{\msq^2}
    \bigg(
    40 
    - \frac{160}{9} \lmmtMsq 
    - \frac{40}{3} \lmmtMsq^2 
    + \lmmtMS \Big(
    \frac{160}{9} 
    + \frac{40}{3} \lmmtMsq
    \Big) 
    - \frac{80}{3} \z{2}
    \bigg)
    \notag\\&
    + \frac{A_t}{\Msusy} \bigg(
    -\frac{416}{27} 
    - \frac{364}{27} \lmumt 
    - \frac{100}{3} \lmumt^2 
    - \frac{304}{9} \lmmtMS^2 
    + \frac{200}{9} \lmmtMsq 
    - 20 \lmmtMsq^2 
    \notag\\&
    + \lmmtMS \Big(
    -\frac{628}{27} 
    - \frac{400}{9} \lmumt 
    + \frac{80}{3} \lmmtMsq
    \Big) 
    - 40 \z{2} 
    + \frac{106}{3} \z{3}
    \bigg)
    \bigg\}
    + \mathcal{O}\left(\frac{\Msusy^4}{\msq^4}\right)
    \bigg]\,,
%  \end{aligned}
\nonumber\\
%\end{align}
%\\
%\displaybreak[2]\\
%\begin{align}
  \hat\Sigma_{\phi_2} 
  &= \frac{
    G_F \Mt^4 \sqrt{2} } {\pi^2 \sin^2\beta}
  \bigg[
  \frac{3}{2} \lmmtMS 
  + \afourpi 
  \bigg(
  4 
  + \Big(
  4 
  + 16 \lmumt
  \Big) \lmmtMS 
  + 4 \lmmtMS^2 
  + \frac{A_t}{\Msusy} \Big(
  4 
  + 8 \lmumt 
  + 4 \lmmtMS
  \Big)
  \bigg) 
  \notag\\&
  + \left(\afourpi\right)^{\!2} \negthickspace
%{\allowdisplaybreaks
%  \begin{aligned}[t]
    \bigg\{{}
%&
%{}&
    \frac{2764}{9} 
    - \frac{116}{27} \lmumt 
    - \frac{136}{3} \lmumt^2 
    + \bigg(
 -\frac{644}{9} 
    + \frac{164}{3} \lmumt
    \bigg) \lmmtMS^2 
    \notag\\&
    + 24 \lmmtMS^3 
    + \frac{400}{3} \lmmtMsq 
    - \frac{200}{3} \lmmtMsq^2 
    - \frac{20}{3} \lmmtMsq^3 
    - 120 \z{2} 
    - 80 \lmmtMsq \z{2} 
    + \frac{8}{3} \z{3} 
    \notag\\&
    - \bigg(
    \frac{2216}{27} 
    + \frac{644}{9} \lmumt 
    - \frac{328}{3} \lmumt^2 
    - 40 \lmmtMsq 
    - 20 \lmmtMsq^2 
    - 40 \z{2} 
    + 16 \z{3}
    \bigg) \lmmtMS 
    \notag\\&
    + \frac{\Msusy^2 }{\msq^2}
%    \begin{aligned}[t]
      \bigg({}
%&
      \frac{42356}{225} 
      + 8 \lmmtMS^2 
      - \frac{2128}{45} \lmmtMsq 
      - \frac{176}{3} \lmmtMsq^2 
%      \notag\\&
      + \Big(
      \frac{3928}{45} 
      + \frac{152}{3} \lmmtMsq
      \Big) \lmmtMS 
      - \frac{400}{3} \z{2}
      \bigg)
%    \end{aligned}
    \notag\\&
    + \frac{A_t \Msusy}{\msq^2}
%    \begin{aligned}[t]
      \bigg({}
%&
      -80 
      + \lmmtMS \Big(
      -\frac{320}{9} 
      - \frac{80}{3} \lmmtMsq
      \Big) 
      + \frac{320}{9} \lmmtMsq 
      + \frac{80}{3} \lmmtMsq^2 
      + \frac{160}{3} \z{2}
      \bigg)
%    \end{aligned}
    \notag\\&
    + \frac{A_t}{\Msusy}
%    \begin{aligned}[t]
      \bigg({}
%&
      \frac{832}{27} 
      + \frac{728}{27} \lmumt 
      + \frac{200}{3} \lmumt^2 
      + \frac{608}{9} \lmmtMS^2 
%      \notag\\&
      + \lmmtMS \Big(
      \frac{1256}{27} 
      + \frac{800}{9} \lmumt 
      - \frac{160}{3} \lmmtMsq
      \Big) 
      \notag\\&
      - \frac{400}{9} \lmmtMsq 
      + 40 \lmmtMsq^2 
      + 80 \z{2} 
      - \frac{212}{3} \z{3}
      \bigg)
%    \end{aligned}
    \notag\\&
    + \frac{A_t^2}{\Msusy^2}
%    \begin{aligned}[t]
      \bigg({}
%&
      -\frac{349}{9} 
      + \frac{32}{9} \lmumt 
      + \frac{32}{9} \lmumt^2 
      + \Big(
      \frac{56}{9} 
      + \frac{64}{9} \lmumt
      \Big)  \lmmtMS 
%      \notag\\&
      + \frac{32}{9} \lmmtMS^2 
      + \frac{94}{3} \z{3}
      \bigg)
      \bigg\}
 \notag\\&
      + \mathcal{O}\left(\frac{\Msusy^4}{\msq^4}\right)
      \bigg]\,,
%    \end{aligned}
%  \end{aligned}
%}
 \end{align}
}
$\!\!$with $m_t=m_t(\mu_r)$, $\Msusy=\Msusy(\mu_r)=
\mstop{1}(\mu_r)=\mstop{2}(\mu_r)=\mgluino(\mu_r)$,
$\lmumt=\ln(\mu_r^2/m_t^2)$, $\lmmtMS=\ln(m_t^2/m_{\rm SUSY}^2)$ and
$\lmmtMsq=\ln(m_t^2/m_{\tilde{q}}^2)$ where $\mu_r$ is the
renormalization scale.

\subsubsection{\label{sec:mhphen} Phenomenological analysis}

In order to quantify  the phenomenological significance of the
three-loop contributions ,
it is interesting to investigate the dependence of $M_h$ on 
SUSY parameters. In the following, we adopt the `` modified $m_h^{\rm max}$
'' scenario as defined in Ref~\cite{Pak:2012xr}. The relevant 
MSSM parameters  for our analysis are the  top squark masses
$m_{\tilde{t}_1}=370$~GeV and $m_{\tilde{t}_2}=1045$~GeV, the gluino
mass $m_{\tilde{g}}=860$~GeV, the squark  mass scale
$m_{\tilde{q}}=1042$~GeV, 
the top  trilinear coupling $A_t=1500$~GeV and the mass of the pseudoscalar
Higgs $M_A= 1000$~GeV.
% and finally $\tan\beta=20$.

\begin{figure}
  \centering
\hspace*{-1.cm}
    \includegraphics[width=0.85\textwidth]{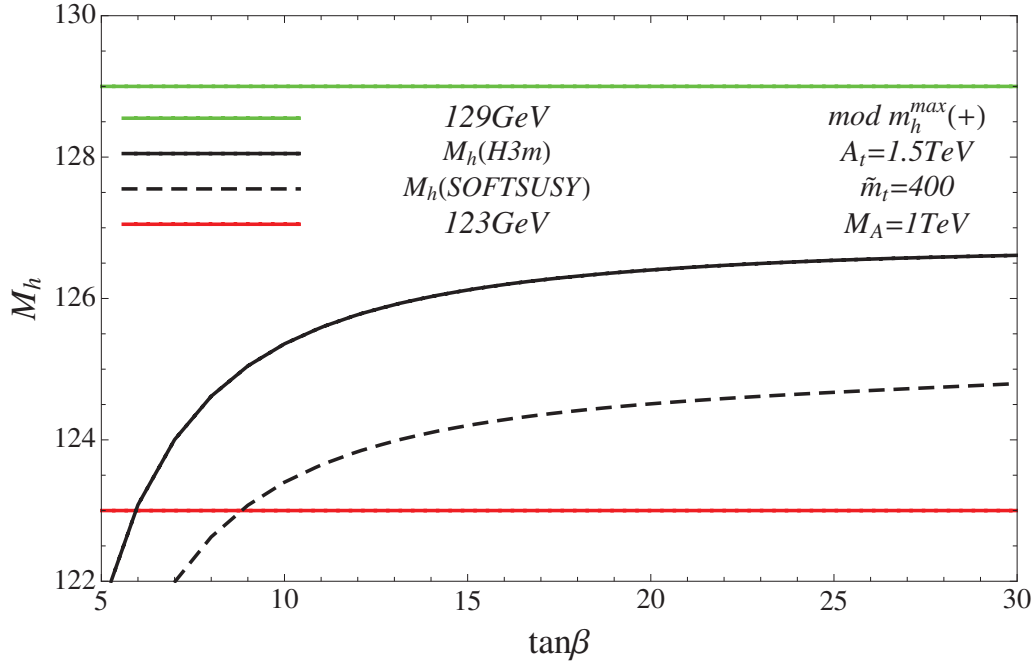}
  \caption{\label{fig::mhmt} The light Higgs boson mass as a function of
    $\tan\beta$ to three-loop accuracy from
     Ref.~\cite{Pak:2012xr}}. 
\end{figure}

In Fig.~\ref{fig::mhmt} from Ref.~\cite{Pak:2012xr} is shown the
comparison between the two- (dashed line)  and three-loop (full line) 
 predictions for the Higgs boson mass as a function of $\tan\beta$ parameter.
As can be read from the plot, the genuine three-loop corrections amount
to around $2$~GeV for the given mass spectrum, independently of the
value of $\tan\beta$. Let us remind the reader that the
experimental accuracy on $M_h$ expected at the LHC is almost an order of
magnitude smaller. It is also worth mentioning that the three-loop
corrections are positive and increase the predicted value for $M_h$ 
beyond $125$~GeV.   For increasing gluino and third generation
squark masses, the light  Higgs boson mass becomes larger  
and values
well above 120~GeV can be reached.

 We can infer from the above analysis that
for a precise comparison with the experimental data expected from the
LHC experiments, the three-loop corrections are indispensable.  Moreover, the
MSSM predictions can easily accommodate  a light Higgs boson mass in the
range between $125<M_h<127$~GeV as observed in the current experiments
at the LHC.

%%%%%%%%%%%%%%%%%%%%%%%%%%%%%%%%%%%%%%%%%%%%%%%%%%%%%%%%%%%%%%%%
\section{\label{sec:hdecay}Hadronic Higgs production and decay in susy models}
\setcounter{equation}{0}
\setcounter{figure}{0}
\setcounter{table}{0}

After the discovery of the new scalar particle with the mass around
$125$~GeV the most important question to be answered is whether it is indeed
the Higgs boson  predicted by the SM or it has another origin. 
%Possible th
%is  just the lightest 
%particle from the five Higgs particles of the MSSM or  wheather it is
%not a fundamental particle at all and  has a composite nature. 
To obtain
the answer one has to study in detail  the interaction properties of
this new scalar 
 with the SM particles. This task requires the comparison of
 the theory predictions for  the production cross
sections  and the  decay rates  of the  
newly discovered scalar particle  with the   
experimental data. In 
 most of the cases, a precision  of the theory predictions at the percent level
 is required in order to cope with the experimental accuracy. 
 This implies that  
  radiative corrections even at next-to-next to leading order (NNLO)
  have to be taken into account.

In the current section we concentrate on the radiative corrections up to
NNLO to the hadronic Higgs production and decays within the MSSM. It
turns out that  in
most of the cases only the NNLO  SUSY-QCD corrections  have to be taken
into account. If available in the literature, also the NNLO top-Yukawa   
corrections, expected to be the next dominant contributions, will be discussed. 
As the exact analytic calculations are not always feasible, several
theoretical methods  employing  phenomenologically  well motivated
simplifying assumptions will be presented.

\subsection{\label{sec:eft} Effective field theory formalism}
 In this section we want to derive the effective field theory formalism
 (EFT) following the method 
of operator product expansion (OPE) introduced by
Wilson~\cite{Wilson:1969zs}\footnote{For a pedagogical overview of the
  method see also Ref~\cite{coll}.}. The main idea is again to
disentangle the long  and short distance physics from each other.
Precisely, the long distance physics is described by local operators
constructed from light degrees of freedom ${\cal O}_i$, whereas the
effects of the heavy degrees of freedom are absorbed into coefficient
functions of the operators. For  QCD  the relevant
local operators have dimension four. Their renormalization and  the
issue of operator mixing under renormalization 
 have been studied in detail in
the literature~\cite{KlubergStern:1975hc,Nielsen:1975ph,Spiridonov:1984}.
For all processes studied in this section, the low energy
effective theory is QCD with five active flavours supplemented with a
light Higgs boson.   
 For completeness, we briefly
review the main results concerning the renormalization of the local
dimension four operators below.

In the following we assume for simplicity that the fundamental theory
is described  by  the SUSY-QCD particle content together with the two
Higgs doublets of the MSSM, $\phi_i\,,$ with $ i=1,2$. 
 The corresponding
interactions are described   by the following Lagrangian

\begin{eqnarray}
{\cal L}&=&{\cal L}_{\rm QCD}+ {\cal L}_{\rm SQCD} 
+\sum_{i=1,2}{\cal L}_{{\rm q}\phi_i} 
 + \sum_{i=1,2}{\cal L}_{{\rm \tilde{q}}\phi_i}\,
\end{eqnarray}
where 
\begin{eqnarray}
{\cal L}_{{\rm q}\phi_i}=-\sum_{q=1}^6 \frac{m_q}{v} g_q^{\phi_i}
 \bar{q} q \phi_i \quad \mbox{and} \quad
{\cal L}_{{\rm \tilde{q}}\phi_i}=-\sum_{q=1}^6\sum_{r,k=1,2} 
\frac{m_q}{v} g_{\tilde{q};kr}^{\phi_i}\tilde{q}^\star_{k}\tilde{q}_r \phi_i\,.
\end{eqnarray}
${\cal L}_{\rm QCD}+ {\cal L}_{\rm SQCD}$ denotes the supersymmetric
 extension of the full QCD Lagrangian with six quark flavours.
 The couplings $g_q^{\phi_i}$ and
 $g_{\tilde{q};kr}^{\phi_i}$ are defined in
 Table~\ref{tab::yukawa_coeff}, where
    $v=\sqrt{v_1^2+v_2^2}$, with
 $v_i\,,i=1,2$, is obtained from the vacuum expectation values of the two Higgs
 doublets of the MSSM. The fields
$\tilde{q}_i\,,$ with $i=1,2$, denote as before the squark mass eigenstates, while
 $\theta_q$ stands for the mixing angle defined through:
\begin{eqnarray}
\sin 2\theta_q=\frac{2 m_q X_q}{m_{\tilde{q}_1}^2-m_{\tilde{q}_2}^2}\,,\quad 
X_q = A_q-\mu_{\rm SUSY} \bigg\{\begin{array}{ll}
\tan\beta\,, &\mbox{ for  down-type quarks}\\
\cot\beta\,,  &\mbox{ for  up-type quarks}
\end{array}  \,,
\label{eq::mixangle}
\end{eqnarray}
where $A_q$ is  the trilinear coupling and $\mu_{\rm SUSY} $    the
Higgs-Higgsino bilinear coupling.

\begin{table}
  \begin{center}
    \begin{tabular}{|l||c|c|c|c|}
\hline
f&$g_q^{\phi_1}$&$g_{\tilde{q};11}^{\phi_1}$&$g_{\tilde{q};12}^{\phi_1}=
g_{\tilde{q};21}^{\phi_1}$&$g_{\tilde{q};22}^{\phi_1}$\\
\hline
up&$0$&$-\mu S_q/S_\beta$&$-\mu  C_q/S_\beta$ &$\mu S_q/S_\beta$\\
\hline
down&$1/C_\beta$& $(2 m_q+A_q S_q)/C_\beta$&$
A_q C_q/C_\beta$&$(2 m_q-A_q S_q)/C_\beta$\\
\hline
\hline
f&$g_q^{\phi_2}$&$g_{\tilde{q};11}^{\phi_2}$&$
g_{\tilde{q};12}^{\phi_2}=g_{\tilde{q};21}^{\phi_2}$&$
g_{\tilde{q};22}^{\phi_2}$\\
\hline
up&$1/S_\beta$& $(2 m_q+A_q S_q)/S_\beta$&$A_q C_q/S_\beta$&$
(2 m_q-A_q S_q)/S_\beta$\\
\hline
down&$0$&$-\mu S_q/C_\beta$&$-\mu  C_q/C_\beta$ &$\mu S_q/C_\beta$\\
\hline
    \end{tabular}
    \caption{\label{tab::yukawa_coeff}Yukawa coupling coefficients for up
      and down type quark and squark, where
$S_q=\sin 2\theta_q$ and $C_q=\cos 2 \theta_q$, and $S_\beta=\sin \beta$ 
and $C_\beta=\cos  \beta$.}
  \end{center}
\end{table}

We assume further  the mass of the lightest Higgs boson
$h$ to be much
 smaller than the mass of the top quark and of the SUSY particles, as well as
  all the other Higgs bosons. In this case, the physical phenomena
 at low energies can be described by an 
effective theory containing  five  quark flavours and the light Higgs
\begin{eqnarray}
{\cal L}\longrightarrow {\cal L}_Y^{\rm eff} + {\cal L}_{\rm
  QCD}^{(5)}\,,
\end{eqnarray}
where $ {\cal L}_{\rm QCD}^{(5)}$ denotes the Lagrangian of QCD with five active
 flavours.\\
  At leading order in the heavy masses, the  effective Lagrangian
 ${\cal L}_Y^{\rm eff}$
can be written as a linear combination of three physical, gauge independent
 operators~\cite{Spiridonov:1984,Chetyrkin:1997un} constructed from the
 light degrees of freedom
\begin{eqnarray}
{\cal L}_Y^{\rm eff}  = -\frac{h^{(0)}}{v^{(0)}}\left[
C_1^0{\cal O}_1^0 + \sum_{q}\left( C_{2q}^0{\cal O}_{2q}^0
 + C_{3q}^0{\cal O}_{3q}^0\right)
\right]\,,
\label{eq::eft}
\end{eqnarray}
where the coefficient functions $C_i\,, i=1,2q,3q$, parametrize the
 effects of the heavy particles on the low-energy phenomena.
 The superscript
 $0$ labels bare quantities. The three
 operators are defined as
\begin{eqnarray}
{\cal O}_1^0 &=& (G_{\mu,\nu}^{0,\prime,a})^2\,,\nonumber\\
{\cal O}_{2q}^0 &=& m_q^{0,\prime}\bar{q}^{0,\prime}
q^{0,\prime}\,,\nonumber\\
{\cal O}_{3q}^0 &=& \bar{q}^{0,\prime}(i\,/\!\!\!\! D^{0,\prime}
-m_q^{0,\prime})q^{0,\prime}\,,
\label{eq::ops}
\end{eqnarray}
where $G_{\mu,\nu}^{0,\prime,a}$ and $ D_{\mu}^{0,\prime}$ are the
gluon field strength tensor and the covariant derivative, respectively. 
The primes label the quantities in the effective theory. 
The relations between the
parameters and fields in the full and effective theories have been
derived in section~\ref{sec::dec}. The explicit formulae
can be read off from Eqs.~(\ref{eq::bare_dec}). 
 The operator ${\cal O}_{3q}$
 vanishes by the fermionic equation
 of motion and it will not contribute to physical observables.
 Thus, the last term in Eq.~(\ref{eq::eft})  might be omitted,
 once the coefficients $C_1^0, C_{2q}^0$ are determined.

For  convenience of the reader we reproduce  the results for the
renormalization constants of the 
 operators ${\cal O}_1^0$ and ${\cal O}_{2q}^0$ that are of interest 
\begin{eqnarray}
&&{\cal O}_1= Z_{11} {\cal O}_1^0 +Z_{12} {\cal O}_{2q}^0\,,\qquad
 {\cal O}_{2q}= Z_{22}{\cal O}_{2q}^0\,,\quad \mbox{where} \nonumber\\
&&Z_{11}=
\left(1-\frac{\pi}{\alpha_s^\prime}\frac{\beta(\alpha_s^\prime)}
{\epsilon}\right)^{-1}
, \,\,\, Z_{12}=-\frac{4\gamma_m(\alpha_s^\prime)}{\epsilon}
\left(1-\frac{\pi}{\alpha_s^\prime}\frac{\beta(\alpha_s^\prime)}
{\epsilon}\right)^{-1},
\, Z_{22}=1\,,\\
&&C_1= Z_{11}^{-1} C_1^0 \,,\qquad\qquad\qquad C_{2q}=C_{2q}^0-
\frac{Z_{12}}{Z_{11}} C_1^0 \,.
\end{eqnarray}
In the above equations the beta function and quark mass anomalous
dimension $\gamma_m$ refer to QCD with $n_l=5$ active flavours evaluated
in the 
\msbar{} scheme. They are needed up to three-loop order
 and have been given explicitly in section~\ref{sec:dred}.

The renormalized coefficient functions and operators are finite but not
renormalization group (RG) invariant. In Ref.~\cite{Chetyrkin:1996ke},
 a redefinition
of the coefficient functions and operators was introduced so that they
are separately renormalization group invariant. The RG invariant operators are defined as follows
\begin{eqnarray}
{\cal O}_g&=&-\frac{2\pi}{\beta_0^{(5)}}\left(
\frac{\pi \beta^{(5)}}{2\alpha_s^{(5)}}{\cal O}_1-2 \gamma_m^{(5)}\sum_q
{\cal O}_{2q}
\right)\,,\nonumber\\
{\cal O}_q&=&{\cal O}_{2q}\,,
\label{eq:giop}
\end{eqnarray}
where the superscript $(5)$ marks that there are five active quarks 
to be considered in the formulas for the beta function and the mass
anomalous dimension $\gamma_m$.
Accordingly, the associated coefficient functions are given by
\begin{eqnarray}
C_g&=&-\frac{\alpha_s^{(5)}\beta_0^{(5)}}{\pi^2\beta^{(5)}} C_1\,,
\nonumber\\
C_q&=&\frac{4\alpha_s^{(5)}\gamma_m^{(5)}}{\pi \beta^{(5)}}C_1 + C_{2q}\,.
\label{eq:gicoeff}
\end{eqnarray}
This procedure allows us to choose independent
renormalization scales for coefficient functions and operators. In
practice, one chooses $\mu\approx M_h$ 
for the
renormalization scale of the operators and  $\mu\approx \tilde{M}$
(where $\tilde{M}$ denotes an averaged mass for the heavy 
supersymmetric particles) for the coefficient functions. Thus, 
 Eq.~(\ref{eq:giop}) is to be evaluated at a low scale   $\mu\approx M_h$, whereas
 Eq.~(\ref{eq:gicoeff}) is to be utilized at a high scale $\mu\approx \tilde{M}$.

For the computation of the Higgs production and decay rates, it is however more convenient
to re-express the effective Lagrangian in terms of the operators ${\cal
  O}_1$ and ${\cal O}_{2q}$. However now, one keeps the separation of the
scales for operators and coefficient functions as given in
Eqs.~(\ref{eq:giop}) and (\ref{eq:gicoeff}).
The new coefficient functions read~\cite{Chetyrkin:1996ke} 
\begin{eqnarray}
{\cal C}_1(\tilde{M}, M_h)&=&
\frac{\alpha_s^\prime(\tilde{M})\beta^{(5)}(\alpha_s^\prime(M_h))}
{\alpha_s^\prime(M_h)\beta^{(5)}(\alpha_s^\prime(\tilde{M}))}C_1(\tilde{M})=-\frac{\pi^2\beta^{(5)}(\alpha_s^\prime(M_h))}{[\alpha_s^\prime(M_h)]^2\beta_0^{(5)}}C_g(\tilde{M})
 \,,\nonumber\\
{\cal C}_2(\tilde{M}, M_h)&=& \frac{4 \alpha_s^\prime(\tilde{M})}{\pi
  \beta^{(5)}(\alpha_s^\prime(\tilde
  M))}[\gamma_m^{(5)}(\alpha_s^\prime(\tilde{M})) 
-\gamma_m^{(5)}(\alpha_s^\prime(M_h))]C_1(\tilde{M})+   
C_{2q}(\tilde{M})\,,
\label{eq::cfrg}
\end{eqnarray}
 The explicit  computation of the
coefficient functions will be 
discussed in detail in the next section.  

\subsection{\label{sec:coefffc}Computation of the coefficient functions $C_1$ and $C_{2q}$}
To calculate the coefficient functions one has to consider appropriate
Green functions in the full and the effective theory and relate them via
the decoupling relations. For example, 
 the amputated Green function
involving the $q \bar{q}$ pair and the zero-momentum insertion of the
operator ${\cal O}_h$ which mediates the couplings to
the light Higgs boson $h$ contains both coefficient functions $C_{2q}$
and $C_{3q}$.   
\begin{eqnarray}
\Gamma_{\bar{q}q{\cal O}_h}^{0}(p,-p)&=&i^2\int {\rm d}x {\rm d}y e^{i
  p (x-y)}\langle T q^{0}(x) \bar{q}^0(y){\cal
  O}_h(0)\rangle^{\rm 1PI}\nonumber\\
&=& -\zeta_2^{(0)}\int {\rm d}x {\rm d}y
 e^{i p (x-y)}\langle T q^{\prime,0}(x)
\bar{q}^{\prime,0}(y)( C_{2q}{\cal O}_{2q} +C_{3q}{\cal
  O}_{3q})\rangle^{\rm 1PI}\,,
\end{eqnarray}
where $p$ is the outgoing momentum of the quark and we label the
quantities in the effective theory with a prime. 

 Upon decomposition of 
 the Green function $\Gamma_{\bar{q}q{\cal O}_h}^{0}$ into its scalar and
vector components and taking the limit $p\to 0$, one obtains for the
coefficient function $C_{2q}$ the following expression 
\begin{eqnarray}
C_{2q}^{0}&=&\frac{\Gamma_{\bar{q}q{\cal
      O}_h; s}^{0,h}(0,0)}{1-\Sigma_s^{0,h}(0)}
+\frac{\Gamma_{\bar{q}q{\cal O}_h;
    v}^{0,h}(0,0)}{1+\Sigma_v^{0,h}(0)}\,.
\label{eq::c2mssm}
\end{eqnarray}
The quantities $\Sigma_v^{0,h}(0)$ and $\Sigma_s^{0,h}(0)$ have been
defined in Eq.~(\ref{eq::letse}). The superscript $h$ in the above
equation marks that only the hard parts of the Green functions survive
when one sets the external momenta to zero $p^2=p_h^2=0$.
%Sample diagrams that contribute to the 
%calculation of the Green function $\Gamma_{\bar{q}q{\cal O}_h}$ in SUSY
%QCD at one-and two-loop order can be found in Fig.~\ref{fig::c2diag}.

From the technical point of view, to separate the vector and scalar
contributions to the vertex Green function $\Gamma_{\bar{q}q{\cal O}_h}$
one has to perform a naive Taylor expansion up to linear order in the 
external momenta carried by quarks. After the projection on vector and
scalar parts, the external momenta can be set to zero. Nevertheless, the
light  Higgs mass 
approximation $M_h^2=p_h^2\approx 0$ can be applied from the very beginning, which
 implies that the quark momenta can be chosen to be equal. As a consequence,
  vertex diagrams are reduced to
two-point functions with vanishing external momenta, that can be further
mapped to vacuum integrals.

Similarly,  one can compute  the coefficient function $C_1$ via the 
Green function formed by the coupling of the operators ${\cal O}_h$
to two gluons
\begin{eqnarray}
\delta^{ab}\Gamma_{GG{\cal O}_h}^{0,\mu\nu}(p_1,p_2)&=&i^2\int {\rm d}x {\rm d}y e^{i
  (p_1\cdot x+p_2\cdot y)}\langle T G^{0,a,\mu}(x) G^{0,b,\nu}(y){\cal
  O}_h(0)\rangle^{\rm 1PI}\,,
\nonumber\\
&=&\delta^{ab}(-g^{\mu\nu} p_1\cdot p_2+p_1^\nu p_2^\mu) \Gamma_{GG{\cal
    O}_h}^{0}(p_1,p_2)\,,
\label{eq:grfcvert}
\end{eqnarray}
where $p_1$ and $p_2$ denote the outgoing momenta of the gluons with the
colour indices $a$ and $b$.
 One can show that~\cite{Chetyrkin:1996ke,Steinhauser:2000ry} the
 coefficient $C_1$ is given by the following relation
\begin{eqnarray}
C_1^0&=&-\frac{1}{4}\frac{1}{\zeta_3^0}\Gamma_{GG{\cal O}_h}^{0}(0,0)
\nonumber\\
&=&-\frac{1}{4}\frac{1}{\Pi^{0,h}(0)}\left(\frac{g_{\mu\nu} p_1\cdot
    p_2-p_{1,\nu}p_{2,\mu}-p_{1,\mu}p_{2,\nu}}{(d-2)( p_1\cdot
    p_2)^2}\Gamma_{GG{\cal O}_h}^{0,\mu\nu}(p_1,p2)
\right)\bigg|_{p_1^2=p_2^2=0}\,,
\label{eq::c1mssm}
\end{eqnarray}
where $d$ denotes as usual the number of space-time dimensions
in dimensional regularization scheme and $\Pi^{0,h}(0)$ has been defined
in Eq.~(\ref{eq::letse}). Let us 
    mention at this point that the projector given in
    Eq.~(\ref{eq::c1mssm}) projects out the coefficient of the term
    proportional to $g^{\mu\nu}$ in Eq.~(\ref{eq:grfcvert}). To
    explicitly verify the transversality of the Green function
    $\Gamma_{GG{\cal O}_h}^{0,\mu\nu}(p_1,p_2)$, one needs to compute
    also the coefficient of the Lorentz structure proportional to
    $p_1^\nu p_2^\mu$ using a second  projector  (for the explicit
    formula see for example    Ref.~\cite{Pak:2010cu}).

%Sample diagrams contribution to the Green's function $\Gamma_{GG{\cal
%  O}_h}^{\mu\nu}$ can be found in Fig~(\ref{fig::c1mssmdia}).
%Since the Higgs coupling to gluons is a loop induced interaction, {\it
%  i.e.} the LO contributions correspond to one-loop diagrams. In
%consequence,  the calculation of the NNLO
%corrections to the coefficient $C_1$ requires the computation of three-loop diagrams. Typical
%diagrams contributing at this loop order can be visualized in
%Refs.~\cite{Pak:2010cu,Pak:2012xr}.

 In equation~(\ref{eq::c1mssm}), one has to keep $p_1\ne 0$ and
$p_2\ne 0$ until the projection is applied.
 When only heavy particles are running in the loops, a naive Taylor
 expansion to the linear 
order in the two external momenta is required. After the expansion, the
factor $( p_1\cdot 
    p_2)^2$ in the denominator cancels and the two external momenta can
    be set to zero. In this way the vertex topologies implied in
    Eq.~(\ref{eq::c1mssm}) are reduced to vacuum
    integrals. Nevertheless, when light particles are present in the
    loops, {\it e.g.} bottom quarks, a naive   Taylor expansion is not
    enough and one has to perform
    an asymptotic expansion. In this case the resulting Feynman
    integrals can be decomposed into massive vacuum integrals and vertex
    integrals with external momenta satisfying $p_1^2=p_2^2=0$ and
    $2p_1\cdot p_2 = M_h^2$, and light quark masses present in the
    loops. Up to now, the light quark mass effects have been evaluated
    at NLO in Refs.~\cite{Degrassi:2010eu,Harlander:2010wr}, which
    requires the computation of two-loop massive vacuum integrals and
    1-loop vertex integrals.

As explained above the computation of the coefficient functions $C_1$
and $C_{2q}$  involves  vacuum integrals with several mass
scales. Up to two-loop order such integrals are known
exactly~\cite{Davydychev:1992mt}. However, the three-loop multi-scale
integrals are not 
known and the computation of the coefficient $C_1$ at NNLO can be
performed only for specific mass hierarchies between the SUSY particles,
that requires application of the asymptotic expansion method (for details
see Refs.~\cite{Pak:2012xr,Kant:2010tf}).

In SM, the coefficient functions $C_1$ and $C_{2q}$ are known up to
the third order in perturbation theory. The first order QCD corrections to
$C_1$ have been computed in Refs.~\cite{Dawson:1990zj, Djouadi:1991tka,
  Inami:1982xt}, while the same order 
contribution to   $C_{2q}$ vanishes in the SM. The second order QCD
corrections to the coefficients $C_1$ and $C_2$ can be found in
Ref.~\cite{Chetyrkin:1996ke}. The leading Yukawa corrections to the
coefficient functions have been evaluated in
Ref.~\cite{Steinhauser:1998cm}.  For the coefficient function $C_1$ 
the fourth order QCD corrections have been computed
recently~\cite{Schroder:2005hy}. Using the low-energy theorem, the
authors of Ref~\cite{Baikov:2006ch} computed even  the fifth order QCD corrections to
the coefficient $C_1$  up to   contributions originating in the
$n_l$-dependent part of the five-loop QCD beta function, that are
currently not known.  

In the MSSM, the coefficient functions $C_1$ and $C_{2q}$ are known at
the NNLO. The NLO corrections to
$C_1$ have been computed within SUSY-QCD for the first time in
Refs.~\cite{Harlander:2003bb,Harlander:2004tp} 
and confirmed analytically~\cite{Degrassi:2008zj}
and numerically~\cite{Anastasiou:2008rm} (see also
Ref.~\cite{Muhlleitner:2008yw}). In
Refs.~\cite{Muhlleitner:2006wx,Bonciani:2007ex} the squark loop
contributions to 
Higgs boson production in the MSSM have
been computed without assuming any mass hierarchy. In SUSY models with
large values of $\tan \beta$, the radiative corrections due to the bottom
sector can become  large and they have been computed analytically at NLO
in Refs.~\cite{Degrassi:2010eu,Harlander:2010wr} and confirmed
numerically in Ref.~\cite{Anastasiou:2008rm}.  For the coefficient function
$C_{2q}$ the NLO  SUSY-QCD and top Yukawa corrections   are known
analytically since quite some time~\cite{Carena:1999py}.  The dominant
($\tan \beta$ enhanced) NNLO SUSY
QCD  and top Yukawa corrections  to $C_{2b}$  have been computed in
Ref.~\cite{Noth:2008tw}. The SUSY
QCD contributions have been  confirmed analytically in~\cite{Mihaila:2010mp}.

For completeness, we display here the one-loop order coefficients  $C_1$ and $C_{2b}$  
 providing also ${\cal O}(\epsilon)$ terms that are necessary for the
 higher order calculations.

\begin{eqnarray}
C_1&=&-\frac{\alpha_s}{3\pi}\Bigg\{+\frac{\sin\alpha}{\cos\beta}
\Bigg[\frac{M_t^2\mu_{\rm SUSY} X_t}{4\Mstu^2\Mstd^2\tan\beta}
-\epsilon \frac{M_t\mu_{\rm SUSY}\sin
  2\theta_t}{8\tan\beta}\left(\frac{\lMstu}{\Mstu^2} -
  \frac{\lMstd}{\Mstd^2}\right) \Bigg]
\nonumber\\
&&-\frac{\cos\alpha}{\sin\beta}\Bigg[
\frac{4\Mstu^2\Mstd^2 + \Mstu^2 M_t^2 + \Mstd^2 M_t^2 - A_t M_t^2
  X_t}{4\Mstu^2\Mstd^2}\nonumber\\
&&
+\epsilon \frac{A_t M_t 
\sin 2\theta_t}{8}\left(\frac{\lMstu}{\Mstu^2} -
    \frac{\lMstd}{\Mstd^2}\right) +
\epsilon\frac{M_t^2}{4}\left(\frac{4 \lMt}{M_t^2} 
+ \frac{\lMstu}{\Mstu^2} + \frac{\lMstd}{\Mstd^2}\right)
\Bigg]
\Bigg\}\,,
\label{eq::c11l}
\end{eqnarray}
\begin{eqnarray}
C_{2b}&=&-\frac{\sin\alpha}{\cos\beta}\frac{1+\frac{\alpha_s}{2\pi} C_F
A_b \Mgl\bigg[F_1(\Msbu^2,\Msbd^2,\Mgl^2)+\epsilon
F_2(\Msbu^2,\Msbd^2,\Mgl^2)\bigg]}{1+\frac{\alpha_s}{2\pi} C_F
X_b \Mgl\bigg[F_1(\Msbu^2,\Msbd^2,\Mgl^2)+\epsilon
F_2(\Msbu^2,\Msbd^2,\Mgl^2)\bigg]}
\nonumber\\
&&
+\frac{\cos\alpha}{\sin\beta}
\frac{\frac{\alpha_s}{2\pi} C_F
(-\mu_{\rm SUSY}\tan\beta) \Mgl\bigg[F_1(\Msbu^2,\Msbd^2,\Mgl^2)+\epsilon
F_2(\Msbu^2,\Msbd^2,\Mgl^2)\bigg]}{1+\frac{\alpha_s}{2\pi} C_F
X_b \Mgl\bigg[F_1(\Msbu^2,\Msbd^2,\Mgl^2)+\epsilon
F_2(\Msbu^2,\Msbd^2,\Mgl^2)\bigg]}
\,,
\end{eqnarray}
where the functions $F_1$ and $F_2$ are defined through
\begin{eqnarray}
F_1(x_,y_,z_)&=&-\frac{x y
  \ln\frac{y}{x}+yz\ln\frac{z}{y}+zx\ln{x}{z}}{(x-y)(y-z)(z-x)}\,,\nonumber\\
F_2(x_,y_,z_)&=&\frac{1}{(x-y)(y-z)(z-x)}
\Bigg[x y
  \ln\frac{y}{x}(1+\frac{1}{2}\ln\frac{\mu^2}{\sqrt{xy}})
\nonumber\\
&&
+yz\ln\frac{z}{y}(1+\frac{1}{2}\ln\frac{\mu^2}{\sqrt{yz}})
+zx\ln{x}{z}(1+\frac{1}{2}\ln\frac{\mu^2}{\sqrt{xz}})\Bigg]\,.
\end{eqnarray}
The corresponding expression for up type quarks can be easily obtained by
replacing $\sin\alpha $ with $\cos \alpha$ and $\sin\beta$ with $\cos
\beta$ and vice versa.

The approach outlined above has the advantage that it simplifies significantly 
  the calculation, once the limit $M_h^2=p_h^2\approx 0$ is
  applied. The validity of this approximation  has
  been proved within the 
  SM at the NNLO~\cite{Harlander:2009mq,Pak:2009dg}\footnote{For the SM,
    it is known
  as the infinite top quark mass approximation.}. Since the SUSY
  particle masses are expected to 
  be considerably heavier than the top quark mass, we expect that this
  approximation holds  in the MSSM even with higher accuracy.

\subsubsection{Low Energy Theorem}
A second possibility to compute the coefficient functions is to relate
them via the Low Energy Theorem (LET) to vacuum polarization and quark
self energy corrections. This approach resides heavily on the fact that
the momenta carried by the Higgs boson can be set to zero. In this case,
it was
shown (within the SM) that the amplitude of a process containing $(N+1)$
external particles from
which, one is a Higgs boson with vanishing momenta, can be computed from
the amplitude with  $N$ external particles, obtained in the absence of
the Higgs external leg~\cite{Ellis:1975ap}:
\begin{eqnarray}
\lim_{p_h\to 0} \Gamma^{h, A_1, A_2,\ldots,
  A_N}(p_h,p_{A_1},p_{A_1},\ldots, P_{A_N}) &=& \frac{\partial}{\partial
  v }  \Gamma^{ A_1, A_2,\ldots,
  A_N}(p_{A_1},p_{A_1},\ldots, P_{A_N}) \,,
\label{eq::letgen}
\end{eqnarray}  
where $v$ denotes the vacuum expectation value (VEV) of the theory. Beyond  tree level, all
kinematic parameters must be considered as bare quantities. For certain
special theories and renormalization schemes the above equation holds
even for renormalized parameters (for details see
Ref.~\cite{Kilian:1995tra}). Within QCD  all order formulae relating the
coefficient functions of dimension four operators with the decoupling
coefficients for the strong coupling and the quark masses  have been
derived~\cite{Chetyrkin:1997un}.  Within the MSSM,
Eq.~(\ref{eq::letgen}) has to be generalized to the case where two
Higgs fields acquire VEVs. Nevertheless, it has been
proved~\cite{Degrassi:2008zj,Mihaila:2010mp,Kurz:2012ff} that within
SUSY-QCD the coefficient functions $C_1$ and $C_{2q}$  can be
derived  up to NNLO from the decoupling coefficients $\zeta_s$ and $\zeta_{m_q}$
through the following relations:
\begin{eqnarray}
C_1^0 &=& (-\sin\alpha \hat{D}^0_{\phi_1} +\cos \alpha
\hat{D}^0_{\phi_2})\ln \zeta_s^0\equiv  \hat{D}^0_{h}\ln
\zeta_s^0\,,\nonumber\\
C_{2q}^0 &=& (-\sin\alpha \hat{D}^0_{\phi_1} +\cos \alpha
\hat{D}^0_{\phi_2})\ln \zeta_{m_q}^0\equiv  \hat{D}^0_{h}\ln
\zeta_{m_q}^0\,.
\end{eqnarray}
As usual, the superscript $0$ labels bare quantities. The operators
$\hat{D}^0_{\phi_i}$, with $i=1,2$,  contain the derivatives w.r.t. the two VEVs of the
MSSM. They have been derived using the field dependent definitions of
quark and  squark masses and  mixing angles in
Ref~\cite{Degrassi:2008zj}. However, for the computation of the
coefficient function $C_1$ at the NNLO, also the dependence of the
\eps-scalar mass on the VEVs through the loop induced Higgs-\eps-scalar
coupling has to be taken into 
account~\cite{Kurz:2012ff}. 
As can be understood from the Eqs.~(21) and (22) in
Ref.~\cite{Degrassi:2008zj}  the dominant contributions to the the
differential operators originate from the pure SUSY-QCD terms. For
exemplification and to fix the normalization, we reproduce
 here the terms corresponding to the
third generation quarks keeping only the linear terms  in  bottom quark
masses\footnote{Please note the sign difference in the definition of
  parameter $\mu_{\rm SUSY}$ between  Ref.~\cite{Degrassi:2008zj} and
   Refs~\cite{Mihaila:2010mp,Kurz:2012ff}.}
\begin{eqnarray}
\hat{D}_{\phi_1}&=&\frac{1}{\cos\beta}( m_b   A_b {\cal F}_b +m_b {\cal G}_b)
-\frac{1}{\sin\beta} m_t
\mu_{\rm SUSY}\sin 2 \theta_t {\cal F}_t\,,\nonumber\\
\hat{D}_{\phi_2}&=& \frac{1}{\cos\beta} (-m_b\mu_{\rm SUSY} {\cal F}_b)
+ \frac{1}{\sin\beta} (m_t A_t \sin 2 \theta_t {\cal F}_t
+2 m_t^2 {\cal G}_t)\,, \quad \mbox{with}\nonumber\\
{\cal F}_b&=& \frac{2}{\Msbu^2-\Msbd^2} (1-\Sqq)
\frac{\partial}{\partial \Sq}\,,\quad {\cal G}_b = \frac{\partial}{\partial m_b}\,,\nonumber\\
{\cal F}_t&=& \frac{\partial}{\partial \Mstu^2}-
\frac{\partial}{\partial \Mstd^2}
+\frac{2}{\Mstu^2-\Mstd^2} \frac{ (1-\Stq)}{\St}
\frac{\partial}{\partial \St}\,,\nonumber\\
{\cal G}_t&=&\frac{\partial}{\partial \Mstu^2}
+\frac{\partial}{\partial \Mstd^2}
+\frac{\partial}{\partial m_t^2}\,.
\label{eq::derivop}
\end{eqnarray}
On the r.h.s. of the above equations, all parameters are the bare
ones. We omitted the superscript ``0'' to avoid clumsy notation.
For large values of $\tan \beta$ the dominant contributions to the
coefficient functions, {\it i.e.}
the terms proportional to $\mu_{\rm  SUSY}\tan\beta$, are generated
through the term containing the derivative ${\cal F}_b$  in
$\hat{D}_{\phi_2}$. Taking into account the
 parametric dependence of the quark self energy $\Sigma^{0,h}$ on masses and mixing angles, 
one can easily derive    these  contributions 
 from the terms proportional to $\sin 2\theta_b$ in $\Sigma_s^{0,h}$.

\subsection{Hadronic Higgs decays}

\begin{figure}[t]
\begin{center}
\epsfxsize=.6\textwidth
\epsffile{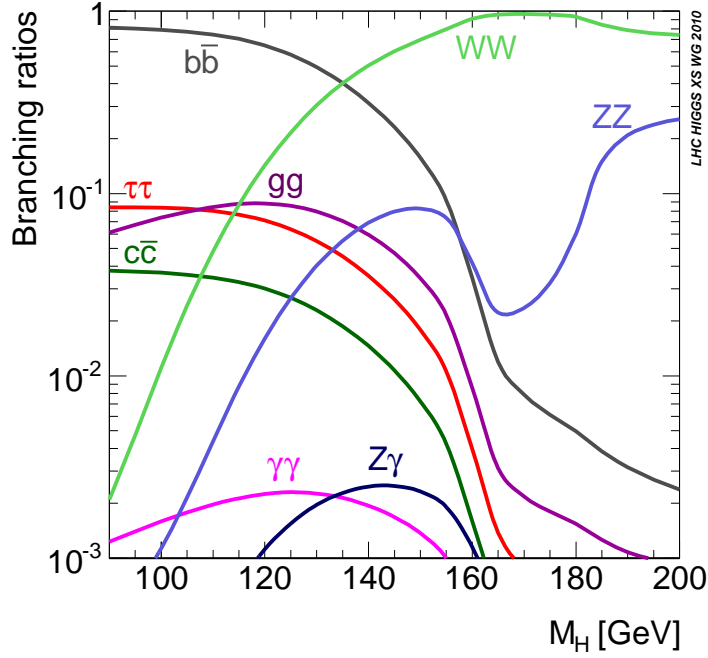}
\end{center}
\vspace*{-0.4cm}
\caption{\label{fig::brlo} Higgs boson branching ratios in the SM at the
LO from  Ref.~\cite{higgswg}.}
\end{figure}

In this section we study the phenomenological applications of the
computations discussed above.  We concentrate on the calculation within
the MSSM 
of the total decay rate into hadrons $\Gamma(h\to \mbox{hadrons})$, that
is composed 
of the partial decay widths
into quarks $\Gamma(h\to q\bar{q})$ and  gluons $\Gamma(h\to gg)$.
Although, the channel $\Gamma(h\to b\bar b)$  gives the dominant contributions  to  the
total Higgs   decay
rate,  it was not used among the Higgs discovery channels at the LHC, due to its
huge background.
Nevertheless, it has a big impact on  all branching ratios and is an
 important channel for
the identification of the Higgs properties. Precisely, the uncertainties
on the partial decay width $\Gamma(h\to b\bar b)$ 
translate into significant systematic errors for all the other
non-leading branching ratios.  
 For illustration we show in Figs.~{\ref{fig::brlo}}
from Ref.~\cite{higgswg} the branching ratios of the Higgs boson in the
SM at
the LO. For precise analysis they have to be complemented by genuine  SM
radiative corrections together with  corrections due to the
supersymmetric particles, that can be  embedded in the decoupling
coefficients as 
discussed in the previous section.  

 Starting from the effective Lagrangian~(\ref{eq::eft}) one 
can derive the following formula for the total decay width into hadrons
\begin{eqnarray}
\Gamma(h\to  \mbox{hadrons})&=&(1+\bar{\delta}_u)^2\bigg \{
\sum_{q}\Gamma^{(0)}_{q\bar{q}}\bigg[
(1+\Delta_{22}) {\cal C}_2^2+ \Delta_{12}{\cal C}_1{\cal C}_2
\bigg]\nonumber\\
&& + \Gamma^{(0)}_{gg}(1+\Delta_{11}) {\cal C}_1^2 \bigg \}\,,
\label{eq::tdw}
\end{eqnarray}
where the coefficient functions ${\cal C}_1$ and ${\cal C}_2$ have been
defined in Eq.~(\ref{eq::cfrg}).

At the lowest order in  perturbation theory, the first line corresponds
to  $\Gamma(h\to q\bar{q})$, whereas the second one stands for $\Gamma(h\to
gg)$. At higher orders, however, the splitting of Eq.~(\ref{eq::tdw})
into the decay widths to fermions and gluons is not straightforward
anymore, due to the occurrence of diagrams contributing to  both
channels.\\
The LO expressions for the  branching ratios are given by 
\begin{eqnarray}
\Gamma^{(0)}_{q\bar{q}} &=& \frac{N_c G_F M_h m_q^2}{4\pi\sqrt{2}}
\left(1-\frac{4m_q^2}{M_h^2}\right)^{3/2}\,\quad \mbox{and}
\quad\Gamma^{(0)}_{gg}= 
\frac{N_c C_F G_F M_h^3}{\pi\sqrt{2}}\,,
\label{eq::born}
\end{eqnarray} 
where $G_F$ denotes the Fermi constant. As is well
known~\cite{Braaten:1980yq,Sakai:1981gr,Inami:1982xt}, the large
logarithms of the type 
 $\ln(M_h^2/m_q^2)$ 
can be resummed by taking $m_q$ in Eq.~(\ref{eq::born}) to be the
\msbar{} mass $m_q^{\msbarmath}(\mu)$ evaluated at the scale
$\mu=M_h$.

 The coefficients $\Delta_{11},\, \Delta_{12},\, \Delta_{22}$ 
describe the low-energy physics. Therefore, they have to be computed in the effective
theory and are independent
of the heavy masses.
 Using the method of  operators
described in the previous section, they can be related via the optical
theorem to the absorptive parts of the scalar correlators $\Pi_{jk}$
\begin{eqnarray}
\Delta_{jk}=\frac{1}{M_h}\rm{Im}( \Pi_{jk})= \frac{1}{M_h}\rm{Im}\left(i
  \int \rm{d}x\, e^{i p x}\, \langle 0|T[{\cal O}_j(x) {\cal 
  O}_k(0)]  | 0 \rangle \bigg|_{p^2=M_h^2}\right)\,,\quad j,k=1,2\,,
\end{eqnarray}
where $p$ is the momentum of the external Higgs boson.
They have been computed  within SM  up to three-loop order\footnote{See
  Ref.~\cite{Steinhauser:2002rq} for a 
  comprehensive review on this topic.}. For the analysis discussed in
this section, their one- and two-loop QCD corrections are required. 
The two-loop QCD contributions to the coefficients $\Delta_{22}$ and
$\Delta_{11}$ are given by~\cite{Gorishnii:1990zu,Chetyrkin:1997sg}.
{\allowdisplaybreaks
\begin{align}
%\begin{eqnarray}
\Delta_{22} &=
%&
\frac{\alpha_s^{\prime}(\mu)}{\pi}\left(\frac{17}{3} + 2
  \ln\frac{\mu^2}{M_h^2}\right)\nonumber\\
%&
&+
\left(\frac{\alpha_s^{\prime}(\mu)}{\pi}\right)^2\bigg[
\frac{10801}{144}-\frac{19}{2}\zeta(2)-\frac{39}{2}
\zeta(3)+\frac{106}{3}\ln\frac{\mu^2}{M_h^2}
+ \frac{19}{4}\ln^2\frac{\mu^2}{M_h^2}
\nonumber\\
%&
&-n_l\left(\frac{65}{24}-\frac{1}{3}\zeta(2)-\frac{2}{3}
\zeta(3)+\frac{11}{9}\ln\frac{\mu^2}{M_h^2}
+ \frac{1}{6}\ln^2\frac{\mu^2}{M_h^2}\right)
\bigg]\,,\\
%\end{eqnarray}
%\begin{eqnarray}
\Delta_{11} &=
%&
\frac{\alpha_s^{\prime}(\mu)}{\pi}\bigg[\frac{73}{4} + \frac{11}{2}
  \ln\frac{\mu^2}{M_h^2} -n_l\left(\frac{7}{6}+\frac{1}{3}
    \ln\frac{\mu^2}{M_h^2}\right)\bigg]\nonumber\\ 
%&
&+
\left(\frac{\alpha_s^{\prime}(\mu)}{\pi}\right)^2\bigg[
\frac{37631}{96}-\frac{363}{8}\zeta(2)-\frac{495}{8}
\zeta(3)+\frac{2817}{16}\ln\frac{\mu^2}{M_h^2}
+ \frac{363}{16}\ln^2\frac{\mu^2}{M_h^2}
\nonumber\\
%&
&-n_l\left(\frac{7189}{144}-\frac{11}{2}\zeta(2)-\frac{5}{4}
\zeta(3)+\frac{263}{12}\ln\frac{\mu^2}{M_h^2}
+ \frac{11}{4}\ln^2\frac{\mu^2}{M_h^2}\right)\nonumber\\
%&
&+n_l^2\left(\frac{127}{108}-\frac{1}{6}\zeta(2)+\frac{7}{12}\ln\frac{\mu^2}{M_h^2} 
+ \frac{1}{12}\ln^2\frac{\mu^2}{M_h^2}\right)
\bigg]\,,
%\end{eqnarray}
\end{align}
}
with $\zeta(x)$ being the Riemann's zeta function.

The additional QCD correction  $\Delta_{12}$ is  generated  through double-triangle
topologies. It was first computed in
Ref.~\cite{Chetyrkin:1996ke} and it reads
\begin{eqnarray}
 \Delta_{12} &=& \frac{\alpha_s^{\prime}(\mu)}{\pi} {C_F} \left(-19
 +6\zeta(2)
-\ln^2\frac{m_q^2}{M_h^2} -6 \ln\frac{\mu^2}{M_h^2}\right)\,.
\end{eqnarray}
The universal corrections $\bar{\delta}_u$ of ${\cal
  O}(\alpha_s^n x_t)$,
 where $x_t=(\alpha_t/4\pi)^2=G_F M_t^2/(8\pi^2\sqrt{2})$, with
 $\alpha_t$ the top-Yukawa coupling, contain
the contributions from the renormalization of the Higgs wave function
and the vacuum expectation value~\cite{Kwiatkowski:1994cu}. It is given by
 \begin{eqnarray}
\bar{\delta}_u &=& x_t\left[\frac{7}{2} +
  \frac{\alpha_s^{\prime}(\mu)}{\pi}\left(\frac{19}{3} -2\zeta(2)
+7\ln\frac{\mu^2}{M_t^2} \right) +{\cal O}(\alpha_s^2)
\right]\,.
\end{eqnarray}
Now, we are in a position to interpret the phenomenological significance
of Eq.~(\ref{eq::tdw}). In the following section we concentrate on the numerical effects of the
radiative corrections to the hadronic Higgs  decay.

\subsubsection{Numerical analysis}

The  SM input parameters are the
strong coupling constant at the Z-boson mass scale
$\alpha_s(M_Z)=0.1184$~\cite{Bethke:2009jm}, the top quark pole mass
$M_t=173.1$~GeV~\cite{Nakamura:2010zzi} and the running bottom quark mass in the
\msbar{} scheme $m_b(m_b)=4.163$~GeV~\cite{Chetyrkin:2009fv}. For the 
supersymmetric parameters  we adopted the corresponding values of
the ``modified $m_h^{\rm max}$''
 scenario as described in section~\ref{sec:mhphen} (for details see Ref.~~\cite{Pak:2012xr}).

\begin{figure}[h]
  \begin{center}
 %   \begin{tabular}{c}
      \epsfig{figure=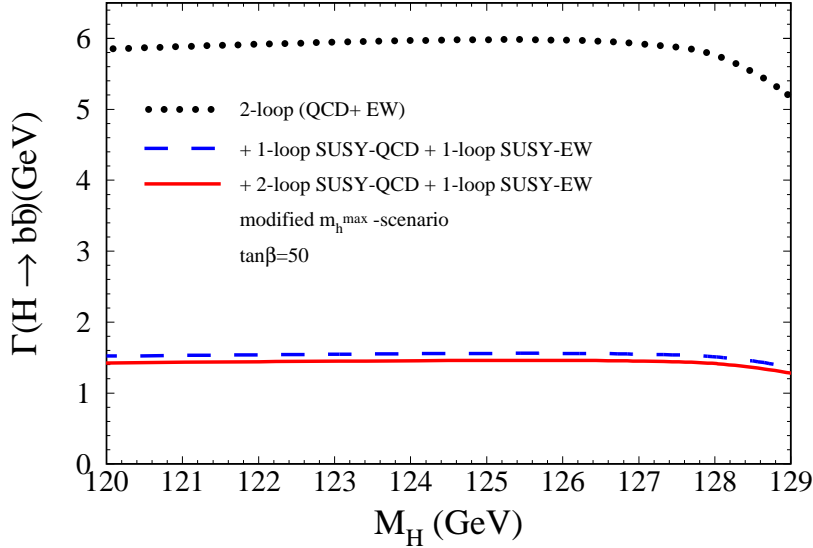,width=.78\textwidth}
%      \\(a)\\
%      \epsfig{figure=figs/plot_Gammagenuine_ew_Mh_gluophobic_tbeta20a.eps,width=.6\textwidth}
%      \\(b)
%    \end{tabular}
    \parbox{14.cm}{
      \caption[]{\label{fig::gamma}\sloppy  $\Gamma(h\to b\bar{b})$
% normalized to the Born results given in Eq.~(\ref{eq::born}) 
 for  the
``modified $m_h^{max}$''   scenario as a function of
 $M_h$. The dotted line displays
 the two-loop QCD and  electroweak corrections together with two-loop corrections to the
 Higgs boson propagator. The dashed and solid lines depict in addition 
the one- and two-loop SUSY-QCD vertex corrections, respectively.          
        }}
  \end{center}
\end{figure}

In Fig.~\ref{fig::gamma} we focus on the decay
channel $h\to b\bar{b}$ and display the decay width   as
 a function of the Higgs boson mass $M_h$.
  We chose in this case $\tan\beta=50$. The two-loop genuine QCD and
  electroweak  corrections ({\i.e.} computed in the effective theory) to
  the  process  $h\to b\bar{b}$ , as well as the two-loop SUSY-QCD 
 corrections to the Higgs boson mass are depicted by the dotted line.
 More precisely, they are derived from  Eq.~(\ref{eq::tdw}), where
 the coefficient functions
 ${\cal C}_1 \, \mbox{and}\, {\cal C}_2$ are set to their tree-level values. 
The additional SUSY-QCD vertex
 corrections parametrized through the coefficient functions 
 ${\cal C}_1 \, \mbox{and}\, {\cal C}_2$ 
 are represented at the one- and two-loop order by the dashed and solid lines,
 respectively. We also take into account the one-loop SUSY-EW corrections to
 the coefficient
 function ${\cal C}_2$ and fix their renormalization scale at
 $\mu_{\rm SEW}=(\Mstu+\Mstd+\mu_{\rm SUSY})/15$, for which the two-loop 
 SUSY-EW corrections
 become negligible~\cite{Noth:2008tw}. The genuine two-loop corrections
 are negligible. Nevertheless, their are essential tools for  the proof of
 the convergence of the perturbative  expansion.\\ 
  The large one-loop SUSY-QCD radiative
 corrections to $\Gamma(h\to b\bar{b})$ have only a relatively small impact 
on the branching ratio $BR(h\to b\bar{b})$, but they can have a large impact
on $BR(h\to \tau^+\tau^-)$. For sufficiently large $\tan\beta$ 
and $\mu_{\rm SUSY}$, the measurement of  $BR(h\to \tau^+\tau^-)$ 
can provide information about the  distinction between
 the SM and MSSM predictions.

The gluonic Higgs decay rate can be directly measured only at $e^+e^-$
colliders. At hadron colliders, they can be measured only indirectly
with rather bad  accuracy of the order of $20$\%. As it has been shown,
the  genuine 
SUSY-QCD corrections to the gluonic Higgs decay are 
rather small~\cite{Spira:1995rr}. For the experimental analysis relevant
at the LHC they can be neglected with respect to the standard
quark contributions to the hadronic decay rate. The QCD corrections are
known in the SM up to the 
NNNLO~\cite{Schroder:2005hy,Baikov:2006ch} in the heavy-top-mass
limit.~\footnote{Here, the mass of the Higgs boson is assumed to be much
  smaller than the mass of the
top quark.} Even the mixed QCD-electroweak
corrections at the three-loop level are 
known~\cite{Steinhauser:1998cm} in the same approximations. 
The genuine NLO   SUSY-QCD corrections have been evaluated in
Refs.~\cite{Harlander:2004tp,Degrassi:2008zj} and amount to about $-5$\%
from the QCD corrections at NLO.

A much more interesting Higgs decay channel from the perspective of the
ongoing experiments conducted at the LHC is the rare $h\to \gamma\gamma$
channel. In this case the coupling of the Higgs to photons is mediated
by loops containing electrically charged particles. If the masses of the
particles inside  loops are generated through the Higgs mechanism, as in
the case of the SM, the  couplings to the Higgs boson grow with the
masses, balancing the decrease due to rising loop masses. If the masses
of the particles are generated by different mechanisms, as is the case
in SUSY, the effect of the heavy particles on the $h\gamma\gamma$
coupling is in general small.\\
 In SM with the Higgs boson mass of about $125$~GeV
only the top quark and the W boson effectively contribute and they
interfere destructively. The radiative corrections are well under
control. The QCD contributions are known up to
NNLO~\cite{Steinhauser:1996wy} and the  electroweak corrections to
NLO~\cite{Aglietti:2004nj}. 
The SUSY-QCD corrections to $\Gamma(h\to\gamma\gamma)$ are known with
the same accuracy as in the case of $\Gamma(h\to
gg)$. The NLO corrections have been computed in
Refs.~\cite{Spira:1995rr,Harlander:2005rq} and  the NNLO
contributions can be found in  Ref.~\cite{kurz:2012}.
Also for this channel, the SUSY corrections are
small  as compared to the SM ones.

\subsubsection{Mass corrections to hadronic Higgs decays}
\begin{figure}
\begin{center}
    \begin{tabular}{ccc}
%      \leavevmode
%      \epsfxsize=.95\textwidth
\vspace*{-0.2em}
      \includegraphics[width=.28\linewidth]{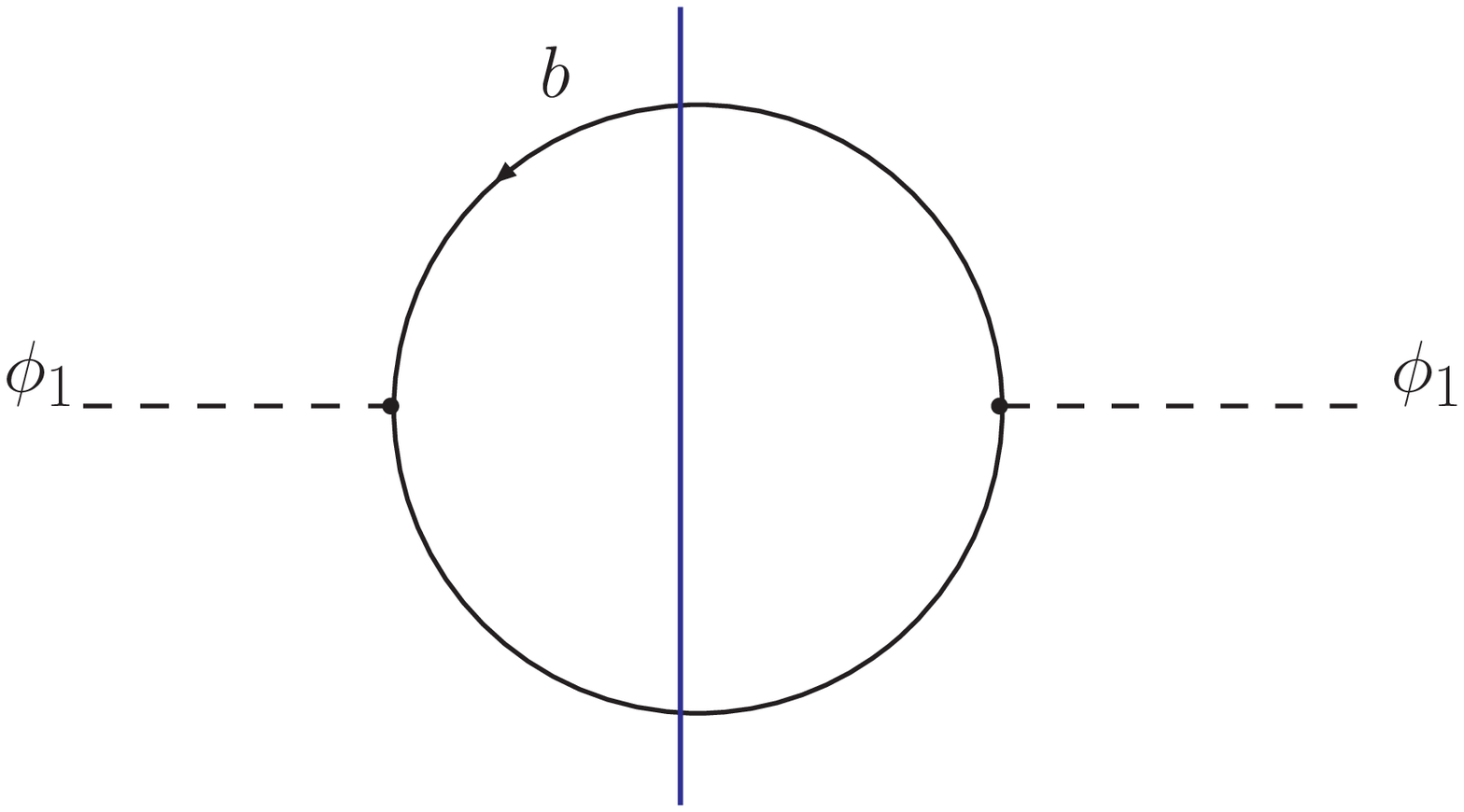}&
      \includegraphics[width=.29\linewidth]{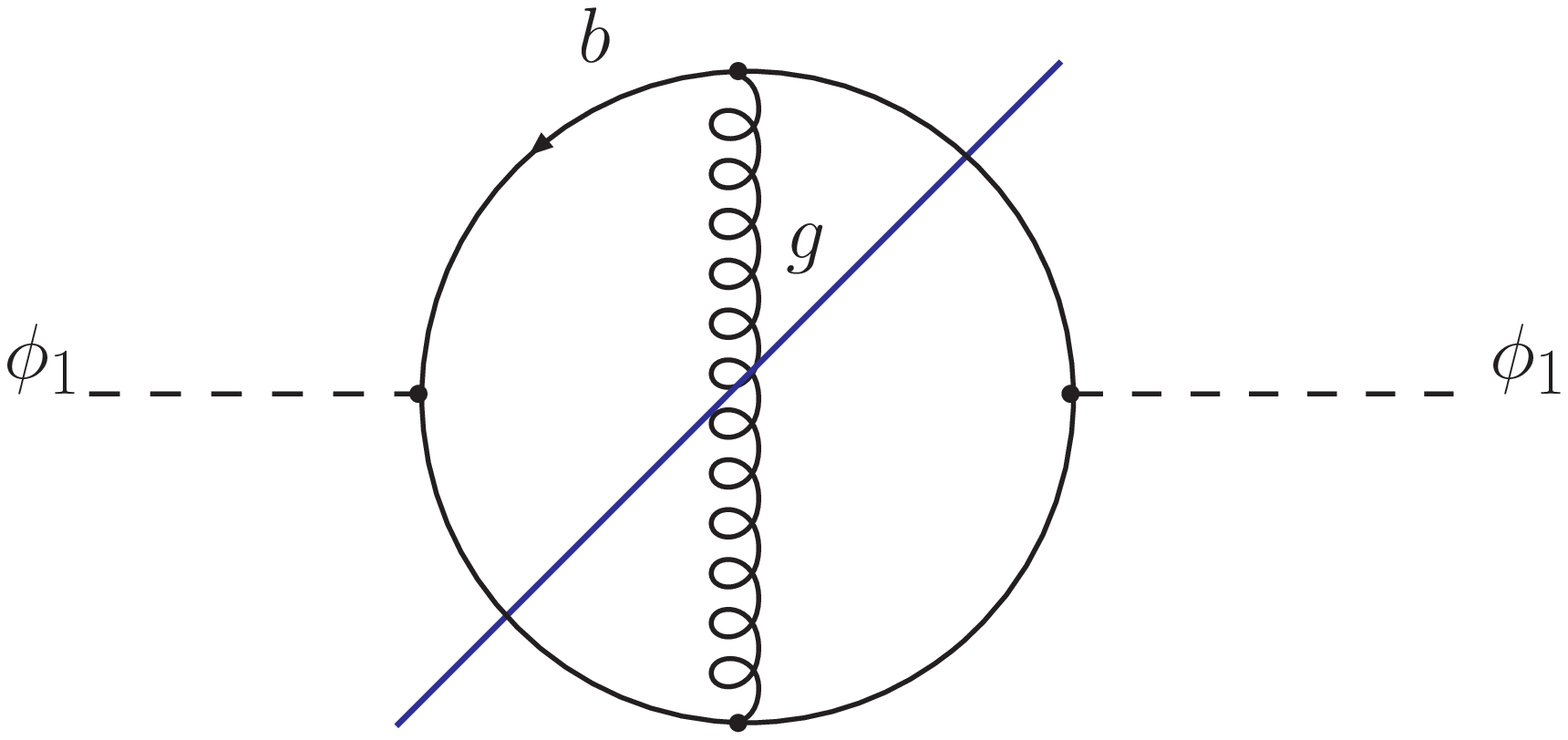}&
\includegraphics[width=.29\linewidth]{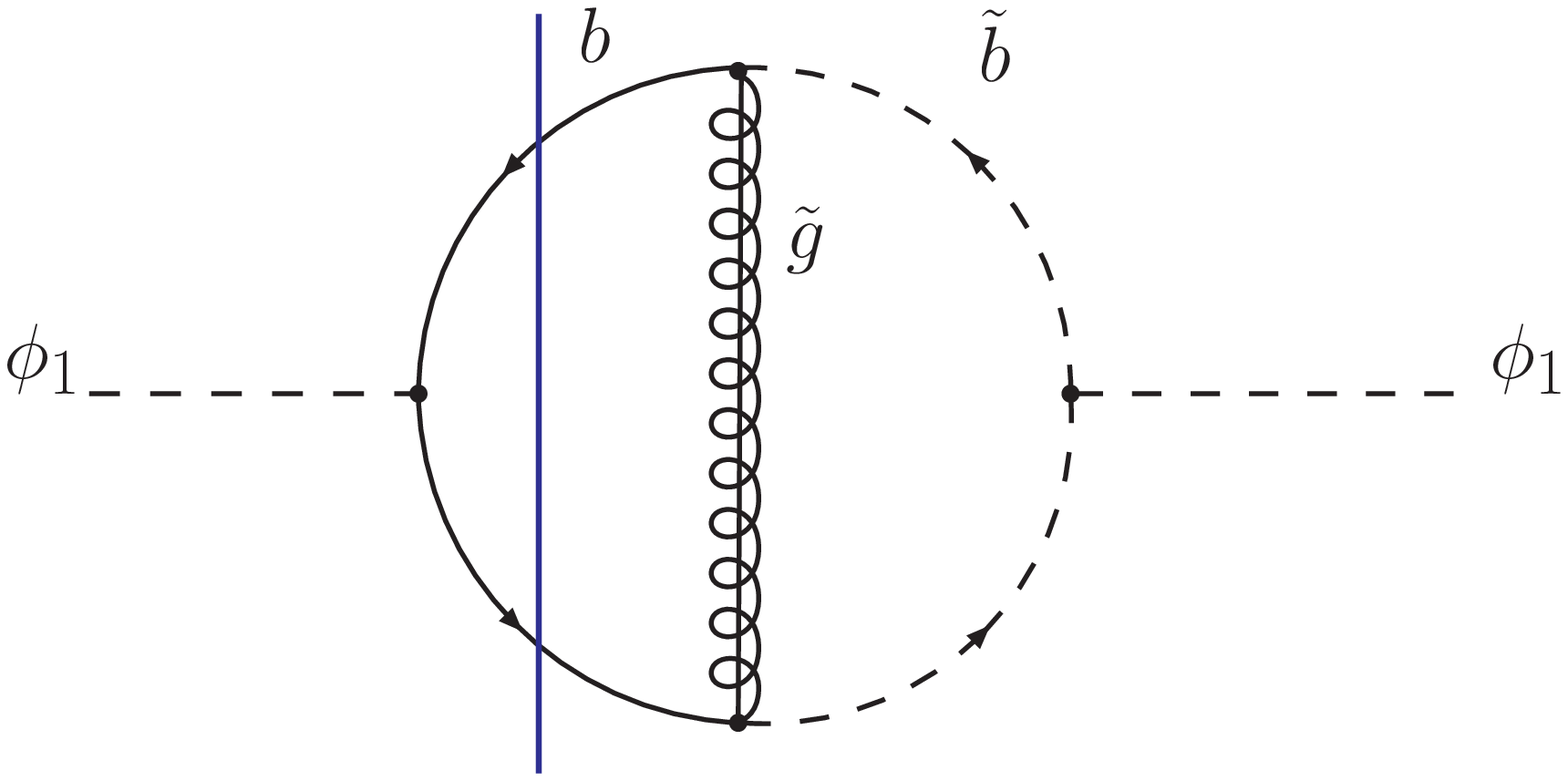}
\\
(a)&(b)&(c)\\
\includegraphics[width=.29\linewidth]{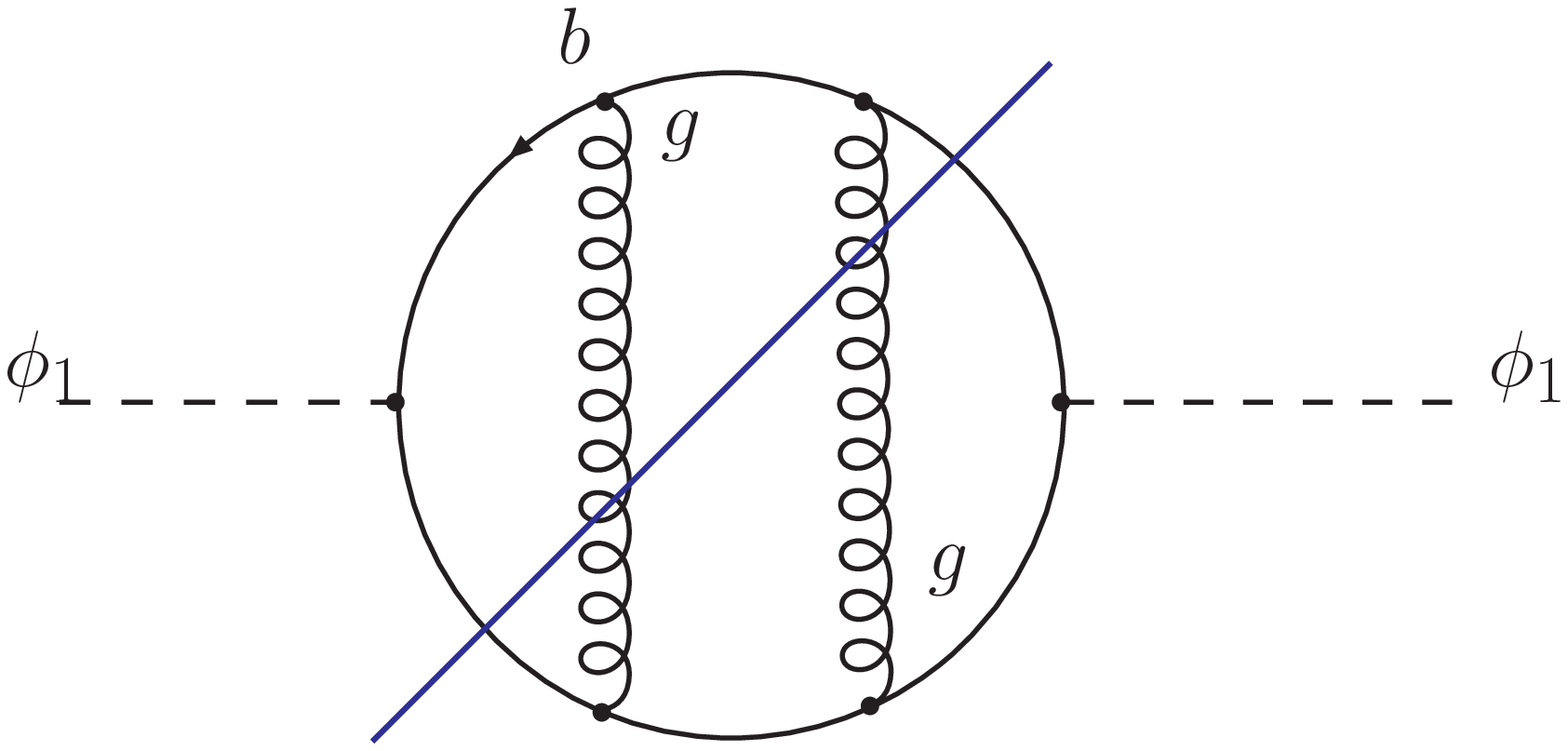}&
\includegraphics[width=.29\linewidth]{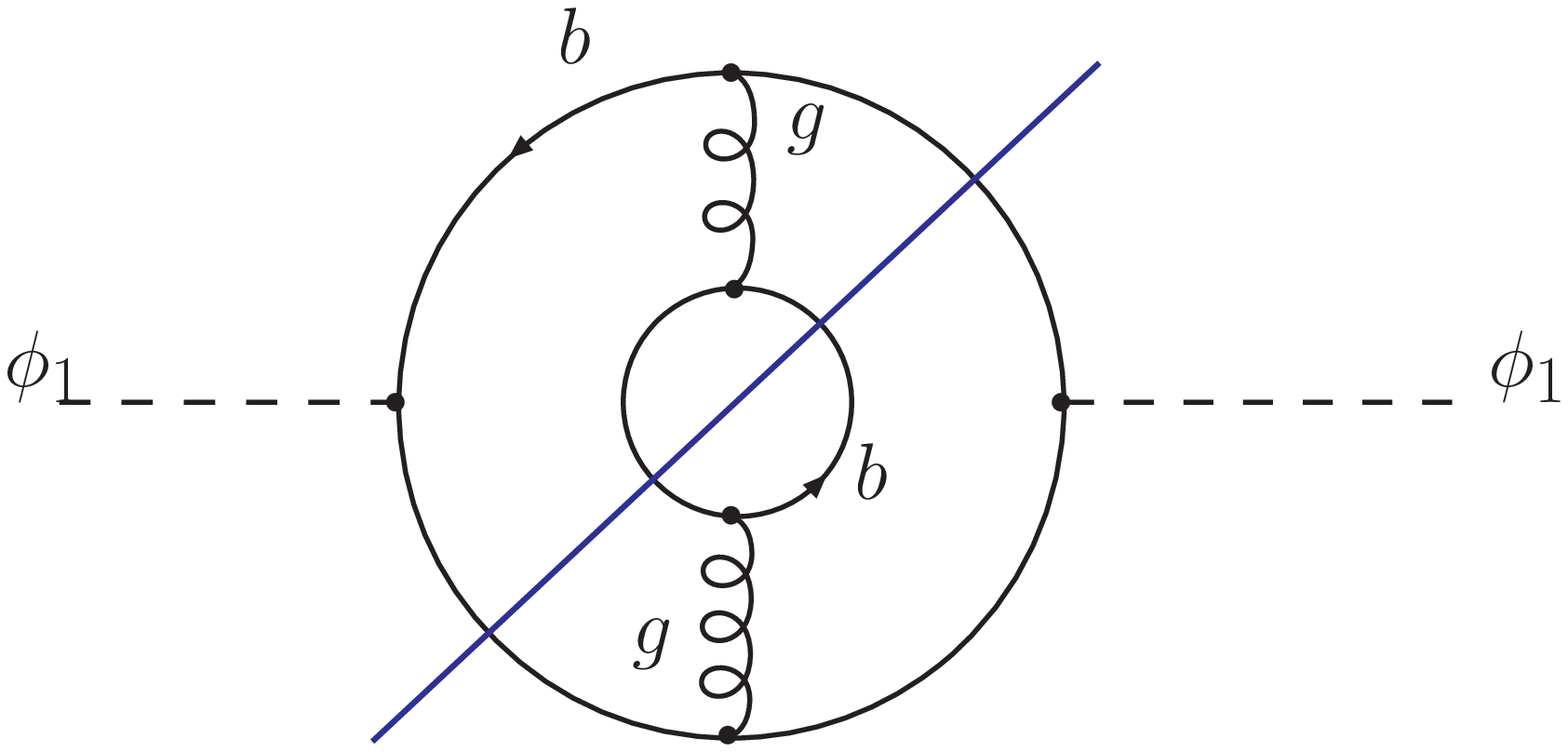}&
%\vspace*{0.5em}
\includegraphics[width=.31\linewidth]{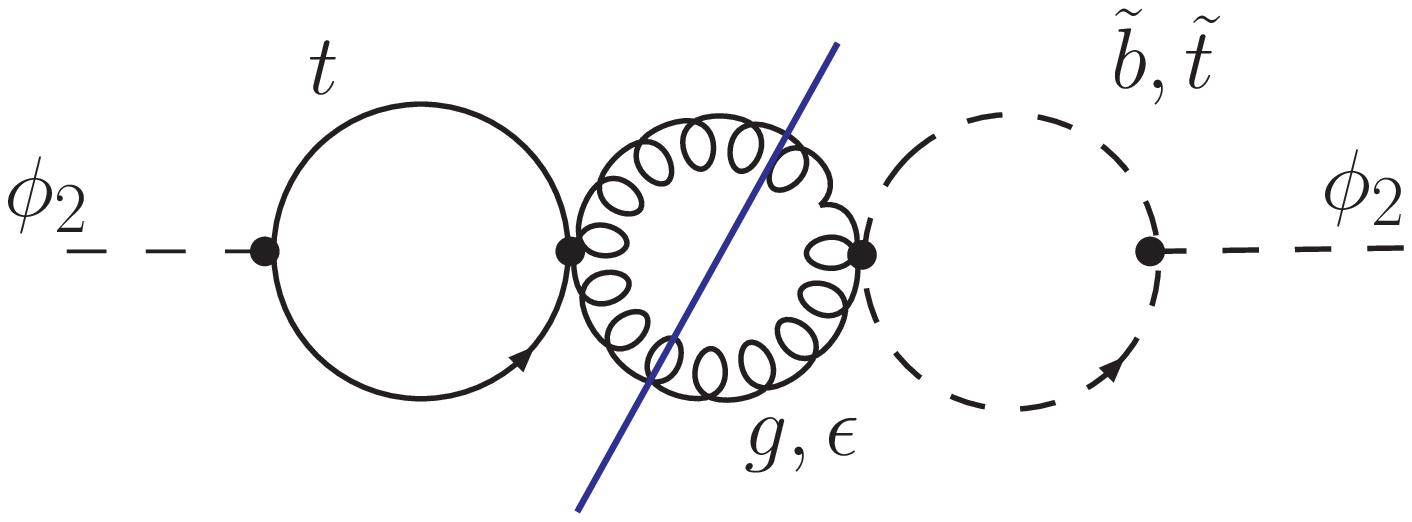}
\\
(d)&(e)&(f)\\
\includegraphics[width=.30\linewidth]{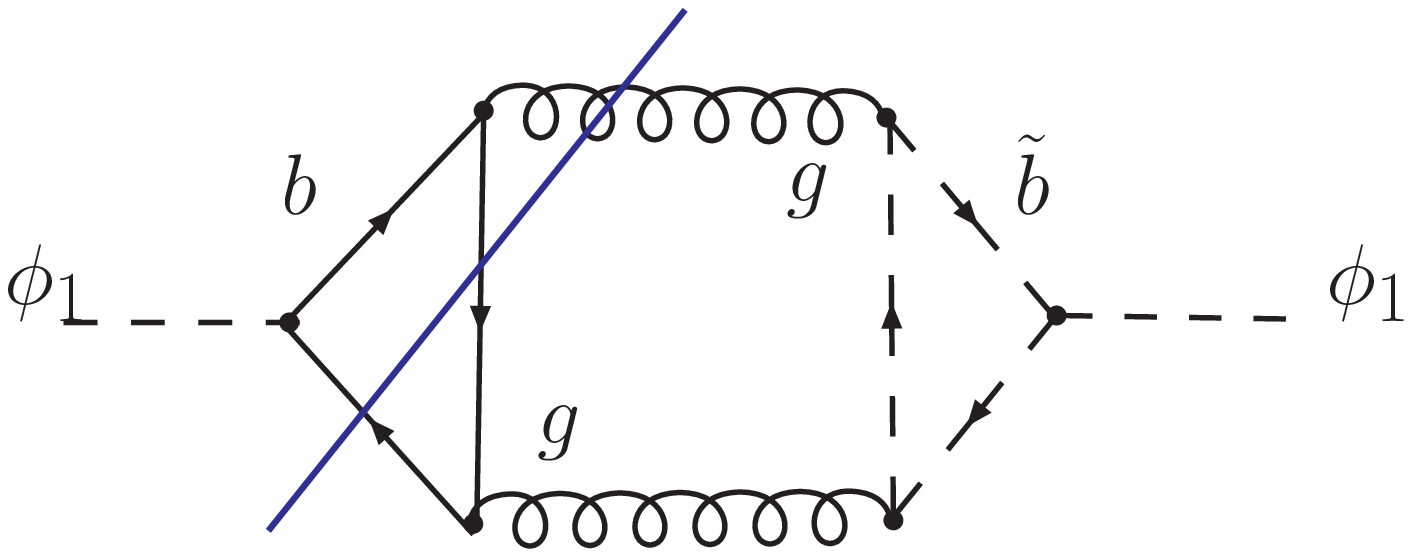}&
\includegraphics[width=.30\linewidth]{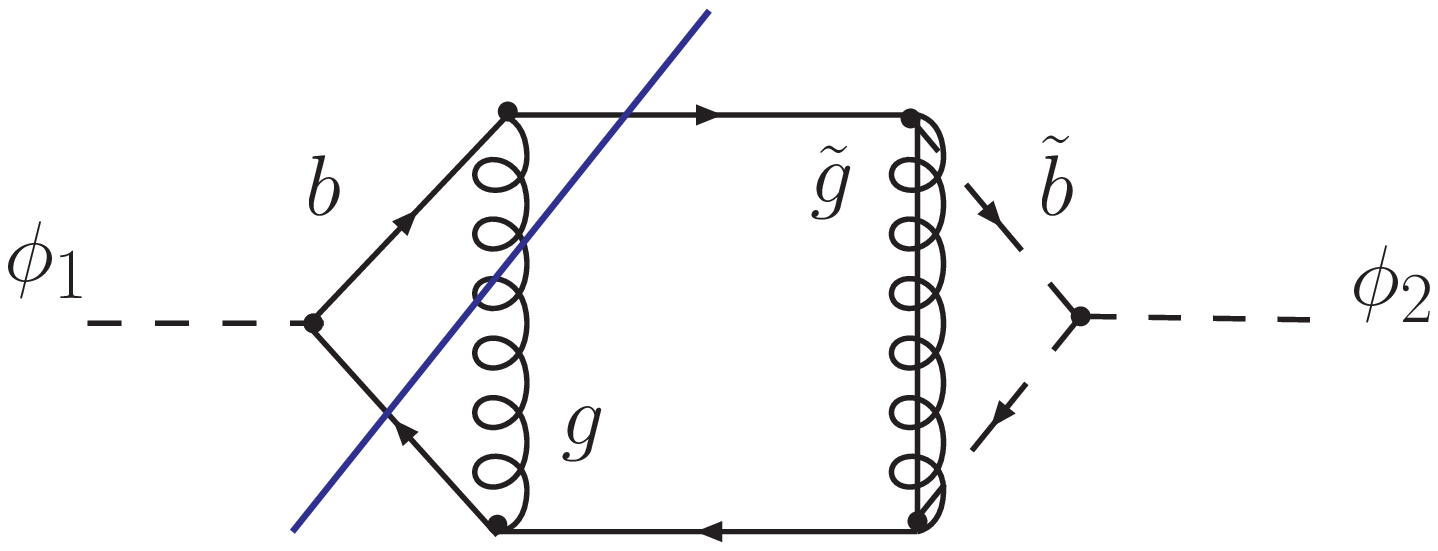}&
\includegraphics[width=.30\linewidth]{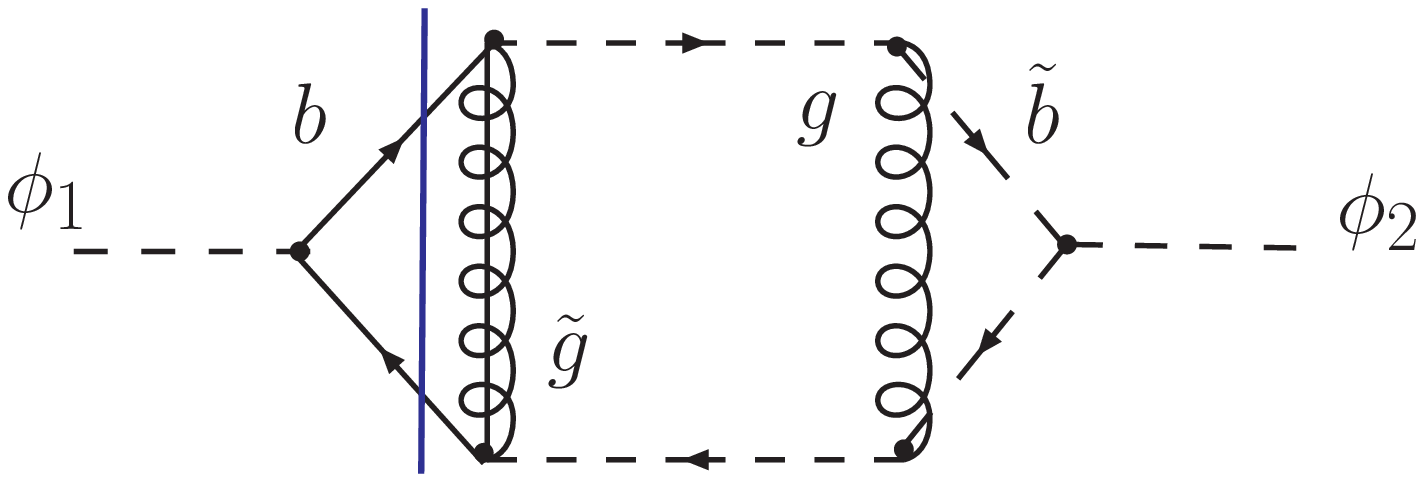}
%\includegraphics[width=.21\linewidth]{figs/p2top23loopa.eps}&
%\vspace*{0.5em}
%\includegraphics[width=.21\linewidth]{figs/p2top23loopb.eps}
\\
(g)&(h)&(i)
%      \epsffile{figs/toydr1loop.eps}& \epsffile{figs/toydr2loopa.eps}&&&&
    \end{tabular}
  \end{center}
  \caption{\label{fig::cutrules}
    One-, two- and three-loop Feynman diagrams contributing to the
    Higgs boson  propagator in SUSY-QCD. Dashed lines denote
    Higgs bosons, whereas oriented dashed lines represent the squarks.
    For the other particles we use the same convention as before.
          }
\end{figure}

For an intermediate  Higgs mass  of about $125$~GeV it is
legitimate to investigate the quality of the approximation  discussed in the
previous section. For  accurate results, one has to take into
consideration in Eq.~(\ref{eq::eft}) also operators of dimension six and
higher, that are suppressed at least by a factor $M_h^2/M_t^2$.
 However, the application of higher dimensional operators in the context
 of SUSY is quite tedious. A more
familiar method for this purpose is to use the optical theorem.
Hereby, one has to consider  corrections to the Higgs boson self-energy
 $\Pi_h(q^2)$. The imaginary part of this quantity provides us with 
the total decay rate of the Higgs boson
\begin{eqnarray}
\Gamma_h&=&\frac{1}{M_h}\rm{Im} \Pi_h(M_h^2)
\label{eq::gamtot}
\end{eqnarray}
According to the Cutkosky cut rules, non-vanishing contributions to the
imaginary part of the
Higgs boson self energy will provide only those diagrams, that can be
cut in such a way that all resulting final state particles can be set
simultaneously on their mass-shell. Sample diagrams contributing to  the
hadronic decay rate can be seen in Fig.~{\ref{fig::cutrules}}.

The imaginary parts originate from
the $i \epsilon$-prescription for on-shell propagators. In the 
results obtained using DRED they are embedded in complex logarithms
occurring in the $\epsilon$-expansion of the expression 
\begin{eqnarray}
\left(\frac{-\mu^2}{q^2+ i\epsilon}\right)^\epsilon &=& 1-\epsilon
\log\left(\frac{-q^2-
    i\epsilon}{\mu^2}\right)+\frac{1}{2}\epsilon^2\log^2\left(\frac{-q^2- i\epsilon}{\mu^2}\right)+{\cal O}(\epsilon^3)\,.
\end{eqnarray}    
After setting the external momenta on the Higgs mass shell $q^2=M_h^2$,
one obtains further
\begin{eqnarray}
\log\left(\frac{-q^2-
    i\epsilon}{\mu^2}\right)&=&\log\left(\frac{M_h^2}{\mu^2}\right) - i \pi\,.
\end{eqnarray}  
The analytic calculation of the three-loop diagrams  contributing
to $\Gamma_h$ in SUSY-QCD is not yet
possible. Nevertheless, for fixed mass hierarchies between the occurring
particles,  the method of  asymptotic expansion can be successfully
applied.  For illustration, we consider a degenerate SUSY mass spectrum
satisfying the following inequality with respect to the SM particle
masses
\begin{eqnarray}
m_q\ll M_h\ll m_t \ll M_{\rm S}\equiv \Mgl = m_{\tilde{q}} \,.
\label{eq::mashierch}
\end{eqnarray} 
Similar with the computation of three-loop SUSY-QCD corrections to the
light Higgs bosons mass, also in this calculation one has to make an
additional Taylor expansion of bottom squark propagator in bottom squark mass
differences $\Delta_b $ defined like
\begin{eqnarray}
\Delta_b&=& \frac{\Msbu^2-\Msbd^2}{\Msbu^2}\,.
\end{eqnarray}
This procedure allows to correctly take into account the contributions
generated by the bottom squark mixing angle renormalization.\\
  In the following we consider the same renormalization scheme as in
section~\ref{subsec:mh3l}. The results for $\Gamma_h$
including the dominant  mass corrections at ${\cal O}(\alpha_s^2)$
read~\cite{kleine:2010} 
{\allowdisplaybreaks
\begin{align}
%\begin{eqnarray}
\Gamma_h &=
\Gamma_{qq}^{(0)}\left(\frac{\sin\alpha}{\cos\beta}\right)^2\bigg\{1+\frac{4}{3}\aspi
\bigg[ \frac{19}{4} + \frac{3}{2}\lnMh -\frac{1}{2} \Ls +
\left(-\frac{15}{2} - 9 \lnMh + 3 \Ls\right)\frac{m_b^2}{M_h^2} 
+  \frac{5}{12}\frac{m_b^2}{\Ms^2}
 \nonumber\\
&
+ \frac{1}{15}\frac{m_b^2M_h^2}{ \Ms^4} - \frac{A_b -\MUH\cot\alpha}{\Ms}
 \left(\frac{1}{2} + \frac{1}{12}\frac{m_b^2}{\Ms^2}
   +\frac{1}{24}\frac{M_h^2}{ \Ms^2}\right)   
  \bigg]+\left(\aspi \right)^2 \bigg[ 
\frac{14093}{216}
 + \frac{541}{18} \lnMh
\nonumber\\ 
&
 + \frac{47}{12}\lnMh^2 - \frac{559}{36} \Ls - \frac{10}{3} \lnMh \Ls + 
 \frac{35}{36} \Ls^2 - \frac{11}{9} \Lt - \frac{1}{3}\lnMh \Lt +
 \frac{1}{6}\Lt^2 - \frac{97}{6} \zeta(3)
\nonumber\\ 
&
 + 
 \left(\frac{107}{675} + \frac{2}{45} \lnMh - \frac{2}{45}
   \Lt\right)\frac{ M_h^2}{m_t^2} +  
\left(-\frac{529}{88200} - \frac{1}{420}\lnMh +\frac{1}{420}\Lt\right)
\frac{M_h^4}{m_t^4}
\nonumber\\ 
&
 + \left(\frac{7}{108} + \frac{1}{9}\Ls - \frac{1}{9}\Lt \right)
 \frac{m_t^2}{\Ms^2} 
+ \left(\frac{5821}{16200} + \frac{17}{135} \lnMh - \frac{17}{135} \Ls
\right)\frac{M_h^2}{\Ms^2} 
\nonumber\\ 
&
 + \frac{ (A_b-\MUH\cot\alpha)^2}{\Ms^2} \left(\frac{1}{9}
 + \frac{1}{54} \frac{M_h^2}{ \Ms^2}\right )
 + 
 \frac{A_b-\MUH\cot\alpha}{\Ms}\bigg[-\frac{119}{18} - \frac{4}{3} \lnMh
 - \frac{1}{18}\Ls 
\nonumber\\ 
&
+    \left(-\frac{7}{54} - \frac{1}{18}\Ls + \frac{1}{18}\Lt \right)\frac{
     m_t^2}{\Ms^2} 
 + \left(-\frac{62}{81} 
- \frac{1}{9}\lnMh - \frac{1}{216}\Ls\right)\frac{M_h^2}{\Ms^2}\bigg] 
\nonumber\\ 
&
 + \frac{\tan\beta}{\sin\alpha}\bigg[
    -\frac{28}{9}
 - \frac{2}{3} \lnMh + \frac{2}{3} \Lt 
+ \frac{5}{54}\frac{ X_t m_t^2}{ \Ms^2}
 + \left(-\frac{2011}{24300} - \frac{41}{1620} \lnMh + \frac{41}{1620}
   \Lt \right) \frac{M_h^2}{m_t^2}
\nonumber\\ 
&
 + \left(-\frac{28307}{4762800} - \frac{47}{22680} \lnMh +
   \frac{47}{22680} \Lt\right)\frac{ M_h^4}{m_t^4} +  
 \left(-\frac{85}{54} - \frac{1}{3}\lnMh +
   \frac{1}{3}\Ls\right)\frac{m_t^2}{\Ms^2}
 - \frac{1}{27}\frac{M_h^2}{\Ms^2} 
\nonumber\\ 
&
- \frac{7}{3240}\frac{ M_h^4}{m_t^2 \Ms^2}
\bigg]
\bigg]\bigg \}
+\Gamma_{gg}^{(0)}\left(\frac{\cos\alpha}{\sin\beta}\right)^2
\left(\aspi\right)^2 \bigg[\frac{1}{144} +\frac{1}{144}\frac{m_t^2}{ \Ms^2} + 
\frac{7}{8640} \frac{M_h^2}{ m_t^2} 
+ \frac{7}{17280}\frac{M_h^2}{ \Ms^2} 
\nonumber\\
&
+  
 \frac{169}{2073600} \frac{M_h^4}{ m_t^4} + \frac{1}{24192}
 \frac{M_h^4}{m_t^2 \Ms^2}
\bigg]+ {\cal O}\left( \frac{M_h^4}{ \Ms^4}, \frac{M_t^4}{
    \Ms^4},\frac{M_h^6}{ m_t^6}\right) \,.
%\end{eqnarray}
\end{align}
}
$\!\!$For a light Higgs mass $M_h= 125$~GeV and SUSY masses of about $1$~TeV,
$\tan\beta = 40$ and SM parameters chosen as in the previous sections
the mass corrections at NLO and NNLO amount to below one percent from
the dominant 
contribution ({\it i.e.} computed in the EFT) at the corresponding order
in perturbation theory. They are beyond the reach of the LHC accuracy, but they
might be of phenomenological interest at a future linear collider.

\subsection{Hadronic Higgs production}
\begin{figure}[t]
\begin{center}
   \begin{tabular}{l}
%      \leavevmode
%\hspace*{-0.5cm}
%\centering
%      \epsfxsize=.95\textwidth
      \epsfig{figure=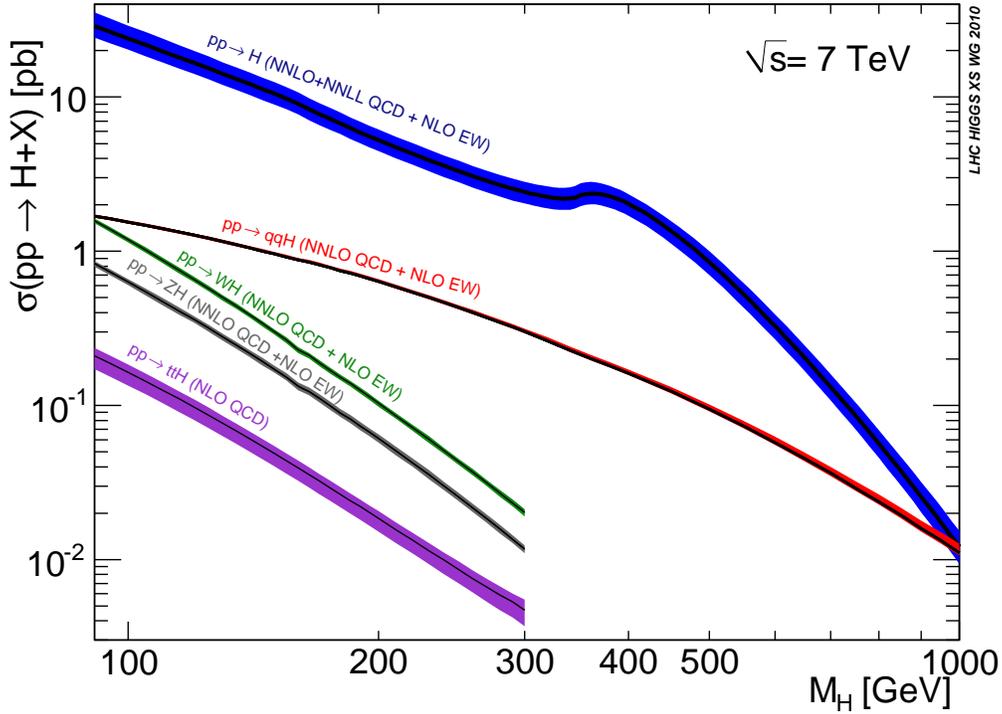,width=36em}
    \end{tabular}
  \end{center}
  \caption{\label{fig:hxslhc}
    Higgs production cross sections at the LHC for $\sqrt{s}=8$~TeV
    together with the uncertainties from the missing higher order
    corrections and  
   the parton density functions  from Ref.~\cite{higgswg}.      }
\end{figure}

During the last years, a lot of effort has been devoted to  precise
predictions for Higgs production at hadron colliders (for reviews, see
Refs.~\cite{Djouadi:2005gi,Harlander:2007zz,Dittmaier:2011ti}). They
constituted basic ingredients for the discovery  of the new scalar
particle at the LHC.  The main production
channel for the  SM Higgs boson at the LHC is the loop-induced gluon-fusion
channel.   For illustration, we reproduce in Fig~\ref{fig:hxslhc} from
Ref.~\cite{higgswg} the
theoretical predictions for the main Higgs production channels  together
with the uncertainties due to missing higher order corrections and 
 to the uncertainties  on the parton density functions (PDFs). 

An important application of the Higgs discovery 
is to constrain  the parameters of theories predicting physics beyond
the SM. This is also the case for SUSY theories. Given the high
sensitivity of the Higgs observables (its  mass, production cross
sections and decay rates) on the  parameters of the top sector in the 
MSSM, one can derive lower bounds for top squark masses and set
constraints for their mixing angle. For this task
 one needs, among other ingredients,  precise predictions for
the Higgs production cross section, including even NNLO SUSY-QCD corrections.
As discussed in section~\ref{sec:coefffc}, exact analytic
 calculations at this order in perturbation theory are not yet
 feasible. Instead one has to use the EFT approach together with the
 method of asymptotic expansions. In the SM, it was
 shown~\cite{Harlander:2009mq,Harlander:2009bw,Harlander:2009my,Pak:2009dg,Pak:2009bx,Pak:2011hs} that the exact result for the
 hadronic cross section for intermediate Higgs masses ($M_h< 2 m_t$) 
 is approximated to better than 1\% level by  
the result derived with EFT approach, if the full top mass dependence at
LO is factored out.\\
 Following the same reasoning for the case of the
MSSM, one can write the hadronic cross section
$\sigma\equiv\sigma(pp\to h+X)$ as as a function of the hadronic
center-of-mass energy $\sqrt{s}$. It reads~\cite{Harlander:2004tp}
\begin{eqnarray}
\sigma(z)= \rho_0 \sigma_0 \left(-3\pi\frac{ C_g(\mu_h)}{c_1^{(0)}}\right)^2
\bigg[\Sigma^{(0)}(z) + \frac{\alpha_s(\mu_s)}{\pi}\Sigma^{(1)}(z) 
+ \left(\frac{\alpha_s(\mu_s)}{\pi}\right)^2\Sigma^{(2)}(z)+
\cdots\bigg]\,,
\label{eq::hadxgen}
\end{eqnarray} 
%where the factorisation of short and  long distance contributions
%according to the EFT formalism is explicit. 
The exact LO
contribution, denoted here $\sigma_0$, is factored out, as discussed
above. The higher order corrections are computed within the EFT approach
and the separation of   short and  long distance contributions is
explicit in  Eq.~(\ref{eq::hadxgen}). For a better convergence of the
perturbative expansion and to avoid the occurrence of large logarithms,
one  makes use of   
 scale separation  as discussed in
 section~\ref{sec:eft}. Thus, the coefficient functions $C_g$ and 
 $c_1^{(0)}$ that contain the radiative corrections due to heavy
 particles are evaluated at a heavy scale of the order of the SUSY particle masses
 $\mu_h = {\cal O}(\tilde{M})$.  The partonic cross sections $\Sigma^{(n)}(z)$ are computed
 at a low-scale of the order of the Higgs mass $\mu_s = {\cal
   O}(M_h)$. The individual building blocks in Eq.~(\ref{eq::hadxgen})
 are discussed below.

The normalization coefficient $\rho_0$ is given by
\begin{eqnarray}
\rho_0&=&\frac{G_F [\alpha_s(\mu_S)]^2}{288\pi\sqrt{2}}\,,
\end{eqnarray}
where the presence of the strong coupling evaluated at the low energy scale
$\mu_s$ is due to the use of renormalization group invariant operators and coefficient
functions as given in Eqs.~(\ref{eq:giop}) and (\ref{eq::cfrg}).

$\sigma_0$ contains the exact dependence on all
 masses and momenta at the LO.
Its analytic expression is known for quite long time. For convenience of
the reader, we reproduce it here, in the normalization of 
Ref.~\cite{Pak:2012xr}
\begin{eqnarray}
\sigma_0&=&\bigg|
\frac{3}{2}\frac{\cos\alpha}{\sin\beta}\bigg\{
A(\tau_t) +\sum_{i=1,2} (-1)^i\bigg[
\frac{\sin(2\theta_t)}{2}\left(\tan\alpha+\frac{1}{\tan\beta}\right)\frac{m_t\mu_{\rm
  SUSY}}{2m_{\tilde{t}_i}^2}
\nonumber\\
&&+\frac{m_t^2}{8m_{\tilde{t}_i}^2}\left(\sin^2(2\theta_t)\frac{m_{\tilde{t}_1}^2-m_{\tilde{t}_2}^2}{m_t^2}-4(-1)^i\right) 
\bigg]\tilde{A}(\tau_{\tilde{t}_i})
\bigg\}+{\cal O}\left(\frac{M_Z^2}{m_{\tilde{t}_i}^2}\right)\bigg|^2\,,
\label{eq::sigma0}
\end{eqnarray}
with
\begin{eqnarray}
\tau_i&=& \frac{4 m_i^2}{M_h^2}\,,\quad A(\tau)= \tau
[1+(1-\tau)f(\tau)]\,,\quad \tilde{A}(\tau)= \tau(1-\tau
f(\tau))\,,\nonumber\\
f(\tau)&=&\left\{\begin{array}{ll}
\arcsin^2(1/\sqrt{\tau})\,,& \tau\ge 1\,,\\
-\frac{1}{4}\left(\ln \frac{1+\sqrt{1-\tau}}{1-\sqrt{1-\tau}}-i\pi
\right)^2\,,& \tau<1\,.
\end{array}\right.
\end{eqnarray}
The coefficient $c_1^{0}$ is defined
through the one-loop relation 
\begin{eqnarray}
c_1^{0} &=& -\frac{3\pi}{\alpha_s}  C_1^{({\rm 1-loop})}\,.
\end{eqnarray}  
Its SUSY-QCD part can be read of directly from
Eq.~(\ref{eq::c11l}). The coefficient $c_1^{0}$
is factored out because it is already contained in the LO contribution
$\sigma_0$ as can be easily understood from Eq.~(\ref{eq::sigma0}). Indeed, 
in the limit of  light Higgs masses $M_h\ll m_t, m_{\tilde{t}},
m_{\tilde{g}}$  and neglecting mass suppressed contributions of the order
of ${\cal O}(M_h^2/m_t^2)$, ${\cal O}(M_h^2/m_{\tilde{t}}^2)$, and ${\cal O}(M_Z^2/m_{\tilde{t}}^2)$ the LO
contribution $\sigma_0$ takes 
the form~\footnote{We adopt here the normalization of  Ref.~\cite{Pak:2012xr}.} 
\begin{eqnarray}
\sigma_0\to  \left| c_1^{(0)}\right|^2\,.
\end{eqnarray} 
The coefficient $C_g$ was defined in Eq.~(\ref{eq:gicoeff}) and has to
be evaluated at the heavy scale. Let us point out that the factor
$-3\pi C_g(\mu_h)/c_1^{0}$ expanded in the strong coupling
$\alpha_s(\mu_h)$ takes the form 
\begin{eqnarray}
-3\pi\frac{ C_g(\mu_h)}{c_1^{0}} = 1
+\frac{\alpha_s(\mu_h)}{\pi} c_g^{(1)}+
\left(\frac{\alpha_s(\mu_h)}{\pi}\right)^2 c_g^{(2)} + \cdots\,,
\end{eqnarray}
where the coefficients $ c_g^{(i)}$, with $i=1,2$, are known, once 
the coefficient $C_1$ is computed up to the NNLO.

Finally, $\Sigma^{(n)}(z)$ is defined through the convolution  
\begin{eqnarray}
\Sigma^{(n)}(z) &=& \sum_{i,j\in\{q\bar{q}g\}} \int_z^1 {\rm d} x_1
\int_{z/x_1}^1 {\rm d} x_2 f_{i/p}(x_1)
f_{j/p}(x_2)\hat{\Sigma}_{ij}^{(n)}\left(\frac{z}{x_1x_2}\right)\,, \quad
z\equiv \frac{M_h^2}{s}\,,
\end{eqnarray}
of $f_{j/p}(x)$  the density of parton $i$ inside the proton  and
$\hat{\Sigma}_{ij}^{(n)}(x)$ the partonic cross section expanded up to the $n$-th order in
$\alpha_s(\mu_s)$ and computed in the effective-theory approach. 
% $s$ is the hadronic center-of-mass energy.
 At the LO, it reads
\begin{eqnarray}
\hat{\Sigma}_{ij}^{(0)}(x)&=& \delta_{ig}\delta_{jg}\delta(1-x)\,.
\end{eqnarray}
The NLO and NNLO contributions contain the real and virtual corrections
associated with the operator ${\cal O}_1$ and its mixture with the
operator ${\cal O}_2$.
Since they are computed within the effective
theory,  they can be taken over from the   SM computations reported
in Refs.~\cite{Dawson:1990zj,Djouadi:1991tka,Harlander:2009mq,Pak:2011hs}.  

Let us mention, that there is also a third scale present in
 Eq.~(\ref{eq::hadxgen}), namely the factorization scale $\mu_F$
 embedded in  the PDFs. Usually it is chosen to be equal to the
 low-scale $\mu_s$, {\it i.e.} $\mu_F=\mu_s$. The choice of scales plays
 an important role 
 in precision calculations of the hadronic Higgs production cross
 section, especially when particles much heavier than 
 the SM ones are present. We discuss in the next section the
 phenomenological impact of the NNLO corrections.

% In order to avoid occurance of large logarithms in  Eq.~(\ref{eq::hadxgen}),

\subsubsection{Numerical analysis}

\begin{figure}[t]
  \centering
  \begin{tabular}{c}
    \includegraphics[width=.8\linewidth]{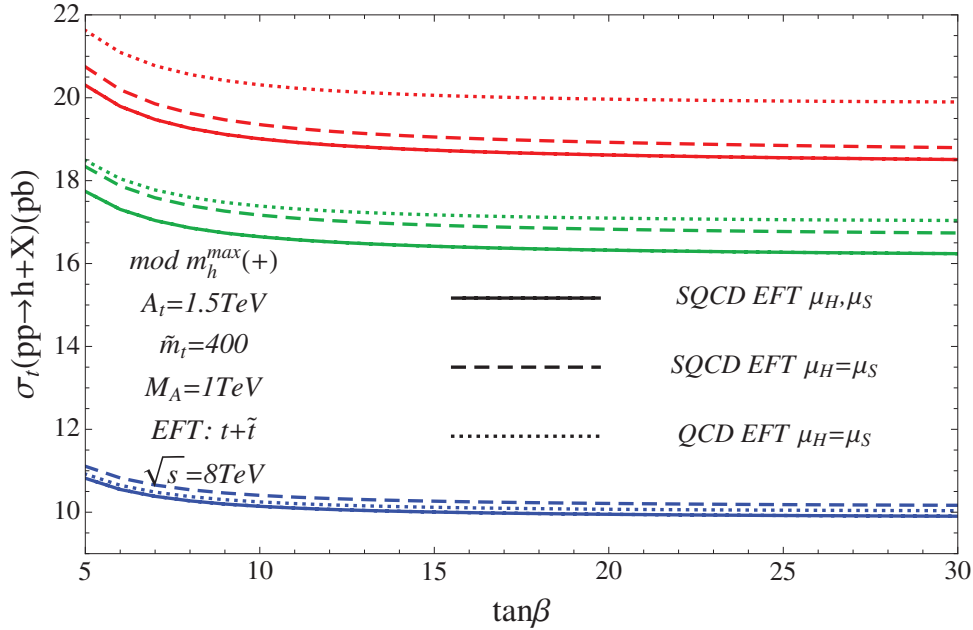}
  \end{tabular}
  \caption[]{\label{fig::sig_tb}The cross section $\sigma_t^{\rm
      SUSY-QCD}(\mu_s,\mu_h)$ as a function of $\tan\beta$ using the parameters
    in  to LO (bottom), NLO (middle) and NNLO
    (top) from Ref.~\cite{Pak:2012xr}.  The dotted line corresponds to the SM and the dashed and solid
    lines to the MSSM.}
\end{figure}

%In the following we discuss the numerical effects of the radiative
%corrections to the hadronic cross section $\sigma(z)$.
 For the numerical analysis  we choose a 
supersymmetric mass spectrum in the so-called `` modified $m_h^{\rm max}$
scenario'' as defined in Ref~\cite{Pak:2012xr}. It is a modification of
the original ``$m_h^{\rm max}$'' scenario~\cite{Carena:2002qg}  such that one 
of the top squarks becomes light and the other one remains heavy, at the
TeV scale. At the
same time Higgs masses as large as $127$~GeV can be achieved. 

%The
%  top trilinear coupling has been set to
%$A_t=1500$~GeV and the mass of the CP-odd Higgs to $M_A=1000$~TeV. In such
%a scenario the supersymmetric radiative corrections to the hadronic
%cross section become sizeable.  

For illustration, we reproduce in   Fig.~\ref{fig::sig_tb} the results
of Ref.~\cite{Pak:2012xr} that 
constitute the most precise prediction for the hadronic cross section in the
gluon fusion channel in the framework of the MSSM. Here,  $\sigma_t^{\rm
  SUSY-QCD}(\mu_s,\mu_h)$ denotes the dominant contribution originating
from the top-sector. From bottom to top, the  LO, NLO and NNLO results
are depicted for  $5\le\tan\beta\le 30$ and choosing 
$m_{\tilde{t}_1}=400$~GeV. The dotted lines represent the SM results. The
solid and the dashed lines show the  MSSM predictions  for two different
scale choices: $\mu_h=M_t$ and $\mu_s=M_h/2$ and $\mu_h=\mu_s=M_h/2$,
respectively. 
The MSSM results are reduced by a few percent as compared to the SM
prediction.  This effect increases when going from LO to NLO and finally
to NNLO where a difference of about 5\% is observed. This behaviour is
specific for  supersymmetric mass spectra containing at least one
light squark of the third generation. For the case when all SUSY
particles are heavy, at the TeV scale, ( the so called decoupling limit)
the  genuine 
SUSY-QCD corrections to the production cross section become
negligible. The fact that the difference between the SM and MSSM
predictions increases when higher order radiative corrections are taken
into account 
can be explained by the occurrence  of many new SUSY contributions.\\  
 As can be seen from
the figure, the effect of  scale choice is not negligible: the results for
$\mu_h=\mu_s$ are in general a few percent above the ones with
$\mu_h\not=\mu_s$. At  NLO, the scale dependence   increases as compared
to the LO case as a consequence of the special organisation of the
perturbative series. Nevertheless, the scale dependence decreases when
going from NLO to NNLO as expected.

%\subsubsection{Numerical analysis}

%\subsection{$h\to g g$  }

%%%%%%%%%%%%%%%%%%%%%%%%%%%%%%%%%%%%%%%%%%%%%%%%%%%%%%%%%%%%%%%%
\section{\label{sec:conclusions}Conclusions}
In this review we report on precision calculations in supersymmetric
theories. They are important ingredients not only for the development of quantum
field theories in general, but they are also required by the current
experimental analyses searching for indirect manifestations of SUSY in collider
experiments at the TeV scale. The latter topic is of utmost importance
for particle physics: the non-observation of any supersymmetric particle
at the TeV scale renders low-energy supersymmetric theories
debatable. Obviously, to prove or disprove a theory for which 
enormous efforts both at theoretical and experimental level have been
devoted over the last four decades, 
is a very complex task. In this review, we concentrate on the indirect
searches for SUSY that can be carried through  precision
tests of 
the gauge coupling unification  hypothesis, the prediction
of a light Higgs boson mass  and the interaction properties of the Higgs
boson with the SM particles. 

It turns  out, that the  hypothesis of gauge coupling unification 
even in the
framework of minimal SUSY SU(5) model cannot be falsified with the
help of currently available experimental data. Let us mention that the
contributions 
at the three-loop order in perturbation theory  are essential in this
analysis. The conclusion drawn from precision calculations
reconfirm earlier results derived from model building arguments.

Furthermore, the theoretical prediction of the light Higgs boson mass
within SUSY  with an
accuracy comparable with the one reached by the ongoing experimental
analyses conducted at the LHC is an  important tool for constraining
the supersymmetric parameter space. For this purpose one needs to
calculate even three-loop Feynman integrals
 involving many different mass scales. At present, an exact analytic
computation is not feasible. Nevertheless, the method of asymptotic
expansion can be applied successfully also   in SUSY theories and
provides us with 
precise  results. Specifically, the lightest Higgs bosons mass within
the MSSM can be 
predicted at present with an accuracy of about $1$~GeV for the 
parameter space of phenomenological interest.

Moreover, after the recent discovery of the Higgs boson at the LHC, the
natural question  
 is whether it
has the characteristics of  the particle predicted by the SM or new
theories are required to describe it. To answer this question from the
perspective of supersymmetric theories, one needs  predictions of the
hadronic Higgs production cross section and its decay rates into SM
particles with the same precision as in the SM.  To achieve such an
accuracy, again multi-loop calculations up to the three-loop order are
required.

Detailed analyses of the data taken or to be taken at the LHC running at
energies up to $14$~TeV are  expected  to provide us with new insights
into the particle physics and hopefully with the answer to the question 
whether low-energy supersymmetry is the right theory to describe the
phenomena at the TeV scale. 

\section*{Acknowledgements}

I am grateful to M.~Steinhauser  and J.H.~K\"uhn for
carefully reading the manuscript and for many valuable comments and
discussions. Furthermore, I would like to thank J.H.~K\"uhn and
M.~Steinhauser for encouraging me to complete this work.
 This work was
supported by the DFG through SFB/TR~9 ``Computational Particle Physics''.

%that in the near future a
%definite answer will be given to the question whether the supersymmetric
%extension 
%%%%%of the SM   is the correct theory to describe the physical phenomena
%left over unexplained by the SM. 

%%%%%%%%%%%%%%%%%%%%%%%%%%%%%%%%%%%%%%%%%%%%%%%%%%%%%%%%%%%%
%%%%%%%%%%%%%%%%%%%%%%%%%%%%%%%%%%%%%%%%%%%%%%%%%%%%%%%%%%%%

\newpage
\begin{appendix}
\renewcommand {\theequation}{\Alph{section}.\arabic{equation}}
\renewcommand {\thefigure}{\Alph{section}.\arabic{figure}}
\renewcommand {\thetable}{\Alph{section}.\arabic{table}}

\section{\label{appA} Group Theory}

We consider a gauge group ${\cal G}$ with generators $R^a$
satisfying the Lie algebra\footnote{Useful sources for some of the
  material in this 
section have included Refs.~\cite{predrag,vanRitbergen:1997va,RSV}.}
\be
\left[R^a, R^b \right] = if^{abc}R^c.
\ee
We work throughout with a  fermion representation consisting of
$N_f$ sets of Dirac fermions or $2 N_f$ sets of two-component fermions, in
irreducible representations with identical Casimir invariants, using $R^a$ to denote
the generators in one such representation. Thus
$R^a R^a$ is proportional to the
unit matrix:
\be
R^a R^a = C_R\cdot I
\ee
For the adjoint representation we have
\be
C_A \delta_{ab} = f_{acd} f_{bcd}.
\ee
$I_2(R)$ is defined by
\be
\Tr[R^aR^b]= I_2 (R)\delta^{ab}.
\ee
Thus we have
\be
C_R d_R = I_2(R) N_A
\ee
where $N_A$ is the number of generators and $d_R$ is the dimensionality
of the representation $R$. Evidently $I_2 (A) = C_A$.
The fully symmetric
tensors $d_R^{abcd}$ and
$d_A^{abcd}$ are defined by
\bea
d_R^{abcd}&=&\frac{1}{6}\Tr[R^{(a}R^bR^cR^{d)}],\nn
d_A^{abcd}&=&\frac{1}{6}\Tr[F^{(a}F^bF^cF^{d)}],
\eea
where
\be
(F^a)^{bc} = if^{bac}
\ee
and
\bea
R^{(a}R^bR^cR^{d)} &=& R^aR^bR^cR^d + R^aR^bR^dR^c
+R^aR^cR^bR^d\nn  &+& R^aR^cR^dR^b +R^aR^dR^bR^c +R^aR^dR^cR^b,
\eea
(similarly for $F^{(a}F^bF^cF^{d)}$).

The additional tensor invariants occurring in
the results  are defined as
\bea
D_3 (RR) &=& d_A^{abc} d_A^{abc}/N_A \nn
D_4 (AA) &=& d_A^{abcd} d_A^{abcd}/N_A \nn
D_4 (RA) &=& d_R^{abcd} d_A^{abcd} /N_A\nn
D_4 (AAA) &=& d_A^{abcd} d_A^{cdef} d_A^{abef}/N_A\nn
D_4 (RAA) &=& d_R^{abcd} d_A^{cdef} d_A^{abef}/N_A.
\eea
In table~\ref{gaminvsa}-\ref{gaminvsc}\ we present results
for the various tensor invariants
for the groups $SU(N)$, $SO(N)$ and $Sp(N)$, when the fermion
representation $R$
is the fundamental representation.

The  canonical choice of $b$ is $b=1$ for all groups,
but sometimes different choices are more convenient~\cite{RSV}.
% as we show in the next
%section.

\begin{table}[ht]
\begin{center}
\begin{tabular}{|c|c|} \hline
&  \\ %\hline
Group & $SU(N)$ \\ \hline
$C_A$& $bN$ \\ %\hline

$C_R$&$b\frac{N^2-1}{2N}$\\%\hline

$I_2(A)$&$bN$\\%\hline

$I_2(R)$&$\frac{b}{2}$\\%\hline

$N_A$ & $N^2-1$ \\ %\hline

$D_4 (AA)$&$\frac{b^4}{24}(N^2+36)N^2$
\\%\hline

$D_4 (RA)$&$\frac{b^4}{48}N(N^2+6)$\\%\hline

$D_4 (RR)$&$\frac{b^4}{96N^2}(18 - 6N^2 + N^4)$
\\ %\hline

$D_4 (AAA)$&$\frac{b^6}{216}N^2(324 + 135N^2 + N^4)$\\%\hline

$D_4 (RAA)$ &$\frac{b^6}{432}N^3(51 + N^2)$ \\ \hline

\end{tabular}
\caption{\label{gaminvsa} $SU(N)$ Group invariants (here $R$ is the
fundamental representation).}
\end{center}
\end{table}

\begin{table}[ht]
\begin{center}
\begin{tabular}{|c|c|} \hline
& \\ %\hline
Group & $SO(N)$ \\ \hline

$C_A$& $b(N-2)$  \\ %\hline

$C_R$&$\frac{b}{2}(N-1)$\\%\hline

$I_2(A)$&$b(N-2)$\\%\hline

$I_2(R)$&$b$\\%\hline

$N_A$ & $\frac{1}{2}N(N-1)$ \\ %\hline

$D_4 (AA)$&$\frac{b^4}{24}(N-2)(-296 + 138N - 15N^2 + N^3)$\\%\hline

$D_4 (RA)$&$\frac{b^4}{24}(N-2)(22 - 7N + N^2)$\\%\hline

$D_4 (RR)$ &$\frac{b^4}{24}(4 - N + N^2)$\\ %\hline

$D_4 (AAA)$&$\frac{b^6}{432}(N-2)(-29440
+23272N- 7018N^2 + 971N^3 - 47N^4 +2N^5)$\\%\hline

$D_4 (RAA)$ & $\frac{b^6}{432}(N-2)(2048 - 1582N
+ 387N^2 - 31N^3 + 2N^4)$\\
\hline

\end{tabular}
\caption{\label{gaminvsb} $SO(N)$ Group invariants (here $R$ is the
fundamental representation).}
\end{center}
\end{table}

\begin{table}[ht]
\begin{center}
\begin{tabular}{|c|c|} \hline
& \\ %\hline
Group & $Sp(N)$ \\ \hline

$C_A$& $b(N+2)$  \\ %\hline

$C_R$&$\frac{b}{4}(N+1)$\\%\hline

$I_2(A)$&$b(N+2)$\\%\hline

$I_2(R)$&$\frac{b}{2}$\\%\hline

$N_A$ & $\frac{1}{2}N(N+1)$ \\ %\hline

$D_4 (AA)$&$\frac{b^4}{384}(N+2)(296 + 138N + 15N^2 + N^3)$\\%\hline

$D_4 (RA)$&$\frac{b^4}{384}(N+2)(22 + 7N + N^2)$\\%\hline

$D_4 (RR)$ &$\frac{b^4}{384}(4 + N + N^2)$\\ %\hline

$D_4 (AAA)$&$\frac{b^6}{27648}(N+2)(29440
+23272N+ 7018N^2 + 971N^3 + 47N^4 +2N^5)$\\%\hline

$D_4 (RAA)$ & $\frac{b^6}{27648}(N+2)(2048 + 1582N
+ 387N^2 + 31N^3 + 2N^4)$\\%\hline
& \\ \hline
\end{tabular}
\caption{\label{gaminvsc} $Sp(N)$ Group invariants (here $R$ is the
fundamental representation).}
\end{center}
\end{table}

%\end{appendix}

%%%%%%%%%%%%%%%%%%%%%%%%%%%%%%%%%%%%%%%%%%%%%%%%%%%%%%%%%%%%
%%%%%%%%%%%%%%%%%%%%%%%%%%%%%%%%%%%%%%%%%%%%%%%%%%%%%%%%%%%%
%\begin{appendix}
%\renewcommand {\theequation}{\Alph{section}.\arabic{equation}}
%\renewcommand {\thefigure}{\Alph{section}.\arabic{figure}}
%\renewcommand {\thetable}{\Alph{section}.\arabic{table}}

\section{\label{appB} Modification of the \drbar{} scheme:
\drbarmod{}}

In the following we provide  analytic expressions for the finite shifts
introduced in the top squark mass counter-terms as compared
 to the \drbar{} scheme. According to the discussion in
 section~\ref{sec:mh}, one can distinguish four cases
 for the mass hierarchies.

Case (i): $\quad\msq \gg \Msti$, $(i=1,2)$
\begin{eqnarray}
  \left(\frac{\Msti^{\overline{\rm MDR}}}{\Msti}\right)^2 &=& 
  1 - \left(\as\right)^2\cf\Nq\TF
  \left(-\frac{1}{2} +\lMsq + \zeta(2) \right)
 \frac{\msq^2}{\Msti^2}
  \,.
  \label{eq:h4shift}
\end{eqnarray}
The label $\Nq=5$ has been introduced for convenience and for the
logarithms the abbreviation $\lMsq=\ln(\mu^2/\msq^2)$ has been introduced.

Case (ii): $\quad\Mstd \gg \Mstu $
\begin{eqnarray}
  \left(\frac{\Mstu^{\overline{\rm MDR}}}{\Mstu}\right)^2 &=& 
  1 - \left(\as\right)^2\cf\TF
  \left(-\frac{1}{4} + \frac{1}{2}\lMstd + \frac{1}{2}\zeta(2) \right)
 \frac{\Mstd^2}{\Mstu^2}
  \,.
\label{eq:h3shift}
\end{eqnarray}
In this equation we have $\lMstd=\ln(\mu^2/\Mstd^2)$.

Case (iii): $\quad\Mgl \gg \Msti$, $(i=1,2)$ and $\msq\gg \Mgl$
\begin{eqnarray}
  \left(\frac{\Msti^{\overline{\rm MDR}}}{\Msti}\right)^2 &=& 
  1+
  \as\cf\left[1 + \lMgl\right]\frac{\Mgl^2}{\Msti^2}
  +
  \left(\as\right)^2\bigg\{
  \cf^2\left[-\frac{11}{4} - \frac{3}{2}\lMgl 
    + \zeta(2)\right]\frac{\Mgl^2}{\Msti^2}
  \nonumber\\
  &+&
  \ca\cf\left[\frac{21}{8} + \frac{7}{2}\lMgl 
    + \frac{9}{8}\lMgl^2 - \frac{1}{4}\zeta(2)\right]\frac{\Mgl^2}{\Msti^2}
  \nonumber\\
  &+&
  \cf\Nt\TF\bigg[-\left(2 + 2\lMgl
      + \frac{3}{4}\lMgl^2\right)\frac{\Mgl^2}{\Msti^2} 
    + \left(1-2\zeta(2)\right)\frac{\Mgl(\Mgl-\Mstd)}{\Msti^2}
\nonumber\\
&+&\left(\frac{1}{4} - \frac{1}{2}\lMstd - \frac{1}{2}\zeta(2) \right)
\frac{\Mstd^2}{\Msti^2} \bigg]
  \nonumber\\        
  &+&\cf\Nq\TF\bigg[\left(-\frac{5}{8} - \frac{3}{4}\lMgl 
    - \frac{5}{4}\lMsq - \frac{3}{2}\lMgl\lMsq + \frac{3}{4}\lMsq^2 
    + \frac{3}{2}\zeta(2)\right)\frac{\Mgl^2}{\Msti^2}
  \nonumber\\
  &+&
  \left(-\frac{43}{36} - \frac{5}{6}\lMsqgl \right) \frac{\Mgl^4}{\msq^2\Msti^2} 
  +
  \left(-\frac{67}{288} - \frac{7}{24}\lMsqgl
  \right)\frac{\Mgl^6}{\msq^4\Msti^2} 
\nonumber\\
&+& \left(+\frac{1}{2} -\lMsq - \zeta(2) \right)
 \frac{\msq^2}{\Msti^2}
  \bigg] 
  \bigg\}
  \,.
  \label{eq:h6shift}
\end{eqnarray}
Here $N_t=1$, $\lMgl=\ln(\mu^2/\Mgl^2)$ and $\lMsqgl=\ln(\msq^2/\Mgl^2)$.

Case (iv): $\quad\Mgl \gg \Mstu$,  and $\msq\approx \Mgl$
\begin{eqnarray}
  \left(\frac{\Msti^{\overline{\rm MDR}}}{\Msti}\right)^2 &=&
  1+ \as\cf\left[1 + \lMgl\right]\frac{\Mgl^2}{\Msti^2}
  +
  \left(\as\right)^2\bigg\{\cf^2\left[-\frac{11}{4} - \frac{3}{2}\lMgl +
    \zeta(2)\right]\frac{\Mgl^2}{\Msti^2}\nonumber\\
  & +& \ca\cf \left[\frac{21}{8} + \frac{7}{2}\lMgl + \frac{9}{8}\lMgl^2
    - \frac{1}{4}\zeta(2)\right]\frac{\Mgl^2}{\Msti^2} \nonumber\\
  &+&
  \cf\Nt\TF\bigg[-\left(2 + 2\lMgl + \frac{3}{4}\lMgl^2\right)
    \frac{\Mgl^2}{\Msti^2} +\left(1 -
      2\zeta(2)\right)\frac{\Mgl(\Mgl-\Mstd)}{\Msti^2}
\nonumber\\
&+& \left(\frac{1}{4} - \frac{1}{2}\lMstd - \frac{1}{2}\zeta(2) \right)
\frac{\Mstd^2}{\Msti^2}  \bigg]
  \nonumber\\
  & +&\cf\Nq\TF\bigg[\left(-\frac{3}{4}\lMgl - \frac{5}{4}\lMsq
    - \frac{3}{2}\lMgl\lMsq + \frac{3}{4}\lMsq^2 +
    \frac{3}{2}\zeta(2)\right)\frac{\Mgl^2}{\Msti^2}
  \nonumber\\
  &-& 4 \zeta(2)\frac{\Mgl(\Mgl-\Mstd)}{\Msti^2}
  -\left(\frac{7}{4}+\lMsq + \zeta(2) \right)\frac{\msq^2}{\Msti^2}
  \bigg]\bigg\}\,.
  \label{eq:h6bshift}
\end{eqnarray}

All the masses on the r.h.s. of the
Eqs.~(\ref{eq:h4shift}), (\ref{eq:h3shift}), (\ref{eq:h6shift}) and
(\ref{eq:h6bshift}) are \drbar{} 
masses. Let us also mention that the above formulae are valid for the
case  $\Mes=0$. The finite shifts given for the cases (iii) and (iv) can
also be used  for other  mass hierarchies like for example $m_{\tilde q}
\gg m_{\tilde t_2} \approx m_{\tilde g} \gg m_{\tilde t_1}$ 
or $m_{\tilde q} \approx m_{\tilde t_2} \approx m_{\tilde g}\gg
m_{\tilde t_1} $. 
%For these particular cases  the shifts for
%$Mstd$ are necessary only at the one-loop order to allow the mixing
%angle to be renormalized in the \drbar{} scheme, whereas the two-loop
%shifts are optional. 

\end{appendix}

%%%%%%%%%%%%%%%%%%%%%%%%%%%%%%%%%%%%%%%%%%%%%%%%%%%%%%%%%%%%


\begin{thebibliography}{99}
%    \input{habil_ref}
%
% habil_ref.tex -- generated by sortref-2.3.6  
% ((C) R. Harlander, http://www.robert-harlander.de/software/)
% on Wed May  1 23:25:45 CEST 2013
%

%1
\bibitem{Glashow:1961tr}
  S.~L.~Glashow,
  Nucl.\ Phys.\  {\bf 22} (1961) 579.
  %%CITATION = NUPHA,22,579;%%

%2
\bibitem{Weinberg:1967tq}
  S.~Weinberg,
  Phys.\ Rev.\ Lett.\  {\bf 19} (1967) 1264.
  %%CITATION = PRLTA,19,1264;%%

%3
\bibitem{Salam:1968rm}
  A.~Salam,
{\it In the Proceedings of 8th Nobel Symposium, Lerum, Sweden, 19-25 May 1968, pp 367-377}.
  %%CITATION = CONFP,C680519,367;%%

%4
\bibitem{GellMann:1981ph}
  M.~Gell-Mann,
  Acta Phys.\ Austriaca Suppl.\  {\bf 9} (1972) 733;
  %%CITATION = APAUA,9,733;%%
\\
  H.~Fritzsch, M.~Gell-Mann and H.~Leutwyler,
  Phys.\ Lett.\ B {\bf 47} (1973) 365.
  %%CITATION = PHLTA,B47,365;%%

%5
\bibitem{Gross:1973id}
  D.~J.~Gross and F.~Wilczek,
  Phys.\ Rev.\ Lett.\  {\bf 30} (1973) 1343;
  %%CITATION = PRLTA,30,1343;%%
  D.~J.~Gross and F.~Wilczek,
  Phys.\ Rev.\ D {\bf 8} (1973) 3633;
  %%CITATION = PHRVA,D8,3633;%%
\\
  H.~D.~Politzer,
  Phys.\ Rev.\ Lett.\  {\bf 30} (1973) 1346.
  %%CITATION = PRLTA,30,1346;%%

%6
\bibitem{Coleman:1967ad}
  S.~R.~Coleman and J.~Mandula,
  Phys.\ Rev.\  {\bf 159} (1967) 1251.
  %%CITATION = PHRVA,159,1251;%%

%7
\bibitem{Haag:1974qh}
  R.~Haag, J.~T.~Lopuszanski and M.~Sohnius,
  Nucl.\ Phys.\  B {\bf 88} (1975) 257.
  %%CITATION = NUPHA,B88,257;%%

%8
\bibitem{Golfand:1971iw}
  Y.~.A.~Golfand and E.~P.~Likhtman,
  JETP Lett.\  {\bf 13} (1971) 323
   [Pisma Zh.\ Eksp.\ Teor.\ Fiz.\  {\bf 13} (1971) 452].
  %%CITATION = JTPLA,13,323;%%

%9
\bibitem{Volkov:1972jx}
  D.~V.~Volkov and V.~P.~Akulov,
  JETP Lett.\  {\bf 16} (1972) 438
   [Pisma Zh.\ Eksp.\ Teor.\ Fiz.\  {\bf 16} (1972) 621].
  %%CITATION = JTPLA,16,438;%%

%10
\bibitem{Wess:1974tw}
  J.~Wess and B.~Zumino,
  Nucl.\ Phys.\ B {\bf 70} (1974) 39.
  %%CITATION = NUPHA,B70,39;%%

%11
\bibitem{Salam:1974yz}
  A.~Salam and J.~A.~Strathdee,
  Nucl.\ Phys.\  B {\bf 76} (1974) 477.

%12
\bibitem{Zumino:1974bg}
  B.~Zumino,
  Nucl.\ Phys.\ B {\bf 89} (1975) 535.
  %%CITATION = NUPHA,B89,535;%%

%13
\bibitem{West:1976wz}
  P.~C.~West,
  Nucl.\ Phys.\ B {\bf 106} (1976) 219.
  %%CITATION = NUPHA,B106,219;%%

%14
\bibitem{Grisaru:1979wc}
  M.~T.~Grisaru, W.~Siegel and M.~Rocek,
  Nucl.\ Phys.\ B {\bf 159} (1979) 429.
  %%CITATION = NUPHA,B159,429;%%

%15
\bibitem{Girardello:1981wz}
  L.~Girardello and M.~T.~Grisaru,
  Nucl.\ Phys.\  B {\bf 194} (1982) 65.
  %%CITATION = NUPHA,B194,65;%%

%16
\bibitem{Gladyshev:2012xq}
  A.~V.~Gladyshev and D.~I.~Kazakov,
  arXiv:1212.2548 [hep-ph].

%17
\bibitem{lepwg}
{\tt http://lepewwg.web.cern.ch/LEPEWWG/}

%18
\bibitem{Ellis:1990wk}
  J.~R.~Ellis, S.~Kelley and D.~V.~Nanopoulos,
  Phys.\ Lett.\  B {\bf 260} (1991) 131.

%19
\bibitem{Amaldi:1991cn}
  U.~Amaldi, W.~de Boer and H.~Furstenau,
  Phys.\ Lett.\  B {\bf 260} (1991) 447.

%20
\bibitem{Langacker:1991an}
  P.~Langacker and M.~x.~Luo,
  Phys.\ Rev.\  D {\bf 44} (1991) 817.

%21
\bibitem{Sofue:2000jx}
  Y.~Sofue and V.~Rubin,
  Ann.\ Rev.\ Astron.\ Astrophys.\  {\bf 39} (2001) 137
  [astro-ph/0010594].

%22
\bibitem{Kaiser:1992ps}
  N.~Kaiser and G.~Squires,
  Astrophys.\ J.\  {\bf 404} (1993) 441.
  %%CITATION = ASJOA,404,441;%%

%23
\bibitem{Kochanek:1994vw}
  C.~S.~Kochanek,
  Astrophys.\ J.\  {\bf 453} (1995) 545
  [astro-ph/9411082].

%24
\bibitem{Eidelman:2004wy}
  S.~Eidelman {\it et al.}  [Particle Data Group Collaboration],
  Phys.\ Lett.\ B {\bf 592} (2004) 1.

%25
\bibitem{atlas:susy}
{\tt  https://twiki.cern.ch/twiki/bin/view/AtlasPublic/SupersymmetryPublicResults}

%26
\bibitem{Angle:2011th}
  J.~Angle {\it et al.}  [XENON10 Collaboration],
  Phys.\ Rev.\ Lett.\  {\bf 107} (2011) 051301
  [arXiv:1104.3088 [astro-ph.CO]].

%27
\bibitem{wmap}
{\tt http://map.gsfc.nasa.gov/}

%28
\bibitem{Bennett:2006fi}
  G.~W.~Bennett {\it et al.}  [Muon G-2 Collaboration],
  Phys.\ Rev.\ D {\bf 73} (2006) 072003
  [hep-ex/0602035].

%29
\bibitem{atlas}
{\tt https://twiki.cern.ch/twiki/bin/view/AtlasPublic/HiggsPublicResults}

%30
\bibitem{cms}
{\tt https://twiki.cern.ch/twiki/bin/view/CMSPublic/PhysicsResults}

%31
\bibitem{Heinemeyer:2004gx}
  S.~Heinemeyer, W.~Hollik and G.~Weiglein,
  Phys.\ Rept.\  {\bf 425} (2006) 265
  [hep-ph/0412214].

%32
\bibitem{Stockinger:2006zn}
  D.~Stockinger,
  J.\ Phys.\ G {\bf 34} (2007) R45
  [hep-ph/0609168].

%33
\bibitem{delAguila:2008iz}
  F.~del Aguila, J.~A.~Aguilar-Saavedra, B.~C.~Allanach, J.~Alwall, Y.~.Andreev, D.~Aristizabal Sierra, A.~Bartl and M.~Beccaria {\it et al.},
  Eur.\ Phys.\ J.\ C {\bf 57} (2008) 183
  [arXiv:0801.1800 [hep-ph]].

%34
\bibitem{Dittmaier:2011ti}
  S.~Dittmaier {\it et al.}  [LHC Higgs Cross Section Working Group Collaboration],
  arXiv:1101.0593 [hep-ph].

%35
\bibitem{Shifman:1986zi}
  M.~A.~Shifman and A.~I.~Vainshtein,
  Nucl.\ Phys.\ B {\bf 277} (1986) 456
   [Sov.\ Phys.\ JETP {\bf 64} (1986) 428]
   [Zh.\ Eksp.\ Teor.\ Fiz.\  {\bf 91} (1986) 723].
  %%CITATION = NUPHA,B277,456;%%

%36
\bibitem{ArkaniHamed:1997mj}
  N.~Arkani-Hamed and H.~Murayama,
  JHEP {\bf 0006} (2000) 030
  [arXiv:hep-th/9707133].
  %%CITATION = JHEPA,0006,030;%%

%37
\bibitem{Wilson:1973jj}
  K.~G.~Wilson and J.~B.~Kogut,
  Phys.\ Rept.\  {\bf 12} (1974) 75.

%38
\bibitem{GellMann:1954fq}
  M.~Gell-Mann and F.~E.~Low,
  Phys.\ Rev.\  {\bf 95}, 1300 (1954).

%39
\bibitem{Callan:1970yg}
  C.~G.~.~Callan,
  Phys.\ Rev.\  D {\bf 2} (1970) 1541.

%40
\bibitem{Symanzik:1970rt}
  K.~Symanzik,
  Commun.\ Math.\ Phys.\  {\bf 18} (1970) 227.
  %%CITATION = CMPHA,18,227;%%

%41
\bibitem{Grisaru:1985tc}
  M.~T.~Grisaru, B.~Milewski and D.~Zanon,
  Phys.\ Lett.\  B {\bf 155} (1985) 357.
  %%CITATION = PHLTA,B155,357;%%s

%42
\bibitem{Hisano:1997ua}
  J.~Hisano and M.~A.~Shifman,
  Phys.\ Rev.\  D {\bf 56} (1997) 5475
  [arXiv:hep-ph/9705417].
  %%CITATION = PHRVA,D56,5475;%%

%43
\bibitem{ArkaniHamed:1998kj}
  N.~Arkani-Hamed, G.~F.~Giudice, M.~A.~Luty and R.~Rattazzi,
  Phys.\ Rev.\  D {\bf 58}, 115005 (1998)
  [arXiv:hep-ph/9803290].
  %%CITATION = PHRVA,D58,115005;%%

%44
\bibitem{Yamada:1994id}
  Y.~Yamada,
  Phys.\ Rev.\  D {\bf 50} (1994) 3537
  [arXiv:hep-ph/9401241].
  %%CITATION = PHRVA,D50,3537;%%

%45
\bibitem{Avdeev:1997vx}
  L.~V.~Avdeev, D.~I.~Kazakov and I.~N.~Kondrashuk,
  Nucl.\ Phys.\  B {\bf 510} (1998) 289
  [arXiv:hep-ph/9709397].
  %%CITATION = NUPHA,B510,289;%%

%46
\bibitem{Kazakov:2000ih}
  D.~I.~Kazakov and V.~N.~Velizhanin,
  Phys.\ Lett.\ B {\bf 485} (2000) 393
  [hep-ph/0005185].
  %%CITATION = HEP-PH/0005185;%%

%47
\bibitem{Novikov:1985rd}
  V.~A.~Novikov, M.~A.~Shifman, A.~I.~Vainshtein and V.~I.~Zakharov,
  Phys.\ Lett.\ B {\bf 166} (1986) 329
   [Sov.\ J.\ Nucl.\ Phys.\  {\bf 43} (1986) 294]
   [Yad.\ Fiz.\  {\bf 43} (1986) 459].
  %%CITATION = PHLTA,B166,329;%%

%48
\bibitem{siegel}
 W.~Siegel,
 Phys.\ Lett.\ B {\bf 84} (1979) 193.

%49
\bibitem{siegelb}
  W.~Siegel,
  Phys.\ Lett.\ B {\bf 94} (1980) 37.

%50
\bibitem{Avdeev:1981vf}
  L.V.~Avdeev, G.A.~Chochia and A.A.~Vladimirov,
  Phys.\ Lett.\ B {\bf 105} (1981) 272.

%51
\bibitem{Avdeev:1982xy}
  L.~V.~Avdeev and A.~A.~Vladimirov,
  Nucl.\ Phys.\  B {\bf 219} (1983) 262.

%52
\bibitem{Jack:1997pa}
  I.~Jack and D.~R.~T.~Jones,
  Phys.\ Lett.\ B {\bf 415} (1997) 383
  [hep-ph/9709364].
  %%CITATION = HEP-PH/9709364;%%

%53
\bibitem{Jack:1998iy}
  I.~Jack, D.~R.~T.~Jones, A.~Pickering,
  Phys.\ Lett.\  {\bf B432 } (1998)  114-119.
  [hep-ph/9803405].

%54
\bibitem{Jack:1996qq}
  I.~Jack, D.~R.~T.~Jones and C.~G.~North,
  Nucl.\ Phys.\  B {\bf 473} (1996) 308
  [arXiv:hep-ph/9603386].

%55
\bibitem{Jack:1996vg}
  I.~Jack, D.~R.~T.~Jones and C.~G.~North,
  Phys.\ Lett.\  B {\bf 386} (1996) 138
  [arXiv:hep-ph/9606323].

%56
\bibitem{Jack:1996cn}
  I.~Jack, D.~R.~T.~Jones and C.~G.~North,
  Nucl.\ Phys.\  B {\bf 486} (1997) 479
  [arXiv:hep-ph/9609325].
  %%CITATION = NUPHA,B486,479;%%

%57
\bibitem{Jack:1994bn}
  I.~Jack, D.~R.~T.~Jones and K.~L.~Roberts,
  Z.\ Phys.\  C {\bf 63} (1994) 151
  [arXiv:hep-ph/9401349].
  %%CITATION = ZEPYA,C63,151;%%

%58
\bibitem{Avdeev:1981ew}
  L.~V.~Avdeev and O.~V.~Tarasov,
  Phys.\ Lett.\  B {\bf 112}, 356 (1982).
  %%CITATION = PHLTA,B112,356;%%

%59
\bibitem{dreg1}
  C.G.~Bollini and J.J.~Giambiagi,
  Nuovo Cim.\ B {\bf 12} (1972) 20.

%60
\bibitem{'tHooft:1972fi}
  G.~'t Hooft and M.~J.~G.~Veltman,
  Nucl.\ Phys.\  B {\bf 44} (1972) 189.

%61
\bibitem{Piguet:1980fa}
  O.~Piguet, K.~Sibold and M.~Schweda,
  Nucl.\ Phys.\ B {\bf 174} (1980) 183.
  %%CITATION = NUPHA,B174,183;%%

%62
\bibitem{wt}
  J.~C.~Ward,
  Phys.\ Rev.\  {\bf 78} (1950) 182;\\
  Y.~Takahashi,
  Nuovo Cim.\  {\bf 6} (1957) 371.

%63
\bibitem{st}
  A.~A.~Slavnov,
  Theor.\ Math.\ Phys.\  {\bf 10} (1972) 99
  [Teor.\ Mat.\ Fiz.\  {\bf 10} (1972) 153].

%64
\bibitem{ds}  
  D.~St\"ockinger,
  JHEP {\bf 0503} (2005) 076
  [arXiv:hep-ph/0503129].

%65
\bibitem{coll}
 J.~Collins, {\it Renormalization}, Cambridge University Press,
 Cambridge 1984.

%66
\bibitem{Korner:1993pv}
  J.~G.~Korner and M.~M.~Tung,
  Z.\ Phys.\  C {\bf 64} (1994) 255.

%67
\bibitem{Misiak:1993es}
  M.~Misiak,
  Phys.\ Lett.\  B {\bf 321} (1994) 113
  [arXiv:hep-ph/9309236].

%68
\bibitem{Kunszt:1993sd}
  Z.~Kunszt, A.~Signer and Z.~Trocsanyi,
  Nucl.\ Phys.\  B {\bf 411} (1994) 397
  [arXiv:hep-ph/9305239].

%69
\bibitem{Smith:2004ck}
  J.~Smith and W.~L.~van Neerven,
  Eur.\ Phys.\ J.\  C {\bf 40} (2005) 199
  [arXiv:hep-ph/0411357].

%70
\bibitem{Altarelli:1980fi}
  G.~Altarelli, G.~Curci, G.~Martinelli and S.~Petrarca,
  Nucl.\ Phys.\  B {\bf 187} (1981) 461.

%71
\bibitem{Capper:1979ns}
  D.~M.~Capper, D.~R.~T.~Jones and P.~van Nieuwenhuizen,
  Nucl.\ Phys.\  B {\bf 167} (1980) 479.

%72
\bibitem{Nicolai:1980km}
  H.~Nicolai and P.~K.~Townsend,
  Phys.\ Lett.\  B {\bf 93} (1980) 111.

%73
\bibitem{Jones:1982zf}
  D.~R.~T.~Jones and J.~P.~Leveille,
  Nucl.\ Phys.\  B {\bf 206} (1982) 473;
  [Erratum-ibid.\  B {\bf 222} (1983) 517].

%74
\bibitem{Adler:1969er}
  S.~L.~Adler and W.~A.~Bardeen,
  Phys.\ Rev.\  {\bf 182} (1969) 1517.

%75
\bibitem{Trueman:1995ca}
  T.~L.~Trueman,
  Z.\ Phys.\  C {\bf 69} (1996) 525
  [arXiv:hep-ph/9504315].

%76
\bibitem{Larin:1993tq}
  S.~A.~Larin,
  Phys.\ Lett.\  B {\bf 303} (1993) 113
  [arXiv:hep-ph/9302240].

%77
\bibitem{Chetyrkin:1998mw}
  K.~G.~Chetyrkin, B.~A.~Kniehl, M.~Steinhauser and W.~A.~Bardeen,
  Nucl.\ Phys.\  B {\bf 535} (1998) 3
  [arXiv:hep-ph/9807241].

%78
\bibitem{Adler:1969gk}
  S.~L.~Adler,
  Phys.\ Rev.\  {\bf 177} (1969) 2426;\\
  J.~S.~Bell and R.~Jackiw,
  Nuovo Cim.\  A {\bf 60} (1969) 47;\\
  W.~A.~Bardeen,
  Phys.\ Rev.\  {\bf 184} (1969) 1848.

%79
\bibitem{Bardeen:1972vi}
  W.~A.~Bardeen, R.~Gastmans and B.~Lautrup,
  Nucl.\ Phys.\  B {\bf 46} (1972) 319.

%80
\bibitem{Chanowitz:1979zu}
  M.~S.~Chanowitz, M.~Furman and I.~Hinchliffe,
  Nucl.\ Phys.\  B {\bf 159} (1979) 225.

%81
\bibitem{Jegerlehner:2000dz}
  F.~Jegerlehner,
  Eur.\ Phys.\ J.\  C {\bf 18} (2001) 673
  [arXiv:hep-th/0005255].

%82
\bibitem{Fleischer:1993ub}
  J.~Fleischer, O.~V.~Tarasov and F.~Jegerlehner,
  Phys.\ Lett.\  B {\bf 319} (1993) 249;\\
  J.~Fleischer, O.~V.~Tarasov, F.~Jegerlehner and P.~Raczka,
  Phys.\ Lett.\  B {\bf 293} (1992) 437.

%83
\bibitem{Freitas:2000gg}
  A.~Freitas, W.~Hollik, W.~Walter and G.~Weiglein,f
  Phys.\ Lett.\  B {\bf 495} (2000) 338
  [Erratum-ibid.\  B {\bf 570} (2003) 260]
  [arXiv:hep-ph/0007091];\\
  A.~Freitas, W.~Hollik, W.~Walter and G.~Weiglein,
  Nucl.\ Phys.\  B {\bf 632} (2002) 189
  [Erratum-ibid.\  B {\bf 666} (2003) 305]
  [arXiv:hep-ph/0202131].

%84
\bibitem{Avdeev:1994db}
  L.~Avdeev, J.~Fleischer, S.~Mikhailov and O.~Tarasov,
  Phys.\ Lett.\  B {\bf 336} (1994) 560
  [Erratum-ibid.\  B {\bf 349} (1995) 597]
  [arXiv:hep-ph/9406363].
  %%CITATION = PHLTA,B336,560;%%

%85
\bibitem{Chetyrkin:1995ix}
  K.~G.~Chetyrkin, J.~H.~Kuhn and M.~Steinhauser,
  Phys.\ Lett.\ B {\bf 351} (1995) 331
  [hep-ph/9502291].

%86
\bibitem{Heinemeyer:2004yq}
  S.~Heinemeyer, D.~Stockinger and G.~Weiglein,
  Nucl.\ Phys.\  B {\bf 699} (2004) 103
  [arXiv:hep-ph/0405255].

%87
\bibitem{Pickering:2001aq}
  A.~G.~M.~Pickering, J.~A.~Gracey and D.~R.~T.~Jones,
  Phys.\ Lett.\  B {\bf 510} (2001) 347
  [Phys.\ Lett.\  B {\bf 512} (2001\ ERRAT,B535,377.2002) 230]
  [arXiv:hep-ph/0104247].

%88
\bibitem{Harlander:2009mn}
  R.~V.~Harlander, L.~Mihaila and M.~Steinhauser,
  Eur.\ Phys.\ J.\ C {\bf 63} (2009) 383
  [arXiv:0905.4807 [hep-ph]].

%89
\bibitem{'tHooft:1973mm}
  G.~'t Hooft,
  Nucl.\ Phys.\  B {\bf 61} (1973) 455.

%90
\bibitem{Bardeen:1978yd}
  W.~A.~Bardeen, A.~J.~Buras, D.~W.~Duke and T.~Muta,
  Phys.\ Rev.\  D {\bf 18} (1978) 3998.

%91
\bibitem{Collins:1974bg}
  J.~C.~Collins,
  Nucl.\ Phys.\  B {\bf 80} (1974) 341; Nucl.\ Phys.\  B {\bf 92} (1975) 477.
  %%CITATION = NUPHA,B80,341;%%

%92
\bibitem{Chetyrkin:1984xa}
  K.~G.~Chetyrkin and V.~A.~Smirnov,
  Phys.\ Lett.\  B {\bf 144} (1984) 419.
  %%CITATION = PHLTA,B144,419;%%

%93
\bibitem{Vladimirov:1979zm}
  A.~A.~Vladimirov,
  Theor.\ Math.\ Phys.\  {\bf 43} (1980) 417
  [Teor.\ Mat.\ Fiz.\  {\bf 43} (1980) 210].

%94
\bibitem{Chetyrkin:1980pr}
  K.~G.~Chetyrkin, A.~L.~Kataev and F.~V.~Tkachov,
  Nucl.\ Phys.\  B {\bf 174} (1980) 345.
  %%CITATION = NUPHA,B174,345;%%

%95
\bibitem{Tarasov:1980au}
  O.~V.~Tarasov, A.~A.~Vladimirov and A.~Y.~Zharkov,
  Phys.\ Lett.\  B {\bf 93} (1980) 429.

%96
\bibitem{Larin:1993tp}
  S.~A.~Larin and J.~A.~M.~Vermaseren,
  Phys.\ Lett.\  B {\bf 303} (1993) 334
  [arXiv:hep-ph/9302208].

%97
\bibitem{Harlander:2006rj}
  R.~Harlander, P.~Kant, L.~Mihaila and M.~Steinhauser,
  JHEP {\bf 0609} (2006) 053
  [arXiv:hep-ph/0607240].
  %%CITATION = JHEPA,0609,053;%%

%98
\bibitem{Chetyrkin:2012rz}
  K.~G.~Chetyrkin and M.~F.~Zoller,
  JHEP {\bf 1206} (2012) 033
  [arXiv:1205.2892 [hep-ph]].
  %%CITATION = ARXIV:1205.2892;%%

%99
\bibitem{Mihaila:2012fm}
  L.~N.~Mihaila, J.~Salomon and M.~Steinhauser,
  Phys.\ Rev.\ Lett.\  {\bf 108} (2012) 151602
  [arXiv:1201.5868 [hep-ph]].

%100
\bibitem{Larin:1991fz}
  S.~A.~Larin, F.~V.~Tkachov and J.~A.~M.~Vermaseren,
preprint NIKHEF-H-91-18 (1991).

%101
 \bibitem{Vermaseren:2000nd}
  J.~A.~M.~Vermaseren,
  arXiv:math-ph/0010025.

%102
\bibitem{Chetyrkin:1997fm}
  K.~G.~Chetyrkin, M.~Misiak and M.~Munz,
  Nucl.\ Phys.\  B {\bf 518} (1998) 473
  [arXiv:hep-ph/9711266].

%103
\bibitem{Chetyrkin:1997dh}
  K.~G.~Chetyrkin,
  Phys.\ Lett.\ B {\bf 404} (1997) 161
  [hep-ph/9703278].
  %%CITATION = HEP-PH 9703278;%%

%104
\bibitem{vanRitbergen:1997va}
  T.~van Ritbergen, J.~A.~M.~Vermaseren and S.~A.~Larin,
  Phys.\ Lett.\ B {\bf 400} (1997) 379
  [hep-ph/9701390].
  %%CITATION = HEP-PH 9701390;%%

%105
\bibitem{Vermaseren:1997fq}
  J.~A.~M.~Vermaseren, S.~A.~Larin and T.~van Ritbergen,
  Phys.\ Lett.\ B {\bf 405} (1997) 327
  [hep-ph/9703284].
  %%CITATION = HEP-PH 9703284;%%

%106
\bibitem{Czakon:2004bu}
  M.~Czakon,
  Nucl.\ Phys.\ B {\bf 710} (2005) 485
  [hep-ph/0411261].
  %%CITATION = HEP-PH 0411261;%%

%107
\bibitem{Curtright:1979mg}
  T.~Curtright,
  Phys.\ Rev.\ D {\bf 21} (1980) 1543.
  %%CITATION = PHRVA,D21,1543;%%

%108
\bibitem{Jones:1980fx}
  D.~R.~T.~Jones,
  Phys.\ Rev.\ D {\bf 22} (1980) 3140.
  %%CITATION = PHRVA,D22,3140;%%

%109
\bibitem{Steinhauser:2000ry}
  M.~Steinhauser,
  Comput.\ Phys.\ Commun.\  {\bf 134} (2001) 335
  [arXiv:hep-ph/0009029].
  %%CITATION = HEP-PH 0009029;%%

%110
\bibitem{Chetyrkin:1996ez}
  K.~G.~Chetyrkin,
  Phys.\ Lett.\  B {\bf 391} (1997) 402
  [arXiv:hep-ph/9608480].

%111
\bibitem{Chetyrkin:1996sr}
  K.~G.~Chetyrkin,
  Phys.\ Lett.\  B {\bf 390} (1997) 309
  [arXiv:hep-ph/9608318].
  %%CITATION = PHLTA,B390,309;%%

%112
\bibitem{Ferreira:1996ug}
  P.~M.~Ferreira, I.~Jack and D.~R.~T.~Jones,
  Phys.\ Lett.\  B {\bf 387}, 80 (1996)
  [arXiv:hep-ph/9605440].
  %%CITATION = PHLTA,B387,80;%%

%113
\bibitem{Bern:2002zk}
  Z.~Bern, A.~De Freitas, L.~J.~Dixon and H.~L.~Wong,
  Phys.\ Rev.\  D {\bf 66} (2002) 085002
  [arXiv:hep-ph/0202271].

%114
\bibitem{Pak:2010cu}
  A.~Pak, M.~Steinhauser and N.~Zerf,
  Eur.\ Phys.\ J.\ C {\bf 71} (2011) 1602
  [arXiv:1012.0639 [hep-ph]].
  %%CITATION = ARXIV:1012.0639;%%

%115
\bibitem{Pak:2012xr}
  A.~Pak, M.~Steinhauser and N.~Zerf,
  JHEP {\bf 1209} (2012) 118
  [arXiv:1208.1588 [hep-ph]].
  %%CITATION = ARXIV:1208.1588;%%

%116
\bibitem{Kant:2009zza}
  P.~Kant, PhD thesis, University of Karlsruhe (2008). 

%117
\bibitem{Tim} D.R.T.~Jones (unpublished) 1979\semi  see also
W.~Siegel, P.K.~Townsend and P.~van Nieuwenhuizen,
Proc. 1980 Cambridge meeting on supergravity, ITP-SB-80-65.

%118
\bibitem{hvand} R.~van Damme and G.~'t~Hooft,
Phys.\ Lett.\ B {\bf 150} (1985) 133.

%119
\bibitem{Jack:1993ws}
  I.~Jack, D.~R.~T.~Jones and K.~L.~Roberts,
  Z.\ Phys.\  C {\bf 62} (1994) 161
  [arXiv:hep-ph/9310301].

%120
\bibitem{Steinhauser:2002rq}
  M.~Steinhauser,
  Phys.\ Rept.\  {\bf 364} (2002) 247
  [hep-ph/0201075].

%121
\bibitem{Jack:2007ni}
  I.~Jack, D.~R.~T.~Jones, P.~Kant and L.~Mihaila,
  JHEP {\bf 0709} (2007) 058
  [arXiv:0707.3055 [hep-th]].
  %%CITATION = JHEPA,0709,058;%%

%122
\bibitem{predrag}\
  P.~Cvitanovic,\
  Phys.\ Rev.\ D {\bf 14} (1976) 1536.\
  %%CITATION = PHRVA,D14,1536;%%\

%123
\bibitem{Dittner:1972hm}
  P.~Dittner,
  Commun.\ Math.\ Phys.\  {\bf 27} (1972) 44.
  %%CITATION = CMPHA,27,44;%%

%124
\bibitem{msw}\
  A.J.~MacFarlane, A.~Sudbery and P.H.~Weisz,\
  Commun.\ Math.\ Phys.\  {\bf 11} (1968) 77.\
  %%CITATION = CMPHA,11,77;%%\

%125
\bibitem{RSV}
T.~van Ritbergen,  A.N.~Schellekens and J.A.M.~Vermaseren,
  Int.\ J.\ Mod.\ Phys.\  A {\bf 14} (1999) 41
  [arXiv:hep-ph/9802376].
  %%CITATION = IMPAE,A14,41;%%

%126
\bibitem{Cheng:1973nv}
  T.~P.~Cheng, E.~Eichten and L.~F.~Li,
  Phys.\ Rev.\  D {\bf 9} (1974) 2259.

%127
\bibitem{Nogueira:1991ex}
  P.~Nogueira,
  J.\ Comput.\ Phys.\  {\bf 105} (1993) 279.
  %%CITATION = JCTPA,105,279;%%

%128
\bibitem{Seidensticker:1999bb}
  T.~Seidensticker,
  [hep-ph/9905298].
  %%CITATION = HEP-PH 9905298;%%

%129
\bibitem{Harlander:1997zb}
  R.~Harlander, T.~Seidensticker and M.~Steinhauser,
  Phys.\ Lett.\ B {\bf 426} (1998) 125
  [hep-ph/9712228].
  %%CITATION = HEP-PH 9712228;%%

%130
\bibitem{Harlander:2006xq}
  R.~V.~Harlander, D.~R.~T.~Jones, P.~Kant, L.~Mihaila and M.~Steinhauser,
  JHEP {\bf 0612} (2006) 024
  [arXiv:hep-ph/0610206].

%131
\bibitem{Baikov:2008jh}
  P.~A.~Baikov, K.~G.~Chetyrkin and J.~H.~Kuhn,
  Phys.\ Rev.\ Lett.\  {\bf 101} (2008) 012002
  [arXiv:0801.1821 [hep-ph]].
  %%CITATION = ARXIV:0801.1821;%%

%132
\bibitem{Baikov:2010iw}
  P.~A.~Baikov, K.~G.~Chetyrkin and J.~H.~Kuhn,
  Nucl.\ Phys.\ Proc.\ Suppl.\  {\bf 205-206} (2010) 237
  [arXiv:1007.0478 [hep-ph]].
  %%CITATION = ARXIV:1007.0478;%%

%133
\bibitem{Jack:1998uj}
  I.~Jack, D.~R.~T.~Jones, A.~Pickering,
  Phys.\ Lett.\  {\bf B435 } (1998)  61-66.
  [hep-ph/9805482].

%134
\bibitem{Jones:1983ip}
  D.R.T.~Jones, Phys.\  Lett.\  B {\bf 123} (1983) 45.
   %%CITATION = PHLTA,B123,45;%% 

%135
\bibitem {Novikov:1983ee} 
  V.A.~Novikov, M.A.~Shifman, A.I.~Vainshtein and V.I.~Zakharov, 
  Nucl.\  Phys.\  B {\bf 229} (1983) 407. 
  %%CITATION = NUPHA,B229,407;%%}

%136
\bibitem{Smirnov:2002pj}
  V.~A.~Smirnov,
  ``Applied asymptotic expansions in momenta and masses,''
  Springer Tracts Mod.\ Phys.\  {\bf 177} (2002) 1.
  %%CITATION = STPHB,177,1;%%

%137
\bibitem{Marquard:2007uj}
  P.~Marquard, L.~Mihaila, J.~H.~Piclum and M.~Steinhauser,
  Nucl.\ Phys.\  B {\bf 773} (2007) 1
  [arXiv:hep-ph/0702185].
  %%CITATION = NUPHA,B773,1;%%

%138
\bibitem{Martin:2001vx}
  S.~P.~Martin,
  Phys.\ Rev.\  D {\bf 65} (2002) 116003
  [arXiv:hep-ph/0111209].
  %%CITATION = PHRVA,D65,116003;%%

%139
\bibitem{tevatron}
{\tt http://www.fnal.gov/pub/science/experiments/energy/tevatron/}

%140
\bibitem{Heinemeyer:2004ms}
  S.~Heinemeyer,
  Int.\ J.\ Mod.\ Phys.\  A {\bf 21} (2006) 2659
  [arXiv:hep-ph/0407244].
  %%CITATION = IMPAE,A21,2659;%%

%141
\bibitem{Degrassi:2009yq}
  G.~Degrassi and P.~Slavich,
  Nucl.\ Phys.\  B {\bf 825} (2010) 119
  [arXiv:0907.4682 [hep-ph]].
  %%CITATION = NUPHA,B825,119;%%

%142
\bibitem{Kant:2010tf}
  P.~Kant, R.~V.~Harlander, L.~Mihaila and M.~Steinhauser,
  JHEP {\bf 1008} (2010) 104
  [arXiv:1005.5709 [hep-ph]].
  %%CITATION = JHEPA,1008,104;%%

%143
\bibitem{Novikov:1983uc}
  V.~A.~Novikov, M.~A.~Shifman, A.~I.~Vainshtein and V.~I.~Zakharov,
  Nucl.\ Phys.\ B {\bf 229} (1983) 381.
  %%CITATION = NUPHA,B229,381;%%

%144
\bibitem{Jack:2003sx}
  I.~Jack, D.~R.~T.~Jones and A.~F.~Kord,
  Phys.\ Lett.\  B {\bf 579} (2004) 180
  [arXiv:hep-ph/0308231].\\
  %%CITATION = PHLTA,B579,180;%%
  I.~Jack, D.~R.~T.~Jones and A.~F.~Kord,
  Annals Phys.\  {\bf 316} (2005) 213
  [arXiv:hep-ph/0408128].

%145
\bibitem{Mihaila:2012pz}
  L.~N.~Mihaila, J.~Salomon and M.~Steinhauser,
  Phys.\ Rev.\ D {\bf 86} (2012) 096008
  [arXiv:1208.3357 [hep-ph]].
  %%CITATION = ARXIV:1208.3357;%%

%146
\bibitem{Hermann:2011ha}
  T.~Hermann, L.~Mihaila and M.~Steinhauser,
  Phys.\ Lett.\ B {\bf 703} (2011) 51
  [arXiv:1106.1060 [hep-ph]].
  %%CITATION = ARXIV:1106.1060;%%

%147
\bibitem{Hollik:2002mv}
  W.~Hollik, E.~Kraus, M.~Roth, C.~Rupp, K.~Sibold and D.~Stockinger,
  Nucl.\ Phys.\ B {\bf 639} (2002) 3
  [hep-ph/0204350].\\
  %%CITATION = HEP-PH/0204350;%%
  W.~Hollik, E.~Kraus and D.~Stockinger,
  Eur.\ Phys.\ J.\ C {\bf 23} (2002) 735
  [hep-ph/0007134].
  %%CITATION = HEP-PH/0007134;%%

%148
\bibitem{Jack:1994rk}
  I.~Jack, D.~R.~T.~Jones, S.~P.~Martin, M.~T.~Vaughn and Y.~Yamada,
  Phys.\ Rev.\  D {\bf 50} (1994) 5481
  [arXiv:hep-ph/9407291].
  %%CITATION = PHRVA,D50,5481;%%

%149
\bibitem{Politzer:1973fx}
  H.~D.~Politzer,
  Phys.\ Rev.\ Lett.\  {\bf 30} (1973) 1346.
  %%CITATION = PRLTA,30,1346;%%

%150
\bibitem{Jones:1974mm} 
  D.~R.~T.~Jones,
  Nucl.\ Phys.\ B {\bf 75}, 531 (1974).
  %%CITATION = NUPHA,B75,531;%%

%151
\bibitem{Tarasov:1976ef}
  O.~V.~Tarasov and A.~A.~Vladimirov,
  Sov.\ J.\ Nucl.\ Phys.\  {\bf 25} (1977) 585
  [Yad.\ Fiz.\  {\bf 25} (1977) 1104].
%%CITATION = SJNCA,25,585;%%

%152
\bibitem{Caswell:1974gg}
  W.~E.~Caswell,
  Phys.\ Rev.\ Lett.\  {\bf 33} (1974) 244.
  %%CITATION = PRLTA,33,244;%%

%153
\bibitem{Egorian:1978zx}
  E.~Egorian and O.~V.~Tarasov,
  Teor.\ Mat.\ Fiz.\  {\bf 41} (1979) 26
   [Theor.\ Math.\ Phys.\  {\bf 41} (1979) 863].
  %%CITATION = TMFZA,41,26;%%

%154
\bibitem{Jones:1981we}
  D.~R.~T.~Jones,
  Phys.\ Rev.\  D {\bf 25} (1982) 581.
  %%CITATION = PHRVA,D25,581;%%

%155
\bibitem{Fischler:1981is}
  M.~S.~Fischler and C.~T.~Hill,
  Nucl.\ Phys.\  B {\bf 193} (1981) 53.
  %%CITATION = NUPHA,B193,53;%%

%156
\bibitem{Machacek:1983tz}
  M.~E.~Machacek and M.~T.~Vaughn,
  Nucl.\ Phys.\  B {\bf 222} (1983) 83; Nucl.\ Phys.\  B {\bf 236}
  (1984) 221; Nucl.\ Phys.\  B {\bf 249} (1985) 70.

%157
\bibitem{Jack:1984vj}
  I.~Jack and H.~Osborn,
  Nucl.\ Phys.\  B {\bf 249} (1985) 472.
  %%CITATION = NUPHA,B249,472;%%

%158
\bibitem{Steinhauser:1998cm}
  M.~Steinhauser,
  Phys.\ Rev.\ D {\bf 59} (1999) 054005
  [hep-ph/9809507].

%159
\bibitem{Bednyakov:2012rb}
  A.~V.~Bednyakov, A.~F.~Pikelner and V.~N.~Velizhanin,
  JHEP {\bf 1301} (2013) 017
  [arXiv:1210.6873 [hep-ph]].
  %%CITATION = ARXIV:1210.6873;%%

%160
\bibitem{Nakamura:2010zzi}
  K.~Nakamura {\it et al.}  [Particle Data Group Collaboration],
  J.\ Phys.\ G G {\bf 37} (2010) 075021.
  %%CITATION = JPHGB,G37,075021;%%

%161
\bibitem{Abbott:1980hw}
  L.~F.~Abbott,
  Nucl.\ Phys.\ B {\bf 185} (1981) 189.
  %%CITATION = NUPHA,B185,189;%%

%162
\bibitem{Denner:1994xt}
  A.~Denner, G.~Weiglein and S.~Dittmaier,
  Nucl.\ Phys.\ B {\bf 440} (1995) 95
  [hep-ph/9410338].
  %%CITATION = HEP-PH/9410338;%%

%163
\bibitem{Peskin:1995ev}
  M.~E.~Peskin and D.~V.~Schroeder, {\it An Introduction to quantum field theory},
  Reading, USA: Addison-Wesley (1995) 842 p.

%164
\bibitem{Bethke:2009jm}
  S.~Bethke,
  Eur.\ Phys.\ J.\ C {\bf 64} (2009) 689
  [arXiv:0908.1135 [hep-ph]].
  %%CITATION = ARXIV:0908.1135;%%

%165
\bibitem{Harlander:2005wm}
  R.~Harlander, L.~Mihaila and M.~Steinhauser,
  Phys.\ Rev.\ D {\bf 72} (2005) 095009
  [hep-ph/0509048].
  %%CITATION = HEP-PH/0509048;%%

%166
\bibitem{Bauer:2008bj}
  A.~Bauer, L.~Mihaila and J.~Salomon,
  JHEP {\bf 0902} (2009) 037
  [arXiv:0810.5101 [hep-ph]].
  %%CITATION = ARXIV:0810.5101;%%

%167
\bibitem{Appelquist:1974tg}
  T.~Appelquist and J.~Carazzone,
  Phys.\ Rev.\  D {\bf 11} (1975) 2856.

%168
\bibitem{Chetyrkin:1997un}
  K.~G.~Chetyrkin, B.~A.~Kniehl and M.~Steinhauser,
  Nucl.\ Phys.\ B {\bf 510} (1998) 61
  [hep-ph/9708255].

%169
\bibitem{Schroder:2005hy}
  Y.~Schroder and M.~Steinhauser,
  JHEP {\bf 0601} (2006) 051 [hep-ph/0512058];\\
  K.~G.~Chetyrkin, J.~H.~Kuhn and C.~Sturm,
  Nucl.\ Phys.\  B {\bf 744} (2006) 121 [hep-ph/0512060].
  %%CITATION = NUPHA,B744,121;%%

%170
\bibitem{Bednyakov:2007vm}
  A.~V.~Bednyakov,
  Int.\ J.\ Mod.\ Phys.\  A {\bf 22} (2007) 5245.

%171
\bibitem{Bednyakov:2009wt}
  A.~V.~Bednyakov,
  Int.\ J.\ Mod.\ Phys.\  A {\bf 25} (2010) 2437.

%172
\bibitem{Noth:2008tw}
  D.~Noth and M.~Spira,
  Phys.\ Rev.\ Lett.\  {\bf 101} (2008) 181801 [arXiv:0808.0087 [hep-ph]];\\
  D.~Noth and M.~Spira,
  JHEP {\bf 1106} (2011) 084
  [arXiv:1001.1935 [hep-ph]].

%173
\bibitem{Kurz:2012ff}
  A.~Kurz, M.~Steinhauser and N.~Zerf,
  JHEP {\bf 1207} (2012) 138
  [arXiv:1206.6675 [hep-ph]].
  %%CITATION = ARXIV:1206.6675;%%

%174
\bibitem{Jack:1994kd}
  I.~Jack and D.~R.~T.~Jones,
  Phys.\ Lett.\  B {\bf 333} (1994) 372
  [arXiv:hep-ph/9405233].

%175
\bibitem{Bednyakov:2002sf}
  A.~Bednyakov, A.~Onishchenko, V.~Velizhanin and O.~Veretin,
  Eur.\ Phys.\ J.\ C {\bf 29} (2003) 87
  [hep-ph/0210258].
  %%CITATION = HEP-PH/0210258;%%

%176
\bibitem{Pierce:1996zz}
  D.~M.~Pierce, J.~A.~Bagger, K.~T.~Matchev and R.~j.~Zhang,
  Nucl.\ Phys.\  B {\bf 491}, 3 (1997)
  [arXiv:hep-ph/9606211].
  %%CITATION = NUPHA,B491,3;%%

%177
\bibitem{Harlander:2004tp}
  R.~V.~Harlander and M.~Steinhauser,
  JHEP {\bf 0409} (2004) 066
  [hep-ph/0409010].
  %%CITATION = HEP-PH/0409010;%%

%178
\bibitem{Martin:2005qm}
  S.~P.~Martin and D.~G.~Robertson,
  Comput.\ Phys.\ Commun.\  {\bf 174} (2006) 133
  [hep-ph/0501132].

%179
\bibitem{Shifman:1978zn}
  M.~A.~Shifman, A.~I.~Vainshtein and V.~I.~Zakharov,
  Phys.\ Lett.\  B {\bf 78} (1978) 443.

%180
\bibitem{Carena:1999py}
  M.~Carena, D.~Garcia, U.~Nierste and C.~E.~M.~Wagner,
  Nucl.\ Phys.\ B {\bf 577} (2000) 88 
  [hep-ph/9912516].

%181
\bibitem{Harlander:2007wh}
  R.~V.~Harlander, L.~Mihaila and M.~Steinhauser,
  Phys.\ Rev.\  D {\bf 76} (2007) 055002
  [arXiv:0706.2953 [hep-ph]].

%182
\bibitem{Mihaila:2009bn}
  L.~Mihaila,
  Phys.\ Lett.\ B {\bf 681} (2009) 52
  [arXiv:0908.3403 [hep-ph]].

%183
\bibitem{Stockinger:2011gp}
  D.~Stockinger and P.~Varso,
  Comput.\ Phys.\ Commun.\  {\bf 183} (2012) 422
  [arXiv:1109.6484 [hep-ph]].

%184
\bibitem{Bethke:2006ac}
  S.~Bethke,
  Prog.\ Part.\ Nucl.\ Phys.\  {\bf 58} (2007) 351
  [hep-ex/0606035].

%185
\bibitem{AguilarSaavedra:2005pw}
  J.~A.~Aguilar-Saavedra, A.~Ali, B.~C.~Allanach, R.~L.~Arnowitt, H.~A.~Baer, J.~A.~Bagger, C.~Balazs and V.~D.~Barger {\it et al.},
  Eur.\ Phys.\ J.\ C {\bf 46} (2006) 43
  [hep-ph/0511344].

%186
\bibitem{Ghodbane:2002kg}
  N.~Ghodbane and H.~U.~Martyn,
  in {\it Proc. of the APS/DPF/DPB Summer Study on the Future of
    Particle Physics (Snowmass 2001) } ed. N.~Graf, 
  [arXiv:hep-ph/0201233].

%187
\bibitem{Dimopoulos:1981zb}
  S.~Dimopoulos and H.~Georgi,
  Nucl.\ Phys.\  B {\bf 193} (1981) 150.
  %%CITATION = NUPHA,B193,150;%%

%188
\bibitem{Sakai:1981gr}
  N.~Sakai,
  Z.\ Phys.\  C {\bf 11} (1981) 153.
  %%CITATION = ZEPYA,C11,153;%%

%189
\bibitem{Goto:1998qg}
  T.~Goto and T.~Nihei,
  Phys.\ Rev.\  D {\bf 59} (1999) 115009
  [arXiv:hep-ph/9808255].
  %%CITATION = PHRVA,D59,115009;%%

%190
\bibitem{Murayama:2001ur}
  H.~Murayama and A.~Pierce,
  Phys.\ Rev.\  D {\bf 65}, 055009 (2002)
  [arXiv:hep-ph/0108104].
  %%CITATION = PHRVA,D65,055009;%%

%191
\bibitem{Bajc:2002bv}
  B.~Bajc, P.~Fileviez Perez and G.~Senjanovic,
  Phys.\ Rev.\  D {\bf 66} (2002) 075005
  [arXiv:hep-ph/0204311].
  %%CITATION = PHRVA,D66,075005;%%

%192
\bibitem{EmmanuelCosta:2003pu}
  D.~Emmanuel-Costa and S.~Wiesenfeldt,
  Nucl.\ Phys.\  B {\bf 661} (2003) 62
  [arXiv:hep-ph/0302272].
  %%CITATION = NUPHA,B661,62;%%

%193
\bibitem{Wiesenfeldt:2004qz}
  S.~Wiesenfeldt, PhD thesis, University of Hamburg (2004).
  %%CITATION = DESY-THESIS-2004-009;%%

%194
\bibitem{Bajc:2002pg}
  B.~Bajc, P.~Fileviez Perez and G.~Senjanovic,
  arXiv:hep-ph/0210374.
  %%CITATION = HEP-PH/0210374;%%

%195
\bibitem{Yamada:1992kv}
  Y.~Yamada,
  Z.\ Phys.\  C {\bf 60} (1993) 83.
  %%CITATION = ZEPYA,C60,83;%%

%196
\bibitem{Hall:1980kf}
  L.~J.~Hall,
  Nucl.\ Phys.\  B {\bf 178} (1981) 75.
  %%CITATION = NUPHA,B178,75;%%

%197
\bibitem{Weinberg:1980wa}
  S.~Weinberg,
  Phys.\ Lett.\  B {\bf 91} (1980) 51.
  %%CITATION = PHLTA,B91,51;%%

%198
\bibitem{Einhorn:1981sx}
  M.~B.~Einhorn and D.~R.~T.~Jones,
  Nucl.\ Phys.\  B {\bf 196} (1982) 475.
  %%CITATION = NUPHA,B196,475;%%

%199
\bibitem{Hagiwara:1992ys}
  K.~Hagiwara and Y.~Yamada,
  Phys.\ Rev.\ Lett.\  {\bf 70} (1993) 709.
  %%CITATION = PRLTA,70,709;%%

%200
\bibitem{Dedes:1996wc}
  A.~Dedes, A.~B.~Lahanas, J.~Rizos and K.~Tamvakis,
  Phys.\ Rev.\  D {\bf 55} (1997) 2955
  [arXiv:hep-ph/9610271].
  %%CITATION = PHRVA,D55,2955;%%

%201
\bibitem{Hisano:1992jj}
  J.~Hisano, H.~Murayama and T.~Yanagida,
  Nucl.\ Phys.\  B {\bf 402} (1993) 46
  [arXiv:hep-ph/9207279].
  %%CITATION = NUPHA,B402,46;%%

%202
\bibitem{Amsler:2008zzb}
  C.~Amsler {\it et al.}  [Particle Data Group],
  Phys.\ Lett.\  B {\bf 667} (2008) 1.
  %%CITATION = PHLTA,B667,1;%%

%203
\bibitem{Teubner:2010ah}
  T.~Teubner, K.~Hagiwara, R.~Liao, A.~D.~Martin and D.~Nomura,
  Chin.\ Phys.\ C {\bf 34} (2010) 728
  [arXiv:1001.5401 [hep-ph]].
  %%CITATION = ARXIV:1001.5401;%%

%204
\bibitem{Steinhauser:1998rq}
  M.~Steinhauser,
  Phys.\ Lett.\  B {\bf 429} (1998) 158
  [arXiv:hep-ph/9803313].
  %%CITATION = PHLTA,B429,158;%%

%205
\bibitem{Kuhn:1998ze}
  J.~H.~Kuhn and M.~Steinhauser,
  Phys.\ Lett.\  B {\bf 437} (1998) 425
  [arXiv:hep-ph/9802241].
  %%CITATION = PHLTA,B437,425;%%

%206
\bibitem{Fanchiotti:1992tu}
  S.~Fanchiotti, B.~A.~Kniehl and A.~Sirlin,
  Phys.\ Rev.\  D {\bf 48} (1993) 307
  [arXiv:hep-ph/9212285].
  %%CITATION = PHRVA,D48,307;%%

%207
\bibitem{Chetyrkin:2000yt}
  K.~G.~Chetyrkin, J.~H.~K\"uhn and M.~Steinhauser,
  Comput.\ Phys.\ Commun.\  {\bf 133} (2000) 43
  [arXiv:hep-ph/0004189].
  %%CITATION = CPHCB,133,43;%%

%208
\bibitem{Dedes:1998hg}
  A.~Dedes, A.~B.~Lahanas and K.~Tamvakis,
  Phys.\ Rev.\  D {\bf 59} (1999) 015019
  [arXiv:hep-ph/9801425].
  %%CITATION = PHRVA,D59,015019;%%

%209
\bibitem{Chamseddine:1982jx}
  A.~H.~Chamseddine, R.~L.~Arnowitt and P.~Nath,
  Phys.\ Rev.\ Lett.\  {\bf 49} (1982) 970.
  %%CITATION = PRLTA,49,970;%%

%210
\bibitem{Allanach:2001kg}
  B.~C.~Allanach,
  Comput.\ Phys.\ Commun.\  {\bf 143} (2002) 305
  [arXiv:hep-ph/0104145].
  %%CITATION = CPHCB,143,305;%%

%211
\bibitem{Martens:2010nm}
  W.~Martens, L.~Mihaila, J.~Salomon and M.~Steinhauser,
  Phys.\ Rev.\ D {\bf 82} (2010) 095013
  [arXiv:1008.3070 [hep-ph]].

%212
\bibitem{Kobayashi:2005pe}
  K.~Kobayashi {\it et al.}  [Super-Kamiokande Collaboration],
  Phys.\ Rev.\  D {\bf 72} (2005) 052007
  [arXiv:hep-ex/0502026].
  %%CITATION = PHRVA,D72,052007;%%

%213
\bibitem{Nambu:1960xd}
  Y.~Nambu,
  Phys.\ Rev.\ Lett.\  {\bf 4} (1960) 380.
  %%CITATION = PRLTA,4,380;%%

%214
\bibitem{Englert:1964et}
  F.~Englert and R.~Brout,
  Phys.\ Rev.\ Lett.\  {\bf 13} (1964) 321.
  %%CITATION = PRLTA,13,321;%%

%215
\bibitem{Higgs:1964ia}
  P.~W.~Higgs,
  Phys.\ Lett.\  {\bf 12} (1964) 132.
  %%CITATION = PHLTA,12,132;%%

%216
\bibitem{Higgs:1964pj}
  P.~W.~Higgs,
  Phys.\ Rev.\ Lett.\  {\bf 13} (1964) 508.
  %%CITATION = PRLTA,13,508;%%

%217
\bibitem{'tHooft:1971rn}
  G.~'t Hooft,
  Nucl.\ Phys.\  B {\bf 35} (1971) 167.
  %%CITATION = NUPHA,B35,167;%%

%218
\bibitem{Barate:2003sz}
  R.~Barate {\it et al.}  [LEP Working Group for Higgs boson searches and ALEPH and DELPHI and L3 and OPAL Collaborations],
  Phys.\ Lett.\ B {\bf 565} (2003) 61
  [hep-ex/0306033].
  %%CITATION = HEP-EX/0306033;%%

%219
\bibitem{Baak:2012kk}
  M.~Baak, M.~Goebel, J.~Haller, A.~Hoecker, D.~Kennedy, R.~Kogler, K.~Moenig and M.~Schott {\it et al.},
  Eur.\ Phys.\ J.\ C {\bf 72} (2012) 2205
  [arXiv:1209.2716 [hep-ph]].
  %%CITATION = ARXIV:1209.2716;%%

%220
\bibitem{:2012nc}
    [TEVNPH (Tevatron New Phenomena and Higgs Working Group) and CDF and D0 Coll],
  [arXiv:1203.3774 [hep-ex]].
  %%CITATION = ARXIV:1203.3774;%%

%221
\bibitem{Ellis:2009tp}
  J.~Ellis, J.~R.~Espinosa, G.~F.~Giudice, A.~Hoecker and A.~Riotto,
  Phys.\ Lett.\ B {\bf 679} (2009) 369
  [arXiv:0906.0954 [hep-ph]].
  %%CITATION = ARXIV:0906.0954;%%

%222
\bibitem{EliasMiro:2011aa}
  J.~Elias-Miro, J.~R.~Espinosa, G.~F.~Giudice, G.~Isidori, A.~Riotto and A.~Strumia,
  Phys.\ Lett.\  B {\bf 709} (2012) 222
  [arXiv:1112.3022 [hep-ph]].
  %%CITATION = PHLTA,B709,222;%%

%223
\bibitem{Bezrukov:2012sa}
  F.~Bezrukov, M.~Y.~Kalmykov, B.~A.~Kniehl and M.~Shaposhnikov,
  JHEP {\bf 1210} (2012) 140
  [arXiv:1205.2893 [hep-ph]].

%224
\bibitem{Degrassi:2012ry}
  G.~Degrassi, S.~Di Vita, J.~Elias-Miro, J.~R.~Espinosa, G.~F.~Giudice, G.~Isidori and A.~Strumia,
  JHEP {\bf 1208} (2012) 098
  [arXiv:1205.6497 [hep-ph]].
  %%CITATION = ARXIV:1205.6497;%%

%225
\bibitem{Ellis:1990nz}
  J.~R.~Ellis, G.~Ridolfi and F.~Zwirner,
  Phys.\ Lett.\  B {\bf 257} (1991) 83;
  Phys.\ Lett.\ B {\bf 262} (1991) 477.
  %%CITATION = PHLTA,B257,83;%%

%226
\bibitem{Okada:1990vk}
  Y.~Okada, M.~Yamaguchi and T.~Yanagida,
  Prog.\ Theor.\ Phys.\  {\bf 85} (1991) 1.
  %%CITATION = PTPKA,85,1;%%

%227
\bibitem{Haber:1990aw}
  H.~E.~Haber and R.~Hempfling,
  Phys.\ Rev.\ Lett.\  {\bf 66} (1991) 1815.
  %%CITATION = PRLTA,66,1815;%%

%228
\bibitem{Allanach:2002nj}
  B.~C.~Allanach {\it et al.},
in {\it Proc. of the APS/DPF/DPB Summer Study on the Future of Particle Physics (Snowmass 2001) } ed. N.~Graf,
  Eur.\ Phys.\ J.\  C {\bf 25} (2002) 113
  [arXiv:hep-ph/0202233].
  %%CITATION = EPHJA,C25,113;%%

%229
\bibitem{Chankowski:1991md}
  P.~H.~Chankowski, S.~Pokorski and J.~Rosiek,
  Phys.\ Lett.\  B {\bf 274} (1992) 191.
  %%CITATION = PHLTA,B274,191;%%

%230
\bibitem{Brignole:1992uf}
  A.~Brignole,
  Phys.\ Lett.\  B {\bf 281}, 284 (1992).
  %%CITATION = PHLTA,B281,284;%%

%231
\bibitem{Dabelstein:1994hb}
  A.~Dabelstein,
  Z.\ Phys.\  C {\bf 67}, 495 (1995)
  [arXiv:hep-ph/9409375].
  %%CITATION = ZEPYA,C67,495;%%

%232
\bibitem{Allanach:2004rh}
  B.~C.~Allanach, A.~Djouadi, J.~L.~Kneur, W.~Porod and P.~Slavich,
  JHEP {\bf 0409} (2004) 044
  [arXiv:hep-ph/0406166].
  %%CITATION = JHEPA,0409,044;%%

%233
\bibitem{Frank:2006yh}
  M.~Frank, T.~Hahn, S.~Heinemeyer, W.~Hollik, H.~Rzehak and G.~Weiglein,
  JHEP {\bf 0702} (2007) 047
  [arXiv:hep-ph/0611326].
  %%CITATION = JHEPA,0702,047;%%

%234
\bibitem{Heinemeyer:2007aq}
  S.~Heinemeyer, W.~Hollik, H.~Rzehak and G.~Weiglein,
  Phys.\ Lett.\  B {\bf 652} (2007) 300
  [arXiv:0705.0746 [hep-ph]].
  %%CITATION = PHLTA,B652,300;%%

%235
\bibitem{Carena:2000yi}
  M.~S.~Carena, J.~R.~Ellis, A.~Pilaftsis and C.~E.~M.~Wagner,
  Nucl.\ Phys.\ B {\bf 586} (2000) 92
  [hep-ph/0003180].
  %%CITATION = HEP-PH/0003180;%%

%236
\bibitem{Martin:2002wn}
  S.~P.~Martin,
  Phys.\ Rev.\  D {\bf 67} (2003) 095012
  [arXiv:hep-ph/0211366].
  %%CITATION = PHRVA,D67,095012;%%

%237
\bibitem{Martin:2007pg}
  S.~P.~Martin,
  Phys.\ Rev.\  D {\bf 75} (2007) 055005
  [arXiv:hep-ph/0701051].
  %%CITATION = PHRVA,D75,055005;%%

%238
\bibitem{Harlander:2008ju}
  R.~V.~Harlander, P.~Kant, L.~Mihaila and M.~Steinhauser,
  Phys.\ Rev.\ Lett.\  {\bf 100} (2008) 191602
  [Phys.\ Rev.\ Lett.\  {\bf 101} (2008) 039901]
  [arXiv:0803.0672 [hep-ph]].
  %%CITATION = PRLTA,101,039901;%%

%239
\bibitem{Heinemeyer:1998yj}
  S.~Heinemeyer, W.~Hollik and G.~Weiglein,
  Comput.\ Phys.\ Commun.\  {\bf 124} (2000) 76
  [arXiv:hep-ph/9812320].
  %%CITATION = CPHCB,124,76;%%

%240
\bibitem{Degrassi:2002fi}
  G.~Degrassi, S.~Heinemeyer, W.~Hollik, P.~Slavich and G.~Weiglein,
  Eur.\ Phys.\ J.\  C {\bf 28} (2003) 133
  [arXiv:hep-ph/0212020].
  %%CITATION = EPHJA,C28,133;%%

%241
\bibitem{Hahn:2009zz}
  T.~Hahn, S.~Heinemeyer, W.~Hollik, H.~Rzehak and G.~Weiglein,
  Comput.\ Phys.\ Commun.\  {\bf 180} (2009) 1426.
  %%CITATION = CPHCB,180,1426;%%

%242
\bibitem{Lee:2003nta}
  J.~S.~Lee, A.~Pilaftsis, M.~S.~Carena, S.~Y.~Choi, M.~Drees, J.~R.~Ellis and
  C.~E.~M.~Wagner,
  Comput.\ Phys.\ Commun.\  {\bf 156} (2004) 283
  [arXiv:hep-ph/0307377].
  %%CITATION = CPHCB,156,283;%%

%243
\bibitem{Lee:2007gn}
  J.~S.~Lee, M.~Carena, J.~Ellis, A.~Pilaftsis and C.~E.~M.~Wagner,
  Comput.\ Phys.\ Commun.\  {\bf 180} (2009) 312
  [arXiv:0712.2360 [hep-ph]].
  %%CITATION = CPHCB,180,312;%%

%244
\bibitem{h3m}
  {\tt http://www-ttp.particle.uni-karlsruhe.de/Progdata/ttp10/ttp10-23/}

%245
\bibitem{FeynHiggs}
  {\tt http://www.feynhiggs.de/}

%246
\bibitem{Schroder:2005db}
  Y.~Schroder and M.~Steinhauser,
  Phys.\ Lett.\ B {\bf 622} (2005) 124
  [hep-ph/0504055].

%247
\bibitem{Chetyrkin:2006bj}
  K.~G.~Chetyrkin, M.~Faisst, J.~H.~Kuhn, P.~Maierhofer and C.~Sturm,
  Phys.\ Rev.\ Lett.\  {\bf 97} (2006) 102003
  [hep-ph/0605201].

%248
\bibitem{Boughezal:2006xk}
  R.~Boughezal and M.~Czakon,
  Nucl.\ Phys.\ B {\bf 755} (2006) 221
  [hep-ph/0606232].

%249
\bibitem{Degrassi:2001yf}
  G.~Degrassi, P.~Slavich and F.~Zwirner,
  Nucl.\ Phys.\ B {\bf 611} (2001) 403
  [hep-ph/0105096].

%250
\bibitem{Wilson:1969zs}
  K.~G.~Wilson,
  Phys.\ Rev.\  {\bf 179} (1969) 1499.
  %%CITATION = PHRVA,179,1499;%%

%251
\bibitem{KlubergStern:1975hc}
  H.~Kluberg-Stern and J.~B.~Zuber,
  Phys.\ Rev.\ D {\bf 12} (1975) 3159.
  %%CITATION = PHRVA,D12,3159;%%

%252
\bibitem{Nielsen:1975ph}
  N.~K.~Nielsen,
  Nucl.\ Phys.\ B {\bf 97} (1975) 527.
  %%CITATION = NUPHA,B97,527;%%

%253
\bibitem{Spiridonov:1984}
  V.~P.~Spiridonov, Report No. INR P-0378, Moscow, 1984.

%254
\bibitem{Chetyrkin:1996ke}
  K.~G.~Chetyrkin, B.~A.~Kniehl and M.~Steinhauser,
  Nucl.\ Phys.\ B {\bf 490} (1997) 19
  [hep-ph/9701277].

%255
\bibitem{Degrassi:2010eu}
  G.~Degrassi and P.~Slavich,
  JHEP {\bf 1011} (2010) 044
  [arXiv:1007.3465 [hep-ph]].

%256
\bibitem{Harlander:2010wr}
  R.~V.~Harlander, F.~Hofmann and H.~Mantler,
  JHEP {\bf 1102} (2011) 055
  [arXiv:1012.3361 [hep-ph]].

%257
\bibitem{Davydychev:1992mt}
  A.~I.~Davydychev and J.~B.~Tausk,
  Nucl.\ Phys.\ B {\bf 397} (1993) 123.
  %%CITATION = NUPHA,B397,123;%%

%258
\bibitem{Dawson:1990zj}
  S.~Dawson,
  Nucl.\ Phys.\ B {\bf 359} (1991) 283.
  %%CITATION = NUPHA,B359,283;%%

%259
\bibitem{Djouadi:1991tka}
  A.~Djouadi, M.~Spira and P.~M.~Zerwas,
  Phys.\ Lett.\ B {\bf 264} (1991) 440.
  %%CITATION = PHLTA,B264,440;%%

%260
\bibitem{Inami:1982xt}
  T.~Inami, T.~Kubota and Y.~Okada,
  Z.\ Phys.\ C {\bf 18} (1983) 69.

%261
\bibitem{Baikov:2006ch}
  P.~A.~Baikov and K.~G.~Chetyrkin,
  Phys.\ Rev.\ Lett.\  {\bf 97} (2006) 061803
  [hep-ph/0604194].

%262
\bibitem{Harlander:2003bb}
  R.~V.~Harlander and M.~Steinhauser,
  Phys.\ Lett.\ B {\bf 574} (2003) 258
  [hep-ph/0307346].

%263
\bibitem{Degrassi:2008zj}
  G.~Degrassi and P.~Slavich,
  Nucl.\ Phys.\ B {\bf 805} (2008) 267
  [arXiv:0806.1495 [hep-ph]].

%264
\bibitem{Anastasiou:2008rm}
  C.~Anastasiou, S.~Beerli and A.~Daleo,
  Phys.\ Rev.\ Lett.\  {\bf 100} (2008) 241806
  [arXiv:0803.3065 [hep-ph]].

%265
\bibitem{Muhlleitner:2008yw}
  M.~Muhlleitner, H.~Rzehak and M.~Spira,
  JHEP {\bf 0904} (2009) 023
  [arXiv:0812.3815 [hep-ph]].

%266
\bibitem{Muhlleitner:2006wx}
  M.~Muhlleitner and M.~Spira,
  Nucl.\ Phys.\ B {\bf 790} (2008) 1
  [hep-ph/0612254].

%267
\bibitem{Bonciani:2007ex}
  R.~Bonciani, G.~Degrassi and A.~Vicini,
  JHEP {\bf 0711} (2007) 095
  [arXiv:0709.4227 [hep-ph]].

%268
\bibitem{Mihaila:2010mp}
  L.~Mihaila and C.~Reisser,
  JHEP {\bf 1008} (2010) 021
  [arXiv:1007.0693 [hep-ph]].

%269
\bibitem{Harlander:2009mq}
  R.~V.~Harlander and K.~J.~Ozeren,
  JHEP {\bf 0911} (2009) 088
  [arXiv:0909.3420 [hep-ph]].

%270
\bibitem{Pak:2009dg}
  A.~Pak, M.~Rogal and M.~Steinhauser,
  JHEP {\bf 1002} (2010) 025
  [arXiv:0911.4662 [hep-ph]].

%271
\bibitem{Ellis:1975ap}
  J.~R.~Ellis, M.~K.~Gaillard and D.~V.~Nanopoulos,
  Nucl.\ Phys.\  B {\bf 106} (1976) 292;
\\
  M.~A.~Shifman, A.~I.~Vainshtein and V.~I.~Zakharov,
  Phys.\ Lett.\  B {\bf 78} (1978) 443;
\\
  M.~A.~Shifman, A.~I.~Vainshtein, M.~B.~Voloshin and V.~I.~Zakharov,
  Sov.\ J.\ Nucl.\ Phys.\  {\bf 30} (1979) 711
  [Yad.\ Fiz.\  {\bf 30} (1979) 1368];
\\
  A.~I.~Vainshtein, V.~I.~Zakharov and M.~A.~Shifman,
  Sov.\ Phys.\ Usp.\  {\bf 23} (1980) 429
  [Usp.\ Fiz.\ Nauk {\bf 131} (1980) 537];
\\
  B.~A.~Kniehl and M.~Spira,
  Z.\ Phys.\  C {\bf 69} (1995) 77 [hep-ph/9505225];
\\
  M.~Spira, A.~Djouadi, D.~Graudenz and P.~M.~Zerwas,
  Nucl.\ Phys.\  B {\bf 453} (1995) 17 [hep-ph/9504378].

%272
\bibitem{Kilian:1995tra}
  W.~Kilian,
  Z.\ Phys.\ C {\bf 69} (1995) 89
  [hep-ph/9505309].

%273
\bibitem{higgswg}
{\tt https://twiki.cern.ch/twiki/bin/view/LHCPhysics/CrossSections}

%274
\bibitem{Braaten:1980yq}
  E.~Braaten and J.~P.~Leveille,
  Phys.\ Rev.\ D {\bf 22} (1980) 715.

%275
\bibitem{Gorishnii:1990zu}
  S.~G.~Gorishnii, A.~L.~Kataev, S.~A.~Larin and L.~R.~Surguladze,
  Mod.\ Phys.\ Lett.\ A {\bf 5} (1990) 2703.

%276
\bibitem{Chetyrkin:1997sg}
  K.~G.~Chetyrkin, B.~A.~Kniehl and M.~Steinhauser,
  Phys.\ Rev.\ Lett.\  {\bf 79} (1997) 2184
  [hep-ph/9706430].

%277
\bibitem{Kwiatkowski:1994cu}
  A.~Kwiatkowski and M.~Steinhauser,
  Phys.\ Lett.\ B {\bf 338} (1994) 66
   [Erratum-ibid.\ B {\bf 342} (1995) 455]
  [hep-ph/9405308].

%278
\bibitem{Chetyrkin:2009fv}
  K.~G.~Chetyrkin, J.~H.~Kuhn, A.~Maier, P.~Maierhofer, P.~Marquard,
  M.~Steinhauser and C.~Sturm, 
  Phys.\ Rev.\  D {\bf 80} (2009) 074010
  [arXiv:0907.2110 [hep-ph]].

%279
\bibitem{Carena:2002qg}
  M.~S.~Carena, S.~Heinemeyer, C.~E.~M.~Wagner and G.~Weiglein,
  Eur.\ Phys.\ J.\  C {\bf 26} (2003) 601
  [hep-ph/0202167].

%280
\bibitem{Guasch:2003cv}
  J.~Guasch, P.~Hafliger and M.~Spira,
  Phys.\ Rev.\ D {\bf 68} (2003) 115001
  [hep-ph/0305101].

%281
\bibitem{Spira:1995rr}
  M.~Spira, A.~Djouadi, D.~Graudenz and P.~M.~Zerwas,
  Nucl.\ Phys.\ B {\bf 453} (1995) 17
  [hep-ph/9504378].

%282
\bibitem{Steinhauser:1996wy}
  M.~Steinhauser,
  In *Tegernsee 1996, The Higgs puzzle* 177-185
  [hep-ph/9612395].

%283
\bibitem{Aglietti:2004nj}
  U.~Aglietti, R.~Bonciani, G.~Degrassi and A.~Vicini,
  Phys.\ Lett.\ B {\bf 595} (2004) 432
  [hep-ph/0404071];\\
  F.~Fugel, B.~A.~Kniehl and M.~Steinhauser,
  Nucl.\ Phys.\ B {\bf 702} (2004) 333
  [hep-ph/0405232];\\
  G.~Degrassi and F.~Maltoni,
  Nucl.\ Phys.\ B {\bf 724} (2005) 183
  [hep-ph/0504137].

%284
\bibitem{Harlander:2005rq}
  R.~Harlander and P.~Kant,
  JHEP {\bf 0512} (2005) 015
  [hep-ph/0509189].

%285
\bibitem{kurz:2012}
A.~Kurz, 
 Diploma thesis, Karlsruhe Institute of Technology (2010). 

%286
\bibitem{kleine:2010}
J.~Kleine, 
 Diploma thesis, Karlsruhe Institute of Technology (2010). 

%287
\bibitem{Djouadi:2005gi}
  A.~Djouadi,
  Phys.\ Rept.\  {\bf 457} (2008) 1
  [hep-ph/0503172];\\
  A.~Djouadi,
  Phys.\ Rept.\  {\bf 459} (2008) 1
  [hep-ph/0503173].

%288
\bibitem{Harlander:2007zz}
  R.~Harlander,
  J.\ Phys.\ G {\bf 35} (2008) 033001.

%289
\bibitem{Harlander:2009bw}
  R.~V.~Harlander and K.~J.~Ozeren,
  Phys.\ Lett.\ B {\bf 679} (2009) 467
  [arXiv:0907.2997 [hep-ph]].

%290
\bibitem{Harlander:2009my}
  R.~V.~Harlander, H.~Mantler, S.~Marzani and K.~J.~Ozeren,
  Eur.\ Phys.\ J.\ C {\bf 66} (2010) 359
  [arXiv:0912.2104 [hep-ph]].

%291
\bibitem{Pak:2009bx}
  A.~Pak, M.~Rogal and M.~Steinhauser,
  Phys.\ Lett.\ B {\bf 679} (2009) 473
  [arXiv:0907.2998 [hep-ph]].

%292
\bibitem{Pak:2011hs}
  A.~Pak, M.~Rogal and M.~Steinhauser,
  JHEP {\bf 1109} (2011) 088
  [arXiv:1107.3391 [hep-ph]].

    \end{thebibliography}
\end{document}